%% file: LightBaryonSpectroscopy.tex
\documentclass[12pt,hidelinks,twoside]{article}
\usepackage{epsfig}

\usepackage{lscape}
\usepackage{rotating}

\usepackage{booktabs} 
\usepackage{multirow} 
\usepackage{longtable}
\usepackage{array}

\usepackage{tabularx}

\usepackage{hyperref}
\usepackage{amsmath}
\usepackage{amssymb}
\usepackage{xcolor}
\usepackage{bm}
\usepackage{wasysym}
\usepackage{bbm}
\usepackage[percent]{overpic}
\usepackage[backend=biber,style=numeric,sorting=none]{biblatex}

\usepackage{theorem}

\theoremstyle{break}
\newtheorem{CompEx}{Problem}

\newcolumntype{C}{>{$}c<{$}}
\newcolumntype{L}{>{$}l<{$}}
\usepackage{xargs} 

\usepackage{todonotes}

\newcommand\blfootnote[1]{%
	\begingroup
	\renewcommand\thefootnote{}\footnote{#1}%
	\addtocounter{footnote}{-1}%
	\endgroup
}

\newcommand{\EPJA}{With kind permission of The European Physical Journal (EPJ). }
\newcommand{\APS}[2]{Reprinted figure with permission from #1. Copyright #2 by the American Physical Society. }
\newcommand{\FewBody}[2]{Reprinted by permission from Few-Body Systems: #1, Copyright #2. }

\newcommand{\be}{\begin{equation}}
\newcommand{\ee}{\end{equation}}
\newcommand{\bea}{\begin{eqnarray}}
\newcommand{\eea}{\end{eqnarray}}

\begin{document}

\title{ \vspace{1cm} Light Baryon Spectroscopy}
\author{A.\ Thiel$^{1}$, F.~Afzal$^{1}$, Y. Wunderlich$^{1}$.\\
\\
$^1$Helmholtz-Institut f\"ur Strahlen- und Kernphysik,\\ University of Bonn, D-53115 Bonn, Germany
}

\maketitle

\blfootnote{© 2022. This manuscript version is made available under the CC-BY-NC-ND 4.0 license \url{https://creativecommons.org/licenses/by-nc-nd/4.0/}}

\begin{abstract} 
This review treats the advances in \textit{Light Baryon Spectroscopy} of the last two decades, which were mainly obtained by measuring meson-production reactions at photon facilities all over the world. 
We provide a consistent compendium of experimental results, as well as a review of the theoretical methods of amplitude analysis used to analyze the data. The most significant datasets are presented in detail and are listed in combination with a full set of the relevant references. In addition, a brief summary of spin-formalisms, which are ubiquitous in \textit{Light Baryon Spectroscopy}, as well as a review on complete experiments, are provided. The synthesis of the reviewed knowledge is presented in a full interpretation of the new results on the \textit{Light Baryon Spectrum}. 
\end{abstract}
\clearpage
\tableofcontents

\clearpage
\section{Introduction}
\input{Introduction.tex}

\clearpage
\section{Models and Lattice QCD calculations}
\input{Models_LatticeQCD.tex}
\clearpage
\section{Spin-formalisms and important reactions}
\input{Formalisms.tex}

\clearpage
\section{Experiments in Light Baryon Spectroscopy}\label{sec:experiments}
\input{Experiments.tex}

\clearpage
\section{Measurements}
\input{Measurements.tex}

\clearpage
\section{Partial Wave- and Phenomenological Analyses}\label{sec:pwa}
\input{PWA_PhenomenologicalAna.tex}

\clearpage
\section{Interpretation}
\input{Interpretation.tex}

\clearpage

\section{Epilogue}
\input{Conclusion.tex}

\section*{Acknowledgment}
We would like to thank E.~Klempt, B.~Krusche and V.~Metag for their fruitful comments to the manuscript. For feedback on different sections, we acknowledge support from M. D\"oring, Ch. Fischer, M. Manley, D. R\"onchen, A. Sarantsev, I. Strakovsky, A. \v{S}varc and R. Workman.

The work of YW has been funded by the Transdisciplinary Research Area Matter (University of Bonn) as part of the Excellence Strategy of the federal and state governments, during the completion of this paper.

\clearpage
\appendix
\section{Appendix}
\input{appendix.tex}

\clearpage
\printbibliography 


\end{document}

%% file: Introduction.tex

\subsection{Introduction into Baryon Spectroscopy}
Ever since the nucleons were identified as composed objects of quarks, the exact interaction and dynamics between the quarks was a desired topic of investigation. Several different methods about how to shed light on the inside of the nucleon were proposed, one of them is spectroscopy, the analysis of the excitation spectrum of the particles.

The interaction between the quarks is described by quantum chromodynamics (QCD), which is not solvable in the energy regime of the stable hadrons.  To overcome this issue, for example quark models or LatticeQCD calculations can be used to get a prediction on how the excitation spectra are expected to look like. The comparison of the spectra with experimental data becomes an essential task in order to understand the interaction.

\begin{figure}[ht]
\includegraphics[width=\linewidth]{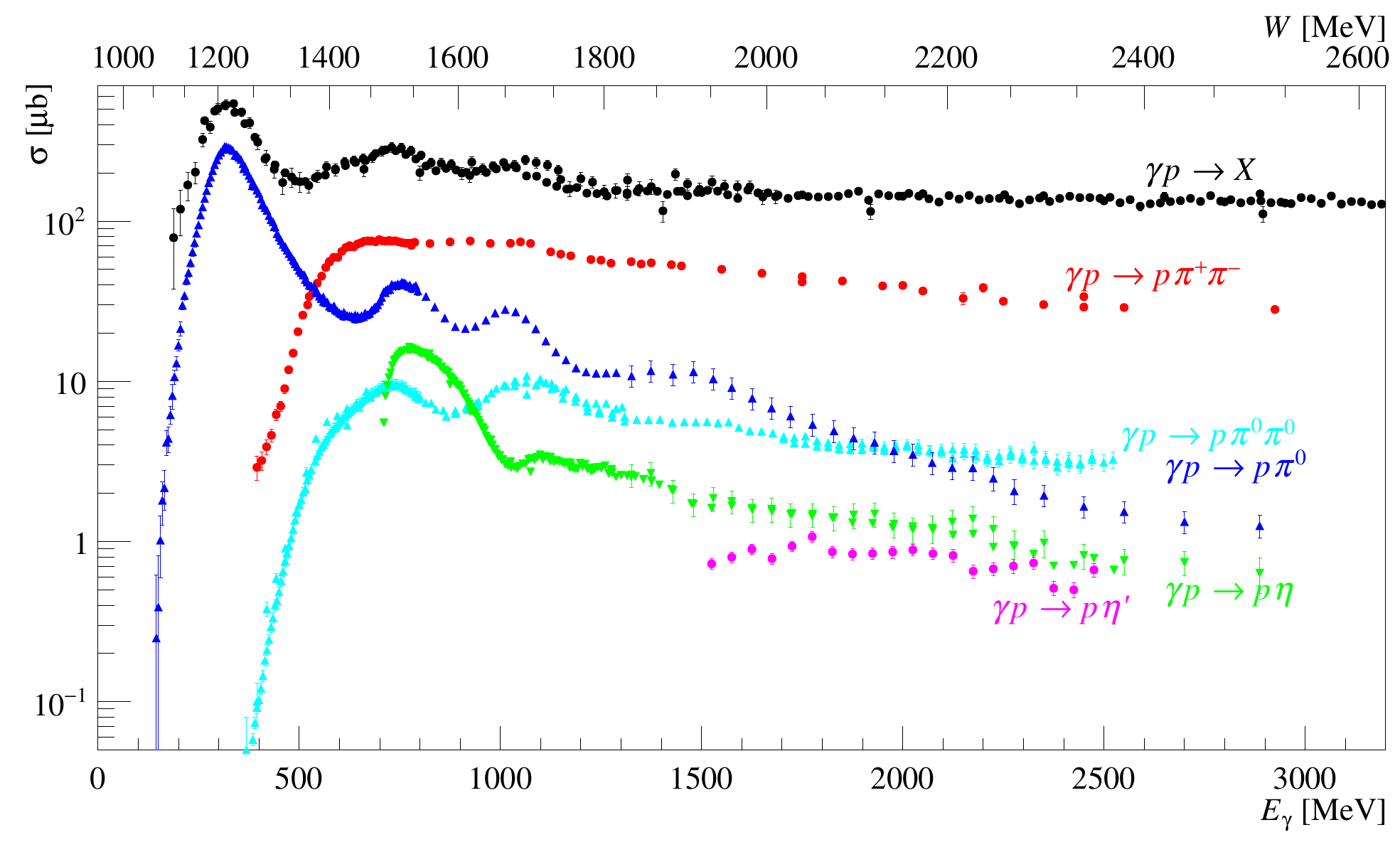}
	\caption{Overview of different cross sections for various final states: black: total cross section \cite{Nakamura:2010zzi}, blue: $p\pi^0$ \cite{Schumann:2010js,Bartholomy:2004uz}, green: $p\eta$ \cite{McNicoll:2010qk,Crede:2009zzb,Crede:2003ax}, red: $p\pi^+\pi^-$ \cite{Braghieri:1994rf,Wu:2005wf,Aachen-Hamburg-Heidelberg-Munich:1975jed}, cyan: $p\pi^0\pi^0$ \cite{Thoma:2007bm, Sokhoyan:2015fra, Schumann:2010js}, magenta: $p\eta'$ \cite{Crede:2009zzb}.}
	\label{fig:Intro:crosssections}
\end{figure}

Nucleons can be excited by different particles, for example pions or photons. The challenging problem here is the determination of the excitation spectrum from the scattering data. In the case of baryon spectroscopy, this is not as simple as for atomic spectroscopy, where the excitation spectrum can be observed as separate lines. The excited states of nucleons manifest themselves in the cross sections as broad distributions, the so-called resonances. These resonances are strongly overlapping, making it difficult to separate them without sophisticated tools. An example is visible in Fig.~\ref{fig:Intro:crosssections}, where cross sections for different final states are compared. At low masses, the $\Delta(1232)$ resonance, the first excited state, can be clearly observed as a peak in the cross section for $p\pi^0$. All other structures are not composites of a single resonant state, but arise due to several overlapping states. Especially at higher masses, the cross section becomes flat. In order to identify the contributing states, partial wave analyses are needed. These analyses allow the extraction of complex amplitudes from the different data sets. The resonant states can be observed as poles in the complex plane and their parameters are compared to the different model predictions.

The scope of this article is to give a recent and broad overview about the topic of \textit{Light Baryon Spectroscopy}. This review is organized as follows: After this short introduction, chapter~\ref{sec:ModelsLatticeQCD} will give an overview about different theoretical predictions and calculations. Here, constituent quark models will be discussed as well as LatticeQCD calculations. In chapter~\ref{sec:formalisms}, a general and broad description of the spin formalisms and their implementation for different final states is given. In addition, the observables, which are of major importance for baryon spectroscopy, are introduced. Chapter~\ref{sec:experiments} focuses on the experimental setups, which are used to extract the different observables in various final states. An overview about these measurements is collected in detail in chapter~\ref{sec:measurements}. Several examples of measurements are presented for different final states alongside with tables containing a collection of all publications since 2000. For a qualified interpretation of these data sets, partial wave and phenomenological analyses are essential. An overview about different methods and analyses is given in chapter~\ref{sec:pwa}. The focus of last chapter, chapter~\ref{sec:interpretation}, lies on the interpretation of the results, where the impact of the measurement of the last 20 years is investigated and the influence of the listing of baryon resonances is analyzed. This review will close with a small outlook on future developments.

\subsection{Comparison to other Review Articles}
Several different review articles have already been published in the past, for example \cite{Hey:1982aj,Krusche:2003ik,Klempt:2009pi,Crede:2013kia,Ireland:2019uwn}. All of these articles had different focuses and gave the state of art at that time. 

In this review, we focus on a detailed listing of the conducted measurement since 2000 and their impact on our current knowledge of the baryon spectrum. This information will be accompanied by a detailed overview of the theoretical models and the partial wave analyses, used for interpretation of the data. Since the field is developing rapidly, this review will provide an overview of the state at the time of writing this, mid of 2021.


%% file: Models_LatticeQCD.tex
\label{sec:ModelsLatticeQCD}

The prediction of the properties of either stable or metastable bound states among quarks (e.g. ground and excited states of the Light Baryon Spectrum) directly from QCD is a highly involved procedure, due to the fact that the formation of such states is an inherently non-perturbative phenomenon. The non-abelian structure of the gauge theory QCD and thereby its property of asymptotic freedom would make the application of perturbation theory in any case difficult if not impossible. Therefore, calculational schemes have to be employed that go beyond perturbation theory. This section is intended to provide some information on and present resulting spectra for four different possible directions of research.

\subsection{Quark models} \label{sec:QuarkModels}

In the quark models as initially proposed by Isgur and Karl~\cite{Isgur:1977ef, Isgur:1978wd}, a hadron is modeled as a bound state of so-called {\it constituent quarks}. This approach has then been carried further by several groups, see for instance~\cite{Capstick:1985xss,Capstick:2000qj,Ferraris:1995ui,Glozman:1996wq,Glozman:1997ag,Loring:2001kv,Loring:2001kx,Giannini:2015zia} and references therein. Constituent quarks are the fundamental ('current'-) quarks from the lagrangian of QCD, but their properties are modified by a surrounding cloud of quarks and gluons. The constituent quarks are thus also sometimes referred to as 'dressed' quarks. QCD itself is not invoked for an \textit{ab initio} description of the interactions among the constituent quarks, due to its complicated nonperturbative behavior. Instead, these interactions are modeled by effective potentials. While a linearly rising confinement-potential is almost universally assumed for the long-range part of the interaction, the differences among models can mostly be traced back to the phenomenological parametrization of the short-ranged interactions of QCD. Once a form of interaction has been fixed, some kind of either differential (i.e. Schr\"{o}dinger-/Dirac-type) or integral (Lippmann-Schwinger-/Bethe-Salpeter-type) equation has to be solved in order to yield the spectrum of ground- and excited states. This makes the calculations technically difficult. 

The degree to which the detailed dynamics of QCD was incorporated into the effective interactions of the constituent quark models is variable for different models and also changed over time. The one-gluon exchange has already been discussed as an effective contribution with a Coulomb-type radial dependence~$\propto 1/r$ to the confining potential~$V_{\text{conf}}$ in the early works by Isgur and Karl~\cite{Isgur:1978wd}. Such Coulomb-type contributions to the potentials are quite common in other models. The modern approach of an effective description of QCD using \textit{chiral dynamics}~\cite{Gasser:1983yg,Gasser:1984gg} has been incorporated into a constituent-quark model for the first time (to our knowledge) in the case of the Graz model by Glozman, Plessas and collaborators~\cite{Glozman:1996wq,Glozman:1997ag}.
In the following, we select the Bonn-model developed by L\"{o}ring and collaborators~\cite{Loring:2001kv,Loring:2001kx} for a more detailed discussion, since it represents a well-known and modern description of the light-baryon spectrum. 

The Bonn-model uses the relativistically covariant Bethe-Salpeter equation for bound states of the three-quark system to calculate all predictions. First results for the nucleon- and $\Delta$-spectra were published in~\cite{Loring:2001kx}. Only up-, down- and strange quarks have been introduced as dynamical quarks. A three-body confinement-kernel has been used as input for the Bethe-Salpeter equation, with a local three-quark potential of the following form: $V_{\mathrm{conf.}}^{(3)} \left( \vec{x}_{1}, \vec{x}_{2}, \vec{x}_{3} \right) = 3 a \hat{\Gamma}_{o} + b \sum_{i<j} \left| \vec{x}_{i} - \vec{x}_{j} \right| \hat{\Gamma}_{s}$~\cite{Loring:2001kx}. This potential introduces two parameters: an offset $a$ and the slope $b$. The operators $\hat{\Gamma}_{o}$ and $\hat{\Gamma}_{s}$ denote Dirac tensor-structures, for which two possible choices have been employed in the original calculation~\cite{Loring:2001kx}, leading to the models $\mathcal{A}$ and $\mathcal{B}$. For the remaining short-range interactions implied by QCD, a two-body potential was chosen based on an effective instanton-induced interaction proposed by 't Hooft~\cite{tHooft:1976snw}, which introduces three additional free parameters. In its original form, the Bonn model has remarkably few free parameters~\cite{Loring:2001kx}: aside from the five parameters mentioned above, only the constituent quark masses $m_{n}$ and $m_{s}$ of the non-strange and strange quarks are further input. Fixing these $7$ parameters from the mass values and splittings of a few well-known resonances, predictions for the whole remainder of the spectrum are obtained~\cite{Loring:2001kx}. 

A possible extension of the Bonn-model has been published more recently in a work by Ronniger and Metsch~\cite{Ronniger:2011td}, where the above-described original model-construction has been modified by an additional quark-flavor dependent interaction. The parametrization of the latter introduces four additional free parameters, two couplings and two effective ranges, which at first increases the number of free parameters to $11$. Upon fixing the effective range of the 't Hooft force to the value obtained in the previous calculation~\cite{Loring:2001kx}, the number of free parameters was reduced to $10$. 

\begin{sidewaysfigure}[p]
\centering
\vspace*{-10pt}
 \includegraphics[width=0.945\textwidth]
      {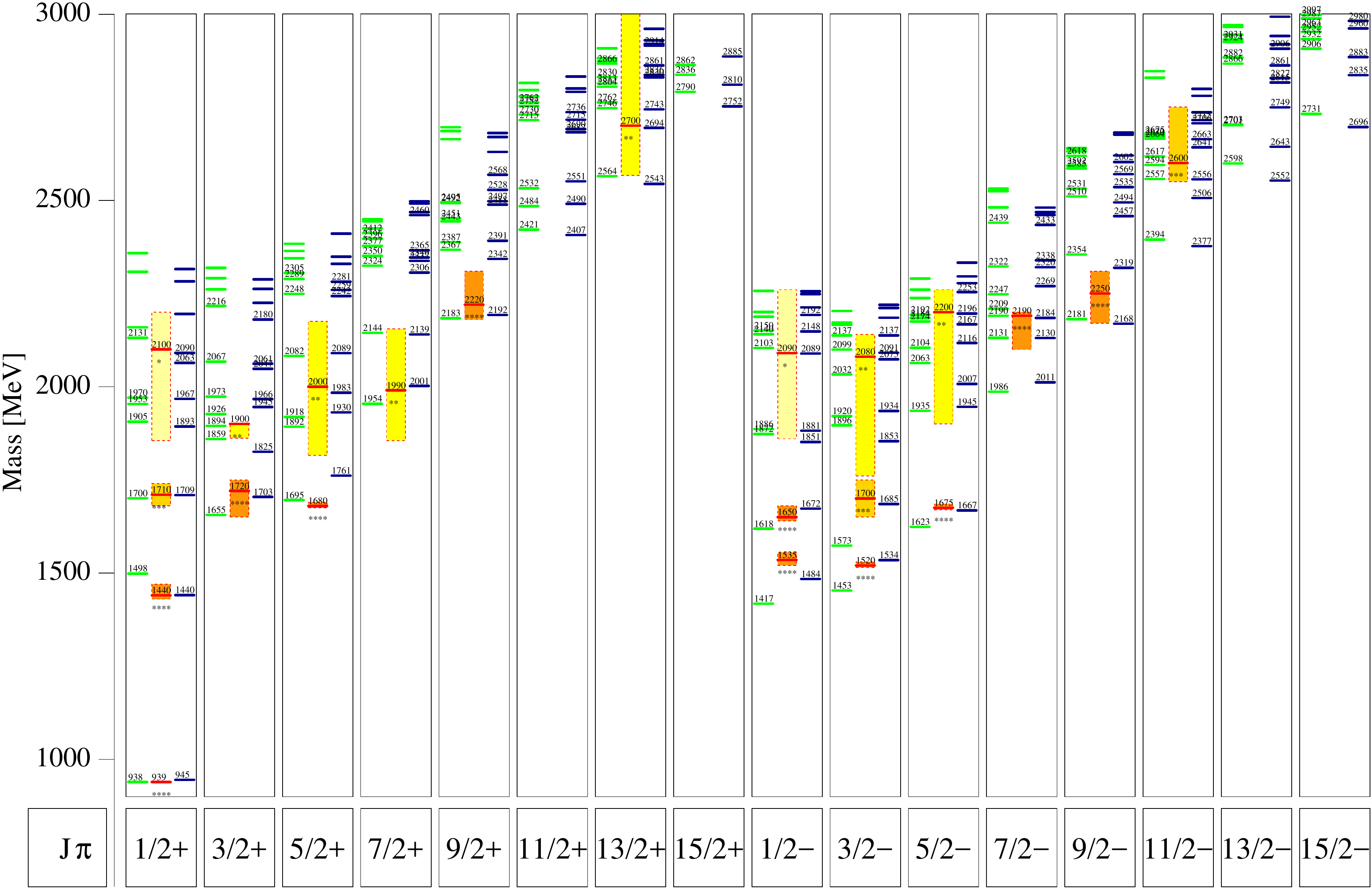}
 \caption{The spectra of nucleon-resonances, calculated in the work by Ronniger and Metsch (model $\mathcal{A}$ (green lines) and model $\mathcal{C}$ (blue lines)), are shown in comparison to experimental data (from the PDG \cite{Nakamura:2010zzi}) in the middle columns ~\cite{Ronniger:2011td}. (Figure provided in high resolution by courtesy of Bernard Metsch.) \EPJA}
 \vspace*{-18pt}
\label{fig:models:NucleonSpectrum}
\end{sidewaysfigure}
%


%
\begin{sidewaysfigure}[p]
\centering
\vspace*{-10pt}
 \includegraphics[width=0.945\textwidth]
      {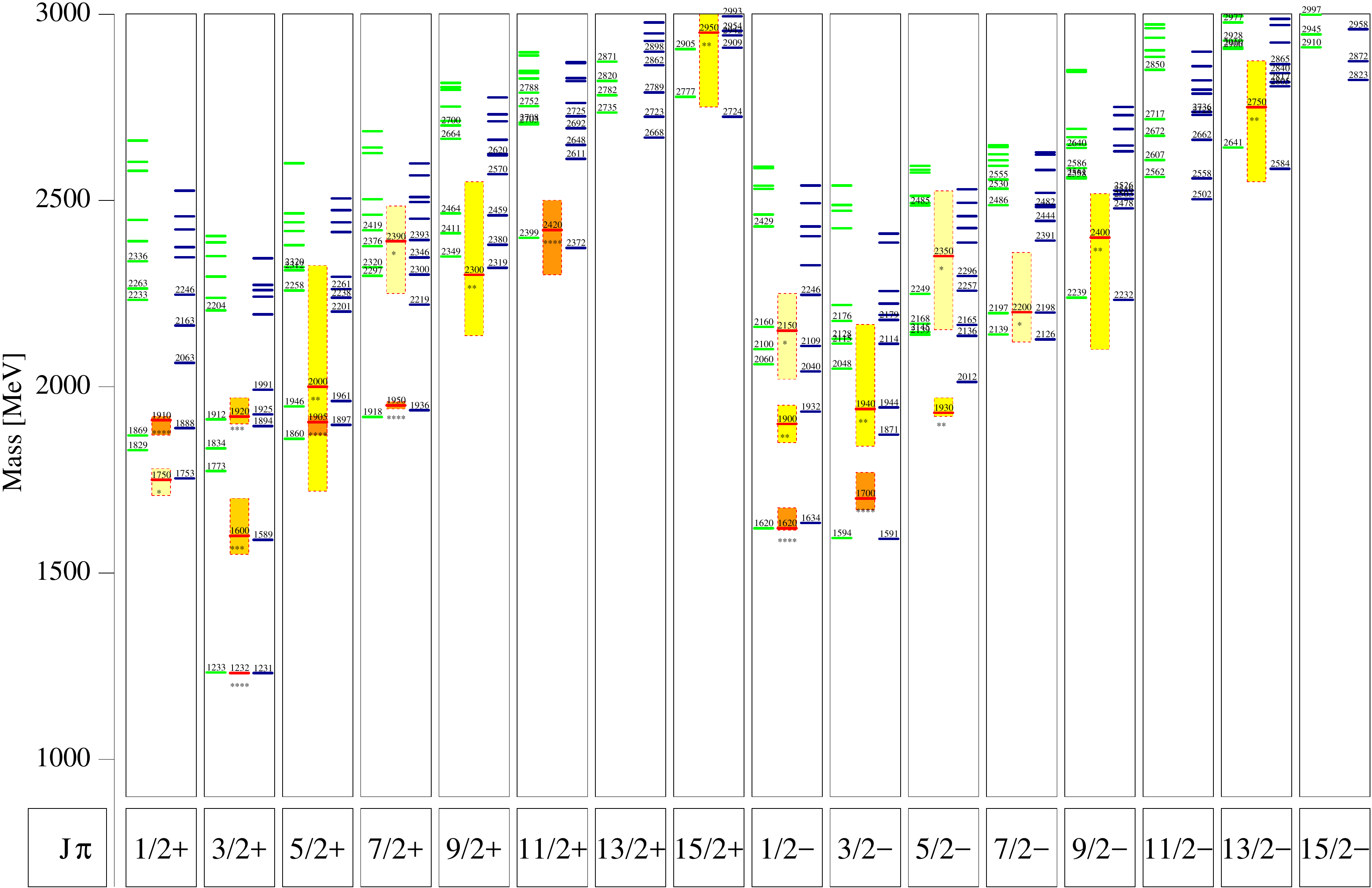}
 \caption{The spectra of $\Delta$-resonances, calculated in the work by Ronniger and Metsch (model $\mathcal{A}$ (green lines) and model $\mathcal{C}$ (blue lines)), are shown in comparison to experimental data (from the PDG \cite{Nakamura:2010zzi}) in the middle columns ~\cite{Ronniger:2011td}. (Figure provided in high resolution by courtesy of Bernard Metsch.) \EPJA}
 \vspace*{-18pt}
\label{fig:models:DeltaSpectrum}
\end{sidewaysfigure}
%


In Figures \ref{fig:models:NucleonSpectrum} and \ref{fig:models:DeltaSpectrum}, the experimental values of baryon resonances from the Review of particle physics~\cite{Nakamura:2010zzi} are compared to results of model $\mathcal{A}$ from the earlier Bonn-model calculation~\cite{Loring:2001kx}, as well as to the newer description involving flavor-dependent forces (model $\mathcal{C}$) from reference~\cite{Ronniger:2011td}. Nucleon resonances with isospin $I = \frac{1}{2}$ are shown in Figure \ref{fig:models:NucleonSpectrum}, whereas Figure \ref{fig:models:DeltaSpectrum} depicts results on $\Delta$-resonances with isospin $I = \frac{3}{2}$. The columns of the spectra are organized according to spin-parity quantum numbers $J^{P}$. The results from model $\mathcal{A}$ are plotted on the left side in each column, results from model $\mathcal{C}$ are on the right side and experimental results from the PDG~\cite{Nakamura:2010zzi} are at the center of each column. The resonance masses predicted from the Bonn model are indicated by a line in each case. The experimental values are plotted as shaded bars which indicate the experimental mass uncertainty. The PDG star-rating~\cite{Nakamura:2010zzi} is also indicated in the figures. 

Upon consideration of Figures \ref{fig:models:NucleonSpectrum} and \ref{fig:models:DeltaSpectrum}, one can see hints of the so-called \textit{missing-resonance problem}: many more states are generically predicted by quark models than have been measured until now, predominantly in the higher mass-region.  
This problem has been one of the main motivations for the performance of further baryon-spectroscopy measurements using electromagnetic probes, such as photoproduction experiments~(cf. descriptions in sections~\ref{sec:formalisms:SingleMesonPhotoproduction},~\ref{sec:measurements:SingleMesonPhotoproduction} and experiments described in chapter~\ref{sec:exp}), in addition to the earliest experiments which used pion beams (cf. sections~\ref{sec:formalisms:PiNScattering} and~\ref{sec:measurements:PionNucleonScattering}). 

The experimental light-baryon spectrum exhibits further intricate features, the description of which has presented challenges to the constituent-quark models. The first feature is given by the fact that the Roper resonance $N(1440) \frac{1}{2}^{+}$ is found experimentally to be situated lower in mass compared to the $\frac{1}{2}^{-}$ states of the nucleon spectrum (cf. colored bars in Figure~\ref{fig:models:NucleonSpectrum}). This ordering of the Roper resonance relative to the $\frac{1}{2}^{-}$ states came out right in the Bonn-model only after Ronniger and Metsch introduced a flavor-dependent interaction (compare states for models~$\mathcal{A}$ and~$\mathcal{C}$ in Figure~\ref{fig:models:NucleonSpectrum}). At least within the context of an instanton-based model such as the Bonn-model, a flavor-dependence seems to be essential for achieving this effect. One can find further substantiation of the importance of flavor-dependent forces in the hypercentral constituent-quark model (see section 4.2 of reference~\cite{Giannini:2015zia}), due to the fact that the correct relative ordering of the Roper resonance is only achieved there upon introducing an ‘isospin-dependent’ interaction term.

A further intricate feature is given by the large mass-splitting between the $J^{P} = \frac{1}{2}^{+}$ and $\frac{3}{2}^{+}$ states in the nucleon spectrum (cf. Figure~\ref{fig:models:NucleonSpectrum}). This effect can be traced back, using basic phenomenological reasoning, to baryon spin- and ﬂavor- wave functions which are both antisymmetric with respect to the exchange of two quarks (see section 3.4 of reference~\cite{Klempt:2012fy}). At least in the context of this phenomenological interpretation, an instanton-induced interaction (as for instance given in the Bonn-model) is rather crucial for a description of the mass splitting, since such interactions only act between pairs of quarks that induce an antisymmetry of the overall wave-function under their exchange.



\subsection{Lattice QCD calculations} \label{sec:LatticeQCD}

The approach of Lattice QCD originates from a seminal work by Wilson~\cite{Wilson:1974sk}. The discussion given in the following is based on a didactical account given in volume II of the book by Aitchison and Hey~\cite{AitchisonHey}. In lattice QCD, the path integrals defining correlation functions in QCD are solved numerically on a discretized spacetime-lattice, where in most cases euclidean spacetime is employed~\footnote{Euclidean spacetime is reached from Minkowski spacetime by the substitution $t \rightarrow \tau:= i t$, i.e.~by going to imaginary time. In this way, the rapidly oscillating integrands of path-integrals become exponentially damped and the path-integrals themselves thus well-defined. After a calculation has been done, one can always analytically continue from euclidean back to Minkowski spacetime.}. It should be mentioned that here the full path-integral is solved and no perturbative expansion is applied. In an idealized situation, arbitrarily large lattices would be possible with arbitrarily fine lattice-spacing $a$, which would facilitate a numerical solution of QCD very close to the so-called {\it continuum-limit}. In practical cases however, simulations on lattices with $16^{4}$ sites are already very CPU intensive~\cite{AitchisonHey} and thus only possible on supercomputers. Practical calculations are therefore usually performed away from the continuum-limit and sophisticated methods have to be used to extrapolate the lattice-results back to the continuum-results. 

In order to avoid so-called 'finite-size effects', the lattices themselves have to be large enough such that the simulated objects, e.g.~nucleons, which have themselves the spatial dimension $R$, fit well inside. In case the lattice itself becomes too large, one would approach the limit $R \simeq a$ and the discrete nature of the lattice would lead to simulation errors. The ideal situation is expressed by the following estimate~\cite{AitchisonHey}:
\begin{equation}
 a \ll R \sim \frac{1}{m} \ll L = N a \mathrm{,} \label{eq:IdealLatticeEstimate}
\end{equation}
where $m$ is the mass of the considered object and a lattice of side-length $L$ with $N$ lattice-sites is employed in the calculation. Furthermore, a finite non-zero lattice-spacing $a$ yields a lower bounds on the masses of hadrons that can be studied: this effect is quantified by the pion-mass $m_{\pi}$. 

\begin{figure}[ht]
\centering
\hspace*{0pt}
 \includegraphics[width=0.95\textwidth]%
      {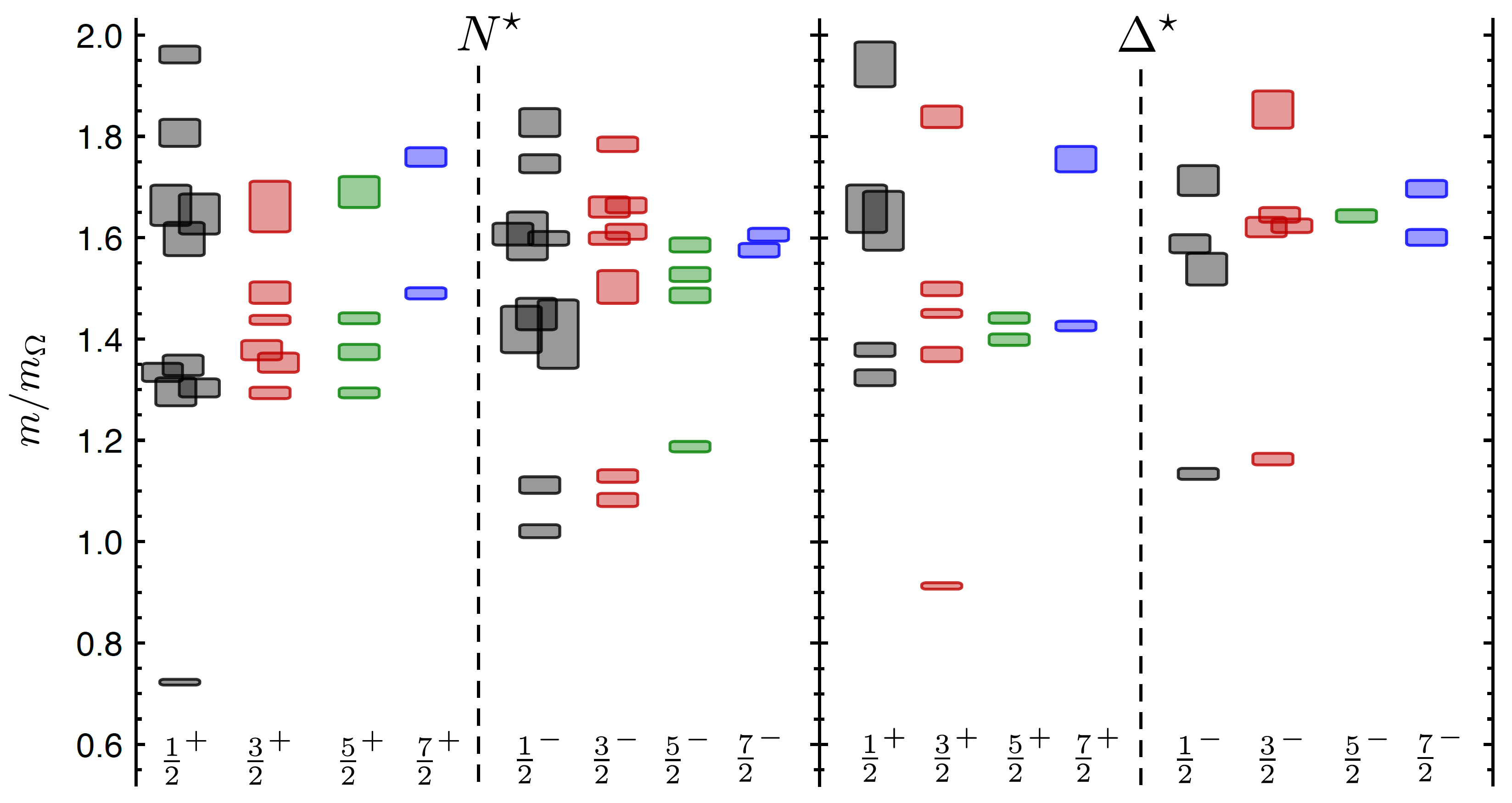}
\vspace*{0pt}
 \caption{The nucleon- (left) and $\Delta$-states (right) from the Lattice-QCD calculation by Edwards and collaborators~\cite{Edwards:2011jj} are depicted here. The spin-parity quantum numbers $J^{P}$ are given and the masses of the states are provided in units of the mass of the $\Omega^{-}$-baryon calculated in the same lattice-QCD simulation (for comparison: the PDG-value is $m_{\Omega} \simeq 1672$ $\mathrm{MeV}$~\cite{Zyla:2020zbs}). The lattice-spacing chosen for the calculation corresponds here to the unphysical pion-mass $m_{\pi}=396$ $\mathrm{MeV}$. The continuous bands of values are an artifact of the spin-identification procedure for the states~\cite{Edwards:2011jj}. \APS{\cite{Edwards:2011jj}}{2011}}
\label{fig:models:LatticeNucleonSpectra}
\end{figure}

An early Lattice-QCD calculation of excited nucleon- and $\Delta$-states has been performed by Edwards and collaborators~\cite{Edwards:2011jj}. The results are depicted in Figure \ref{fig:models:LatticeNucleonSpectra} for an unphysical pion-mass of $m_{\pi}=396$ $\mathrm{MeV}$. 
The pion-mass is still far from a realistic value, i.e. far away from the physical one $m_{\pi} = 140$ $\mathrm{MeV}$. Furthermore, the excited states evaluated in this Lattice-QCD calculation are not resonances in the sense that they are stable particles. I.e. not even the decays of high-lying states, which would lead them to dynamically acquire a resonance width, are modeled in the calculation. Still, this Lattice-QCD result can be compared to the results of the Bonn constituent-quark model~\cite{Loring:2001kx} shown in Figures \ref{fig:models:NucleonSpectrum} and \ref{fig:models:DeltaSpectrum}. One observes already a quite striking resemblance between the lattice- and quark-model spectra. In particular for the low-lying states the patterns of levels look similar. However, the relative level-ordering can be quite different between both calculational approaches. In particular, the relative ordering of the Roper resonance~$N(1440) \frac{1}{2}^{+}$ compared to the $\frac{1}{2}^{-}$ states of the nucleon spectrum is not correctly reproduced by the Lattice-QCD calculation of reference~\cite{Edwards:2011jj} (compare Figure~\ref{fig:models:LatticeNucleonSpectra} with experimental and quark-model states given in Figure~\ref{fig:models:NucleonSpectrum}). In the discussion of quark models (cf. section~\ref{sec:QuarkModels}), a quark-flavor dependence of the interactions was found to be crucial for reproducing this relative level-ordering. Apparently, a pure Lattice-QCD simulation cannot reproduce this effect. However, at least the large shift of the proton-mass relative to the $\frac{3}{2}^{+}$ states of the nucleon is reproduced correctly in the Lattice-QCD calculation (see Figure~\ref{fig:models:LatticeNucleonSpectra}). 

Comparing the Lattice-QCD results further with the experimental and quark-model results shown in Figures~\ref{fig:models:NucleonSpectrum} and~\ref{fig:models:DeltaSpectrum}, it becomes apparent that the \textit{missing-resonance problem} is also implied by Lattice QCD. However, one should keep in mind the following two important additional sources of error in the Lattice-QCD calculation. First, the bands representing higher-mass resonances are anyways at least in part quite large and overlapping (cf. Figure~\ref{fig:models:LatticeNucleonSpectra}) due to the spin-identification procedure. Second, the baryon spectra evaluated in reference~\cite{Edwards:2011jj} showed still some dependence on the value of the pion-mass~$m_{\pi}$, or equivalently on the employed lattice-spacing~$a$. Such discretization effects lead to additional uncertainties, in particular in the higher-mass region.

A new calculation of the low-lying positive-parity $\Delta$ and $N$ spectra has been published recently \cite{Khan:2020ahz}. Here, pion masses of $m_\pi = 358$~MeV and $278$~MeV were used to calculate the excitation states. Remarkably, the extracted spectra exhibit a counting of states similar to the most popular quark models. Additionally, hybrid states are naturally emerging in the spectra, which further increases the number of states. This also points in the direction that a systematic search for baryonic states with gluonic degrees of freedom might be important.

\subsection{Dyson-Schwinger and Bethe-Salpeter equation (DSE/BSE-) approach} \label{sec:DysonSchwinger}

A different Ansatz for the calculation of non-perturbative QCD phenomena, which is complementary to lattice QCD (sec.~\ref{sec:LatticeQCD}) in the sense that it does not require a discretization of spacetime, is given by the methodology collectively denoted as Dyson-Schwinger and Bethe-Salpeter equations (DSE/BSE). In this approach, hadronic bound-states are obtained as solutions of intricate recursive integral-equations in momentum space. Although this may not seem dissimilar to some of the quark models mentioned in Section~\ref{sec:QuarkModels}, the difference in the case of DSE/BSE is that the interaction kernels, which are then iterated to all orders via the recursive bound-state equations, are taken directly from QCD and not from effective QCD-inspired potentials. However, in order to make the calculations tractable, the interaction-vertices of QCD have in most cases to be replaced by approximate effective interactions (cf. the discussion further below). This approach has been successfully promoted in recent years by Roberts, Fischer, Eichmann, and others~\cite{Roberts:1994dr,Maris:2003vk,Fischer:2006ub,Eichmann:2009qa,Eichmann:2016hgl,Eichmann:2016yit,Eichmann:2016nsu,Sanchis-Alepuz:2017jjd,Eichmann:2018adq}.

\begin{figure}[ht]
\centering
\hspace*{0pt}
 \includegraphics[width=0.95\textwidth]%
      {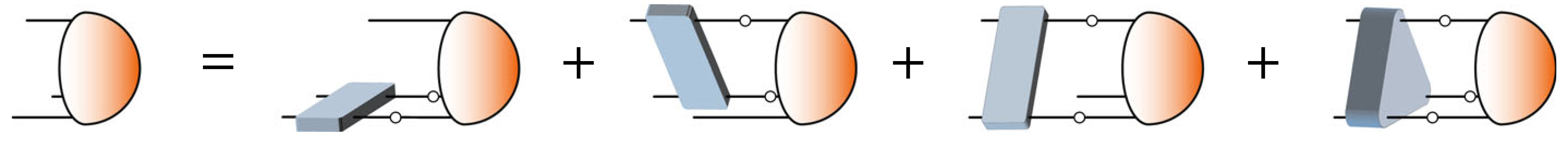}
\vspace*{0pt}
 \caption{The $3$-body Faddeev-equation is illustrated diagrammatically (figure taken from reference~\cite{Eichmann:2018adq}). The two- and three-body interactions between the valence quarks (gray square blobs) are iterated to all orders to obtain the $3$-quark Faddeev amplitude (orange rounded blob). Quark propagators are denoted as solid lines.  \FewBody{\cite{Eichmann:2018adq}}{2019}}
\label{fig:models:ThreeBodyFaddeev}
\end{figure}

For the evaluation of baryons as three-quark bound-states, the correct integral-equation is the covariant $3$-body Faddeev equation~\cite{Eichmann:2009qa,Eichmann:2016hgl,Eichmann:2016yit}, which is illustrated in Figure~\ref{fig:models:ThreeBodyFaddeev}. The Faddeev equation is an exact equation in QCD and it iterates the irreducible kernels describing two- and three-body interactions between the valence quarks to all orders. The solutions are the $3$-body Faddeev amplitudes, which once obtained are used to extract physical properties of baryonic states.

In all solution-approaches for the Faddeev equation employed by the Graz-Giessen-Lisboa- (GGL-) group~\cite{Eichmann:2016hgl,Eichmann:2016yit,Eichmann:2018adq}, the term containing contributions from irreducible three-body forces (i.e. the rightmost term on the right-hand-side in Figure~\ref{fig:models:ThreeBodyFaddeev}) is set equal to zero as a first step. This approximation is justified by consideration of the leading terms in a skeleton expansion of this three-body contribution~\cite{Sanchis-Alepuz:2017jjd}. For instance, the first term is given by a dressed $3$-gluon vertex, which vanishes due to color algebra. The covariant $3$-body Faddeev equation can then be solved using only the two-body kernels as input, utilizing sophisticated numerical methods~\cite{ARPACKGuide,Sanchis-Alepuz:2017jjd}. The numerical treatment of this full three-body problem, which amounts to the solution of eigenvalue problems for matrices of extremely large dimensions, is very CPU intensive. The high computing-power required, in combination with unfavorable analytic properties of the quark propagator, prevent calculations above $2$~GeV~\cite{Eichmann:2016hgl,Eichmann:2016nsu}. Therefore, the GGL group provides only the ground- and first few excited states using this full three-body approach.

An alternative approach for the solution of the Faddeev equation (Fig.~\ref{fig:models:ThreeBodyFaddeev}) is given by the quark-diquark approximation~\cite{Hellstern:1997pg,Oettel:1998bk,Eichmann:2016hgl}. This method proceeds by first re-formulating the whole Faddeev equation (with irreducible three-body contributions again set to zero beforehand) in an equivalent form, where the two-body kernels are eliminated for formal $2 \rightarrow 2$ $T$-matrices to describe the quark-quark interaction~\cite{Eichmann:2016hgl}. In a then ensuing simplifying approximation, these $T$-matrices are expanded into separable quark-diquark correlations (including diquark-systems with scalar, pseudoscalar, vector and axialvector quantum numbers), where the hierarchy of the expansion is organized according to the mass-scales of the diquarks. Using this quark-diquark approximation, the Faddev equation reduces to a coupled system of effective two-body bound-state equations for the quark-diquark system, which have a mathematical form similar to the usual two-body Bethe-Salpeter equations~\cite{Eichmann:2016hgl}. The effective two-body equations can then be solved for the quark-diquark Bethe-Salpeter amplitude which encodes the physical properties of baryonic states. However, this procedure needs the fully dressed quark propagator, diquark Bethe-Salpeter amplitudes and diquark propagators as input, which need to be calculated beforehand~\cite{Eichmann:2016hgl,Eichmann:2018adq}. The quark-diquark method is numerically more tractable and thus allows to make predictions for the whole spectrum. It has to be stressed that the full three-body approach described above, in combination with the quark-diquark approximation, provides a theoretical tool for a systematic study of the question to which degree baryons are predominantly quark-diquark systems~\cite{Eichmann:2018adq}.

Both solution methods for the Faddeev equation need input information on the irreducible two-body kernels, in particular on the quark-gluon vertex. One could in principle take the fully dressed quark-gluon vertex directly from QCD. However, since this object has a very complicated spin-tensor structure~\cite{Eichmann:2016yit}, calculations would become impossibly extensive. One possibility of a simplification is here to replace the quark-gluon vertex by its leading spin-structure, which is proportional to the Dirac matrix $\gamma^{\mu}$. This replacement is called the \textit{rainbow-ladder approximation}~\cite{Eichmann:2016hgl,Eichmann:2016yit}, and it implies an effective running coupling constant~$\alpha (k^{2})$ for the quark-gluon interaction, which depends on the squared gluon-momentum~$k^{2}$ only. Although it is in principle possible to evaluate~$\alpha (k^{2})$ self-consistently from QCD, this function is often modeled. Authors from the GGL group regularly use the parametrization for~$\alpha (k^{2})$ initially proposed by Maris and Tandy~\cite{Maris:1999nt}.

\begin{figure}[htb]
\centering
\hspace*{0pt}
 \includegraphics[width=0.999\textwidth]%
      {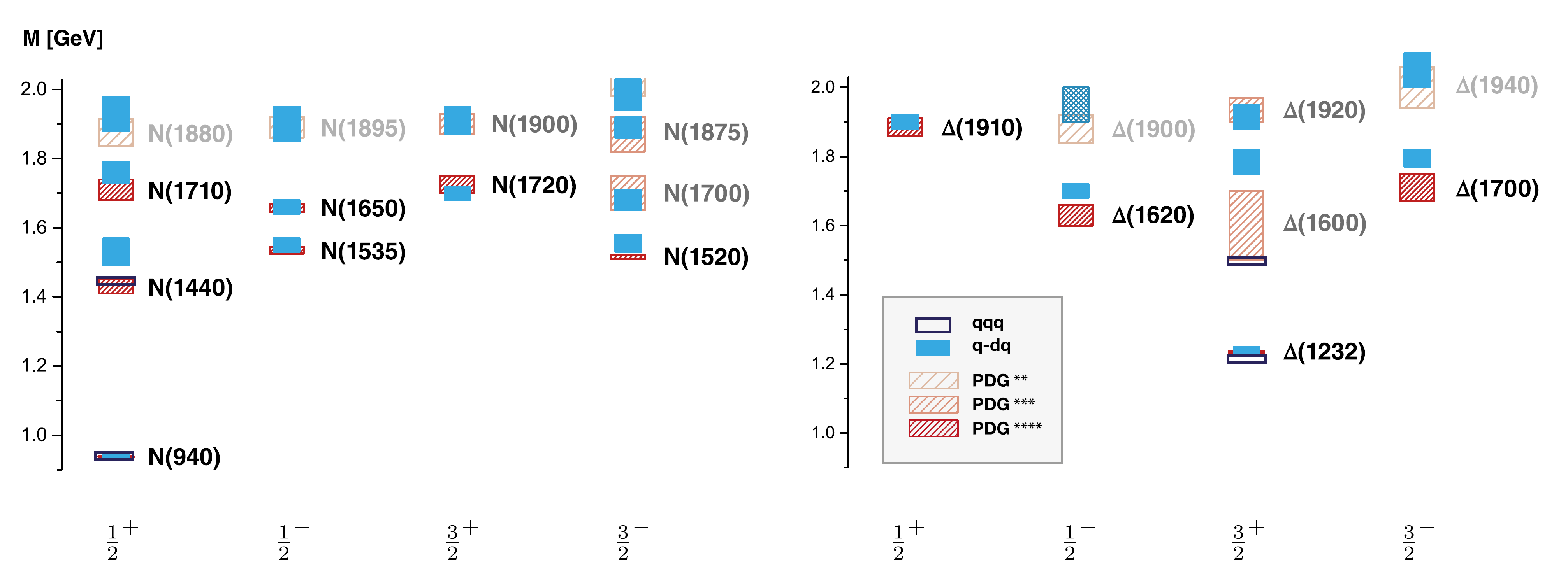}
\vspace*{0pt}
 \caption{Predictions for the nucleon- and $\Delta$-spectra, as evaluated from the rainbow-ladder approximation, are shown for the~$J^{P} = \frac{1}{2}^{\pm}$ and~$J^{P} = \frac{3}{2}^{\pm}$ states (figure taken from reference~\cite{Eichmann:2018adq}). The open boxes indicate results obtained from solution of the full $3$-body Faddeev equation, while the filled boxes denote mass-values resulting from the quark-diquark approximation (the quark-diquark interaction was modified slightly in order to mimic effects 'beyond rainbow-ladder', as described in the main text). The PDG mass-values~\cite{ParticleDataGroup:2016lqr} and their experimental uncertainties are indicated as hatched filled boxes. The widths of the 3-body and quark-diquark results indicate a systematic error inherent to the solution-procedure for the hadronic bound-state equations. \FewBody{\cite{Eichmann:2018adq}}{2019}}
\label{fig:models:DSEBSENucleonDeltaSpectra}
\end{figure}

Results for the Light Baryon Spectrum, i.e. nucleon- and $\Delta$-states, evaluated in the DSE/BSE-method~\cite{Eichmann:2016hgl,Eichmann:2018adq} are depicted in Figure~\ref{fig:models:DSEBSENucleonDeltaSpectra}. The results from the three-body approach and 'pure' (i.e. unmodified) quark-diquark calculations turn out to agree well with each other, as well as with experimental values, in certain 'good' channels: the $J^{P} = \frac{1}{2}^{+}$ channel in case of the nucleon and the $J^{P} = \frac{3}{2}^{+}$ channel for the $\Delta$. In order to get the agreement between quark-diquark calculations and experimentally determined resonances to the quality depicted in Figure~\ref{fig:models:DSEBSENucleonDeltaSpectra}, the GGL group had to modify the strength of the Maris/Tandy-interaction via a further constant prefactor, which was however only applied to the interaction in the 'bad' channels of the pseudoscalar and vector diquarks~\cite{Eichmann:2016hgl,Eichmann:2018adq}. The new prefactor was then adjusted to the experimental mass-splitting among the $\rho$- and $a_{1}$ meson. This modified quark-diquark calculation, which mimics effects 'beyond rainbow-ladder'~\cite{Eichmann:2016hgl}, yields then much improved results in all Light-Baryon channels, which are depicted in Figure~\ref{fig:models:DSEBSENucleonDeltaSpectra}. In any case, the good results for the Light Baryon Spectrum illustrated two facts. First, the validity of dropping the last term in the Faddeev equation (Figure~\ref{fig:models:ThreeBodyFaddeev}) has been verified a posteriori. Second, the experimental two-, three- and four-star baryon resonances from the PDG~\cite{ParticleDataGroup:2016lqr} are well reproduced by assuming baryons as dominated by quark-diquark correlations, at least within the context of the DSE/BSE-method described here. More specifically, this means that within the DSE/BSE-formalism it is perfectly justified to treat the baryons as effective $2$-body bound states. The only thing which is really necessary in order to reliably reproduce the measured spectrum of resonances is the modification of the Maris/Tandy-interaction described above, which is a rather ‘ad hoc’ approach for the introduction of effects that go beyond the lainbow-ladder approximation. Thus, the final result shows that approaching the full spin-structure of the quark-gluon vertex posed by QCD, which is effectively what one does when going beyond rainbow-ladder (cf. reference~\cite{Eichmann:2016yit}), seems to be more important for a good description of the light-baryon spectra than genuine $3$-body effects. This is a quite striking result.

It is worthwhile to further compare the DSE/BSE-results shown in Figure~\ref{fig:models:DSEBSENucleonDeltaSpectra} to the spectra evaluated in the Bonn constituent-quark model (Figures~\ref{fig:models:NucleonSpectrum} and~\ref{fig:models:DeltaSpectrum}) as well as to those originating from Lattice QCD (Figure~\ref{fig:models:LatticeNucleonSpectra}). The results of the DSE/BSE-framework are only comparable to the Lattice QCD calculation in an immediate way for the few low-lying states which were actually evaluated as true 3-body bound-states within the former approach (states marked as black boxes in Figure~\ref{fig:models:DSEBSENucleonDeltaSpectra}). The reason for this is that the Lattice QCD calculation from reference~\cite{Edwards:2011jj} is a true $3$-body calculation, at least in the sense that $3$-quark operators are defined and evaluated on lattices of different spacings and at there are at no point any assumptions made regarding quark-diquark configurations. Thus, the states we restrict our attention to for the comparison are the $N(940) \frac{1}{2}^{+}$, $N(1440)\frac{1}{2}^{+}$, the $\Delta(1232)\frac{3}{2}^{+}$ and the $\Delta(1600)\frac{3}{2}^{+}$. For these states, the agreement between the BSE/DSE- and Lattice QCD results is quite fair, at least with regard to the order of magnitude of the masses of the states, as well as to their relative separation (the comparison is made more difficult by the fact that Figure~\ref{fig:models:LatticeNucleonSpectra} is gauged in units of~$m_{\Omega}$). For the remainder of the spectrum, one can state directly that the relative level-ordering of the Roper resonance compared to the $N^{\ast}$ states with $J^{P} = \frac{1}{2}^{-}$ is described very well by both the Bonn constituent-quark model discussed in section~\ref{sec:QuarkModels} and by the DSE/BSE-framework discussed here (cf. Figure~\ref{fig:models:DSEBSENucleonDeltaSpectra}), while this feature is completely missed by the Lattice QCD calculation. The large mass-splittings between the $\frac{1}{2}^{+}$ and $\frac{3}{2}^{+}$ states in the $N^{\ast}$ spectrum are correctly described by all three approaches.


Predictions for the spectra of baryons with strangeness have also been evaluated using the DSE/BSE-approach~\cite{Sanchis-Alepuz:2014sca,Eichmann:2018adq}. In this case, the calculations indeed predict many states which have not yet been seen in experiment, for instance in the predicted $\Omega$-spectrum~\cite{Eichmann:2018adq}. Other predictions, such as those for the $\Sigma$- and $\Xi$-hyperons, closely resemble superpositions of the multiplets for the already calculated Light Baryon Spectra (Figure~\ref{fig:models:DSEBSENucleonDeltaSpectra}), at least with regard to their general structure.

%


The DSE/BSE-method already had impressive successes concerning the description of baryonic states, but there still exist room for improvement. First of all, the bound states predicted in the calculations presented here are stable and cannot decay to open hadronic channels~\cite{Eichmann:2018adq} (similarly to the lattice QCD calculation presented in section~\ref{sec:LatticeQCD}), which would cause them to dynamically acquire a width. Such decays can in principle be calculated within DSE/BSE, but then the decay mechanisms would have to be fed back into the BSEs in order to determine the resonance widths. This iterative procedure cannot be technically realized for baryons at the present moment, but first investigations have been performed for mesons~\cite{Santowsky:2020pwd,Williams:2018adr}. A further open challenge of the DSE/BSE is to systematically calculate effects 'beyond rainbow-ladder'~\cite{Eichmann:2016hgl,Eichmann:2016yit,Eichmann:2018adq}. This problem also poses tremendous technical difficulties, although first investigations are currently underway~\cite{Sanchis-Alepuz:2015qra}.

\subsection{Holographic QCD} \label{sec:HolographicQCD}

The \textit{Anti-de Sitter/Conformal Field Theory- (AdS/CFT-) correspondence}~\cite{Maldacena:1997re,Aharony:1999ti}, more generally referred to as \textit{gauge/gravity duality}~\cite{Ammon:2015wua}, is a conjectured duality (i.e. an equivalence) between quantum field theories with gauge symmetries on the one side and gravity theories in higher-dimensional spacetime (i.e. in the 'bulk') on the other side. A necessary constraint is here that the gauge theory has to be defined on the boundary spacetime of its gravitational dual, which marks the gauge/gravity duality as a particular realization of the \textit{holographic principle}. The whole approach was initialized in a celebrated work by Maldacena~\cite{Maldacena:1997re}. The most well-studied example is certainly the duality between $\mathcal{N} = 4$ Super Yang-Mills theory in $3+1$ dimensional spacetime and type IIB superstring theory defined on a direct-product space of five-dimensional Anti-de Sitter spacetime and a five-sphere, i.e. on~$\text{AdS}_{5} \times S^{5}$~\cite{Aharony:1999ti}. For more detailed and pedagogical introductions to the subject of gauge/gravity duality, see references~\cite{Aharony:1999ti,Ammon:2015wua}. 

In the AdS/CFT-correspondence, there exists a precise 'dictionary' relating operators of the gauge theory to fields of the higher-dimensional string theory, thus providing a precise relation on how physical processes in one theory can be computed in terms of the dual theory. Furthermore, in a particular limit a strong coupling of the gauge theory implies a weak coupling on the string theory side~\cite{Ammon:2015wua}. The latter property makes the gauge/gravity duality very attractive as a possible solution-Ansatz for strong QCD, since in case a gravitational dual to QCD were known, non-perturbative QCD would indeed be solvable in terms of computations in the weakly coupled dual gravitational theory. This idea set the research area of \textit{Holographic QCD} (or \textit{AdS/QCD})~\cite{deTeramond:2008ht,Brodsky:2014yha} in motion. Important contributions to this subject came from de Teramond, Brodsky and others~\cite{deTeramond:2005su,deTeramond:2008ht,Gutsche:2011vb,Brodsky:2014yha}. A very detailed report on the approach is given in~\cite{Brodsky:2014yha}.

At the time of this writing the unique, correct dual gravitational quantum-theory of QCD is not known, in particular since QCD has a much lower degree of overall symmetry compared to the above-mentioned $\mathcal{N} = 4$ Super Yang-Mills theory. Still, it is possible to build semiclassical models in Holographic QCD, using the following 'bottom up' approach~\cite{Brodsky:2014yha}: one starts with the known quantum field theory defined on the boundary and then searches for a corresponding classical gravity theory, ideally as simple as possible (i.e. an effective theory). The classical dual gravity-theory should reproduce the most important properties of the quantum field theory at the boundary (This approach of using a classical gravity-theory on the AdS side is called the 'weak form' of gauge/gravity duality in section~5.1 of reference~\cite{Ammon:2015wua}.). There exist convincing heuristic arguments for the fact that the energy scales on the gauge theory side map inverse proportionally to the spatial distance from the boundary into AdS spacetime~\cite{Brodsky:2014yha} (cf. arguments given further below). Therefore, in such models one typically has to limit the spatial range into the AdS space~\cite{Brodsky:2014yha}. We outline in the following two models corresponding to two possible methods of achieving such a spatial limitation: the 'hard-wall' model~\cite{deTeramond:2005su} and the 'soft wall' model~\cite{Gutsche:2011vb}. We also briefly summarize the results both approaches yield for the light-baryon spectrum.

As a preparatory step, we need the following definition: $d+1$-dimensional Anti-de Sitter space~$\text{AdS}_{d+1}$ is a $d + 1$ space with negative constant curvature and a $d$-dimensional flat spacetime boundary. Defining coordinates~$x^{0}, x^{1}, x^{2}, \ldots, x^{(d-1)}, z := x^{d}$, the differential line element (or the \textit{metric tensor}) of Anti-de Sitter space reads
\begin{equation}
  ds^{2} = \frac{R^{2}}{z^{2}}  \left( \eta_{\mu \nu} d x^{\mu} d x^{\nu} - d z^{2} \right)  . \label{eq:models:AdSSpaceLineElement}
\end{equation}
Here,~$\eta_{\mu \nu}$ is the usual $d$-dimensional Minkowski metric, $R$ is the AdS-radius, $z$ is called the AdS coordinate (or holographic coordinate) and the $d$-dimensional boundary space is reached for $z = 0$. Note that the AdS metric~\eqref{eq:models:AdSSpaceLineElement} is invariant under the spatial dilatations~$x^{\mu} \rightarrow \rho x^{\mu}$ and~$z \rightarrow \rho z$~\cite{deTeramond:2005su,Brodsky:2014yha}, which maps to (a part of) the known conformal symmetry of QCD in the limit of massless quarks on the gauge-theory side. On the gauge-theory side, the conformal invariance of QCD is anomalously broken by quantum effects, which leads to confinement setting in roughly at the QCD energy-scale~$\Lambda_{\text{QCD}}$. On the AdS side, one thus has to model confinement by modifying the AdS background-geometry in the large-$z$ infrared region, where $z \sim 1 / \Lambda_{\text{QCD}}$. This is accomplished by the above-mentioned spatial limitation in the semiclassical models of Holographic QCD.

In the hard-wall approach~\cite{deTeramond:2005su,Brodsky:2014yha}, one introduces a sharp cutoff in the infrared region of AdS space, thus limiting considerations to a finite slice~$0 \leq z \leq z_{0}$. Physical states on the AdS side, corresponding to hadronic modes on the gauge theory side, are introduced as normalizable fields (or 'modes')~$\Psi (x,z) = e^{- i P \cdot x} f(z)$ on AdS spacetime. The plane wave defined on the Minkowski coordinates~$x^{\mu}$ represents a free hadron with four-momentum~$P^{\mu}$ and invariant mass~$M^{2} = P_{\mu} P^{\mu}$. One has to impose strict boundary conditions for these fields at the infrared AdS boundary~$z_{0} \sim 1 / \Lambda_{\text{QCD}}$. The solution of the AdS wave-equation (i.e. the eigenvalue problem) for the AdS field~$\Psi (x,z)$, together with imposing the infrared boundary-conditions, determines the mass-spectrum of the hadronic states on the gauge-theory side. The higher spin-states of hadrons on the gauge-theory side are identified with quantum fluctuations around the spin-$0$,~$\frac{1}{2}$,~$1$ and~$\frac{3}{2}$ string-solutions on the AdS side~\cite{deTeramond:2005su}.

\begin{figure}[tbp]
\centering
 \includegraphics[width=0.665\textwidth]%
      {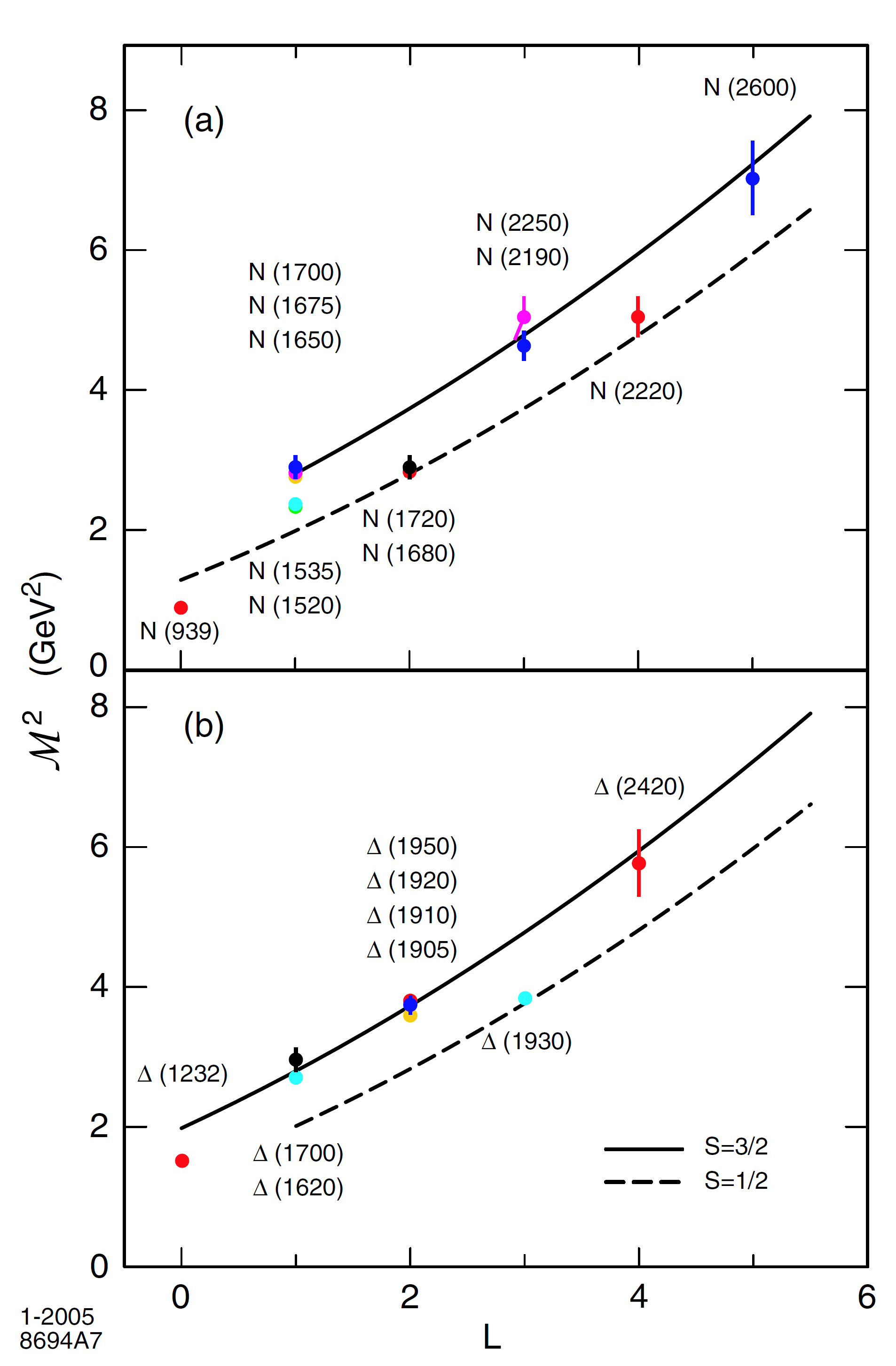}
 \caption{The orbital spectrum for light baryon resonances as evaluated in the hard-wall model is shown for
$\Lambda_{\text{QCD}} = 0.22 \text{ GeV}$. The figure has been taken over identically from reference~\cite{deTeramond:2005su}. Nucleon states are shown in plot (a), while the trajectories for the $\Delta$~states are given in plot (b). \APS{\cite{deTeramond:2005su}}{2005}}
\label{fig:models:HolographicQCDNucleonDeltaSpectra}
\end{figure}

For the determination of the baryon spectrum, the full ten-dimensional Dirac equation has to be solved on the space~$\text{AdS}_{5} \times S^{5}$. Furthermore, baryons are charged under the internal $\text{SU} (4)_{R}$ symmetry of the compact space~$S^{5}$. One can expand the baryon field into eigenfunctions of the Dirac operator on the compact space, with eigenvalues~$\lambda_{\kappa}$, and then the following AdS Dirac equation~\cite{deTeramond:2005su} has to be solved:
\begin{equation}
  \left[  z^{2} \partial_{z}^{2} - d z \partial_{z} + z^{2} M^{2} - \left( \lambda_{\kappa} + \mu \right)^{2} R^{2}  +  \frac{d}{2} \left( \frac{d}{2} + 1 \right) + \left( \lambda_{\kappa} + \mu \right) R \hat{\Gamma} \right]  f(z) = 0   , \label{eq:models:AdSDiracEquation}
\end{equation}
where~$\mu$ is called the AdS mass. The field~$\Psi$ decomposes as~$\Psi (x,z) = e^{- i P \cdot x} f(z)$ and one has the following eigenvalue equation for free Dirac spinors:~$\hat{\Gamma} u_{\pm} = \pm u_{\pm}$. The eigenvalues on $S^{5}$ are~$\lambda_{\kappa} R = \pm \left( \kappa + \frac{5}{2} \right)$, for~$\kappa = 0,1,\ldots$~(this is the general expression on~$S^{d + 1}$, given in~\cite{deTeramond:2005su}, evaluated for~$d = 4$). The normalizable modes on the AdS side, for $\kappa = 0$, are:
\begin{equation}
  \Psi_{\alpha, k} (x,z) = C_{\alpha, k} e^{- i P \cdot x} z^{\frac{5}{2}} \left[ J_{\alpha} \left( z \beta_{\alpha, k} \Lambda_{\text{QCD}} \right)  u_{+} (P) + J_{\alpha + 1} \left( z \beta_{\alpha, k} \Lambda_{\text{QCD}} \right)  u_{-} (P)  \right] , \label{eq:models:SpinOneHalfAdSModes}
\end{equation}
where~$\alpha = 2 + L$, the conformal dimension is~$\Delta = \frac{9}{2} + L$ and $L$ is the orbital angular momentum. The $J_{\alpha}$ are Bessel functions and $\beta_{\alpha, k}$ is the $k$-th zero of the Bessel function. In addition to the spin~$1/2$ Dirac equation, one also has to solve the spin~$3/2$ Rarita-Schwinger equation on AdS spacetime. This leads to additional technical complications, which can however be handled~\cite{deTeramond:2005su}, resulting in a Rarita-Schwinger field~$\Psi_{\mu} (x,z)$. The four-dimensional mass spectrum of the hard-wall model then follows from the boundary conditions~$\Psi^{\pm} (x, z_{0}) = 0$ and~$\Psi^{\pm}_{\mu} (x, z_{0}) = 0$~\cite{deTeramond:2005su}:
\begin{equation}
 M^{+}_{\alpha, k} =  \beta_{\alpha, k} \Lambda_{\text{QCD}} , \hspace*{5pt}  M^{-}_{\alpha, k} =  \beta_{\alpha + 1, k}  \Lambda_{\text{QCD}} . \label{eq:models:AdSHardWallMassSpectrum}
\end{equation}
The internal spin~$S$ of a hadronic state on the gauge-theory side has to match the spin of the string-mode on the dual gravity-theory side~\cite{deTeramond:2005su} (boundary conditions:~$\Psi^{+} (x, z_{0}) = 0$ for~$S = \frac{1}{2}$ and~$\Psi^{-}_{\mu} (x, z_{0}) = 0$ for~$S = \frac{3}{2}$). All these considerations lead to the orbital spectra of nucleon- and $\Delta$ resonances in the hard-wall model, which are shown in Figure~\ref{fig:models:HolographicQCDNucleonDeltaSpectra}, for~$\Lambda_{\text{QCD}} = 0.22 \text{ GeV}$.

%


The agreement of the hard-wall model with the data is already quite good. In particular, the phenomenological \textit{Regge-behavior}~\cite{Klempt:2002cu,Klempt:2012fy} of the baryon resonances, i.e. an approximate arrangement of the resonances on linear \textit{Regge-trajectories} in plots of $M^{2}$ against the orbital angular momentum $L$, is roughly reproduced. However, the precise form of the Regge-trajectories in the hard-wall model is not perfectly linear, but rather slightly konvex~(cf. Figure~\ref{fig:models:HolographicQCDNucleonDeltaSpectra}). In particular, the masses of the lowest states in~$L$ turn out to be slightly too large: a common feature of all hard-wall models for hadronic resonances~\cite{deTeramond:2005su,Brodsky:2014yha}. Furthermore, the hard-wall mass-spectrum~\eqref{eq:models:AdSHardWallMassSpectrum} predicts a new parity degeneracy between the neighboring (i.e. parallel) trajectories in Figure~\ref{fig:models:HolographicQCDNucleonDeltaSpectra}, which can be observed upon shifting the upper (solid) curve one unit of~$L$ to the right. For instance, the $L = 1$ states $N(1650)$, $N(1675)$ and $N(1700)$ are degenerate with the~$L = 2$ states $N(1680)$ and $N(1720)$. Similar parity degeneracies have been suggested, in the form of parity-doublets, to occur due to chiral symmetry restoration in the higher-mass region of the light-baryon spectrum~\cite{Glozman:1999tk}. In summary, the hard-wall model achieves already quite a good determination of the light baryon spectrum, depending on only one single energy-scale:~$\Lambda_{\text{QCD}}$. In order to remedy the above-mentioned problems with the non-linearity of the Regge-trajectories (among others), refined approaches have been developed, one of which will be discussed in the following.

\begin{figure}[tb]
\centering
\hspace*{0pt}
 \includegraphics[width=0.85\textwidth]%
      {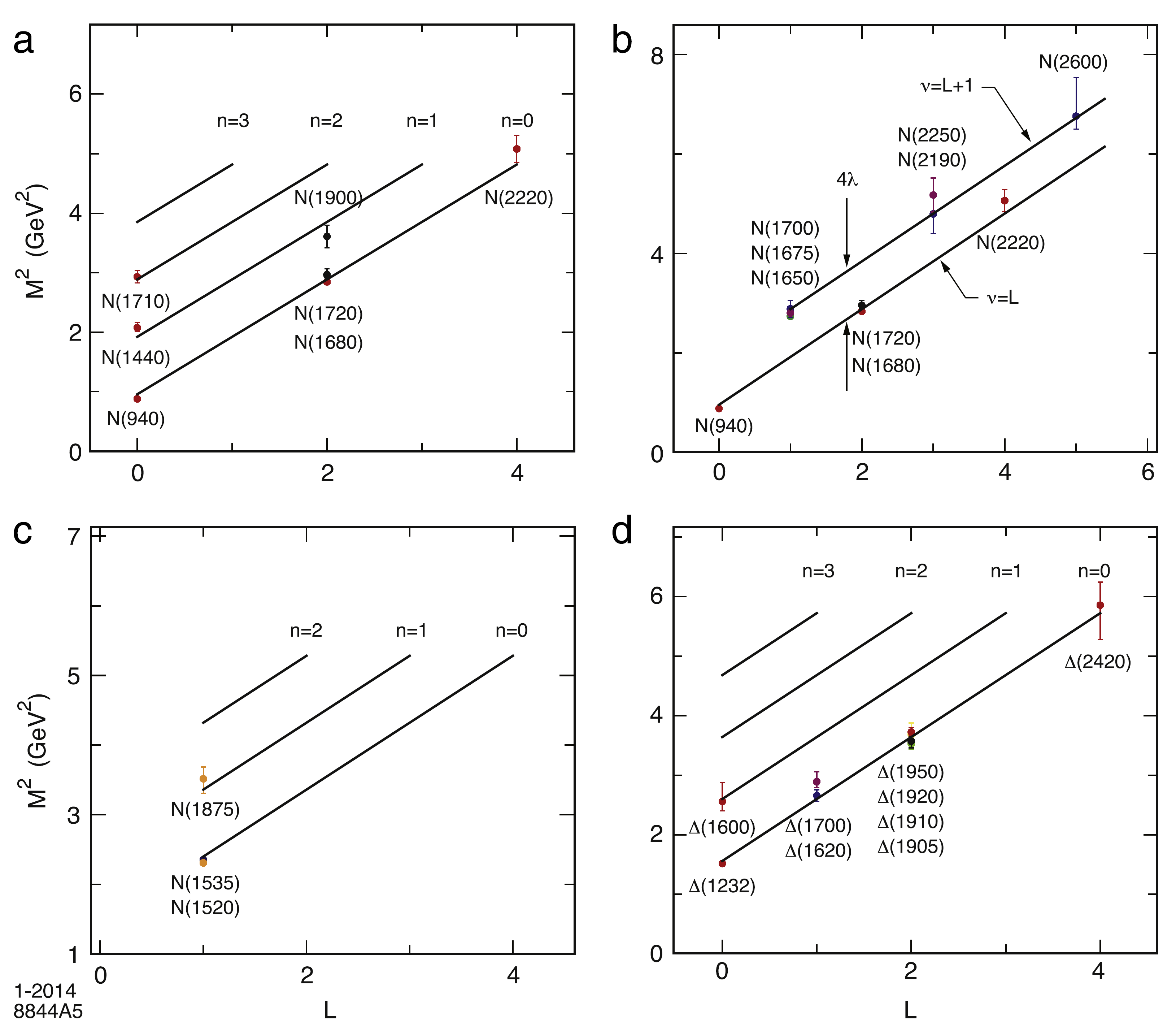}
\vspace*{0pt}
 \caption{The orbital and radial excitation spectra are shown for nucleon- and $\Delta$ resonances as evaluated in the soft-wall model, for values of $\sqrt{\lambda} = 0.49 \text{ GeV}$ (nucleons) and $\sqrt{\lambda} = 0.51 \text{ GeV}$ ($\Delta$'s). The figure has been taken over identically from reference~\cite{Brodsky:2014yha}. The positive-parity spin-$\frac{1}{2}$ nucleons are shown in~(a), the spectrum gap between the negative-parity spin-$\frac{3}{2}$ and the positive-parity spin-$\frac{1}{2}$ nucleons is illustrated in~(b), the negative-parity nucleon states are shown in~(c) and the positive- and negative-parity $\Delta$ states are collected in plot~(d).}
\label{fig:models:HolographicQCDNucleonDeltaSpectraParities}
\end{figure}

In the soft-wall approach~\cite{Gutsche:2011vb,Brodsky:2014yha}, a smooth cutoff at large distances in AdS space is achieved by introducing a so-called \textit{dilaton}-background in the holographic coordinate~$z$. The dilaton field breaks the maximal AdS symmetry and further leads to an effective $z$-dependent curvature in AdS space, which induces conformal symmetry breaking on the QCD (i.e. on the gauge-theory) side of the duality. For half-integer spin fields in an AdS light-front quantization scheme, the dilaton field can be fully absorbed into a field re-definition of the fermions~\cite{Brodsky:2014yha}. Thus, confinement has to be implemented via an effective potential in the AdS field equations. For this purpose, the authors of~\cite{Gutsche:2011vb,Brodsky:2014yha} chose a linear confinement potential
\begin{equation}
  V (\zeta) = \lambda_{F} \zeta   , \label{eq:models:AdSSoftWallConfinementPotential}
\end{equation}
where~$\zeta$ is the so-called light-front invariant impact parameter and the slope~$\lambda_{F}$ defines the only relevant energy-scale in the problem, i.e.~$\sqrt{\lambda_{F}}$. Solving the light-front AdS field-equations for the fermion fields of positive and negative chirality, one finds the following invariant mass-spectrum for the baryon states on the gauge-theory side (the mathematical details are fully written out in reference~\cite{Brodsky:2014yha}):
\begin{equation}
 M^{2} = 4 \lambda_{F} \left( n + \nu + 1 \right) . \label{eq:models:AdSSoftWallMassSpectrum}
\end{equation}
Here,~$n$ is the radial quantum number and~$\nu$ is called the light-front orbital angular momentum, which can be related to the ordinary orbital angular momentum~$L$ of the baryons via phenomenological assignments (depending on the parity and internal spin of the states, cf. Table~5.5 of reference~\cite{Brodsky:2014yha}). The variable~$\lambda_{F}$ can be set equal to the Regge slope for mesons, i.e. ~$\lambda_{F}= \lambda$, due to universality arguments. The lowest possible stable light-baryon state, the proton, fixes the energy scale to~$\sqrt{\lambda} \simeq 0.5~\text{GeV}$ by setting~$n = 0$ and~$\nu = 0$ in equation~\eqref{eq:models:AdSSoftWallMassSpectrum}. The resulting orbital- and radial spectra of the nucleon- and $\Delta$ states in the soft-wall model are shown in Figure~\ref{fig:models:HolographicQCDNucleonDeltaSpectraParities}.

The orbital Regge-behavior of the trajectories resulting from the soft-wall model is now perfectly linear, a clear improvement compared to the hard-wall model (cf. Figure~\ref{fig:models:HolographicQCDNucleonDeltaSpectra}). In particular, the masses of the lowest states in~$L$ have correct values. The radial (i.e. $n$-dependent) spectra also show a linear slope in the soft-wall model, a fact which is not accounted for at all in the hard-wall approach~\cite{Brodsky:2014yha}. However, the choice of the linear effective confinement potential~\eqref{eq:models:AdSSoftWallConfinementPotential} is mainly motivated by the fact that it results in linear Regge-trajectories~\cite{Brodsky:2014yha}. 

The Roper resonance~$N (1440)$ and the~$N(1710)$ can be explained in the soft-wall model as the first and second radial excited states of the proton (cf. plot~(a) in Figure~\ref{fig:models:HolographicQCDNucleonDeltaSpectraParities}). The model is also successful in explaining parity degeneracies, which have been conjectured elsewhere~\cite{Glozman:1999tk}, just as in the case of the hard-wall model. The scale~$\lambda$ determines not only the slope of the Regge trajectories, but it does double duty by also setting the gap (of size~$4 \lambda$) in the spectrum between the positive-parity spin~$1/2$ and the negative-parity spin~$3/2$ nucleon states (see plot~(b) in Figure~\ref{fig:models:HolographicQCDNucleonDeltaSpectraParities}), which is a remarkable achievement. The soft-wall model furthermore predicts the positive- and negative-parity $\Delta$ states to lie on the same trajectory, which which has been observed for some resonances in experimental data (cf. plot~(d) in Figure~\ref{fig:models:HolographicQCDNucleonDeltaSpectraParities}).


%

%% file: Formalisms.tex
\label{sec:formalisms}

In Light Baryon Spectroscopy, particles in the intermediate states always have half-integer spins. In order to produce such intermediate states, the conservation of the total angular momentum implies that at least one particle in the initial- as well as the final-state has to have non-vanishing half-integer spin. Thus, in Baryon spectroscopy (other than in Meson Spectroscopy), one necessarily deals with reactions among particles with spin.

In this section, we first collect some generic features of the formalisms for such reactions with spin. Also, we give a brief overview on the important topic of so-called complete experiments, which has received some increased attention over the last few years. Then, we briefly discuss the most important reactions for Baryon Spectroscopy, one after the other. Since the photoproduction of single pseudoscalar mesons is most prominently discussed in this review, special emphasis is put on this particular reaction.

This section is mainly meant to help the reader in understanding the numerous datasets and results for amplitudes discussed in later chapters.

\subsection{Generic features of formalisms for reactions among particles with spin}
\label{sec:formalisms:GenericFeatures}

\subsubsection{Decomposition of the full transition-matrix into spin-amplitudes}
\label{sec:formalisms:SpinAmplitudes}

For every reaction considered in this review, it is possible to write a decomposition of the matrix element $\mathcal{T}_{fi} = \left< f \right| \hat{T} \left| i \right>$ of the transition-matrix (or $T$-matrix) into so-called {\it spin-amplitudes}. In order to do this, one collects all the four-momenta and spin-operators for the particles in the initial- and final states and finds a set of Dirac-structures that are Lorentz-invariant and furthermore respect other symmetries that have to hold in the considered process (for instance, gauge-invariance in the case of photoproduction). The number of such structures is always finite. Then, one writes a general expansion of $\mathcal{T}_{fi}$ into these kinematical structures and every expansion-coefficient of each particular structure is a so-called spin-amplitude.

A lot of times, the thus-obtained decompositions are given in the literature with the kinematical variables restricted to the center-of-mass (CMS-) frame (cf. appendix~\ref{sec:appendix:Kinematics}). For instance, in case of the reaction of Pion-Nucleon (or Spin-$0$ Spin-$1/2$) scattering (more detail given in section~\ref{sec:formalisms:PiNScattering} below), the general decomposition of the $T$-matrix reads~\cite{Hoehler84,Anisovich:2013tij}:
\begin{equation}
 \mathcal{T}_{fi} (W, \theta) = \frac{q}{2 \pi} \chi_{m_{s_{f}}}^{\dagger} \left[ G (W, \theta) + i \left( \vec{\sigma} \cdot \hat{\vec{n}} \right) H (W, \theta) \right] \chi_{m_{s_{i}}}  . \label{eq:formalisms:TMatrixPiN}
\end{equation}
Here, $q$ is the modulus of the final-state CMS $3$-momentum $\vec{q}$, $\chi_{m_{s_{i,f}}}$ are Pauli spinors that describe the states of the spin-$\frac{1}{2}$ particles in the initial and final states, $\vec{\sigma}$ is a vector-operator collecting the usual Pauli-matrices and the vector $\hat{\vec{n}}$ defines the normal of the {\it reaction plane}\footnote{The reaction plane is spanned by the CMS $3$-momenta $\vec{k}$ and $\vec{q}$ (cf. appendix~\ref{sec:appendix:Kinematics}) and is defined for each $2$-body reaction. Thus, the vector $\hat{\vec{n}}$ is defined as: $\hat{\vec{n}} := \left( \vec{k} \times \vec{q} \right) / \left| \vec{k} \times \vec{q} \right|$.}. The phase-space of Pion-Nucleon scattering can be described fully using the two variables $W$ (the CMS-energy) and $\theta$ (CMS scattering-angle), since it is just a $2$-body reaction (see appendix~\ref{sec:appendix:Kinematics}).

Another important example is given by single pseudoscalar-meson photoproduction ($\gamma N \rightarrow \varphi B$, cf. section~\ref{sec:formalisms:SingleMesonPhotoproduction}). Here, the decomposition into spin-momentum structures reads~\cite{Chew:1957tf}:
\begin{equation}
 \mathcal{T}_{fi} = \frac{4 \pi W}{\sqrt{m_{N}m_{B}}} \chi_{m_{s_{f}}}^{\dagger} \left[  i \vec{\sigma} \cdot \hat{\epsilon} F_{1} + \vec{\sigma} \cdot \hat{q}
   \vec{\sigma} \cdot \left( \hat{k} \times \hat{\epsilon} \right)  F_{2}  + i \vec{\sigma} \cdot \hat{k}  \hat{q} \cdot \hat{\epsilon}  F_{3} + i \vec{\sigma} \cdot \hat{q}  \hat{q} \cdot \hat{\epsilon} F_{4}  \right] \chi_{m_{s_{i}}} . \label{eq:formalisms:TMatrixPhotoprod}
\end{equation}
Here, $\hat{k}$ and $\hat{q}$ are normalized CMS three-momenta in the initial- and final state, while $\hat{\epsilon}$ is the normalized polarization vector of the photon. The four spin-amplitudes $F_{i} = F_{i} (W, \theta)$ of single meson photoproduction are called CGLN-amplitudes. We stress again that the decomposition~\eqref{eq:formalisms:TMatrixPhotoprod} only holds for one pseudoscalar meson~$\varphi$ ($J^{P} = 0^{-}$) and one spin-$1/2$ baryon~$B$ ($J^{P} = \frac{1}{2}^{+}$) in the final state.

Regardless of the process under consideration, a full knowledge of all spin-amplitudes, including all of their respective phases, at every point in phase-space fully characterizes the process. However, the decompositions described in this section are still all fully model-independent. The modeling of the $T$-matrix using more physical input is the subject of chapter~\ref{sec:PWA}.

The examples shown in equations~\eqref{eq:formalisms:TMatrixPiN} and~\eqref{eq:formalisms:TMatrixPhotoprod} correspond to a conventional choice of specifying kinematic variables in the CMS-frame and quantizing spins along the $z$-axis. Moreover, different bases (or systems of amplitudes) exist as well ,which mostly correspond to a different scheme of spin-quantization. Important systems of amplitudes encountered in the literature are:
\begin{itemize}
 \item[-] {\it Invariant amplitudes:} while the examples~\eqref{eq:formalisms:TMatrixPiN} and~\eqref{eq:formalisms:TMatrixPhotoprod} have been written in CMS-kinematics, a general expansion into Lorentz-invariant Dirac-structures, such as described in the beginning of this section, is defined in terms of so-called {\it invariant amplitudes}. These types of spin-amplitudes have to be invariant, since all the individual Dirac-structures are Lorentz-invariant and the fully $T$-matrix element has to be Lorentz-invariant as well. \\
 It has been argued that invariant amplitudes are free of kinematical singularities~\cite{Mikhasenko:2017rkh}, thus they are naturally well-suited for physical models that implement the analyticity-constraint very precisely (cf. section~\ref{sec:PWA:SMatrixPrinciples}). Moreover, invariant amplitudes are well-suited for matching to Quantum Field Theoretical calculations. However, the equations connecting amplitudes to observables are often very complicated in this basis.
 \item[-] {\it Helicity amplitudes:} In this type of amplitude-system, one employs the {\it helicities} 
 of the initial- and final-state particles in the CMS for spin-quantization, as opposed to the spin-$z$ components which are central to examples like equations~\eqref{eq:formalisms:TMatrixPiN} and~\eqref{eq:formalisms:TMatrixPhotoprod}. This helicity formalism is widely used in the literature, since it has become the canonical formalism to define partial-wave decompositions (see section~\ref{sec:formalisms:PWExpansion}).
 \item[-] {\it Transversity amplitudes:} In order to define amplitudes in the so-called {\it transversity basis}, the choice of the axis of spin-quantization is changed once more to the normal of the reaction-plane. Thus, transversity amplitudes are straightforwardly defined for $2$-body reactions, while their generalization to $2 \rightarrow n$-reactions is generally non-trivial (cf. section~\ref{sec:formalisms:MultiMesonPhotoproduction}). \\
 Since the equations connecting amplitudes to observables generally take the most concise and elegant form in the transversity basis, this type of basis is most commonly used to discuss complete experiments (cf. section~\ref{sec:formalisms:CompleteExperiments}).
\end{itemize}
The different systems for spin-amplitudes mentioned above can always be transformed into each other via linear and invertible relations (the coefficients of these relations may carry kinematic dependencies, but don't have to). Therefore, the number $N_{\mathcal{A}}$ of spin-amplitudes describing a particular process remains unchanged. E.g., Pion-Nucleon scattering is always described by two amplitudes, while for photoproduction one always has four amplitudes. 

For any reaction, the number of relevant spin-amplitudes can be simply counted. This result is yielded quickly in the helicity-formalism. One just has to multiply the spin-multiplicities $\bm{n}_{1}, \bm{n}_{2}, \ldots, \bm{n}_{n}$ of all the initial- and final-state particles of the reaction:
\begin{equation}
  N_{\mathcal{A}} = \bm{n}_{1} \hspace*{0.5pt} \bm{n}_{2} \hspace*{0.5pt} \bm{n}_{3} \hspace*{0.5pt} \ldots  \hspace*{0.5pt} \bm{n}_{n} . \label{eq:formalisms:NumberOfSpinAmplitudes}
\end{equation}
For a $2$-body reaction~'$1 + 2 \rightarrow 3 + 4$', a simple parity-relation among the helicity- (or transversity-) amplitudes can be invoked~(cf.~\cite{Roberts:2004mn}), which reduces the number of non-redundant amplitudes~\eqref{eq:formalisms:NumberOfSpinAmplitudes} further by a factor of~$1/2$.

\subsubsection{Polarization observables}
\label{sec:formalisms:PolObservables}

For a $2 \rightarrow n$ meson-production reaction, such as all those considered in this review, the basic experimental source of information is given by the differential cross section, accessed via count-rates measured in the detectors. However, the spins which are necessarily present in reactions used for Light Baryon Spectroscopy give access to many different cross sections. One can either prepare the spins of the particles in the initial state (i.e. {\it polarize} these particles), or measure certain preferred spin-states for final-state particles, although the former task is more easily accomplished experimentally than the latter.

In most cases, a generic {\it polarization observable} $\mathcal{O}$ is given as an {\it asymmetry} among differential cross sections for different polarization-configurations, i.e. (with possible kinematical pre-factors $\bm{c}$):
\begin{equation}
  \mathcal{O} = \bm{c} \left[ \left( \frac{d \sigma}{d \Omega} \right)^{a} - \left( \frac{d \sigma}{d \Omega} \right)^{b} \right]  . \label{eq:formalisms:GenericObservableAsymmetry}
\end{equation}
These asymmetries often appear as normalized by the unpolarized differential cross section, but they don't have to. The super-scripts '$a$' and '$b$' appearing in equation~\eqref{eq:formalisms:GenericObservableAsymmetry} represent a short-notation for the possible polarization-configurations, which depending on the reaction under consideration could be either single-, double- or triple-polarization configurations, at least for all the reactions considered in this review. The spin-states of unpolarized initial-state particles have been implicitly averaged over in equation~\eqref{eq:formalisms:GenericObservableAsymmetry}, while for unpolarized final-state particles the spin-states have to be summed.

Considering the definitions of polarization observables in terms of spin-amplitudes, one encounters for each spin-reaction that the former are {\it bilinear hermitean forms} of the latter. When restricting to transversity-amplitudes $b_{i}$, this means that the observables $\mathcal{O}^{\alpha}$ of a generic process take the form (see the work by Chiang and Tabakin~\cite{Chiang:1996em} which substantiates this claim; possible kinematical pre-factors are suppressed in this expression): 
\begin{equation}
 \mathcal{O}^{\alpha} = \sum_{i,j = 1}^{N} b_{i}^{\ast} \Gamma^{\alpha}_{ij} b_{j} , \hspace*{2,5pt} \alpha = 1, \ldots, N_{\mathcal{A}}^{2} . \label{eq:formalisms:GeneralDefinitionPolarisationObservable}
\end{equation}
The matrices $\Gamma^{\alpha}$ compose a basis of the $N_{\mathcal{A}}^{2}$-dimensional vectorspace of hermitean complex $N_{\mathcal{A}} \times N_{\mathcal{A}}$-matrices~\cite{Chiang:1996em,Wunderlich:2020umg}. The observables in the left-hand-side of equation~\eqref{eq:formalisms:GeneralDefinitionPolarisationObservable}, as well as the amplitudes on the right-hand-side, depend on all the variables necessary to parametrize the phase-space of the considered reaction, which generally are $3 (2 + n) - 10$ variables for a $2 \rightarrow n$-process (cf. appendix~\ref{sec:appendix:Kinematics}). For a $2 \rightarrow 2$-process, one would just have $(W, \theta)$. The $\Gamma$-matrices do not carry kinematical dependencies. Particular examples for the general definition~\eqref{eq:formalisms:GeneralDefinitionPolarisationObservable} are provided by the algebra of observables in Pion-Nucleon scattering (cf. Table~\ref{tab:formalisms:PiNTrObs} in section~\ref{sec:formalisms:PiNScattering}) or by the observables in single-meson photoproduction (see Table~\ref{tab:formalisms:PhotoproductionObservables} in section~\ref{sec:formalisms:SingleMesonPhotoproduction}).

Note that due to the bilinear structure of equation~\eqref{eq:formalisms:GeneralDefinitionPolarisationObservable}, a set of polarization observables can only determine the amplitudes $b_{i}$ up to one unknown overall phase $\phi$ (i.e. one phase for all amplitudes)~\cite{YannickDiploma,YannickPhD,Svarc:2020cic}. In other words, only moduli and relative-phases of the $b_{i}$ can be determined. This unknown phase can, based on the defining equation~\eqref{eq:formalisms:GeneralDefinitionPolarisationObservable} alone, carry any dependence on the $3 (2 + n) - 10$ kinematical phase-space variables.

Further consideration of the structure of equation~\eqref{eq:formalisms:GeneralDefinitionPolarisationObservable} directly reveals why the measurement of polarization observables is so important for Light Baryon Spectroscopy: polarization observables give experimental access to many different {\it interference terms} among amplitudes. The unpolarized cross section on the other hand is always strictly a sum of squares: $\left( \frac{d \sigma}{d \Omega} \right)_{0} = (1/N_{\mathcal{A}}) \sum_{i} \left| b_{i} \right|^{2}$.
Thus, this single observable simply cannot yield enough information to determine all amplitude up to an overall phase. The higher the degree of the complications due to spin is for a particular reaction, the more spin-amplitudes are introduced and, correspondingly, the larger the amount of different observables to be measured becomes for this particular process. This fact makes the unraveling of the Light Baryon Spectrum experimentally very demanding.

As can be seen, the total number of $N_{\mathcal{A}}^{2}$ accessible observables by far outweighs the real degrees of freedom given by $N_{\mathcal{A}}$ complex amplitudes (minus one unknown overall phase), especially for the more complicated reactions. Thus, one may expect that not all observables have to be measured in order to extract the full amount of physical information. This type of optimization question directly leads to the subject of so-called {\it complete experiments}, which is treated in section~\ref{sec:formalisms:CompleteExperiments}.

\subsubsection{Expansion into partial waves}
\label{sec:formalisms:PWExpansion}

While the general model-independent expansions into spin-amplitudes discussed in previous sections are already quite useful, in Light Baryon Spectroscopy one wishes to gain information on {\it resonances}. Thus, one has to transform once more to a different set of amplitudes that allow for a (pre-) selection of resonance-contributions according to the spin-parity quantum numbers of the resonance. This feat is accomplished by transforming the transition-matrix to the {\it partial-wave basis}~\cite{Jacob:1959at,MartinSpearman}.

The partial-wave expansion of the $T$-matrix element $\mathcal{T}_{fi}$ of a particular $2\rightarrow2$-reaction among particles with spin can be most easily and elegantly written in the helicity-formalism. It reads~\cite{MartinSpearman}
\begin{equation}
  \mathcal{T}_{\mu_{1} \mu_{2}, \lambda_{1} \lambda_{2}} (s,t) = e^{i (\lambda - \mu) \phi} \sum_{j = \text{max} (|\lambda|,|\mu|)}^{\infty} (2 j + 1) \mathcal{T}^{j}_{\mu,\lambda} (s) \hspace*{1pt} d^{j}_{\mu,\lambda} (\theta)   , \label{eq:formalisms:General2BodyHelicityPWExpansion} 
\end{equation}
where $\lambda_{1,2}$ and $\mu_{1,2}$ are the helicities of the particles in the initial- and final states, $\mu := \mu_{1} - \mu_{2}$ and $\lambda := \lambda_{1} - \lambda_{2}$ are relative helicities, $\phi$ is the azimuthal angle of the final-state CMS three-momentum $\vec{q}$ and the $d^{j}_{\mu,\lambda}$ are the  Wigner $d$-functions\footnote{These functions form irreducible representations of the rotation-group and they are listed in many books on scattering theory (for instance~\cite{MartinSpearman,BransdenMoorhouse}).}. The infinite series runs over all possible total angular momentum quantum numbers $j$. The functions $\mathcal{T}^{j}_{\mu,\lambda} (s)$ are the partial waves (or $T$-matrix elements in the partial-wave basis) and they carry a pure energy-dependence via the Mandelstam variable $s$~(appendix~\ref{sec:appendix:Kinematics}). The expansion~\eqref{eq:formalisms:General2BodyHelicityPWExpansion} can be inverted for the $\mathcal{T}^{j}_{\mu,\lambda}$ using a suitable projection integral over the angular variable $\cos \theta$ (see~\cite{MartinSpearman,BransdenMoorhouse}).

The generalization of expression~\eqref{eq:formalisms:General2BodyHelicityPWExpansion} to processes with $3$ or more final-state particles is non-trivial. 
For $3$ or more final-state particles, the kinematical phase-space is at least $5$-dimensional, or higher. Furthermore, this space is necessarily spanned by at least $2$ or more relevant angular variables. The angular dependencies then have to be eliminated using the partial-wave expansion. However, since the parametrization of such a higher-dimensional phase-space is in most cases highly convention-dependent and thus non-unique, so is the set of occurring angles. Thus, model-independent partial-wave expansions for $3$ or more final-state particles are highly involved nested expansions, which are furthermore generally non-unique. For the example of two-meson photoproduction, we quote an expression in section~\ref{sec:formalisms:MultiMesonPhotoproduction}.

Still, the concept of the partial-wave expansion has a high significance for at least three reasons:
\begin{itemize}
 \item[(i)] One can extract the partial waves themselves from the data in so-called {\it truncated partial-wave analyses (TPWAs)}. In this procedure, expansions like~\eqref{eq:formalisms:General2BodyHelicityPWExpansion} are truncated at some maximal angular momentum $j_{\text{max}}$ (or at some maximal orbital angular momentum $\ell_{\text{max}}$, respectively) and the angular distributions of a set of observables are analyzed in order to extract the partial waves themselves as parameters. \newline
 For $2$-body reactions, this procedure takes place at single isolated bins in energy and is therefore also termed a {\it single-energy fit}. Fitted partial waves are then in some sense an equivalent representation of the data and may be analytically continued using model-assumptions in order to search for resonance-poles (cf. section~\ref{sec:PWA}). 
 \item[(ii)] The partial-wave series leads to the possibility to do {\it moment analysis} (or {\it angular analysis}) on the observables themselves, without yet extracting partial waves. Expressions such as~\eqref{eq:formalisms:General2BodyHelicityPWExpansion} can be used in order to derive a separation of energy- and angular variables already at the level of observables. Such a separation is then achieved by a finite Legendre-expansion of the angular dependence, defined by a finite set of {\it Legendre-moments}. Such angular analyses are simple but very useful procedures. We elaborate this further below for the example of single-meson photoproduction (see section~\ref{sec:formalisms:SingleMesonPhotoproduction}).
 \item[(iii)] The partial-wave amplitudes provide the natural basis to formulate phenomenological models for the extraction of resonance-parameters (cf. section~\ref{sec:PWA}). Additional physical assumption are used in order to parameterize the partial-waves as energy-dependent functions. Therefore, this procedure is sometimes called {\it energy-dependent fit}.
\end{itemize}
For these reasons, and due to the fact that the most important reactions considered in the following are all (quasi-) two-body reactions, one can state without exaggeration that partial waves are central theoretical constructs for Light Baryon Spectroscopy.

\subsubsection{Complete experiments}
\label{sec:formalisms:CompleteExperiments}

As already mentioned at the end of section~\ref{sec:formalisms:PolObservables}, there remains the question whether or not all $N_{\mathcal{A}}^{2}$ observables have to be measured for a particular process. This leads naturally to the search for so-called {\it complete experiments}~\cite{Barker:1975bp,Chiang:1996em}. Since this is an important question for the optimization of spin-measurements in Light Baryon Spectroscopy, the subject of complete experiments has received quite some recent attention in the literature~\cite{Keaton:1995pw, Keaton:1996pe, Ireland:2010bi, YannickDiploma, Vrancx:2013pza, Nys:2015kqa, Nys:2016uel, WorkmanEtAl2017, YannickPhD, Nakayama:2018yzw, Wunderlich:2020umg}.

\begin{figure}[ht]
\begin{overpic}[width=0.99\textwidth,trim=0 95 0 65,clip]%
      {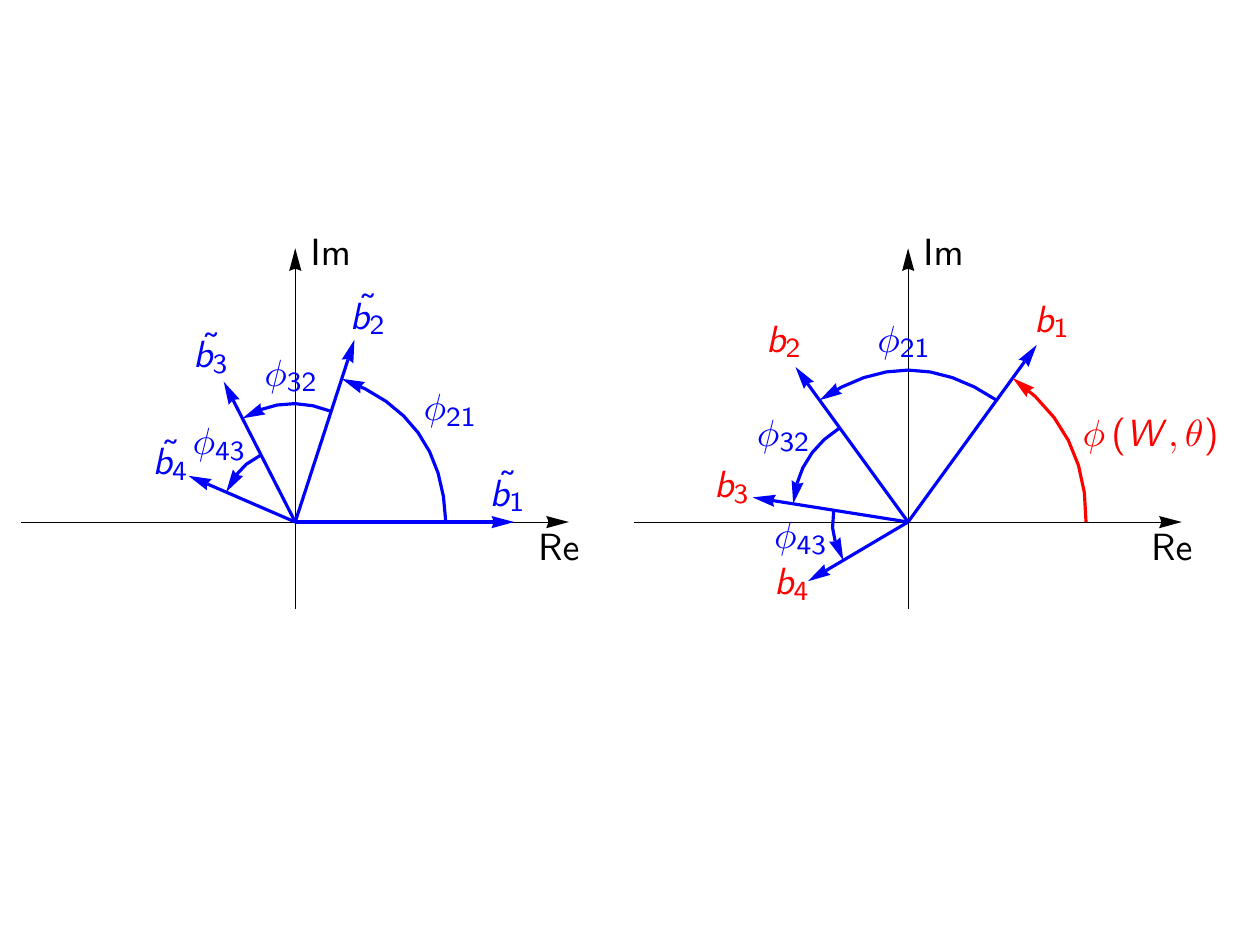}
\end{overpic}
\caption{An arrangement of amplitudes in the complex plane is shown for the example of $4$ complex transversity-amplitudes in single-meson photoproduction~(cf. section~\ref{sec:formalisms:SingleMesonPhotoproduction}). The figure is taken over identically from reference~\cite{Svarc:2020cic}. Left: The reduced amplitudes $ \tilde{b}_{i} $, defined by the phase-contraint $\text{Im} \left[ b_{1} \right] = 0$ $\&$ $\text{Im} \left[ b_{1} \right] > 0$, are shown. Three relative-phase angles are indicated. The CEA can directly and uniquely yield such reduced amplitudes. Right: The \textit{true} solution for the actual transversity amplitudes $ b_{i} $ is shown, which is obtained from the reduced amplitudes via a rotation by the overall phase $ \phi \left(W, \theta \right) $. All quantities drawn in red, i.e. $ \phi \left(W, \theta \right) $ as well as the $ b_{i} $, cannot be calculated directly from the CEA.}
\label{fig:formalisms:CompareReducedAmplitudesToActualAmplitudes}
\end{figure}

We first consider complete experiments for the unique extraction of the full spin-amplitudes, a procedure termed {complete experiment analysis (CEA)}~(cf.~\cite{Svarc:2020cic}) from now on. Furthermore, everything written in the following holds for the academic case of vanishing measurement uncertainty. The simplest formulation of the considered problem is the following:
\begin{CompEx}[CEA: amplitude formulation]
In order to construct a complete experiment for a particular spin-reaction, the task is to select from the entire set of $N_{\mathcal{A}}^{2}$ polarization observables (cf. equation~\eqref{eq:formalisms:GeneralDefinitionPolarisationObservable}) given for this reaction a possibly minimum subset that allows for an unambiguous extraction of the $N_{\mathcal{A}}$ complex amplitudes describing the process.
\end{CompEx}
The term unambiguous means here a unique determination of all moduli and relative-phases for the $N_{\mathcal{A}}$ amplitudes, with one overall phase remaining undetermined (see Figure~\ref{fig:formalisms:CompareReducedAmplitudesToActualAmplitudes}). Furthermore, this extraction of amplitudes has to take place at each point in phase-space individually. \newline
Additionally to the complete experiment problem stated in terms of amplitudes, there exists a slightly different formulation due to reference~\cite{Tiator:2011tu} that concentrates more on the measurement aspect. It reads as follows (see also~\cite{YannickDiploma}):
\begin{CompEx}[CEA: measurement formulation]
A complete experiment is a set of measurements that is sufficient to predict all other possible experiments. Regarding polarization experiments, this means that a subset of all existing polarization observables has to be chosen that is capable of determining all the remaining observables.
\end{CompEx}

These two formulations of the central problem in the CEA are equivalent. However, the second formulation makes more obvious why complete experiments are so valuable for Light Baryon Spectroscopy: a complete experiment represents the least amount of work necessary in order to automatically complete the whole database for a particular reaction!

In order to obtain a unique solution of the CEA, one needs at least $2 N_{\mathcal{A}}$ carefully selected observables. While one can find compelling heuristic arguments for the fact that $2 N_{\mathcal{A}}$ is indeed the universally correct minimal size of a complete experiment (cf. the introduction of reference~\cite{Wunderlich:2020umg} as well as the first footnote of~\cite{Keaton:1996pe}), for an individual process one always has to carefully analyze the equations defining the observables~\eqref{eq:formalisms:GeneralDefinitionPolarisationObservable} in order to determine which sets are in fact complete. For Pion-Nucleon scattering, the analysis is very simple and contained in section~\ref{sec:formalisms:PiNScattering}. For single-meson photoproduction, early results by Barker, Donnachie and Storrow~\cite{Barker:1975bp} have been corrected in a seminal work by Chiang and Tabakin~\cite{Chiang:1996em}, while the latter work has been put on a rigorous mathematical basis recently by Nakayama~\cite{Nakayama:2018yzw}. Two-meson photoproduction has been analyzed in a first study by Arenhoevel and Fix~\cite{Arenhoevel:2014dwa}, while a more recent study has first worked out the complete sets of minimal length~\cite{Kroenert:2020ahf}. For vector meson photoproduction, brief considerations can be found in the work by Pichkowsky, Savkli and Tabakin~\cite{Pichowsky:1994gh}, but a more thorough treatment is to our knowledge lacking at the moment. The problem of a general solution of the CEA for an arbitrary number of amplitudes $N_{\mathcal{A}}$ has been shed new light upon in a recent study~\cite{Wunderlich:2020umg}. However, there it was not possible to get the length of the complete sets derived from the given rules down to the minimum of~$2 N_{\mathcal{A}}$ for all the considered reactions. The possibility of reducing the number of observables in a complete set further down towards~$2 N_{\mathcal{A}}$, using new graphical techniques for in principle any reaction, was further investigated in a more recent work~\cite{Wunderlich:2021xhp}. We will give some more details on the complete sets for each of the specific reactions treated in section~\ref{sec:formalisms:MostImportantReactions}, whenever appropriate. 

In contrast to the CEA elaborated above, one can also ask for complete experiments in the context of a TPWA. One can formulate this as follows~(cf.~\cite{YannickPhD,Grushin})
\begin{CompEx}[Complete experiment: TPWA]
In order to construct a complete experiment, the problem is to find and utilize a (possibly minimum) set of Legendre moments, determined from the angular distributions of a corresponding (also possibly minimum) set of observables that permits an unambiguous determination of the partial waves contributing in the considered energy area bound by a value $ W_{\mathrm{max}} $ (i.e. up to some $j_{\text{max}}$ or $ \ell_{\mathrm{max}} $, respectively). \newline
The word unambiguous in this case means only unambiguous up to an overall phase for all the partial waves not set to zero by the truncation.
\end{CompEx}
One should note in addition that the complete experiment problem in the case of an 'exact' TPWA truncated at some finite $j_{\text{max}}$ (or $\ell_{\text{max}}$) is certainly a mathematical idealization due to the complete neglect of higher partial waves. The latter are certainly not zero in more physically motivated models (see section~\ref{sec:PWA}). Still, the result on complete experiments in a TPWA can serve as a useful guideline for measurements in the low-energy area. Thus, this problem has received some recent attention~\cite{Wunderlich:2013iga,WorkmanEtAl2017,YannickPhD}.

Furthermore, note that partial waves cannot be simply obtained from the results of a CEA via projection integrals, due to the unknown angular dependence of the overall phase in the CEA (cf. Figure~\ref{fig:formalisms:CompareReducedAmplitudesToActualAmplitudes}). Some proposals have been put forward on how to extract information on the overall phase from actual measurements~\cite{Goldberger:1963,Ivanov:2012na}, but at the time of this writing none of these proposed experiments are realistic. Thus, to obtain information on partial waves, a TPWA is definitely necessary.

A surprising fact is given by the result that in all cases of processes where the TPWA was studied, the number of observables needed for a complete TPWA reduces when compared to the CEA~\cite{Omelaenko,Grushin,Wunderlich:2013iga,WorkmanEtAl2017,YannickPhD}. The reason for this is at present not fully clear, but it certainly has to have something to do with the fact that the TPWA utilizes full angular distributions, instead of just isolated points in angle.
We give more information on complete sets in the TPWA for Pion-Nucleon scattering (section~\ref{sec:formalisms:PiNScattering}), single-meson photoproduction (section~\ref{sec:formalisms:SingleMesonPhotoproduction}) and also cite some recent results in two-meson photoproduction (section~\ref{sec:formalisms:MultiMesonPhotoproduction}) further below.

Finally, we mention that although all considerations given above hold for the case of vanishing uncertainties, the mathematical results have also been tested numerically for data with errors in some recent studies. For the CEA, relevant studies are given in~\cite{Ireland:2010bi, Vrancx:2013pza, Nys:2015kqa, Nys:2016uel} while for the TPWA, effects of uncertainties have been studied in selected chapters of the PhD-thesis~\cite{YannickPhD}. One can summarize the results by stating that in many cases of realistic errors, the mathematically derived complete sets have to be enlarged in order to restore a unique solution for the amplitudes.

The search for complete experiments is certainly of vital importance for Light Baryon Spectroscopy. The results of the CEA and the TPWA as stated in this section are model-independent, in contrast to the phenomenological partial-wave analysis models discussed in section~\ref{sec:PWA}. Furthermore, the results of a CEA or a TPWA for a single-channel problem, even in case the set of observables under consideration is complete, still leave one overall phase undetermined (either one phase for all spin-amplitudes, or one for all partial waves). This missing phase-information cannot be extracted from such model-independent analyses of one single reaction. In order to do this, unitarity-constraints applied in coupled-channel analyses have proven useful (cf. section~\ref{sec:PWA:Unitarity}).

\subsection{The most important reactions for Light Baryon Spectroscopy}
\label{sec:formalisms:MostImportantReactions}

Here, we give more information on the model-independent amplitude descriptions of the most important reactions considered in Light Baryon Spectroscopy. The discussion is ordered according to the encountered degree of the complications due to spin.

\subsubsection{Pion-Nucleon Scattering}
\label{sec:formalisms:PiNScattering}

In the following, we outline the description of $2 \rightarrow 2$ scattering with one pseudoscalar spin-$0$-particle (meson) and one spin-$\frac{1}{2}$-particle (Baryon) in the initial- as well as final state. The most important example is certainly Pion-Nucleon scattering $\pi N \longrightarrow \pi N$, which has historically been one of the first reactions used to study Baryon resonances. However, the formalism applies to a larger class of Pion-induced reactions important for Light Baryon Spectroscopy.

Pion-Nucleon scattering is generally described by $N_{\mathcal{A}} = 2$ amplitudes (equation~\eqref{eq:formalisms:NumberOfSpinAmplitudes}), for which one can take either the amplitudes $G (W, \theta)$ and $H (W, \theta)$ (cf. equation~\eqref{eq:formalisms:TMatrixPiN}), or the transversity amplitudes $b_{1,2} := \frac{1}{\sqrt{2}} \left( G \pm i H \right)$ (defined at each point in energy and angle $(W, \theta)$). Thus, one considers a reaction with $N_{\mathcal{A}}^{2} = 4$ observables, the definitions of which are collected in Table~\ref{tab:formalisms:PiNTrObs}. These measurable quantities include the unpolarized differential cross section $\sigma_{0}$, the recoil-polarization $\hat{P}$, as well as the so-called spin-rotation parameters $\hat{R}$ and $\hat{A}$~\cite{DeLesquen:1972viv,Supek:1993qa,Hoehler84,Anisovich:2013tij}. Instead of the individual spin-rotation parameters, sometimes the so-called {\it spin-rotation angle} $\beta$ is measured, which is defined as follows~\cite{Anisovich:2013tij,Hoehler84}
\begin{equation}
  \beta = \arctan \left[ - \frac{\hat{R}}{\hat{A}} \right]  . \label{eq:formalisms:PiNSpinRotAngle}
\end{equation}
\begin{table}[bt]
\centering
\begin{tabular}{llr}
\hline \hline Observable & Transversity representation  &  Measurement \\
\hline \\ [-5.5pt]
$ \sigma_{0} $ & $ \left| b_{1} \right|^{2} + \left| b_{2} \right|^{2} $ &   Unpolarized differential CS, \\[+1.5pt]
$ \hat{P} $ & $  \left| b_{1} \right|^{2} - \left| b_{2} \right|^{2} $   & Recoil-polarization, \\ \hline \\[-8.5pt]
$ \hat{R} $ & $ - 2 \hspace*{1pt} \mathrm{Im} \left[ b_{1}^{\ast} b_{2} \right] = - 2 \hspace*{1pt} \left| b_{1} \right| \left| b_{2} \right| \sin \phi_{21} $ &    Spin-rotation- \\[+1.5pt]
$ \hat{A} $ & $ 2 \hspace*{1pt} \mathrm{Re} \left[ b_{1}^{\ast} b_{2} \right] = 2 \hspace*{1pt} \left| b_{1} \right| \left| b_{2} \right| \cos \phi_{21} $ &    parameters. \\
\hline
\hline
\end{tabular}
\caption{The $4$ observables in the process of Pion-Nucleon- (or more generally spin-$0$ spin-$1/2$-) scattering~\cite{DeLesquen:1972viv,Supek:1993qa,Hoehler84} are listed in the transversity representation. The corresponding measurements are indicated. The sign-conventions of reference~\cite{Anisovich:2013tij} are adopted here (phase-space factors are omitted). These bilinear equations can be summarized in terms of Pauli matrices (extended with the identity~$\hat{\sigma}^{0}$): $\mathcal{O}^{\alpha} = \left< b \right| \hat{\sigma}^{\alpha} \left| b \right>$, for~$\alpha = 0, \ldots, 3$. Therefore, this algebra of~$4$ observables is one particular example for the general definition given in equation~\eqref{eq:formalisms:GeneralDefinitionPolarisationObservable} of section~\ref{sec:formalisms:PolObservables}.}
\label{tab:formalisms:PiNTrObs}
\end{table}

The most widely used form of the partial-wave expansion for Pion-Nucleon scattering is in most cases written for the amplitudes $G$ and $H$ (cf.~\cite{Gersten:1969ae,Hoehler84,Anisovich:2013tij}):
\begin{align}
 G (W, \theta) &=  \sum_{\ell = 0}^{\infty} \left\{ (\ell + 1) T_{\ell +} (W) + \ell T_{\ell -} (W) \right\} P_{\ell} (\cos \theta)  , \label{eq:formalisms:GPWExpPiN} \\
 H (W, \theta) &=  \sin \theta \hspace*{1.5pt} \sum_{\ell = 1}^{\infty} \left\{ T_{\ell +} (W) - T_{\ell -} (W) \right\} P_{\ell}^{'} (\cos \theta)  . \label{eq:formalisms:HPWExpPiN}  
\end{align}
Here, $P_{\ell}, P_{\ell}^{'}$ are Legendre-polynomials and derivatives thereof and $\ell$ is the orbital angular momentum between the two particles in the final state. The total angular momentum, which is conserved, can become $J = \left| \ell \pm \frac{1}{2} \right|$ depending on the partial wave $T_{\ell \pm}$. Parity is conserved as well and it takes the value $P = \left( -1 \right)^{\ell + 1}$. Thus, a one-to-one correspondence between resonance quantum-numbers $J^{P}$ and partial waves $T_{\ell \pm}$ is established.

The CEA in Pion-Nucleon scattering needs all $4$ observables for a unique solution. The corresponding argument (taken from the introduction of reference~\cite{Anisovich:2013tij}) can be established rather quickly. The $4$ observables are not completely independent from each other, but rather satisfy the following quadratic constraint (cf. the definitions in Table~\ref{tab:formalisms:PiNTrObs}, or the introduction of reference~\cite{Anisovich:2013tij}):
\begin{equation}
  \sigma_{0}^{2} - \hat{P}^{2} - \hat{R}^{2} - \hat{A}^{2} = 0   . \label{eq:formalisms:PiNFierzRelation}
\end{equation}
This equation has to hold at each point in $(W,\theta)$. It is seen quickly that in case any set of three observables is selected (for instance $\sigma_{0}$, $\hat{P}$ and $\hat{A}$), then the fourth observable is only determined up to a sign-ambiguity. Thus, the completeness condition in the measurement-formulation of section~\ref{sec:formalisms:CompleteExperiments} is not fulfilled. Reference~\cite{Wunderlich:2020umg} contains another more detailed treatment of the CEA for the reaction considered here.


In the academic scenario of a mathematically exact TPWA, the Pion-Nucleon formalism allows for unique solutions with just $3$ instead of all $4$ observables. More precisely, utilizing full angular distributions from the $3$ quantities $\sigma_{0}$, $\hat{P}$ and $\hat{A}$ leads to a unique solution in the TPWA. This result can be inferred from root-decompositions of the polynomial amplitude such as those given in~\cite{Gersten:1969ae,Barrelet:1971pw,VanHorn:1974hu}. However, numerical tests of this statement have, to our knowledge, only been performed in recent private communications~\cite{LotharAlfredPrivComm}.

\subsubsection{Single-Meson Photoproduction}
\label{sec:formalisms:SingleMesonPhotoproduction}
We consider the photoproduction of a single pseudoscalar meson $\varphi$ on a nucleon, i.e. $\gamma N \longrightarrow \varphi B$ (B is called recoil-baryon). The most important special case is certainly Pion-photoproduction: $\gamma N \longrightarrow \pi N$. Since many of the new recent results in Light Baryon Spectroscopy have been obtained from photoproduction data, we outline the (tedious) formalism of this reaction in a bit more detail in the following.

Single-meson photoproduction is described by $N_{\mathcal{A}} = 4$ spin-amplitudes (cf. eq.~\eqref{eq:formalisms:NumberOfSpinAmplitudes}), which could be given for instance in the basis of CGLN-amplitudes $F_{i}$ (cf. eq.~\eqref{eq:formalisms:TMatrixPhotoprod}) or transversity amplitudes $b_{i}$. Correspondingly, $N_{\mathcal{A}}^{2} = 16$ polarization observables are measurable in single-meson photoproduction. The definitions are collected in Table~\ref{tab:formalisms:PhotoproductionObservables}. The observables are divided into four different groups (or 'shape-classes'~\cite{Wunderlich:2020umg}) with $4$ observables in each group. The unpolarized differential cross section $\sigma_{0}$ and the three single-polarization observables $\check{\Sigma}$, $\check{T}$ and $\check{P}$ compose the group $\mathcal{S}$. Furthermore, there exist three groups of double-polarization measurements, namely the beam-target $(\mathcal{BT})$, beam-recoil $(\mathcal{BR})$ and target-recoil $(\mathcal{TR})$ experiments. From these, the groups $\mathcal{BR}$ and $\mathcal{TR}$ are by far the most difficult to access experimentally. Triple-polarization experiments can be defined as well, but the corresponding observables do not yield any additional information in the case of single-meson photoproduction. It becomes apparent upon consideration of the most general expression for the single-meson photoproduction cross section~\cite{Sandorfi:2010uv} that there exist double assignments in the sense that a specific observable can be accessed via several different polarization measurements. For instance, the single-polarization observable~$P$ can be measured either via a single recoil-polarization measurement, or alternatively via a specific beam-target polarization configuration. The assignments of spin-configurations to polarization observables for single-meson photoproduction are further summarized in Table~\ref{tab:formalisms:SandorfiObservableTable}.  

The definitions of the~$16$ observables in terms of transversity amplitudes~$b_{i}$ (cf.~Table~\ref{tab:formalisms:PhotoproductionObservables}) are just a special case of the general definition~\eqref{eq:formalisms:GeneralDefinitionPolarisationObservable} (see section~\ref{sec:formalisms:PolObservables}). As an example, we explicitly evaluate this definition for the observable~$\check{G}$ (for the definition of the corresponding matrix~$\tilde{\Gamma}^{3}$, as well as all remaining matrices for single-meson photoproduction, see Appendix~A.1 of reference~\cite{YannickPhD}):
\begin{align}
\check{G} &=  \mathcal{O}^{3} =  \frac{1}{2} \left< b \right| \tilde{\Gamma}^{3} \left| b \right> = \frac{1}{2} \left[ \begin{array}{cccc}
 b_{1}^{\ast} & b_{2}^{\ast} & b_{3}^{\ast} & b_{4}^{\ast} \end{array}
  \right] \left[ \begin{array}{cccc}
 0 & 0 & -i & 0  \\
 0 & 0 & 0 & -i \\
 i & 0 & 0 & 0 \\
 0 & i & 0 & 0 \end{array} \right] \left[ \begin{array}{c}
 b_{1} \\
 b_{2} \\
 b_{3} \\
 b_{4} \end{array} \right]  \nonumber \\
 &=  \frac{1}{2} \left[ \begin{array}{cccc}
 b_{1}^{\ast} & b_{2}^{\ast} & b_{3}^{\ast} & b_{4}^{\ast} \end{array}
  \right] \left[ \begin{array}{c}
 - i b_{3} \\
 - i b_{4} \\
 i b_{1} \\
 i b_{2} \end{array} \right] = \frac{1}{2} \left[ - i b_{1}^{\ast} b_{3} - i b_{2}^{\ast} b_{4} + i b_{3}^{\ast} b_{1} + i b_{4}^{\ast} b_{2} \right] \nonumber \\
 &= \frac{1}{2i} \left[ - b_{3}^{\ast} b_{1} + b_{3} b_{1}^{\ast} - b_{4}^{\ast} b_{2} + b_{4} b_{2}^{\ast} \right] = \mathrm{Im} \left[ - b_{3}^{\ast} b_{1} - b_{4}^{\ast} b_{2} \right] \mathrm{.} \label{eq:formalisms:SampleCalculationG}
\end{align}
In the following, we outline the formalism for the TPWA in single-meson photoproduction, i.e. a special case of the general methodology described in section~\ref{sec:formalisms:PWExpansion}. To do this, we need the expressions connecting the transversity amplitudes~$b_{i}$ to the CGLN amplitudes~$F_{i}$. The relevant expressions read~\cite{Omelaenko,Wunderlich:2013iga} ($\mathcal{C}$ is a convention-dependent factor, in our case:~$\mathcal{C} = i/\sqrt{2}$):

\begin{align}
	b_{1} \left( W, \theta\right) &= - b_{3} \left( W, \theta\right)
	+ i \mathcal{C} \sin \theta \left[ F_{3} \left( W, \theta\right) e^{- i \frac{\theta}{2}} + F_{4} \left( W, \theta\right) e^{ i \frac{\theta}{2}} \right] \mathrm{,} \label{eq:formalisms:b1BasicForm} \\
	b_{2} \left( W, \theta\right) &= - b_{4} \left( W, \theta\right)
	- i \mathcal{C} \sin \theta \left[ F_{3} \left( W, \theta\right) e^{i \frac{\theta}{2}} + F_{4} \left( W, \theta\right) e^{- i \frac{\theta}{2}} \right] \mathrm{,} \label{eq:formalisms:b2BasicForm} \\
	b_{3} \left( W, \theta\right) &= \mathcal{C} \left[ F_{1} \left( W, \theta\right) e^{- i \frac{\theta}{2}} -  F_{2} \left( W, \theta\right) e^{ i \frac{\theta}{2}} \right] \mathrm{,} \label{eq:formalisms:b3BasicForm} \\
	b_{4} \left( W, \theta\right) &= \mathcal{C} \left[ F_{1} \left( W, \theta\right) e^{i \frac{\theta}{2}} -  F_{2} \left( W, \theta\right) e^{- i \frac{\theta}{2}} \right] \mathrm{.} \label{eq:formalisms:b4BasicForm}
\end{align}
The partial-wave expansion for single-meson photoproduction is in most cases defined for the CGLN-amplitudes, and the rather elaborate expansions read as follows~\cite{Chew:1957tf,Sandorfi:2010uv}:
\begin{align}
	F_{1} \left( W, \theta \right) &= \sum \limits_{\ell = 0}^{\infty} \Big\{ \left[ \ell M_{\ell+} \left( W \right) + E_{\ell+} \left( W \right) \right] P_{\ell+1}^{'} \left( \cos \theta \right) \nonumber \\
	& \quad \quad \quad + \left[ \left( \ell+1 \right) M_{\ell-} \left( W \right) + E_{\ell-} \left( W \right) \right] P_{\ell-1}^{'} \left( \cos \theta \right) \Big\} \mathrm{,} \label{eq:formalisms:MultExpF1} \\
	F_{2} \left( W, \theta \right) &= \sum \limits_{\ell = 1}^{\infty} \left[ \left( \ell+1 \right) M_{\ell+} \left( W \right) + \ell M_{\ell-} \left( W \right) \right] P_{\ell}^{'} \left( \cos \theta \right) \mathrm{,} \label{eq:formalisms:MultExpF2} \\
	F_{3} \left( W, \theta \right) &= \sum \limits_{\ell = 1}^{\infty} \Big\{ \left[ E_{\ell+} \left( W \right) - M_{\ell+} \left( W \right) \right] P_{\ell+1}^{''} \left( \cos \theta \right) \nonumber \\
	& \quad \quad \quad + \left[ E_{\ell-} \left( W \right) + M_{\ell-} \left( W \right) \right] P_{\ell-1}^{''} \left( \cos \theta \right) \big\} \mathrm{,} \label{eq:formalisms:MultExpF3} \\
	F_{4} \left( W, \theta \right) &= \sum \limits_{\ell = 2}^{\infty} \left[ M_{\ell+} \left( W \right) - E_{\ell+} \left( W \right) - M_{\ell-} \left( W \right) - E_{\ell-} \left( W \right) \right] P_{\ell}^{''} \left( \cos \theta \right) \mathrm{.} \label{eq:formalisms:MultExpF4}
\end{align}
These are expansions into Legendre polynomials $P_{\ell}$ and derivatives thereof, where $\ell$ is the orbital angular momentum in the final meson-baryon system. 

\begin{table}[tbh]
 \begin{center}
 \begin{tabular}{llr}
 \hline
 \hline
  Observable  &  $\bm{c} \left( \sigma^{a} - \sigma^{b} \right)$  &  Group ('shape-class')  \\
  \hline
  $\sigma_{0} = \frac{1}{2} \left( \left| b_{1} \right|^{2} + \left| b_{2} \right|^{2} + \left| b_{3} \right|^{2} + \left| b_{4} \right|^{2} \right)$  &  -  &  \\
  $\check{\Sigma} = \frac{1}{2} \left( - \left| b_{1} \right|^{2} - \left| b_{2} \right|^{2} + \left| b_{3} \right|^{2} + \left| b_{4} \right|^{2} \right)$  &  $a = \left( \perp,0,0 \right)$; $b = \left( \parallel,0,0 \right)$  &   $\mathcal{S}$ \\
  $\check{T} = \frac{1}{2} \left( \left| b_{1} \right|^{2} - \left| b_{2} \right|^{2} - \left| b_{3} \right|^{2} + \left| b_{4} \right|^{2} \right)$  &  $a = \left( 0,+y,0 \right)$; $b = \left( 0,-y,0 \right)$  &   \\
  $\check{P} = \frac{1}{2} \left( - \left| b_{1} \right|^{2} + \left| b_{2} \right|^{2} - \left| b_{3} \right|^{2} + \left| b_{4} \right|^{2} \right)$  &  $a = \left( 0,0,+y' \right)$; $b = \left( 0,0,-y' \right)$  &   \\
  \hline
   $\check{E}  = \mathrm{Re} \left[ - b_{3}^{\ast} b_{1} - b_{4}^{\ast} b_{2} \right]$   &  $a = \left( +1,+z,0 \right)$; $b = \left( +1,-z,0 \right)$  &  \\
   $\check{F} = \mathrm{Im} \left[ b_{3}^{\ast} b_{1} - b_{4}^{\ast} b_{2} \right] $  &  $a = \left( +1,+x,0 \right)$; $b = \left( +1,-x,0 \right)$  &  $\mathcal{BT} $ \\
   $\check{G} = \mathrm{Im} \left[ - b_{3}^{\ast} b_{1} - b_{4}^{\ast} b_{2} \right]$   &  $a = \left( +\pi / 4,+z,0 \right)$; $b = \left( +\pi / 4,-z,0 \right)$ &  \\
   $ \check{H} = \mathrm{Re} \left[ b_{3}^{\ast} b_{1} - b_{4}^{\ast} b_{2} \right]$   &  $a = \left( +\pi / 4,+x,0 \right)$; $b = \left( +\pi / 4,-x,0 \right)$ &   \\
   \hline
   $\check{C}_{x'}  = \mathrm{Im} \left[ - b_{4}^{\ast} b_{1} + b_{3}^{\ast} b_{2} \right] $ &  $a = \left( +1,0,+x' \right)$; $b = \left( +1,0,-x' \right)$  &  \\
   $\check{C}_{z'} = \mathrm{Re} \left[ - b_{4}^{\ast} b_{1} - b_{3}^{\ast} b_{2} \right]$  &  $a = \left( +1,0,+z' \right)$; $b = \left( +1,0,-z' \right)$  &  $\mathcal{BR}$   \\
   $\check{O}_{x'} = \mathrm{Re} \left[ - b_{4}^{\ast} b_{1} + b_{3}^{\ast} b_{2} \right] $  &  $a = \left( +\pi / 4,0,+x' \right)$; $b = \left( +\pi / 4,0,-x' \right)$  &  \\
   $\check{O}_{z'} = \mathrm{Im} \left[ b_{4}^{\ast} b_{1} + b_{3}^{\ast} b_{2} \right]$  &  $a = \left( +\pi / 4,0,+z' \right)$; $b = \left( +\pi / 4,0,-z' \right)$  &   \\
   \hline
   $\check{L}_{x'} = \mathrm{Im} \left[ - b_{2}^{\ast} b_{1} - b_{4}^{\ast} b_{3} \right]$  &  $a = \left( 0,+z,+x' \right)$; $b = \left( 0,+z,-x' \right)$  &  \\
   $\check{L}_{z'}  = \mathrm{Re} \left[ - b_{2}^{\ast} b_{1} - b_{4}^{\ast} b_{3} \right]$  &  $a = \left( 0,+z,+z' \right)$; $b = \left( 0,+z,-z' \right)$  &  $\mathcal{TR}$  \\
   $\check{T}_{x'} = \mathrm{Re} \left[ b_{2}^{\ast} b_{1} - b_{4}^{\ast} b_{3} \right]$  &  $a = \left( 0,+x,+x' \right)$; $b = \left( 0,+x,-x' \right)$  &   \\
   $\check{T}_{z'} = \mathrm{Im} \left[ - b_{2}^{\ast} b_{1} + b_{4}^{\ast} b_{3} \right]$ &  $a = \left( 0,+x,+z' \right)$; $b = \left( 0,+x,-z' \right)$  &  \\
   \hline
   \hline
 \end{tabular}
 \end{center}
 \caption{The definitions of the $16$ ob\-serva\-bles in single-meson photoproduction are given in terms of transversity amplitudes $b_{i}$ (cf. reference~\cite{Wunderlich:2013iga}). These bilinear expressions are one particular realization of the general definition~\eqref{eq:formalisms:GeneralDefinitionPolarisationObservable} from section~\ref{sec:formalisms:PolObservables}. The phase-space factor $\rho = q/k$ (cf. appendix~\ref{sec:appendix:Kinematics}) has been suppressed. The polarization-configurations defining the asymmetries (i.e. '$a$' and '$b$') correspond to the usual CMS-frame (unprimed coordinates) and the coordinate system reached from the CMS by a rotation of the $z$-axis to the meson-momentum $\vec{q}$ (primed coordinates).}
 \label{tab:formalisms:PhotoproductionObservables}
\end{table}
\begin{table}[tbh]
\begin{center}
\begin{tabular}{c|c|ccc|ccc|ccccccccc}
\multicolumn{17}{c}{} \\
\hline
\hline
Beam &  & \multicolumn{3}{c|}{Target} & \multicolumn{3}{c|}{Recoil} & \multicolumn{9}{c}{Target + Recoil} \\
  & -& -& -& -& $ x' $ & $ y' $ & $ z' $ & $ x' $ & $ x' $ & $ x' $ & $ y' $ & $ y' $ & $ y' $ & $ z' $ & $ z' $ & $ z' $  \\
  & -& $ x $ & $ y $ & $ z $ & -& -& -& $ x $ & $ y $ & $ z $ & $ x $ & $ y $ & $ z $ & $ x $ & $ y $ & $ z $ \\
\hline
  &  &  &  &  &  &  &  &  &  &  &  &  &  &  \\
unpolarized & $ \sigma_{0} $ &  & $ T $ &  &  & $ P $ &  & $ T_{x'} $ &   & $ L_{x'} $ &  & $ \Sigma $ &  & $T_{z'}$ &  & $L_{z'}$ \\
  &  &  &  &  &  &  &  &  &  &  &  & \\
linearly pol. & $ \Sigma $ & $ H $ & $ P $ & $ G $ & $ O_{x'} $ & $ T $ & $ O_{z'} $ &  $ L_{z'} $  &  $ C_{z'} $  &  $ T_{z'} $  &  $ E $  &  $ \sigma_{0} $  &  $ F $  &  $ L_{x'}$  &  $ C_{x'} $  &  $ T_{x'} $  \\
  &  &  &  &  &  &  &  &  &  &  &  &  &  &  \\
circularly pol. &  & $ F $ &  & $ E $ & $ C_{x'} $ &  & $ C_{z'} $ &  & $ O_{z'} $ &  & $ G $ &  & $ H $ &  & $ O_{x'} $ &  \\
\hline
\hline
\end{tabular} 
\end{center}
\caption{The $16$ polarization observables of single-meson photoproduction are tabulated with regard to the different polarization measurements that make a specific observable accessible. Information on the signs of the observables has been omitted (cf. reference~\cite{Sandorfi:2011nv} for a discussion of different sign-conventions). This Table is a compressed version of a larger one to be found in reference~\cite{Sandorfi:2010uv}.}
\label{tab:formalisms:SandorfiObservableTable}
\end{table}
%


The partial waves $E_{\ell \pm}$ and $M_{\ell \pm}$ are called {\it electric and magnetic multipoles} and they correspond to a final state with total angular momentum $J = \left| \ell \pm \frac{1}{2}  \right|$, resulting from an excitation by an electric or magnetic photon, respectively. This multipole-expansion has turned out to be the standard-convention in the literature to write the photoproduction partial-wave series, though many different basis systems of partial-wave amplitudes can be defined.

A simple one-to-one correspondence between a resonances spin-parity quantum numbers $J^{P}$ and multipoles can be established using the conservation-laws for parity and total angular momentum~\cite{LeukelPhD,YannickPhD}. One has to keep in mind that the parity in the initial state reads $P = (-1)^{L_{\gamma}}$ for an electric photon and $P = (-1)^{L_{\gamma} + 1}$ for a magnetic photon, where $L_{\gamma}$ is the relative orbital angular momentum between photon and nucleon. The selection-rules for multipoles resulting from these considerations are collected in Table~\ref{tab:formalisms:MultipolesMesonToPhotonCorrespondence}.


%
\begin{table}[htb]
\centering
\begin{tabular}{|c|c|c|c|c|c||c|c|c|c|c|c|}
\hline
\multicolumn{2}{|c|}{$ \gamma N $-system} & \multicolumn{4}{c||}{$ \varphi B $-system} & \multicolumn{2}{c}{$ \gamma N $-system} & \multicolumn{4}{|c|}{$ \varphi B $-system} \\
\hline
$ L_{\gamma} $ & $ \mathcal{M} L $ & $ J $ & $ \ell $ & $ \mathcal{M}_{\ell \pm} $ & $ P $ & $ L_{\gamma} $ & $ \mathcal{M} L $ & $ J $ & $ \ell $ & $ \mathcal{M}_{\ell \pm} $ & $ P $ \\
\hline
\hline
 $ 1 $ & $ E1 $ & $ 1/2 $ & $ 0 $ & $ E_{0+} $ & $ - $ &  $ 2 $ & $ E2 $ & $ 3/2 $ & $ 1 $ & $ E_{1+} $ & $ + $ \\
\cline{4 - 6} \cline{10 - 12}
  &  &  &  $ \displaystyle{\not} 1 $ &  &  &  &  &  &  $ \displaystyle{\not} 2 $ &  &  \\
\cline{3 - 6} \cline{9 - 12}
  &  & $ 3/2 $ &  $ \displaystyle{\not} 1 $ &  &  &  &  & $ 5/2 $ &  $ \displaystyle{\not} 2 $ &  &  \\
\cline{4 - 6} \cline{10 - 12}
  &  &  & $ 2 $ & $ E_{2-} $ & $ - $ &  &  &  & $ 3 $ & $ E_{3-} $ & $ + $ \\
\cline{2 - 6} \cline{8 - 12}
  & $ M1 $ & $ 1/2 $ & $ \displaystyle{\not} 0 $ &  &  &  & $ M2 $ & $ 3/2 $ & $ \displaystyle{\not} 1 $ &  &  \\
\cline{4 - 6} \cline{10 - 12}
  &  &  &  $ 1 $ & $ M_{1-} $ & $ + $ &  &  &  & $ 2 $ & $ M_{2-} $ & $ - $ \\
\cline{3 - 6} \cline{9 - 12}
  &  & $ 3/2 $ &  $ 1 $ & $ M_{1+} $ & $ + $ &  &  & $ 5/2 $ & $ 2 $ & $ M_{2+} $ & $ - $ \\
\cline{4 - 6} \cline{10 - 12}
  &  &  & $ \displaystyle{\not} 2 $ &  &  &  &  &  & $ \displaystyle{\not} 3 $ &  &  \\
\hline
\end{tabular}
\caption{The multipoles $ E_{\ell \pm} $, $M_{\ell \pm}$ which can be reached via electromagnetic dipole ($ L_{\gamma} = 1 $) and quadrupole ($ L_{\gamma} = 2 $) excitations, are given in this Table. The selection-rules mentioned in the main text forbid some values for the quantum number $\ell$ (i.e., the slashed ones). This Table is taken over from references~\cite{LeukelPhD,YannickPhD}.}
\label{tab:formalisms:MultipolesMesonToPhotonCorrespondence}
\end{table}

The multipole-series achieves a crucial separation of the dependencies on energy $W$ and angle $\theta$ at the level of amplitudes. When one tries to accomplish the same at the level of {\it observables}, the formalism of {\it (Legendre-) moment analysis} arises, which recently has received attention as a useful analysis-tool for single-meson photoproduction~\cite{Wunderlich:2016imj,YannickPhD,CBELSA/TAPS:2020yam}. Truncating the multipole-series (eqs.~\eqref{eq:formalisms:MultExpF1} to~\eqref{eq:formalisms:MultExpF4}) at some finite $\ell_{\text{max}} \geq 1$ and inserting it into the $16$ observables as written in Table~\ref{tab:formalisms:PhotoproductionObservables} (using the expressions~\eqref{eq:formalisms:b1BasicForm} to~\eqref{eq:formalisms:b4BasicForm} in the process), one recovers the following generic form for any observable $\mathcal{O}^{\alpha}$ (see~\cite{Wunderlich:2016imj,YannickPhD} for more mathematical details):
\begin{align}
\mathcal{O}^{\alpha} \left( W, \theta \right) &= \frac{q}{k} \hspace*{3pt} \sum \limits_{n = \beta_{\alpha}}^{2 \ell_{\mathrm{max}} + \beta_{\alpha} + \gamma_{\alpha}} \left(a_{L}\right)_{n}^{\mathcal{O}^{\alpha}} \left( W \right) P^{\beta_{\alpha}}_{n} \left( \cos \theta \right) \mathrm{,} \hspace*{1.5pt} \alpha = 1,\ldots,16 \hspace*{1.5pt} \mathrm{,}  \label{eq:formalisms:PhotoProdAssocLegParametrization1} \\
\left(a_{L}\right)_{n}^{\mathcal{O}^{\alpha}} \left( W \right) &= \left< \mathcal{M}_{\ell_{\mathrm{max}}} \left( W \right) \right| \left( \mathcal{C}_{L}\right)_{n}^{\mathcal{O}^{\alpha}} \left| \mathcal{M}_{\ell_{\mathrm{max}}} \left( W \right) \right> \mathrm{.} \label{eq:formalisms:PhotoProdAssocLegParametrization2}
\end{align}
The {\it Legendre-moments} $\left(a_{L}\right)_{n}^{\mathcal{O}^{\alpha}}$ turn out to be bilinear hermitean forms of the multipoles\footnote{The multipoles have been collected into the vector $\left| \mathcal{M}_{\ell_{\mathrm{max}}} \right> = \left[ E_{0+}, E_{1+}, M_{1+}, M_{1-}, E_{2+}, E_{2-}, \ldots , M_{\ell_{\mathrm{max}}-} \right]^{T} $.}, defined by certain matrices $\left( \mathcal{C}_{L}\right)_{n}^{\mathcal{O}^{\alpha}}$ (expressions for such matrices are given, for instance, in the appendix of the thesis~\cite{YannickPhD}). While the form given above is generic, the parameters $\beta_{\alpha}$ and $\gamma_{\alpha}$ which specify the expansion for each observable are given in Table~\ref{tab:formalisms:AngularDistributions}.


%
\begin{table}[htb]
\centering
\begin{tabular}{ccc|cccccl||cccc|cccclllll}
\hline
\hline
Type & $ \mathcal{O}^{\alpha} $ & $\alpha$ &  & $ \beta_{\alpha} $ &  & $ \gamma_{\alpha} $ &  & $N_{a}$ & Type & $ \mathcal{O}^{\alpha} $ & $\alpha$ &  & $ \beta_{\alpha} $ &  & $ \gamma_{\alpha} $ &  &  $N_{a}$  \\
\hline
  &  &  &  &  &  &  &  &  &  &  &  &  &  &  &  &  &  \\
  & $ \sigma_{0} $ & $1$ &  & $ 0 $ &  & $ 0 $ &  & $2 \ell_{\text{max}} + 1$ &  & $ \check{O}_{x'} $ & $14$ &  & $ 1 $ &  & $ 0 $ &  &  $2 \ell_{\text{max}} + 1$ \\
 $\mathcal{S}$ & $ \check{\Sigma} $ & $4$ &  & $ 2 $ &  & $ -2 $ &  &  $2 \ell_{\text{max}} - 1$  & $\mathcal{BR}$ & $ \check{O}_{z'} $ & $7$ &  & $ 2 $ &  & $ -1 $ &  &  $2 \ell_{\text{max}}$  \\
  & $ \check{T} $ & $10$ &  & $ 1 $ &  & $ -1 $ &  & $2 \ell_{\text{max}}$ &  & $ \check{C}_{x'} $ & $16$ &  & $ 1 $ &  & $ 0 $ &   &   $2 \ell_{\text{max}} + 1$ \\
  & $ \check{P} $ & $12$ &  & $ 1 $ &  & $ -1 $ &  & $2 \ell_{\text{max}}$ &  & $ \check{C}_{z'} $ & $2$ &  & $ 0 $ &  & $ +1 $ &   &  $2 \ell_{\text{max}} + 2$ \\
\hline
  &  &  &  &  &  &  &  &  &  &  &  &  &  &  &  &  &  \\
  & $ \check{G} $ & $3$ &  & $ 2 $ &  & $ -2 $ &  & $2 \ell_{\text{max}} - 1$ &  & $ \check{T}_{x'} $ & $6$ &  & $ 2 $ &  & $ -1 $ &   &  $2 \ell_{\text{max}}$  \\
 $\mathcal{BT}$ & $ \check{H} $ & $5$ &  & $ 1 $ &  & $ -1 $ &  & $2 \ell_{\text{max}}$ & $\mathcal{TR}$ & $ \check{T}_{z'} $ & $13$ &  & $ 1 $ &  & $ 0 $ &  &  $2 \ell_{\text{max}} + 1$  \\
  & $ \check{E} $ & $9$ &  & $ 0 $ &  & $ 0 $ &  & $2 \ell_{\text{max}} + 1$ &  & $ \check{L}_{x'} $ & $8$ &  & $ 1 $ &  & $ 0 $ &   &  $2 \ell_{\text{max}} + 1$  \\
  & $ \check{F} $ & $11$ &  & $ 1 $ &  & $ -1 $ &  & $2 \ell_{\text{max}}$ &  & $ \check{L}_{z'} $ & $15$ &  & $ 0 $ &  & $ +1 $ &  &  $2 \ell_{\text{max}} + 2$ \\
\hline
\hline
\end{tabular}
\caption{The parameters defining the angular parameterization (\ref{eq:formalisms:PhotoProdAssocLegParametrization1}) of the observables $\mathcal{O}^{\alpha}$ in a \mbox{(Legendre-)} moment analysis are collected here. The notation is the same as in reference~\cite{Wunderlich:2016imj}. The observables are labeled by $\alpha = 1,\ldots,16$.  The number $N_{a}$ of Legendre moments yielded by an observable in each truncation order $\ell_{\text{max}}$ is given as well.}
\label{tab:formalisms:AngularDistributions}
\end{table}
%


This formalism for the moment analysis in single-meson photoproduction is useful mainly due to the following two reasons~\cite{YannickNStar,YannickPhD,Wunderlich:2016imj}:
\begin{itemize}
 \item[(i)]  \underline{$\ell_{\text{max}}$-analysis:} The angular distributions are fitted according to the parameterization given in eq.~\eqref{eq:formalisms:PhotoProdAssocLegParametrization1}, for different (rising) truncation orders $\ell_{\text{max}}$. The goodness-of-fit $\chi^{2}/\text{ndf}$ is compared for different fits, ascending from the lowest possible $\ell_{\text{max}}$. In case the fit-quality is not (yet) satisfactory, one has to increase $\ell_{\text{max}}$. In case a satisfactory fit is obtained, the resulting $\ell_{\text{max}}$ gives, in most cases, already quite a good estimate for the maximal angular momentum that is detectable in the data. Once the $\chi^{2}/\text{ndf}$ for all the fits is plotted versus energy, one can observe 'bumps' whenever new important contributions from higher angular momenta enter the data.
 \item[(ii)] \underline{Model-comparisons:} The Legendre-moments $\left(a_{\ell_{\text{max}}}\right)_{k}^{\mathcal{O}^{\alpha}}$ extracted from the data can be compared to the right hand side of equation~\eqref{eq:formalisms:PhotoProdAssocLegParametrization2}, when the latter is evaluated using multipoles stemming from an energy-dependent phenomenological model (cf. sec.~\ref{sec:PWA}). Such comparisons can be performed switching on/off certain partial waves coming from the model. In this way, one can obtain valuable information on which partial-wave interferences are important. The results of such comparisons can be interpreted in view of important physical effects visible in the data.
\end{itemize}
The solution-theory for the CEA in single-meson photoproduction has been worked out in detail in references~\cite{Barker:1975bp,Keaton:1995pw,Chiang:1996em,Nakayama:2018yzw}. For this process, $2 N_{\mathcal{A}} = 8$ carefully selected observables can uniquely solve the CEA, i.e. yield one unique solution for the four moduli $\left| b_{1} \right|, \ldots, \left| b_{4} \right|$ and three suitable selected relative phases, e.g. $\phi_{21}$, $\phi_{32}$ and $\phi_{43}$ (cf. Figure~\ref{fig:formalisms:CompareReducedAmplitudesToActualAmplitudes}). One overall phase $\phi (W, \theta)$ remains undetermined. One typically assumes the four group-$\mathcal{S}$ observables to be already measured~\cite{Chiang:1996em,Nakayama:2018yzw}, such that the remaining $4$ measurements only have to fix the relative-phases. For these remaining $4$ observables, one finds the simple rule that they must not all be selected from the same class of double-polarization observables. Further rules for the selection of complete sets cannot be distilled here into a short form. Instead, we refer the reader to the detailed mathematical treatment, as well as to the resulting rules, as given in the recent work by Nakayama~\cite{Nakayama:2018yzw}.

The theory of complete experiments in the context of a TPWA for single-meson photoproduction has been treated at length in the publications~\cite{Omelaenko,Grushin,Wunderlich:2013iga,WorkmanEtAl2017,YannickPhD}. The {\it 'reduction'} of complete sets in the TPWA is even more pronounced than in the case of Pion-Nucleon scattering (section~\ref{sec:formalisms:PiNScattering}). The absolutely minimal number of observables for a complete set in the TPWA is $4$, a result which has been found purely numerically by Tiator~\cite{YannickPhD}: there exist $196$ such complete sets of $4$. An example for such a set, which excludes any recoil-polarization observables, is $\left\{ \sigma_{0}, \check{\Sigma}, \check{F}, \check{H} \right\}$. Apart from the complete sets of $4$, there exist also complete sets of $5$ observables for which an algebraic derivation can be given using root-decompositions of the polynomial amplitude~\cite{Omelaenko,Wunderlich:2013iga}. Examples for such theoretically better justified complete sets of $5$ are given by~\cite{Wunderlich:2013iga,YannickPhD}:
\begin{equation}
 \left\{ \sigma_{0}, \check{\Sigma}, \check{T}, \check{P}, \check{F} \right\} \text{ and } \left\{ \sigma_{0}, \check{\Sigma}, \check{T}, \check{P}, \check{G} \right\} . \label{eq:formalism:ExamplesCompleteSetsOf5}
\end{equation}
Finally, we mention that a useful and more detailed theoretical comparison of the procedures of CEA and TPWA for single-meson photoproduction can be found in the appendix of reference~\cite{Svarc:2020cic}.


\subsubsection{Multi-Meson Photoproduction}
\label{sec:formalisms:MultiMesonPhotoproduction}

This section is devoted to multi-meson photoproduction and we consider as a first example the photoproduction of two pseudoscalar mesons on a nucleon: $\gamma N \longrightarrow \varphi_{1} \varphi_{2} B$ (with a recoil-baryon $B$). Important special cases are two-Pion photoproduction as well as Pion-Eta photoproduction. Multi-Meson production reactions are generally particularly interesting for Baryon Spectroscopy in the higher-mass region, since it is generally assumed that heavier resonances tend to decay favorably via cascades into multi-meson final states \cite{Sokhoyan:2015fra,Thiel:2015kuc}. The model-independent amplitude-formalism for two-meson photoproduction, as summarized in the following, was first written and published by Roberts and Oed~\cite{Roberts:2004mn}. 

Due to the three-particle final state, parity-relations among spin-amplitudes can no longer be applied in order to reduce the number of amplitudes (cf. equation~\eqref{eq:formalisms:NumberOfSpinAmplitudes}). Thus, multiplying the spin-multiplicities of all initial- and final-state particles, one arrives at a description with $N_{\mathcal{A}} = 8$ amplitudes. Roberts and Oed have introduced these amplitudes in a helicity- as well as in a transversity-basis~\cite{Roberts:2004mn}. Correspondingly, for two-meson photoproduction there exist $N_{\mathcal{A}}^{2} = 64$ measurable polarization observables, which are collected in Table~\ref{tab:formalisms:2MesonProductionObservables}. In addition to the unpolarized differential cross section $\text{I}_{0}$, two-meson photoproduction allows for the measurement of $9$ single-polarization asymmetries involving beam- ($\mathcal{B}$-), target- ($\mathcal{T}$-) and recoil- ($\mathcal{T}$-) polarization. The notation for these single-spin measurements is $\text{I}^{\mathcal{B}}$, $\text{P}_{\mathcal{T}}$ and $\text{P}_{\mathcal{R}}$, respectively (Table~\ref{tab:formalisms:2MesonProductionObservables}). Furthermore, two-meson photoproduction allows for the extraction of $27$ double-polarization asymmetries, which subdivide into $9$ observables of type $\mathcal{BT}$ (notation: $\text{P}^{\mathcal{B}}_{\mathcal{T}}$), $9$ measurements of type $\mathcal{BR}$ (notation: $\text{P}^{\mathcal{B}}_{\mathcal{R}}$) and $9$ observables of type $\mathcal{TR}$ (notation: $\mathcal{O}_{\mathcal{TR}}$). The remaining $27$ observables correspond to genuine triple-polarization ($\mathcal{BTR}$-) measurements and they are written as $\mathcal{O}^{\mathcal{B}}_{\mathcal{TR}}$ (Table~\ref{tab:formalisms:2MesonProductionObservables}). 


%
\begin{table}[htb]
\centering
\begin{tabular}{CCL}
 \hline \hline
    \text{Category} & \text{Subcategory} & \text{Observables} \\
    \midrule
   &  & \text{I}_{0}\\
 \midrule
\multirow{2}{*}{$\mathcal{B}$} & \mathcal{B}_\text{l} & \text{I}^\text{s}, \text{I}^\text{c}\\
& \mathcal{B}_{\odot} & \text{I}^{\odot}\\
\midrule
\mathcal{T} & & \text{P}_\text{x}, \text{P}_\text{y}, \text{P}_\text{z}\\
\midrule
\mathcal{R} & & \text{P}_\text{x'}, \text{P}_\text{y'}, \text{P}_\text{z'}\\
\hline
\hline
\multirow{2}{*}{$\mathcal{BT}$} & \mathcal{B}_\text{l} \mathcal{T}& \text{P}^\text{s}_\text{x}, \text{P}^\text{s}_\text{y}, \text{P}^\text{s}_\text{z}, \text{P}^\text{c}_\text{x}, \text{P}^\text{c}_\text{y}, \text{P}^\text{c}_\text{z}\\
& \mathcal{B}_{\odot} \mathcal{T}& \text{P}^{\odot}_\text{x}, \text{P}^{\odot}_\text{y}, \text{P}^{\odot}_\text{z}\\
\midrule
\multirow{2}{*}{$\mathcal{BR}$} & \mathcal{B}_\text{l} \mathcal{R} &  \text{P}^\text{s}_\text{x'}, \text{P}^\text{s}_\text{y'}, \text{P}^\text{s}_\text{z'}, \text{P}^\text{c}_\text{x'}, \text{P}^\text{c}_\text{y'}, \text{P}^\text{c}_\text{z'}\\
& \mathcal{B}_{\odot} \mathcal{R} & \text{P}^{\odot}_\text{x'}, \text{P}^{\odot}_\text{y'}, \text{P}^{\odot}_\text{z'}\\
\midrule
\multirow{2}{*}{$\mathcal{TR}$} & & \mathcal{O}_\text{xx'}, \mathcal{O}_\text{xy'},\mathcal{O}_\text{xz'},\mathcal{O}_\text{yx'},\mathcal{O}_\text{yy'},\mathcal{O}_\text{yz'},\mathcal{O}_\text{zx'}\\
 & & \mathcal{O}_\text{zy'},\mathcal{O}_\text{zz'}\\
\hline
\hline
\multirow{6}{*}{$\mathcal{BTR}$} & \mathcal{B}_\text{l} \mathcal{TR} & \mathcal{O}_\text{xx'}^\text{s}, \mathcal{O}_\text{xy'}^\text{s},\mathcal{O}_\text{xz'}^\text{s},\mathcal{O}_\text{yx'}^\text{s},\mathcal{O}_\text{yy'}^\text{s},\mathcal{O}_\text{yz'}^\text{s},\mathcal{O}_\text{zx'}^\text{s}\\
 &  & \mathcal{O}_\text{zy'}^\text{s},\mathcal{O}_\text{zz'}^\text{s}\\
  & & \mathcal{O}_\text{xx'}^\text{c}, \mathcal{O}_\text{xy'}^\text{c},\mathcal{O}_\text{xz'}^\text{c},\mathcal{O}_\text{yx'}^\text{c},\mathcal{O}_\text{yy'}^\text{c},\mathcal{O}_\text{yz'}^\text{c},\mathcal{O}_\text{zx'}^\text{c}\\
 &  & \mathcal{O}_\text{zy'}^\text{c},\mathcal{O}_\text{zz'}^\text{c}\\
  & \mathcal{B}_{\odot} \mathcal{TR} & \mathcal{O}_\text{xx'}^{\odot}, \mathcal{O}_\text{xy'}^{\odot},\mathcal{O}_\text{xz'}^{\odot},\mathcal{O}_\text{yx'}^{\odot},\mathcal{O}_\text{yy'}^{\odot},\mathcal{O}_\text{yz'}^{\odot},\mathcal{O}_{zx'}^{\odot}\\
 &  & \mathcal{O}_\text{zy'}^{\odot},\mathcal{O}_\text{zz'}^{\odot} \\
 \hline \hline
\end{tabular}
\caption{The 64 observables of two-meson photoproduction are listed here. They are further subdivided into eight categories according to the polarization that is needed to measure each observable (i.e. beam- ($\mathcal{B}$), target- ($\mathcal{T}$) and recoil- ($\mathcal{R}$) polarization). The observable $\text{I}_0$ in the first row is the unpolarized differential cross section. We have adopted the original notation by Roberts and Oed~\cite{Roberts:2004mn}, while this Table itself is taken over from reference~\cite{Kroenert:2020ahf}.}
\label{tab:formalisms:2MesonProductionObservables}
\end{table}
%


Equations relating the $64$ observables to the transversity- (or helicity-) amplitudes are not given here due to their length, but can be found in references~\cite{Roberts:2004mn,Kroenert:2020ahf}. Here, we just mention the fact that the observables can be grouped into $8$ 'shape-classes'~\cite{Kroenert:2020ahf}, containing $8$ observables each, similarly to the single-meson photoproduction observables (cf. Table~\ref{tab:formalisms:PhotoproductionObservables}).

The phase-space of two-meson photoproduction, as a $2 \rightarrow 3$-reaction, is generally parameterized by $5$ independent kinematical variables (cf. appendix~\ref{sec:appendix:Kinematics}). The observables and spin-amplitudes introduced above generally depend on all these $5$ variables. Which variables are chosen in detail for the description of the phase-space heavily depends on the employed conventions and therefore this choice is generally highly non-unique over the whole literature. One example would be the choice by Roberts and Oed~\cite{Roberts:2004mn}, i.e.: $\left( s, t, s_{\varphi_{1} \varphi_{2}}, \Omega_{\varphi_{1} \varphi_{2}} = \left\{ \theta_{\varphi_{1} \varphi_{2}}^{\ast}, \phi_{\varphi_{1} \varphi_{2}}^{\ast} \right\} \right)$, where $s$ is the CMS-energy squared over the whole process, $t$ is the square of the four-momentum difference between initial-state nucleon and final-state baryon, $s_{\varphi_{1} \varphi_{2}}$ is the total energy of the meson-pair squared and $\Omega_{\varphi_{1} \varphi_{2}}$ are the polar coordinates specifying the direction of the meson-pair's total three-momentum in the CMS-frame. Another example is given by the parameterization chosen in the works by Fix and Arenhoevel~\cite{Fix:2012ds,Fix:2013hqa}, namely $\left( \omega_{1}, \omega_{2}, \phi_{B}, \Omega_{\gamma} = \left\{ \theta_{\gamma}, \phi_{\gamma} \right\} \right)$. Here, $\omega_{1,2}$ are the energies of the two mesons in the CMS while the angles $\phi_{B}$ and $\Omega_{\gamma}$ describe the final-state baryon's and initial photon's three-momenta in a special choice of coordinates which Fix and Arenhoevel call the {\it rigid-body system}~\cite{Fix:2012ds}. The general problem has become clear: the non-uniqueness of descriptions rises for higher-dimensional phase-spaces and the situation is generally not as clear cut as in the $2$-body case.

Fix and Arenhoevel have also used their rigid-body system in order to propose the first model-independent partial-wave expansion of the full transition amplitude for two-meson photoproduction, which reads~\cite{Fix:2012ds}:
\begin{equation}
 \mathcal{T}_{\nu; \lambda \mu} \left( \omega_{1}, \omega_{2} , \phi_{B}, \Omega_{\gamma} = \left\{ \phi_{\gamma}, \theta_{\gamma} \right\} \right) =  e^{i \left( \lambda - \mu \right) \phi_{\gamma}} e^{- i \nu \phi_{B}} \sum_{J,M} t^{JM}_{\nu;\lambda \mu} \left( \omega_{1}, \omega_{2} \right)  e^{- i M \phi_{\gamma B}} d^{J}_{M,(\lambda - \mu)} \left( \theta_{\gamma} \right) . \label{eq:formalisms:2MesonPWExp}
\end{equation}
In addition to the kinematical variables specified above, this expression contains the helicities $\lambda$, $\mu$ and $\nu$ for the photon, initial nucleon and final-state baryon, respectively. In addition, $\phi_{\gamma B} = \phi_{\gamma} - \phi_{B}$ and $(J,M)$ are angular-momentum quantum numbers characterizing the total angular momentum of the outgoing three-particle system. Expression~\eqref{eq:formalisms:2MesonPWExp} is again an expansion into Wigner-$d$ functions (cf. Eq.~\eqref{eq:formalisms:General2BodyHelicityPWExpansion}). The complex functions $t^{JM}_{\nu;\lambda \mu} \left( \omega_{1}, \omega_{2} \right)$, which depend on the meson-energies $\omega_{1,2}$ only, are the two-meson photoproduction partial-waves which contain complete information on the dynamics of the process. Note that, as mentioned in section~\ref{sec:formalisms:PWExpansion}, the partial-wave expansion for this three-body final state separates out multiple angular variables, the choice of which is however, as mentioned above, non-unique.

The CEA for two-meson photoproduction has first been considered in a work by Arenhoevel and Fix~\cite{Arenhoevel:2014dwa}. In this work, the inverse function theorem was used to derive complete sets composed of $2 N_{\mathcal{A}} = 16$ observables. The drawback of this theorem consists of the fact that the resulting sets are locally free of ambiguities (i.e. they allow for no continuous ambiguities), but not globally (i.e. they might still allows for discrete ambiguities). A more recent study~\cite{Kroenert:2020ahf} has proposed alternative methods for the derivation of complete sets with $16$ observables that are completely free of ambiguities, locally as well as globally. These alternative methods consist of a very specific algebraic approach (based on earlier ideas by Nakayama~\cite{Nakayama:2018yzw}) to directly derive complete sets of minimal length $16$, numerical tests for the completeness of any proposed set, as well as graphical methods originally proposed for the solution of a CEA with an arbitrary number $N_{\mathcal{A}}$ of amplitudes~\cite{Wunderlich:2020umg}. For explicit listings of complete sets in the two-meson photoproduction CEA, we refer the reader to references~\cite{Kroenert:2020ahf,Arenhoevel:2014dwa}.

The problem of a complete experiment in a TPWA for two-meson photoproduction has first (and up to this point only) been studied in reference~\cite{Fix:2013hqa}, using Eq.~\eqref{eq:formalisms:2MesonPWExp} as a starting point. The authors have confined their attention to the simplest possible case of a truncation at $J_{\text{max}} = \frac{1}{2}$, due to the complexity of the equations resulting for higher orders. For the considered truncation-order, the authors find that the following combination of two-meson photoproduction observables is an example for a complete set~\cite{Fix:2013hqa}:
\begin{equation}
 \left\{ \text{I}_{0}, \text{I}^{\odot}, \text{P}_{z'}, \text{P}^{\odot}_{z'}, \text{P}_{x'}, \text{P}^{\odot}_{x'} \right\} .  \label{eq:formalisms_2MesonTPWACompleteSetExample}
\end{equation}
It is interesting to see that again a kind of {\it reduction} was obtained compared to the complete sets in the CEA, which at least have to contain $16$ observables. Furthermore, the authors note in reference~\cite{Fix:2013hqa} not that no triple-polarization observables are needed for a complete set in the TPWA with $J_{\text{max}} = \frac{1}{2}$. This is similar to single-meson photoproduction (see Sec.~\ref{sec:formalisms:SingleMesonPhotoproduction}), where it was possible to avoid recoil-polarization in the TPWA. The solution of the complete-experiment problem in the two-meson TPWA for arbitrary $J_{\text{max}}$ is probably a highly complicated problem and nothing has been published about it, at the time of this writing.

For the photoproduction of higher multi-meson final-states, i.e. three mesons $\gamma N \longrightarrow \varphi_{1} \varphi_{2} \varphi_{3} B$ and above, the spin-structure of the description above does not change, i.e. one still has $8$ amplitudes and $64$ observables. However, the description of the kinematical phase-space gets more complicated drastically, i.e. for processes with $4$-, $5$-, $\ldots$ body final states one generally needs $8$, $11$, $\ldots$ kinematical variables. This also makes the search for a model-independent partial-wave expansion for such processes much more complicated.

\subsubsection{Vector-Meson Photoproduction}
\label{sec:formalisms:VectorMesonPhotoproduction}

In the following, we consider the process of photoproduction of one single massive vector- (i.e. Spin-$1$-) meson $\vec{V}$ on a nucleon: $\gamma N \longrightarrow \vec{V} B$ with the recoil-baryon $B$. The most important special cases are $\omega$-, $\rho$- and $\phi$-photoproduction. The process of vector-meson photoproduction is interesting mainly for the search for higher-mass resonances. Vector-mesons are heavier than pseudoscalars. Thus, in case the coupling of particular higher-lying resonances to vector-mesons would happen to be sizeable, one would have the chance to do spectroscopy for such a state as well, still within the relatively simple kinematical context of a $2$-body reaction. For the model-independent amplitude-description, which is outlined in the following, we mainly follow the work by Pichowsky, Savkli and Tabakin~\cite{Pichowsky:1994gh}.

The massive vector-particle has three helicity-states. Therefore, one arrives at a description with $N_{\mathcal{A}} = 12$ spin-amplitudes (cf. equation~\eqref{eq:formalisms:NumberOfSpinAmplitudes}), which in reference~\cite{Pichowsky:1994gh} are literally defined as one set of $4$ helicity amplitudes $\left\{ H_{1,\lambda_{V}}, \ldots, H_{4,\lambda_{V}} \right\}$ for each of the three possible values $\lambda_{V} = -1,0,+1$ of the vector-meson's helicity. For $\lambda_{V} = 0$, one recovers the description of pseudoscalar meson photoproduction (section~\ref{sec:formalisms:SingleMesonPhotoproduction}) in the helicity-formalism~\cite{Pichowsky:1994gh}.

As a consequence of the description via $N_{\mathcal{A}} = 12$ spin-amplitudes, for vector-meson photoproduction it is possible to measure a vast array of $N_{\mathcal{A}}^{2} = 144$ polarization observables, which are naturally defined with respect to the initial $\gamma N$- (or final $\vec{V} B$-) CMS-coordinates. The observables can be written in the following bilinear-helicity-product form (using a vector $\left| H \right>$ that collects all the $12$ helicity-amplitudes)~\cite{Pichowsky:1994gh}:
\begin{equation}
  \check{\mathcal{O}}^{\alpha \beta} = \pm \frac{1}{2} \left< H \right| \Gamma^{\alpha} \omega^{\beta} \left| H \right> \text{, } \alpha = 1,\ldots,16 \text{, } \beta = 1,\ldots,9  . \label{eq:formalisms:VecMesonObsHelicityProductForm}
\end{equation}
Here, the $16$ $\Gamma^{\alpha}$-matrices comprise the same algebra that also defines the polarization observables of single-meson photoproduction (section~\ref{sec:formalisms:SingleMesonPhotoproduction}), while the $9$ $\omega^{\beta}$-matrices are $3 \times 3$-matrices that act in the vector-meson's helicity-space. The observables~\eqref{eq:formalisms:VecMesonObsHelicityProductForm} are defined in terms of the direct product of these matrices, such that the correct total number of $16 \times 9 = 144$ observables is recovered. In case the two indices $(\alpha, \beta)$ are combined in a multi-index, the form given above is of course nothing else but the generic definition~\eqref{eq:formalisms:GeneralDefinitionPolarisationObservable}.

In more detail, the process of vector-meson photoproduction allows for the measurement of $8$ non-redundant single-spin observables. Apart from the unpolarized cross section $\sigma_{0}$, these are composed of the target-polarization $\check{T}_{N}$, the recoil-polarization $\check{P}_{B}$, the photon's beam-asymmetry $\check{\Sigma}_{\gamma}$, the vector-meson's vector-polarization $\check{P}^{V}_{y}$ as well as three tensor-polarizations for the vector-meson which are typically specified in spherical coordinates: $\check{T}^{V}_{21}$, $\check{T}^{V}_{22}$ and $\check{T}^{V}_{20}$. Furthermore, there exist $51$ non-vanishing double-polarization observables which are written in the following notation by Pichowsky, Savkli and Tabakin~\cite{Pichowsky:1994gh}:
\begin{equation}
  \check{C}^{\bm{I} \bm{J}}_{ij}    .   \label{eq:formalisms:VecMesonDoublePolarizationObs}
\end{equation}
More precisely, there are $5$ non-vanishing beam-target observables ($\bm{I} \bm{J} = \gamma N$), $5$ observables of type beam-recoil ($\bm{I} \bm{J} = \gamma B$), $5$ target-recoil observables ($\bm{I} \bm{J} = N B$), $12$ beam-vector meson observables ($\bm{I} \bm{J} = \gamma V$), $12$ observables of type recoil-vector meson ($\bm{I} \bm{J} = B V$) as well as $12$ target-vector meson observables ($\bm{I} \bm{J} = N V$). The sub-scripts $i,j$ specify the polarization-directions with respect to CMS-coordinates for each observable. The remaining $144 - 51 - 8 = 85$ polarization observables have to be composed of triple- and quadruple-spin observables. In particular, triple-polarization observables yield important experimental information for vector-meson photoproduction, as opposed to single-meson photoproduction (section~\ref{sec:formalisms:SingleMesonPhotoproduction}). For the explicit definitions of all single-, double-, triple- and quadruple-polarization observables in terms of helicity-amplitudes, as well as for more details on this highly elaborate formalism, we again refer to reference~\cite{Pichowsky:1994gh}.

Instead of the idealized process $\gamma N \longrightarrow \vec{V} B$, due to the short lifetime of the vector-mesons one can only measure final-states in the detector generated by vector-meson decays. The latter could be decays into two or three peudoscalar mesons, i.e. $\vec{V} \longrightarrow \varphi_{1} \varphi_{2}$ (e.g. $\rho \longrightarrow \pi \pi$ or $\phi \longrightarrow K \bar{K}$) or $\vec{V} \longrightarrow \varphi_{1} \varphi_{2} \varphi_{3}$ (e.g. $\omega \longrightarrow \pi \pi \pi$), as well as decays into two spin-$\frac{1}{2}$ leptons $\vec{V} \longrightarrow \ell \bar{\ell}$ (e.g. $\rho \longrightarrow e^{+} e^{-}$). Thus, instead of the $144$ polarization observables defined above as rather theoretical quantities in the $\gamma N$- (or $\vec{V} B$-) center-of-mass frame, one instead only has experimental access to the polarization-information of the vector-mesons via their decay angular-distributions. The latter are parameterized by so-called {\it spin-density matrix elements (SDMEs)}~\cite{Schilling:1969um,Kloet:1998js}, which are thus sometimes viewed as the actual fundamental observables in vector-meson photoproduction. Spin-density matrix elements are not defined in the $\gamma N$-CMS, but instead in the rest-frame of the decaying vector meson either the helicity-, Adair-, or Gottfried-Jackson frame~\cite{Schilling:1969um,Kloet:1998js}.

For example, the decay angular-distribution for the vector-meson decay into two pseudoscalar mesons is connected to the spin-density matrix $\rho_{V}$ via the equation~\cite{Schilling:1969um} $W (\cos \theta, \phi, \rho_{V}) \equiv \bm{M} \rho_{V} \bm{M}^{\dagger}$, where $\bm{M}$ is the vector meson's decay-amplitude $\bm{M} := \left< \theta, \phi \right| M \left| \lambda_{V} \right>$, the angular dependence of which is fully controlled by the rotation group. The angles $\theta$ and $\phi$ are here decay-angles in the vector-meson's rest frame. For an experiment with polarized photons, one can decompose the decay angular-distribution further~\cite{Schilling:1969um}: 
\begin{equation}
  W (\cos \theta, \phi, \rho_{V}) = W^{0} (\cos \theta, \phi) + \sum_{\alpha = 1}^{3} P_{\gamma}^{\alpha} W^{\alpha} (\cos \theta, \phi)  , \label{eq:VecMesonDecayAngDistPolarizedPhoton}
\end{equation}
where $P_{\gamma}^{\alpha}$ is the Stokes-vector specifying the polarization of the incoming photon and the individual angular distributions $W^{\alpha} (\cos \theta, \phi) \equiv W^{\alpha} (\cos \theta, \phi, \rho_{V}^{\alpha})$ are defined by the projections $\rho^{\alpha}_{V}$ of the full vector-meson spin-density matrix $\rho_{V}$ onto the photon's spin-space. The matrix elements $\rho^{\alpha}_{\lambda_{V} \lambda_{V}'} = \left( \rho_{V} \right)^{\alpha}_{\lambda_{V} \lambda_{V}'}$ are thus the spin-density matrix elements one is interested in experimentally. Their extraction from data can proceed via the equations~\cite{Schilling:1969um}:
\begin{align}
  W^{0} (\cos \theta, \phi) &=  \frac{3}{4 \pi} \left[ \frac{1}{2} \left( 1 - \rho^{0}_{0 0} \right) + \frac{1}{2} \left( 3 \rho^{0}_{00} - 1 \right) \cos^{2} \theta - \sqrt{2} \text{Re} \left[ \rho^{0}_{10} \right] \sin 2 \theta \cos \phi - \rho^{0}_{1 -1} \sin^{2} \theta \cos 2 \phi \right]  ,  \label{eq:WZeroDefinition} \\
  W^{1} (\cos \theta, \phi) &=  \frac{3}{4 \pi} \left[  \rho^{1}_{11} \sin^{2} \theta + \rho^{1}_{00} \cos^{2} \theta  - \sqrt{2} \rho^{1}_{10} \sin 2 \theta \cos \phi - \rho^{1}_{1 -1} \sin^{2} \theta \cos 2 \phi \right]  ,  \label{eq:WOneDefinition} \\
  W^{2} (\cos \theta, \phi) &=  \frac{3}{4 \pi} \left[ \sqrt{2} \text{Im} \left( \rho^{2}_{10}  \right) \sin 2 \theta \sin \phi + \text{Im} \left( \rho^{2}_{1-1}  \right) \sin^{2} \theta \sin 2\phi \right]  ,  \label{eq:WTwoDefinition} \\
  W^{3} (\cos \theta, \phi) &=  \frac{3}{4 \pi} \left[ \sqrt{2} \text{Im} \left( \rho^{3}_{10}  \right) \sin 2 \theta \sin \phi + \text{Im} \left( \rho^{3}_{1-1}  \right) \sin^{2} \theta \sin 2\phi \right]  .  \label{eq:WThreeDefinition}
\end{align}
The formalism for the decay into a lepton-pair is more involved due to the lepton-spin. Elaborations on this case can be found in reference~\cite{Kloet:1998js}.

While the spin-density matrix elements are defined in the vector-meson's rest frame, the $144$ fundamental polarization observables are naturally defined with respect to the $\gamma N$-CMS. Thus, a matching of the two types of quantities involves a rather involved analysis of the transition form one frame to the other, where kinematic conversion-factors arise upon which one has to keep track carefully. This problem has been elaborated at length in the works by Kloet, Chiang and Tabakin~\cite{Kloet:1998js,Kloet:1999wc}. For instance, for an incoming photon linearly polarized according to $P_{\gamma}^{x} = \pm 1$, one finds the following relation between spin-density matrix elements and beam-vector meson double-polarization observables, which is written as a $3 \times 3$-matrix equation~\cite{Kloet:1998js}:
\begin{equation}
  \rho^{x}  = \frac{1}{3} \left[ \begin{array}{ccc} \check{\Sigma}^{\gamma}_{x} + \sqrt{\frac{1}{2}} \check{C}^{\gamma V}_{x 20} & \frac{3}{2 \sqrt{2}} \left( - i \check{C}^{\gamma V}_{x y}  \right) - \sqrt{\frac{3}{2}} \check{C}^{\gamma V}_{x 21}  &  \sqrt{3} \check{C}^{\gamma V}_{x 22} \\  \frac{3}{2 \sqrt{2}} \left(  i \check{C}^{\gamma V}_{x y}  \right) - \sqrt{\frac{3}{2}} \check{C}^{\gamma V}_{x 21}  & \check{\Sigma}^{\gamma}_{x} - \sqrt{2} \check{C}^{\gamma V}_{x 20} & \frac{3}{2 \sqrt{2}} \left( - i \check{C}^{\gamma V}_{x y}  \right) + \sqrt{\frac{3}{2}} \check{C}^{\gamma V}_{x 21} \\ \sqrt{3} \check{C}^{\gamma V}_{x 22} & \frac{3}{2 \sqrt{2}} \left(  i \check{C}^{\gamma V}_{x y}  \right) + \sqrt{\frac{3}{2}} \check{C}^{\gamma V}_{x 21} &  \check{\Sigma}^{\gamma}_{x} + \sqrt{\frac{1}{2}} \check{C}^{\gamma V}_{x 20} \end{array} \right]    . \label{eq:ExampleRelationSpinDensityToActualObservables}
\end{equation}
Investigating further the connection between spin-density matrix elements and polarization observables, it was found that {\it not all} polarization observables involving vector-meson polarization are accessible by measuring decay angular-distributions of mesonic decays~\cite{Kloet:1998js}. Rather, this type of measurement imposes clear restrictions upon which of the $144$ polarization observables are experimentally accessible at all. In more detail, within the full set of single- and double-polarization observables, one finds $16$ observables that are inaccessible via the mesonic decay-modes. These $16$ observables are composed of the vector-meson's vector-polarization $\check{P}^{V}_{y}$, as well as $5$ certain double-polarization observables from each of the classes of beam-vector meson-, target-vector meson- and recoil baryon-vector meson observables~\cite{Kloet:1998js}. These $16$ observables are only accessible by measuring the lepton-spin in leptonic decay-modes. 

Since vector-meson photoproduction itself is just a $2 \rightarrow 2$-reaction, the simple helicity partial-wave expansion~\eqref{eq:formalisms:General2BodyHelicityPWExpansion} applies here as well and it is used in reference~\cite{Pichowsky:1994gh}. Other than in the cases of Pion-Nucleon Scattering (section~\ref{sec:formalisms:PiNScattering}) and single-meson photoproduction (section~\ref{sec:formalisms:SingleMesonPhotoproduction}), there has not yet emerged a peculiar alternative convention in the literature to write the partial-wave expansion. We mention that the authors of reference~\cite{Pichowsky:1994gh} also define and discuss a formalism for {\it moment analysis} in vector-meson photoproduction, although they propose to expand the angular distributions of the polarization observables into Wigner-$d$ functions and not into Legendre polynomials.

The theoretical CEA in vector-meson photoproduction is an unsolved problem at the time of this writing. Some considerations on this problem, which are however brief and not very conclusive, are offered in the work by Pichowsky, Savkli and Tabakin~\cite{Pichowsky:1994gh}. From the heuristic arguments described in section~\ref{sec:formalisms:CompleteExperiments}, it is clear that a complete set in vector-meson photoproduction, which allows neither for continuous nor for any discrete ambiguities, has to contain at least $2 N_{\mathcal{A}} = 24$ carefully selected observables. How to select these $24$ observables out of the $144$ available ones is a difficult problem which probably demands for a highly elaborate and careful analysis. The approach of Nakayama~\cite{Nakayama:2018yzw} may be applicable here. However, it is probably very hard to extend Nakayama's treatment of a problem with $4$ amplitudes to $12$ amplitudes, since the number of cases to be considered increases exponentially. Another possible road to solve the CEA for vector-meson photoproduction might be to use the graphical methods proposed in the recent study~\cite{Wunderlich:2020umg}. In this case, the difficulty lies in the extremely large number of possible graph-topologies that have to be considered for $N_{\mathcal{A}} = 12$ amplitudes, in order to find all complete sets. Furthermore, as argued in reference~\cite{Wunderlich:2020umg}, the graphical methods will probably result in slightly over-complete sets, i.e. in complete sets containing a few more than $24$ observables (see also reference~\cite{Kroenert:2020ahf}).

The problem of a complete experiment in the context of a TPWA for vector-meson photoproduction has, to our knowledge, not yet been treated. However, here one also may expect a {\it reduction} from $24$ down to a smaller number of observables, due to the use of full angular distributions.

%% file: Experiments.tex
\label{sec:exp}


In the onset of dedicated baryon spectroscopy experiments, a substantial part was contributed by $\pi N$ scattering. These experiments were for example performed with the Crystal Ball detector at the Brookhaven National Lab (BNL), at the Paul-Scherrer-Institute (PSI) or at the ITEP-PNPI in Gatchina. Further details about the experiments can be found in Sec.~\ref{sec:measurements:PionNucleonScattering} and the references therein.

In the last years, the field was dominated by several experiments, which delivered data sets for different final states mostly using photoproduction. An overview of the most important contributors is given in the following sections. The full data sets extracted by these experiments are given in Sec.~\ref{sec:measurements}.

\subsection{GRAAL at ESRF}\label{sec:exp:GRAAL}
The GRenoble Anneau Accélérateur Laser (GRAAL) experiment \cite{Bartalini:2005wx} was located at the European Synchrotron Radiation Facility (ESRF) in Grenoble (France). This experiment was in operation from 1995 to 2008 \cite{Vegna:2013ccx}. Electrons circulated the storage ring with an energy of 6.03 GeV. Photons were produced by Compton backscattering of laser light off the electrons. High polarization degree values of the photons were achieved (96-98\% at the Compton edge, which decreases down to 20\% at 550 MeV \cite{Bartalini:2005wx}) due to the high polarization degree of the laser light. This is an advantageous feature of Compton backscattering in comparison to Bremsstrahlung photon beams.

\begin{figure}[h]
\centering
  \includegraphics[width=0.45\textwidth, angle=-90]{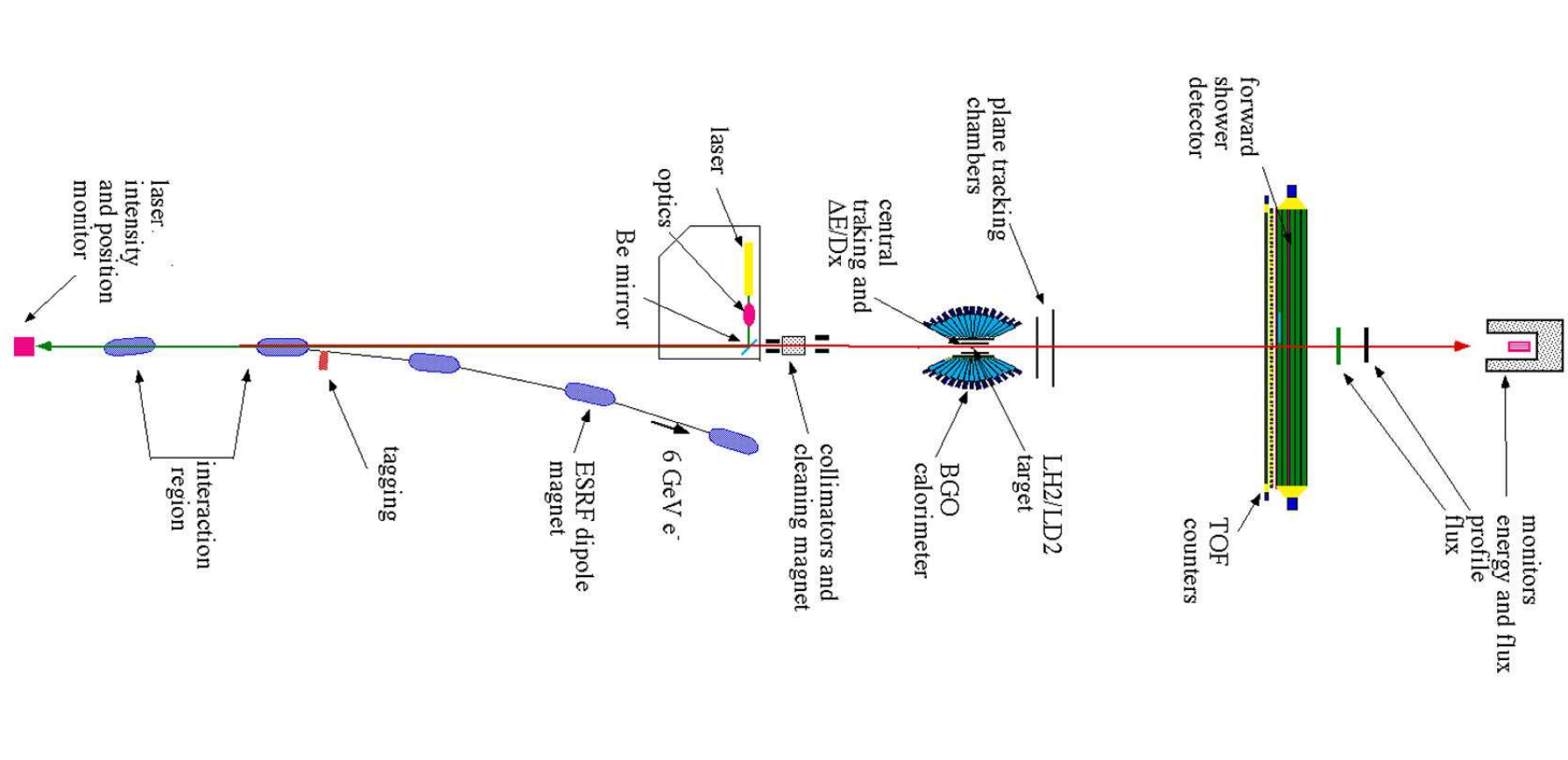}
 \caption{Overview of the GRAAL experiment \cite{Bartalini:2005wx}. \EPJA}
 \label{fig:graalexperiment}
\end{figure}

The photons were tagged in the photon energy range from 550 MeV to 1500 MeV. Either a liquid hydrogen or deuterium target was used during data taking. At the center of the 4$\pi$ LA$\gamma$RANGE detector setup, see Fig.~\ref{fig:graalexperiment}, a BGO calorimeter was used to detect photons. Multi-Wire Proportional Chambers (MWPC) surrounding the target, a plastic scintillator barrel and in forward direction a double plastic scintillator hodoscope and lead-scintillator sandwich ToF wall allowed charge identification and tracking, dE/dx and Time-of-Flight (TOF) measurements. After the end of the GRAAL experiment, the BGO calorimeter has been moved to Bonn and is now part of the BGOOD experiment (see Sec.~\ref{sec:exp:BGOOD}).

\subsection{TAPS, GDH and Crystal Ball/TAPS at MAMI }\label{sec:exp:A2}
The Mainz Microtron (MAMI) accelerator facility is located at the Institute for Nuclear Physics at the University of Mainz \cite{Kaiser:2008zza,Jankowiak:2010iua}. It was one of the first continuous-wave electron accelerator. Since 1990, MAMI-B delivered electrons with energies between 180 - 883 MeV energy and 100 $\mu$A using a sequence of three race-track microtrons. In 2006, a fourth, harmonic double-sided microtron \cite{Kaiser:2008zza} was added for the MAMI-C upgrade, see Fig.~\ref{fig:a2tapsexperiment}, left. It provides unpolarized or longitudinally polarized electrons that can be accelerated in the mentioned different stages to energies of up to 1604 MeV (MAMI-C). Currently, the accelerator facility is expanded by the new linear accelerator (MESA) \cite{Hug:2017ypc}.

The electron beam is used for several nuclear and particle physics experiments. In the A2 hall, real photons are produced through bremsstrahlung off a thin radiator foil and are tagged with the Glasgow tagger \cite{Anthony:1991eq}. The photons can be linearly, circularly or elliptically polarized. In addition, many different targets are available: liquid hydrogen, polarized butanol or dbutanol target \cite{Thomas:2011zza}, active target, solid-state targets (Ca). 
\subsubsection{TAPS experiment}
\begin{figure}[h]
\centering
 \includegraphics[width=0.47\textwidth]{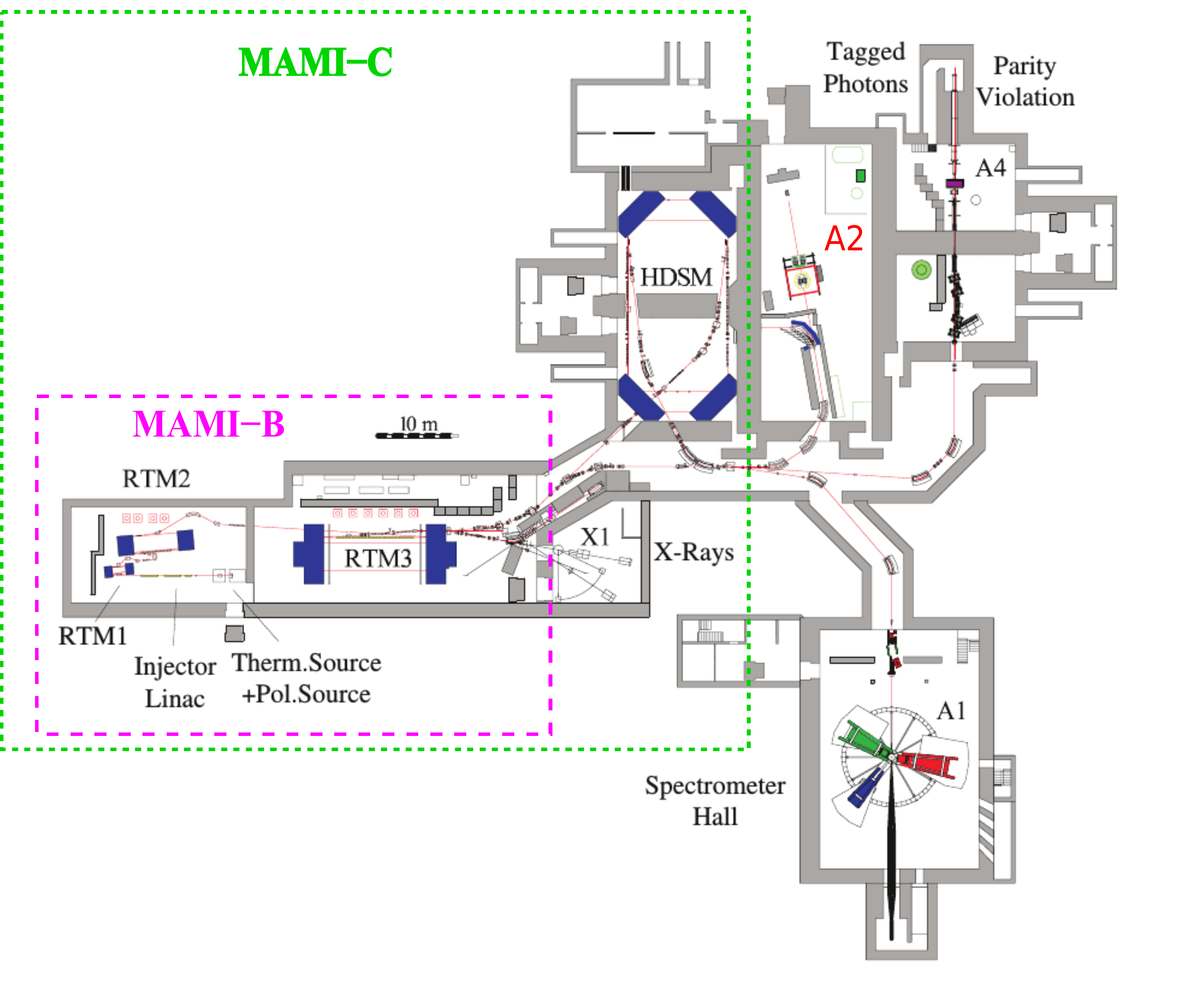}
 \includegraphics[width=0.47\textwidth]{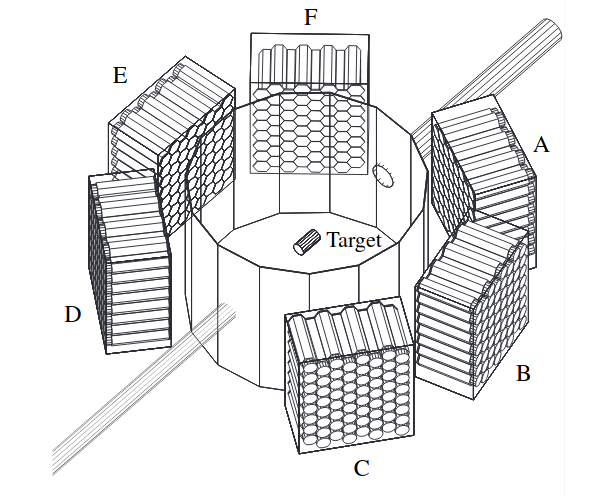}
 \caption{Left: The MAMI accelerator in Mainz \cite{Kaiser:2008zza}. Right: The TAPS experiment as used in 1995/1996 in Mainz \cite{deLeon:2001dnx}. \EPJA}
 \label{fig:a2tapsexperiment}
\end{figure}

First photoproduction reactions were studied in Mainz with the TAPS setup \cite{Schadmand:2000wk, Krusche:1998wm} (see Fig.~\ref{fig:a2tapsexperiment}). The TAPS setup consisted of six detector blocks of BaF$_2$ crystals arranged around the target and one larger forward block. It provided energy, pulse-shape, time of flight and, together with the plastic scintillators in front of the crystals, charge information of the detected particles \cite{Krusche:1998wm}.

\subsubsection{GDH experiment with the DAPHNE detector}
\begin{figure}[h!]
\centering
 \includegraphics[width=0.85\textwidth]{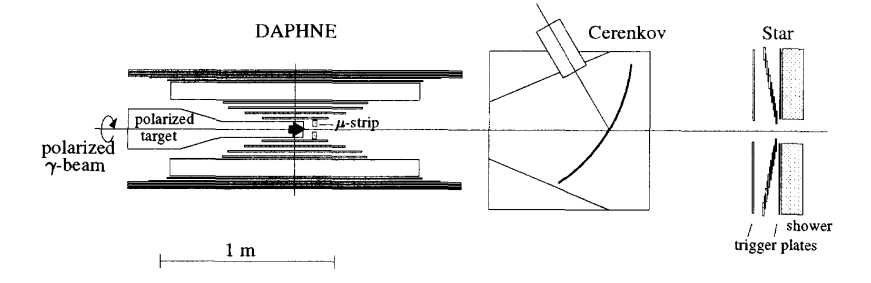}
 \caption{The GDH experiment as used in 2002 in Mainz \cite{Thomas:1999ry,Arends:2000xe}.}
 \label{fig:a2gdhexperiment}
\end{figure}

Furthermore, the GDH experiment, see Fig.~\ref{fig:a2gdhexperiment}, was also used in Mainz from 1998 to 2004 in a collaborative effort with ELSA (see Sec. \ref{sec:exp:ELSA}) to check the Gerasimov-Drell-Hearn sum rule \cite{Ahrens:2004pf}. It comprised mainly of the DAPHNE detector, which consisted of three multi-wire proportional chambers that were enclosed by segments of a plastic scintillator calorimeter and a sandwich of plastic scintillators. It was complemented in forward direction by the scintillator ring detector STAR. This setup was used with a circularly polarized photon beam and a longitudinally polarized target. 

\subsubsection{Crystal Ball/TAPS experiment}
\begin{figure}[h]
\centering
 \includegraphics[width=0.85\textwidth]{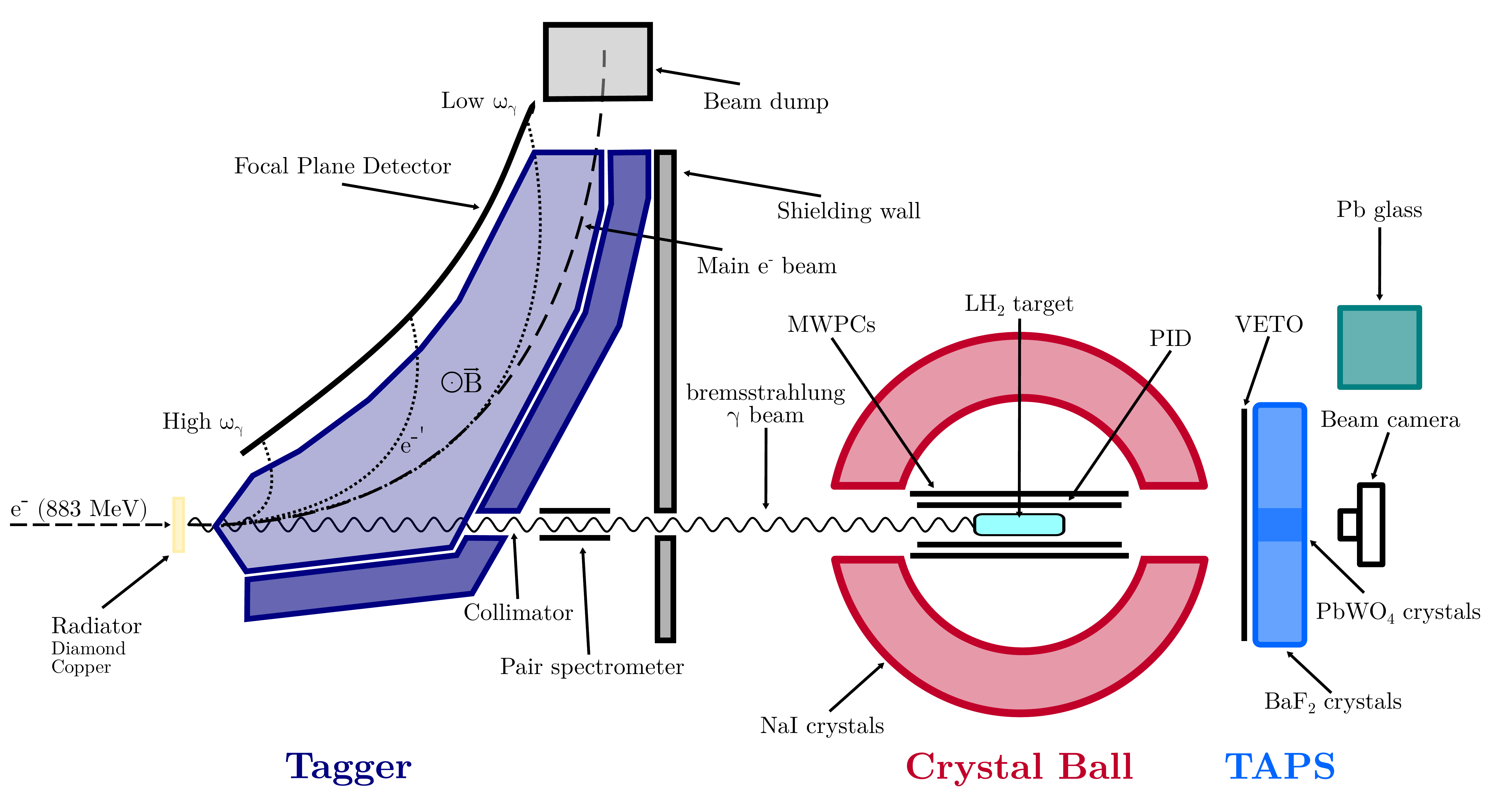}
 \caption{The latest Crystal Ball/TAPS experiment in the A2 hall \cite{PhDMornacchi}.}
 \label{fig:a2setup}
\end{figure}

Currently, the Crystal Ball/TAPS setup is located in the A2 hall (see Fig.~\ref{fig:a2setup}).
The main components of the detector setup are the Crystal Ball and the TAPS electromagnetic calorimeters. In total, 97\% of the full 4$\pi$ angular range is covered. The two multiwire proportional chambers, the PID detector and TAPS vetoes allow the analysis of not only neutral mesons, but also charged mesons, like $\pi^+$.

\subsection{Experiments at the ELSA Accelerator}
\label{sec:exp:ELSA}
The Electron Stretcher Accelerator (ELSA)  \cite{Hillert:2006yb} is located below the Physikalisches Institut of the University of Bonn. In its current setup, it consists of two linear accelerators, a synchrotron ring and a stretcher accelerator and can provide electron beams with energies up to 3.5~GeV. Two different experimental halls are currently in use, which allow experiments with polarized or unpolarized electron beams.
\subsubsection{SAPHIR}
\begin{figure}[h]
	\centering
\includegraphics[width=0.6\textwidth]{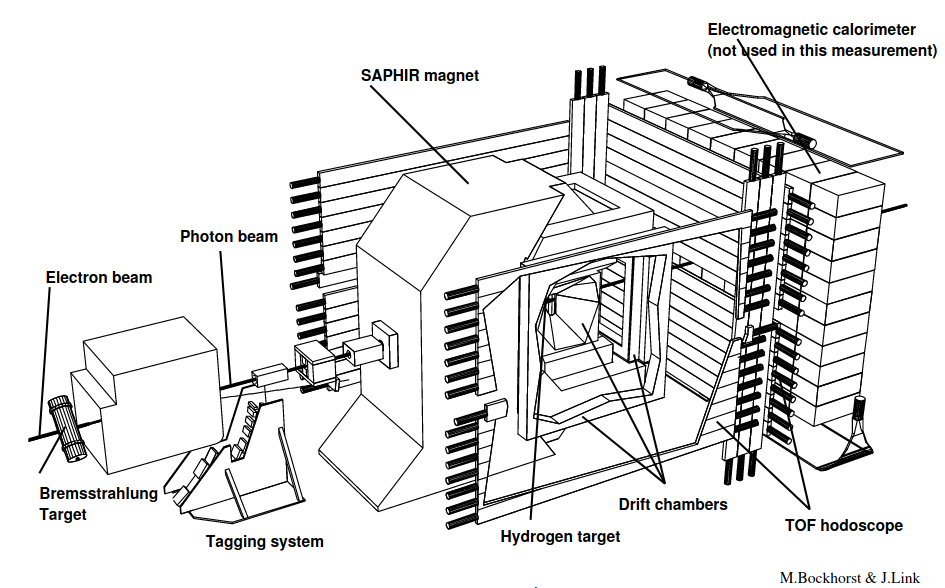}
	\caption{Layout of the SAPHIR experiment at ELSA \cite{Wu:2005wf}. \EPJA}
	\label{fig:saphir}
\end{figure}

The large solid angle magnetic spectrometer SAPHIR (Spectrometer Arrangement for PHoton Induced Reactions) \cite{Schwille:1994vg} was built up at the ELSA accelerator in Bonn and took data between 1993 and 1998. Its setup featured a large magnet with several drift chambers inside. The outside was covered with TOF hodoscopes and in forward direction an electromagnetic calorimeter. The physics program included the measurement of reactions containing strange and vector mesons. An overview of its setup can be seen in Fig.~\ref{fig:saphir}

\subsubsection{CB-ELSA, CB-TAPS and CBELSA/TAPS}\label{sec:exp:CB}
\begin{figure}[h]
\centering
 \includegraphics[width=0.6\textwidth]{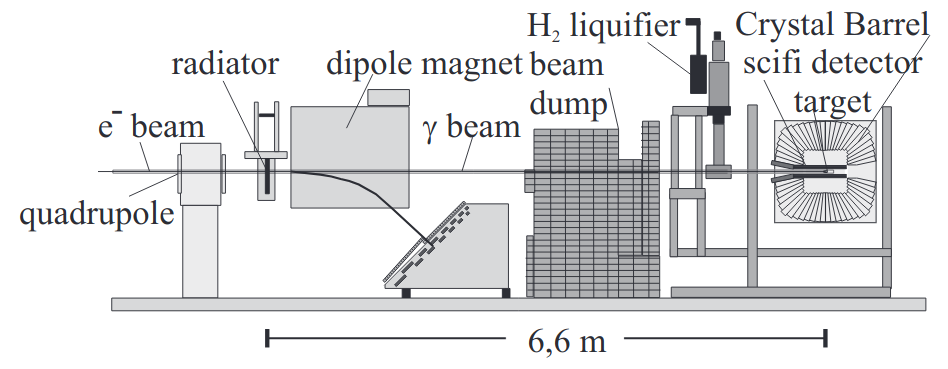}\\
  \includegraphics[width=0.8\textwidth]{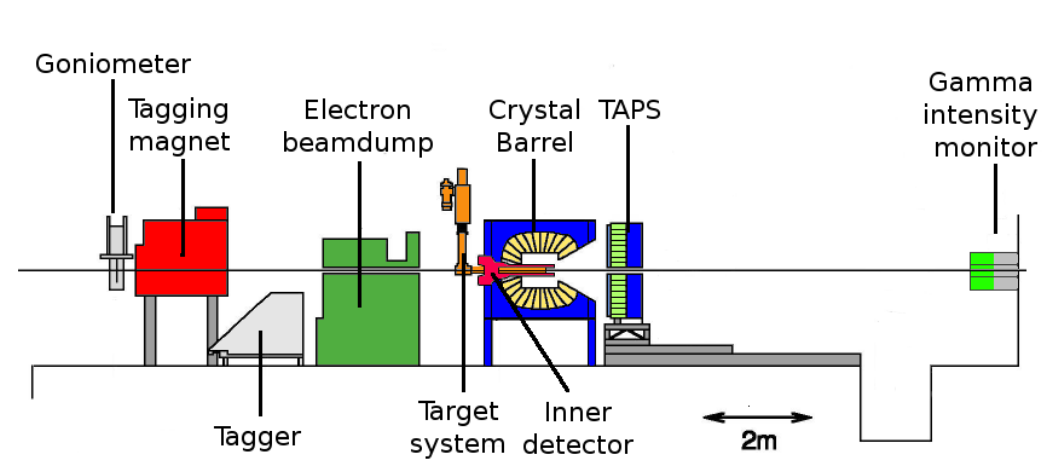}
 \caption{The setup of the CBELSA experiment \cite{Bartholomy:2007zz} (top) starting in 2000. It was extended by the TAPS forward detector for the CBTAPS experiment \cite{Gutz:2014wit} (bottom) and took data until 2003. \EPJA}
 \label{fig:cbelsa}\label{fig:cbtaps}
\end{figure}

After the closure of the LEAR ring at CERN, the Crystal Barrel detector \cite{Aker:1992ny} was moved to the ELSA accelerator in 1997. The electron beam is converted to photons via bremsstrahlung on a radiator target. A tagging system is used to determine the energies of the produced photons. The Crystal Barrel calorimeter, which has a high granularity with over 1300 CsI(Tl) detector modules, is ideally suited for measurements of mesons decaying into photons, in particular high mutiplicty photon final states. The first data sets were taken in 2000 and were mainly focused on the production of single or multiple mesons by a photon beam of up to 3.2~GeV. The complete setup at that time can be found in Fig.~\ref{fig:cbelsa} (top).

Later, the experimental setup was completed by the forward calorimeter TAPS, see Fig.~\ref{fig:cbtaps} (bottom), to provide a higher angular coverage. In addition, linearly polarized photons by utilizing bremsstrahlung off a diamond crystal became available. These led to several new measurements of different polarization observables. This modified setup was also called CB-TAPS experiment.

\begin{figure}[h]
\centering
 \includegraphics[width=0.8\textwidth]{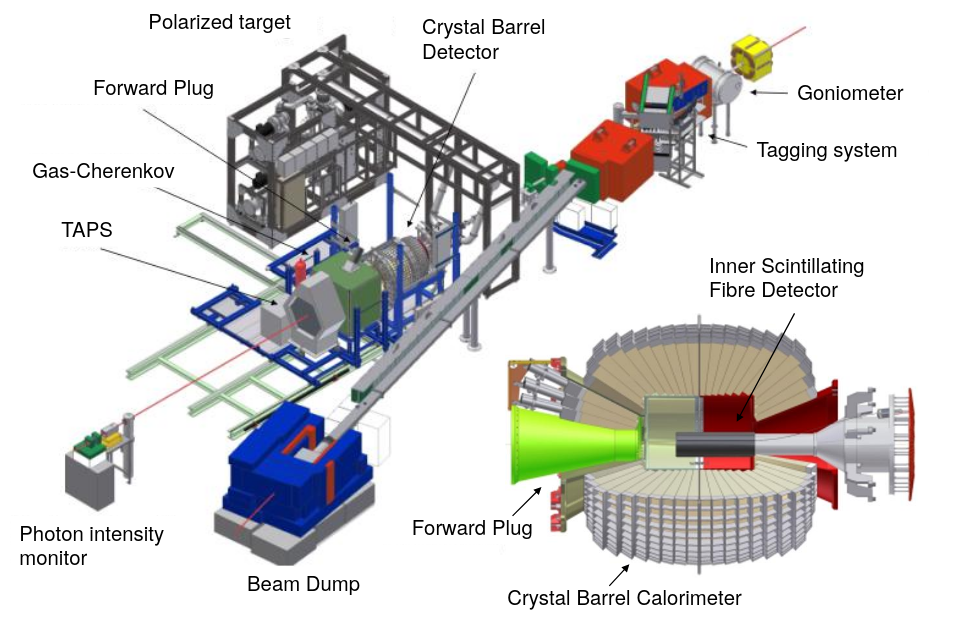}
 \caption{A schematic view of the CBELSA/TAPS experiment \cite{Gottschall:2019pwo}, which is in use since 2006. \EPJA}
 \label{fig:cbelsataps}
\end{figure}

In a third upgrade, the detector system moved to a different experimental area, see Fig.~\ref{fig:cbelsataps}. Here, a new tagging system was installed and circularly polarized photons became available. However, the major difference was the new target system, allowing experiments on polarized protons or neutrons \cite{Dutz:2017kvr}.

An upgrade of the calorimeters was performed starting in 2014 \cite{Urban:2015tma}. The readout was improved to efficiently trigger also on final states containing neutrons \cite{HonischCBUpgrade}.

\subsubsection{BGOOD  }\label{sec:exp:BGOOD}
\begin{figure}[h]
\centering
 \includegraphics[width=0.7\textwidth]{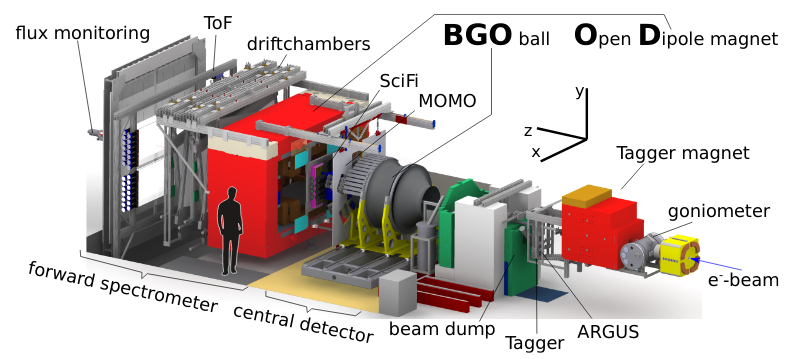}
 \caption{A schematic overview of the BGOOD \cite{Alef:2019imq} experiment at ELSA. \EPJA}
 \label{fig:bgood}
\end{figure}

After the CB-TAPS experiment, see Sec.~\ref{sec:exp:CB}, moved to a different experimental area, a new experiment was built-up: the BGOOD experiment \cite{Alef:2019imq}, see Fig.~\ref{fig:bgood}. The central part comprises the BGO ball from the former GRAAL experiment, see Sec.~\ref{sec:exp:GRAAL}, and is complemented in forward direction by a forward spectrometer. The main feature is the tracking of charged particles in forward direction in addition to a measurement of neutral final states in the central calorimeter. By using linearly or circularly polarized photon beams, it focuses on the extraction of polarization observables mainly in final states containing charged as well as neutral particles.

\subsection{Experiments at the CEBAF Accelerator}\label{sec:exp:CEBAF}
\subsubsection{CLAS}
The CLAS experiment (CEBAF large acceptance spectrometer) \cite{Mecking:2003zu} was located in Hall B at the CEBAF accelerator at the Thomas Jefferson National Accelerator Facility in Newport News, Virginia. 
\begin{figure}[h]
	\centering
	\includegraphics[width=0.45\textwidth]{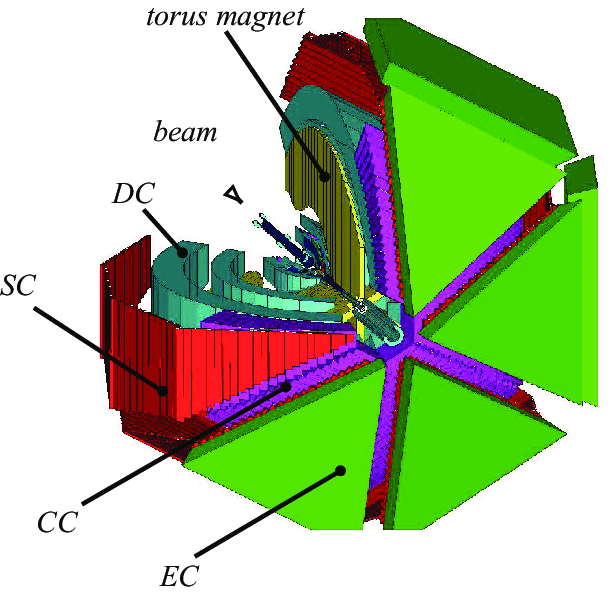}
	\caption{A schematic overview of the central detectors of the CLAS \cite{Bedlinskiy:2014tvi} experiment. \APS{\cite{Bedlinskiy:2014tvi}}{2014}}
	\label{fig:clas}
\end{figure}
The experiment, which is shown in Fig.~\ref{fig:clas}, consisted of a six-coil torodial magnet with drift chambers, Cherenkov counters, time-of-flight scintillators and electromagnetic calorimeters. The detector system is ideally suited for the measurement of charged final states. Among other targets, also two different polarized targets \cite{Keith:2003ca,Bass:2014bez} were available for measurements. A broad range of different experimental topics has been investigated with this setup by using photoproduction as well as electroproduction. 

\subsection{LEPS and LEPS2/BGOegg at SPring-8}\label{sec:exp:LEPS}
Since 2000, hadron physics experiment have been carried out at the Super Photon ring 8-GeV (SPring-8) facility in Harima Science Park City, Japan. 

\subsubsection{LEPS}
\begin{figure}[h]
\centering
 \includegraphics[width=0.4\textwidth]{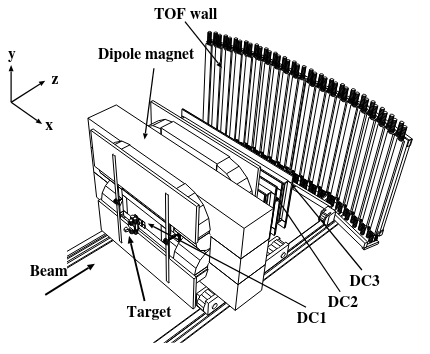}
 \includegraphics[width=0.56\textwidth]{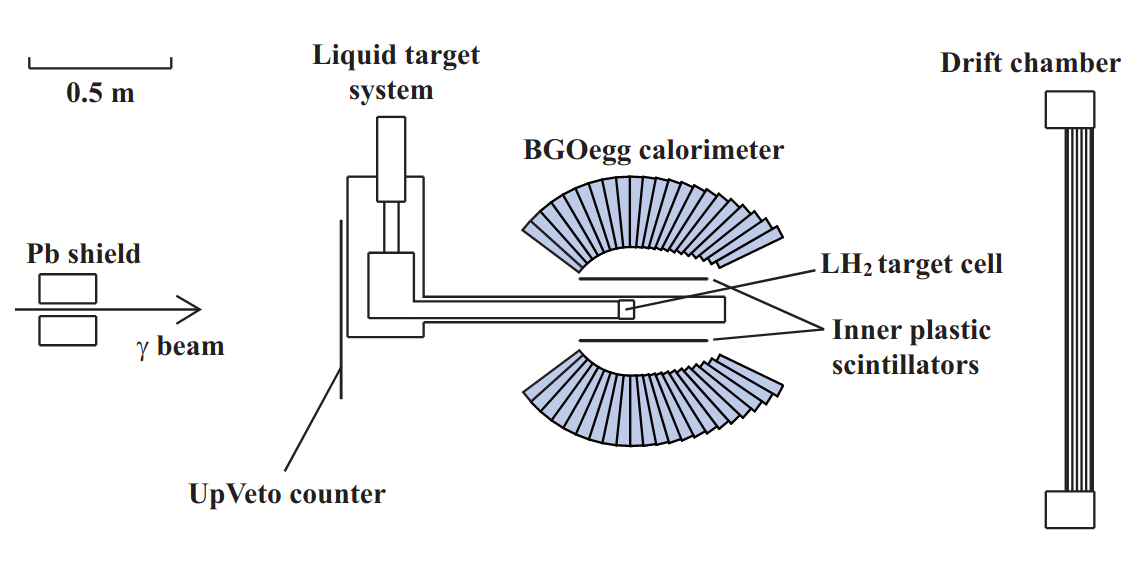}
 \caption{Left: The LEPS forward spectrometer \cite{Sumihama:2005er}. \APS{\cite{Sumihama:2005er}}{2006} Right: The LEPS2/BGOegg experimental setup \cite{Muramatsu:2019sro}. \APS{\cite{Muramatsu:2019sro}}{2019}}
 \label{fig:lepsexperiment}
\end{figure}

At the Laser-Electron-Photon (LEP) beamline, a photon beam is produced via laser-induced backward Compton scattering using electrons from the 8 GeV electron storage ring. Polarizing the laser light allows for high polarization degrees of linearly or circularly polarized photons.
A tagged photon energy range from 1.5 GeV to 2.9 GeV is used by the LEPS experiment \cite{Muramatsu:2013yya}. The setup, which is shown in Fig.~\ref{fig:lepsexperiment} (left), consists of a forward spectrometer with an acceptance of $\pm 20^\circ$ and $\pm10^\circ$ in horizontal and vertical directions, respectively. This experiment is well suited for measuring the momentum and time-of-flight of charged particles with high precision and thus, offers data at extreme angles.

\subsubsection{BGOegg}
In 2011, a new beam line, LEPS2, was constructed at SPring-8, where the BGOegg experiment \cite{Nakano:2017zrr} is located, see Fig.~\ref{fig:lepsexperiment} (right). A tagged photon beam in the energy range from 1.3 to 2.4 GeV is provided in combination with e.g. liquid hydrogen, carbon or copper target. A polarized Hydrogen-Deuteride (HD) target is planned for the future as well \cite{Ohta:2020iwy}. The BGOegg calorimeter covers the polar angular range from 24$^\circ$ to 144$^\circ$. In the forward region, a planar drift chamber (DC) and resistive platechambers (RPC) are used to detect charged particles with excellent time resolution. Recently, the forward region detectors were replaced by another electromagnetic calorimeter \cite{Muramatsu:2020odu}. 

\subsubsection{LEPS2 solenoidal spectrometer}
Apart from the BGOegg experiment, a large solid angle solenoidal spectrometer has been under construction at LEPS2, which was commissioned in 2018-2019. This LEPS2 experiment  \cite{Ryu:2020pjk} consists of a 1-T solenoid magnet, which in combination with the time projection chamber (TPC) and forward drift chambers is used to track charged particles. TOF measurements are performed with the Barrel and forward RPCs.  In addition, aerogel Cherenkov counters (AC) and barrel Pb/scintillator sampling calorimeter (Barrel-$\gamma$) are used as well. Overall, the experiment is optimized to measure charged and neutral final states. 


\subsection{Other facilities and experiments}\label{sec:exp:elph}\label{sec:exp:other}
For photoproduction, different other facilities are available, like for example the Research Center for Electron Photon Science (ELPH) at the Tohoku University. ELPH uses a 90 MeV linac and a Booster STorage ring (BST) to accelerate electrons up to 1.3 GeV. Previous to 2009, the facility was known as Laboratory for Nuclear Science (LNS). Two tagged photon beam lines are available, where the experiments FOREST and NKS2 are located.

The Four-pi Omni-directional Response Extended Spectrometer Trio (FOREST) detector system \cite{Ishikawa:2016kin} consists of three different types of electromagnetic calorimeters. CsI crystals, lead scintillating-fiber modules and lead-glass Cherenkov counters were used to assemble this detector system between 2006 to 2009, covering almost the full polar angular range from 4$^\circ$ to 175$^\circ$. In addition, a scintillation hodoscope was placed in front of the calorimeters to distinguish between charged and neutral particles. A liquid hydrogen or deuterium target is available. This experiment provides a high acceptance for photons, and is well suited for studying neutral mesons. 
%
The Neutral Kaon Spectrometer 2 (NKS2) \cite{Kaneta:2017nit} consists of a dipole magnet with a magnetic field strength of 0.42 T at the center. Two drift chambers are used to track charged particles and an inner and outer hodoscopes provide further TOF measurements. At the center of the experiment a target cell is located, that is filled with liquid deuterium. The NKS2 experiment focuses on the study of strangeness photoproduction. 

In the last years, additional experiments were conducted for baryon spectroscopy, which do not exploit photoproduction. For example the HADES experiment at GSI, Darmstadt, performs $\pi N$ scattering experiments and is therefore able to increase the previously existing data base \cite{Adamczewski-Musch:2020tfm}. For further details of the experimental setup and the measurement, see Sec.~\ref{sec:measurements:PionNucleonScattering} and the references therein.

Another, quite different approach has been performed by the CLEO-c \cite{CLEO:2010fre} and the BESIII \cite{BESIII:2012ssm} experiment. They used $e^+ e^-$ collisions and investigated $\psi(3686)$ decays. Here, they were able to identify several nucleon resonances in different channels, leading to an increased star rating for some of them. For further details see \cite{CLEO:2010fre} and \cite{BESIII:2012ssm}.

%% file: Measurements.tex
\label{sec:measurements}
\subsection{Pion-Nucleon Scattering}
\label{sec:measurements:PionNucleonScattering}
The first steps towards a mapping of the excited nucleon spectrum were performed starting in 1957 utilizing pion nucleon scattering. As an advantage, the number of amplitudes is limited, therefore only a reduced number of observables needs to be measured, see Sec.~\ref{sec:formalisms:PiNScattering}. A vast amount of measurements were conducted, but the largest data sets were collected at LAMPF in Los Alamos, US, TRIUMF in Vancouver, Canada, and PSI (former SIN) in Villigen, Switzerland. Early comparisons to quark models showed a significant lack of experimentally found resonances, which resulted in the well-known problem of the \textit{missing resonances}. 

In recent years, the data base was extended by measurements of different experiments, like the Crystal Ball experiment at BNL, see for example \cite{Prakhov:2005qb,Prakhov:2004zv,Starostin:2005pd}, the PSI  \cite{Breitschopf:2006gn,Meier:2004kn} and by ITEP-PNPI \cite{Alekseev:2005zr,Alekseev:2008cw}. An overview about the complete data sets can be found in the SAID database\footnote{Webpage of the SAID database: http://gwdac.phys.gwu.edu/}.

\begin{figure}[htb]
	\centering
	\includegraphics[width=\textwidth]{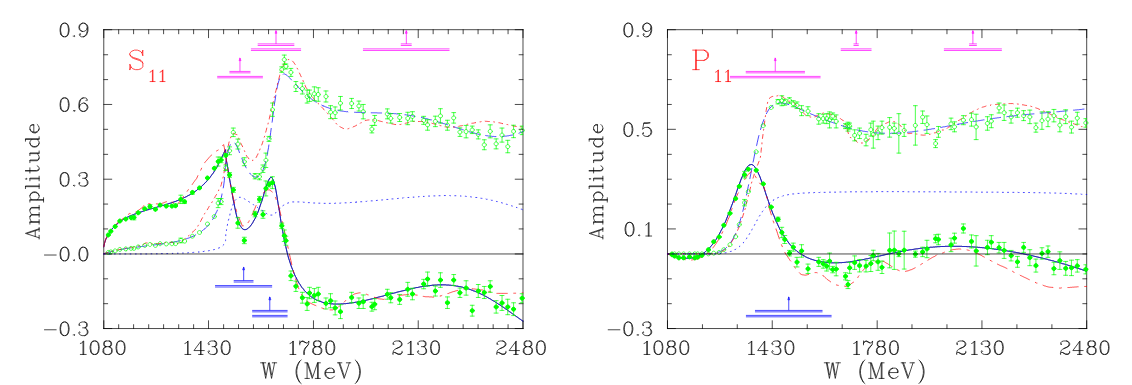}
	\caption{The $S_{11}$ and the $P_{11}$ as extracted by an anaysis of the pion-nucleon scattering data by the SAID group \cite{Arndt:2006bf}. The real parts of the single energy solutions (green) can be seen as filled circles, the imaginary parts as open circles. Figure adapted from \cite{Arndt:2006bf}. }
	\label{fig:M:pionnucleon_partialwaves}
\end{figure}

Different partial wave analysis groups used the measured data sets to compose a full partial wave analysis, see for example the work from Cutkosky et al.~\cite{Cutkosky:1979fy}. Most recent is the analysis of the SAID group \cite{Arndt:2006bf}, which performed a global fit including all data sets, which were available at that time. An example of the $S_{11}$ and the $P_{11}$ partial wave, determined by this analysis can be found in Fig.~\ref{fig:M:pionnucleon_partialwaves}. Their fit method consisted of a set of single energy solutions, which were described by an energy-dependent fit (SP06).
From their fit, they determined several resonance parameters and were the most important contributor to the PDG at that time. Most importantly, the extracted partial waves for pion nucleon scattering by the SAID group are used as input for other partial wave analysis, like the BnGa partial wave analysis, see \ref{sec:PWA:BnGaModel}, which has their focus on the description of the different photo production data sets.

To complete the amplitudes to the high energy region, an analysis was performed by Huang et al.~\cite{Huang:2008nr}. Here, they were interested in the extrapolation to higher energies, above the range with was investigated by the models from SAID \cite{Arndt:2006bf} and the Karlsruhe-Helsinki partial wave analysis \cite{Koch:1980ay,Hoehler84}. In this region, the two-hadron reactions are diffractive and can be described with Regge phenomenology \cite{Huang:2008nr}. In their analysis they found the magnitudes of the helicity non-flip amplitudes not well constrained. However, the magnitudes of the helicity flip amplitudes were considerably better.

In a second analysis, Huang et al.~\cite{Huang:2009pv} analyzed the world data sets of backward pion-nucleon scattering for $\sqrt{s} > 3$~GeV in a Regge Model. They were able to reproduce the polarization asymmetries for $\pi^+p$ and $\pi^-p$ backward elastic scattering within Regge phenomenology.

Another method was used in Mathieu et al.~\cite{Mathieu:2015gxa} to connect the $\pi N$ amplitudes from SAID with the high energy region. Here, they used finite energy sum rules, which provide a stronger constraint than the dispersion relations. By this, they were able to continue the amplitudes to the high energy region.

In addition to the previously mentioned measurements, a reverse measurement of photoproduction has been performed, which is worth mentioning at this point. The reaction $\pi^- p \to n \gamma$ was measured by the Crystal Ball experiment located at BNL \cite{Shafi:2004zn} and used to extract  the differential cross section of the reverse photoproduction reaction. Comparison with photoproduction data showed good agreement, however, due to the closing of this hadron facility, this method was not exploited any further.

Although the contributions to baryon spectroscopy were dominated in the last years by photoproduction measurements, a new opportunity for $\pi N$ scattering became possible with the HADES experiment at GSI \cite{HADES:2017mzn}. There, a $\pi^-$ beam was used to produce final states containing one or two pions \cite{Adamczewski-Musch:2020tfm}. The extracted data sets could be described by the BnGa partial wave analysis. By this, they were able to show that the $N(1520)\frac{3}{2}^-$ resonance dominated the $N\rho$ final states. This was the first measurement of its kind with the HADES spectrometer and it demonstrates that precise $\pi N$ measurements can still be a useful tool to gather information about baryon resonances.

\subsection{Single Meson Photoproduction} \label{sec:measurements:SingleMesonPhotoproduction}
\begin{figure}[h]
\includegraphics[width=\textwidth]{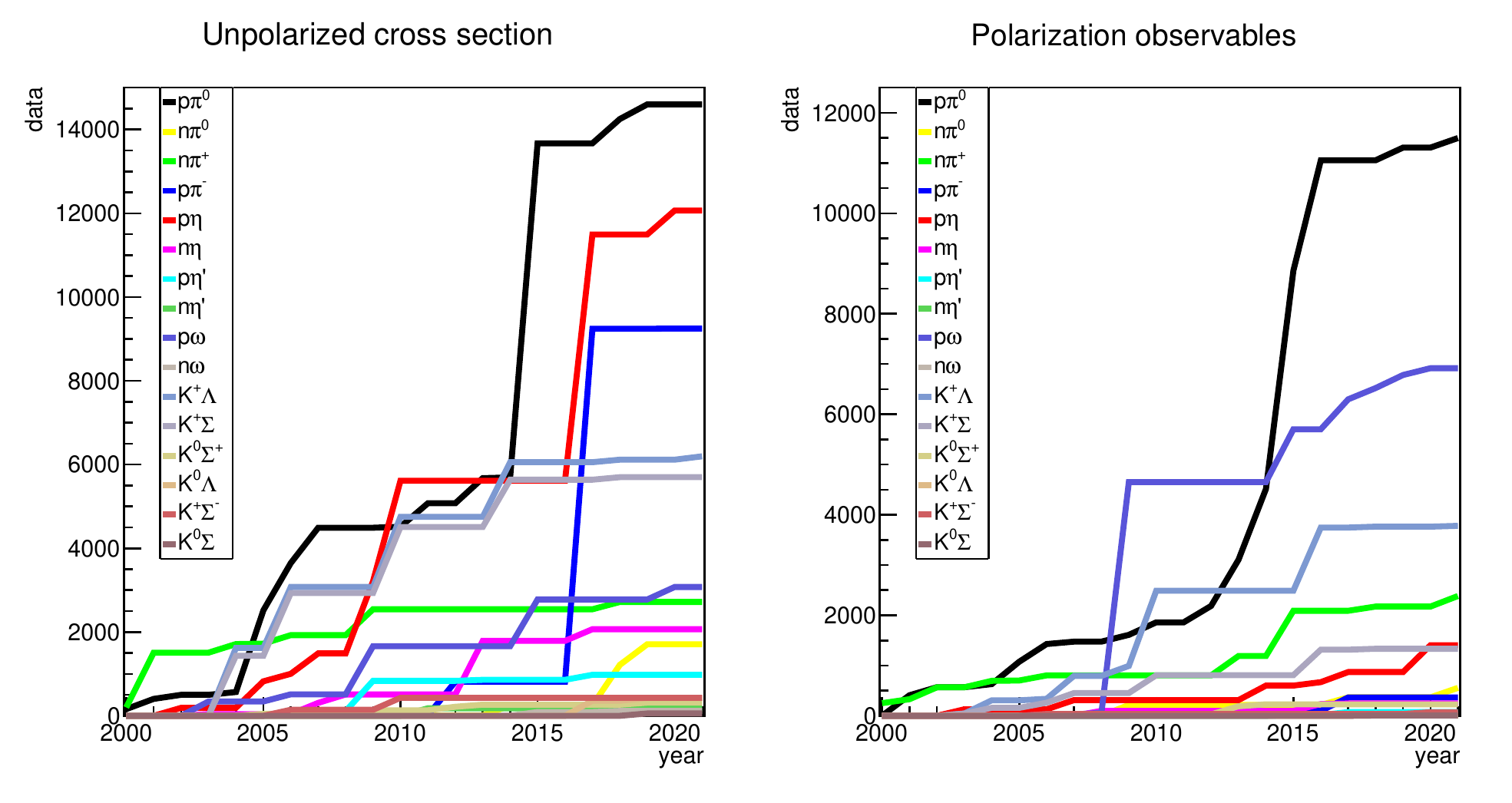}
\caption{Development of the existing database between the years 2000 - 2021 for various final states.  An overview is given for the unpolarized cross section (left) and all polarization observables (right). Note: Including all the existing data prior to the year 2000 will result in an offset on the y-axis for almost all curves.}
\label{fig:measurement:database}
\end{figure}

The aforementioned experimental facilities (see Section \ref{sec:experiments}) were used to conduct measurements of the unpolarized cross section as well as to perform measurements of polarization observables for different photoproduction reactions $\gamma N \to X$. Fig.~\ref{fig:measurement:database} gives an overview of the total number of measured data points within the time period of the last two decades (years 2000 - 2021) for single meson photoproduction reactions. The database was compiled based on the SAID database \cite{SAIDweb}, the database listed in \cite{Ireland:2019uwn} and from publications listed in the Tables \ref{tab:measSinglePion1} - \ref{tab:measStrangeness2}, that were not present in the first two databases. Since the measurement of the polarization observables requires more sophisticated experimental tools like polarized photon beams or polarized targets, it is more challenging to measure them. We therefore show the overview separately for the polarization observables and the unpolarized cross section.

The most datapoints exist for the $p\pi^0$ final state. The ratio between the polarization observables and the unpolarized cross section is, here, quite high with around 80\%. However, most other final states show a large imbalance between the two categories, most notably so for the $p\pi^-$ final state with a ratio of only 4\% and the $p\eta$ final state with a ratio of only 11\%. The possibility of measuring the recoil polarization leads to relatively high ratios for strangeness photoproduction reactions like the $K^+\Lambda$ final state with a ratio of 60\%. It is remarkable that the $p\omega$ final state has more data points for the polarization observables (due to the \textit{SDME} measurements) than for the unpolarized cross section. However, one should keep in mind that a lot more polarization observables are needed for a complete experiment in case of vector mesons in comparison to pseudoscalar mesons. 

In the following subsections, we give an overview of all the available data for each single meson photoproduction channel and highlight exemplary some of the measured data. 

\subsubsection{Pion Photoproduction}
\label{sec:measurements:PionPhotoproduction}
Pion photoproduction can be considered as the best measured final state. First measurements of the $\Delta(1232)$-Resonance were performed for example in 1965 in Bonn \cite{Freytag:1965ifa} and since then a large amount of measurements were published. The measurements performed after 2000 can be found in Tab.~\ref{tab:measSinglePion1} and Tab.~\ref{tab:measSinglePion2}. It directly becomes obvious that most of the experiments concentrated on the reaction $\gamma p \to p \pi^0$, where several papers were published by experiments at the MAMI accelerator in Mainz, Germany or the ELSA accelerator in Bonn, also Germany. Here, it is worth mentioning that not only measurements of cross sections were performed, but also several polarization observables have been extracted. In the following, an overview of the status of measurements for pion photoproduction is given.

\begin{figure}[h]
	\centering
	\includegraphics[width=\textwidth]{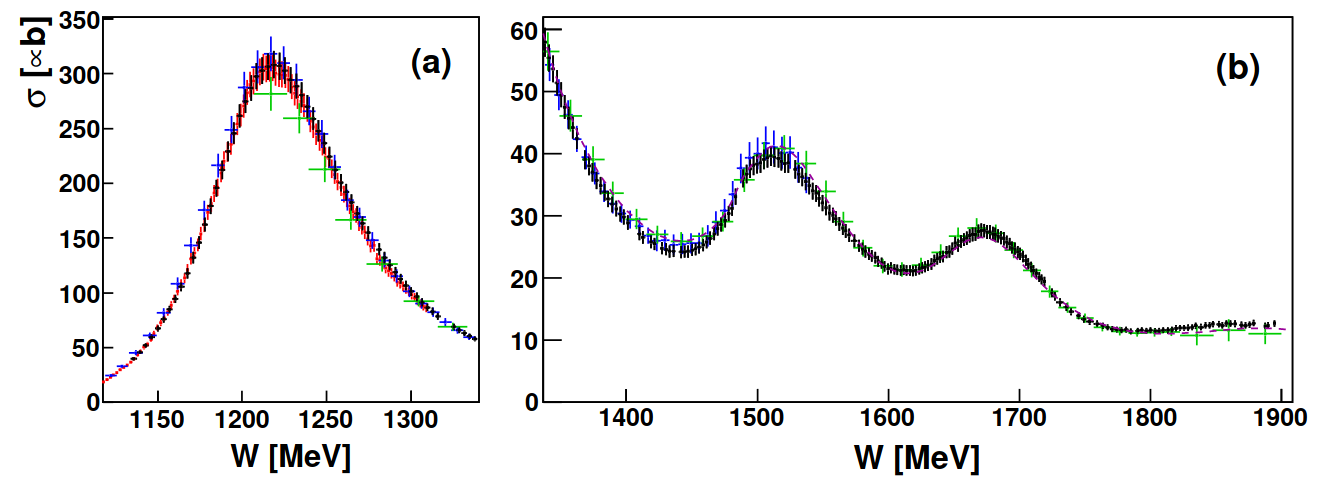}
	\caption{A high precision measurement of the total cross section for $\gamma p \to p\pi^0$ as measured by the Crystal Ball/TAPS collaboration (black points). The data are compared to earlier results of A2 \cite{Hornidge:2012ca} (red points), \cite{Beck:2002tm,Beck:2006ye} (blue points) and CBELSA \cite{Bartholomy:2004uz,vanPee:2007tw} (green points). \cite{Adlarson:2015byy}.}
	\label{fig:M:ppi0_crossection_mami}
\end{figure}

A large amount of the measurements listed in Tab.~\ref{tab:measSinglePion1} and Tab.~\ref{tab:measSinglePion2} are focused on the determination of the total or differential cross section. The cross section was considered to be well known, however, the recent measurements of Crystal Ball/TAPS \cite{Adlarson:2015byy} are noteworthy. In their recent measurement, they were able to cover the energies up to $W=$1.9~GeV with an outstanding precision and energy bins of just about 4~MeV width. The total cross section, covering the $\Delta$ as well as the second and third resonance region, can be seen in Fig.~\ref{fig:M:ppi0_crossection_mami}. With these measurements, signals of higher partial waves became visible already at lower energies. This demonstrates that energy and angular coverage are important as well as a high precision to be able to extract additional information.  However, for such a precise measurement the investigation and handling of systematic uncertainties becomes especially important, see also Wunderlich et al.~\cite{Wunderlich:2016imj}.

\begin{figure}
	\centering
	\includegraphics[width=0.52\textwidth]{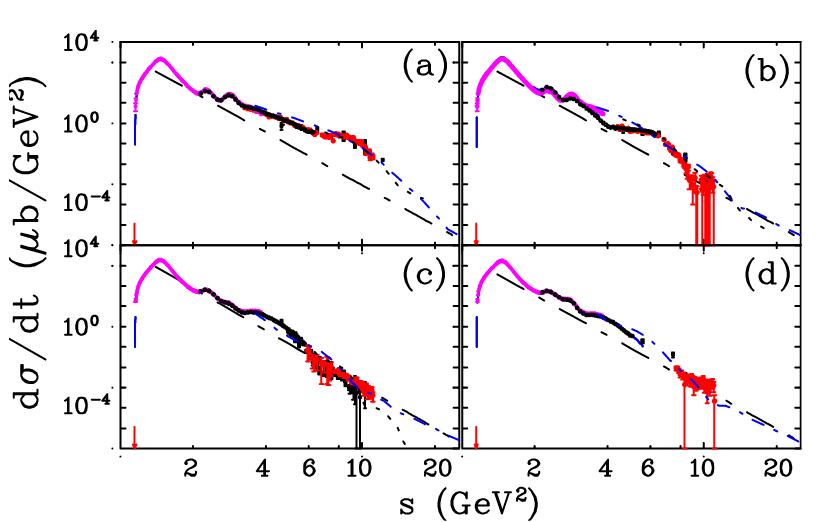}
	\includegraphics[width=0.45\textwidth]{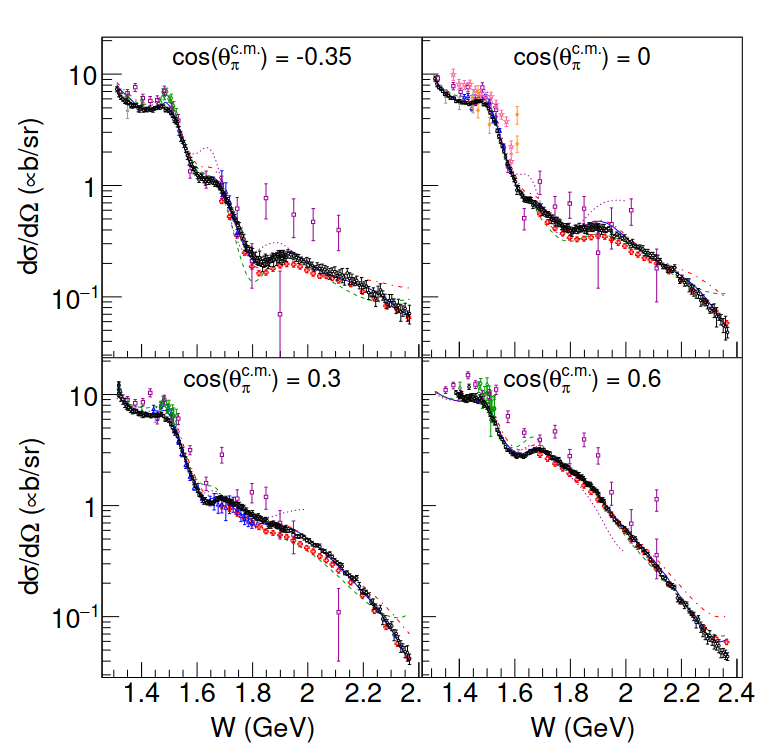}
	\caption{Cross section measurements by the CLAS collaboration. Left: The differential cross section $\frac{d\sigma}{dt}$ for the $p\pi^0$ final state for four different angles: (a) $50^\circ$, (b) $70^\circ$, (c) $90^\circ$ and (d) $110^\circ$. Taken from \cite{Kunkel:2017src}. Right: Cross section data (black circles) for the reaction $\gamma n \to p \pi^-$. Taken from \cite{Mattione:2017fxc}. \APS{\cite{Mattione:2017fxc}}{2017}}
	\label{fig:M:ppi0_crossection_clas}\label{fig:M:ppi-_crossection_clas}
\end{figure}

In another recent measurement of the total cross section the CLAS collaboration \cite{Kunkel:2017src} was able to extend the energy range of the previously measured data up to 5.425~GeV, see Fig.~\ref{fig:M:ppi0_crossection_clas}, left. For this, they investigated the decay mode of the $\pi^0$ into charged particles $\pi^0 \to e^+ e^- \gamma$, even though the Dalitz decay mode has only a branching ratio of about 1\% \cite{Zyla:2020zbs}. The achieved precision is remarkable and it paves the way to measurements at higher energies, which can be performed with the upgraded CEBAF accelerator.

Measurements of reactions off the neutron are much harder to perform compared to reactions off the proton. This fact is also visible in the amount of published datasets in Tab.~\ref{tab:measSinglePion1} and Tab.~\ref{tab:measSinglePion2}. However, it is worth mentioning that several interesting measurements were published within the last years. For example the Crystal Ball/TAPS experiment also measured the cross section for the neutral reaction $\gamma n \to n \pi^0$ and were able to publish two remarkable datasets \cite{Briscoe:2019cyo,Dieterle:2018adj}. These measurements were performed with deuteron target and in order to extract the neutral reaction, a comparison to data off proton targets is necessary. This also allows to shed light on the final state interactions by comparing reactions of bound and of free protons \cite{Dieterle:2018adj}.


Reactions with charged pions can be successfully extracted for example by the CLAS experiment. They were able to extract the cross section over a wide energy range \cite{Dugger:2009pn, Mattione:2017fxc}. An example is shown for the reaction $\gamma n \to p \pi^-$ in Fig.~\ref{fig:M:ppi-_crossection_clas}, right.

As already discussed in detail in Sec.~\ref{sec:formalisms:SingleMesonPhotoproduction}, by using polarized photon beams, polarized targets or by measuring the polarization of the recoiling nucleon, different polarization observables can be extracted. Old measurements mostly concentrated on the measurement of the cross section, but in the last years the data basis of double polarization observables increased tremendously. Here, we will present a few highlights of these measurements.

\begin{figure}[tb]
	\centering
	\includegraphics[width=0.8\textwidth]{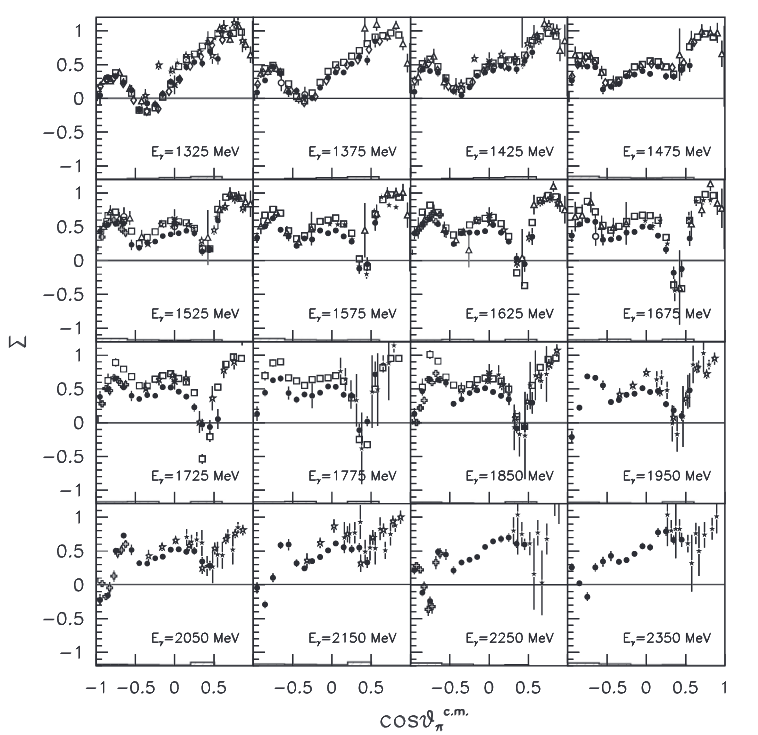}
	\caption{The beam asymmetry $\Sigma$ from the BGOegg experiment (closed circles) \cite{Muramatsu:2019sro} and from CLAS (open squares) \cite{Dugger:2013crn} for the reaction $\gamma p \to p \pi^0$. The other recent measurements are: CBELSA/TAPS (open triangles) \cite{Sparks:2010vb}, GRAAL (open diamond) \cite{Bartalini:2005wx}, LEPS (open cross) \cite{Sumihama:2007qa}. Further details can be found in \cite{Muramatsu:2019sro}. \APS{\cite{Muramatsu:2019sro}}{2019}}
	\label{fig:M:ppi0_beamasymmetry_LEPS}
\end{figure}

As a first example, it is worth mentioning the beam asymmetry measurements by the CLAS collaboration from 2013 \cite{Dugger:2013crn} and BGOegg at LEPS2 from 2019 \cite{Muramatsu:2019sro}. By using a linearly polarized photon beam on an unpolarized proton target, the beam asymmetry could be extracted over a wide energy range with a low statistical uncertainty. The results of both measurements can be found in Fig.~\ref{fig:M:ppi0_beamasymmetry_LEPS}. Both measurements could extract the beam asymmetry with small statistical uncertainties and the agreement is remarkable in most energy bins. Only in the highest energy bins of the CLAS measurement around $E_\gamma = 1700$~MeV systematic deviations are visible, which still need to be understood in detail. However, both experiments used opposite methods to extract the beam asymmetry. For example, for the production of the linearly polarized photon beam, Compton backscattering was utilized for the BGOegg experiment. In contrast, the CLAS experiment used bremsstrahlung off a diamond radiator to produce their linearly polarized photon beam. Therefore the BGOegg experiment can provide data points even at high photon energies. Also the method of identification for the meson is different. While the BGOegg experiment is able to measure both decay photons of the $\pi^0$ and reconstruct the meson, the CLAS experiment can measure the recoiling proton with high accuracy and determine the $\pi^0$ via a missing mass analysis. Since the CLAS detector setup is well-suited to detect charged particles, other final states like $n\pi^-$ can be extracted at the same time and are also published in the same paper \cite{Dugger:2013crn}.

\begin{figure}[tb]
	\centering
	\includegraphics[width=\textwidth]{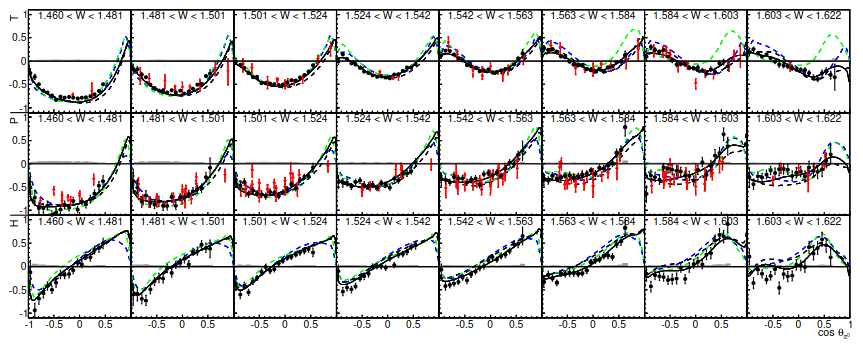}
	\caption{The double polarization observables T, P and H for $\gamma p \to p \pi^0$ measured by the CBELSA/TAPS experiment \cite{Hartmann:2014mya}. For these measurements, a linearly polarized photon beam was used in combination with a transversely polarized target. }
	\label{fig:M:ppi0_TPH_CBELSATAPS}
\end{figure}

The first observable in the era of double polarization measurements was the publication of the observable $G$, which was published by the CBELSA/TAPS collaboration \cite{Thiel:2012yj}. Here, a selection of four energy bins was published and it was demonstrated that the unpolarized carbon content in the polarized target material could be properly handled in the analysis. The full data set was published a few years later \cite{Thiel:2016chx}.

One of the first high statistics measurements of multiple double polarization observables is the publication of the observable $T$, $P$, $H$ by the CBELSA/TAPS collaboration \cite{Hartmann:2015kpa,Hartmann:2014mya}. Here, it was possible with just one experiment to determine three different observables simultaneously (compare for example \cite{Hartmann:2015kpa}). A selection of the measured data is shown in Fig.~\ref{fig:M:ppi0_TPH_CBELSATAPS}. The measurement provided a fine energy binning in combination with an almost complete coverage of the angular range and high statistics. This increases the current data base for these observables by a large factor, which can be seen by the comparison to the previous measurement in Fig.~\ref{fig:M:ppi0_TPH_CBELSATAPS}. It is worth mentioning at this point that the recoil polarization observable $P$ can also be measured by a double polarization measurement using a transversely polarized target and linearly polarized photons, see Sec.~\ref{sec:formalisms:SingleMesonPhotoproduction}. Since the determination of the polarization of the recoiling nucleon is not trivial, a double polarization measurement can provide a better method. 

\begin{figure}[tb]
	\centering
	\includegraphics[width=0.53\textwidth]{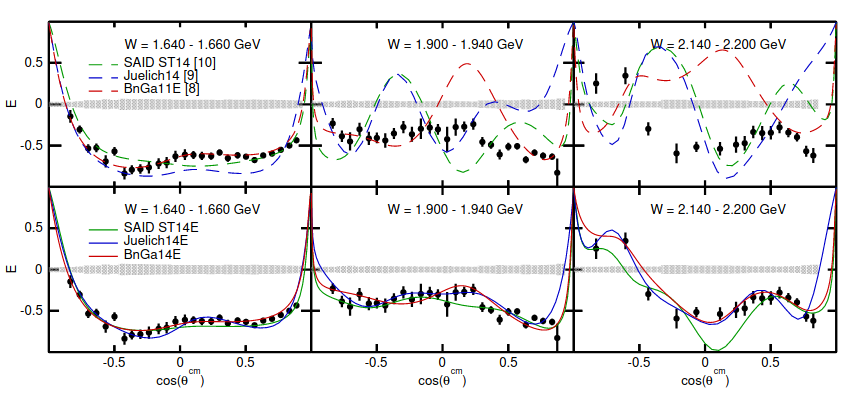}
	\includegraphics[width=0.43\textwidth]{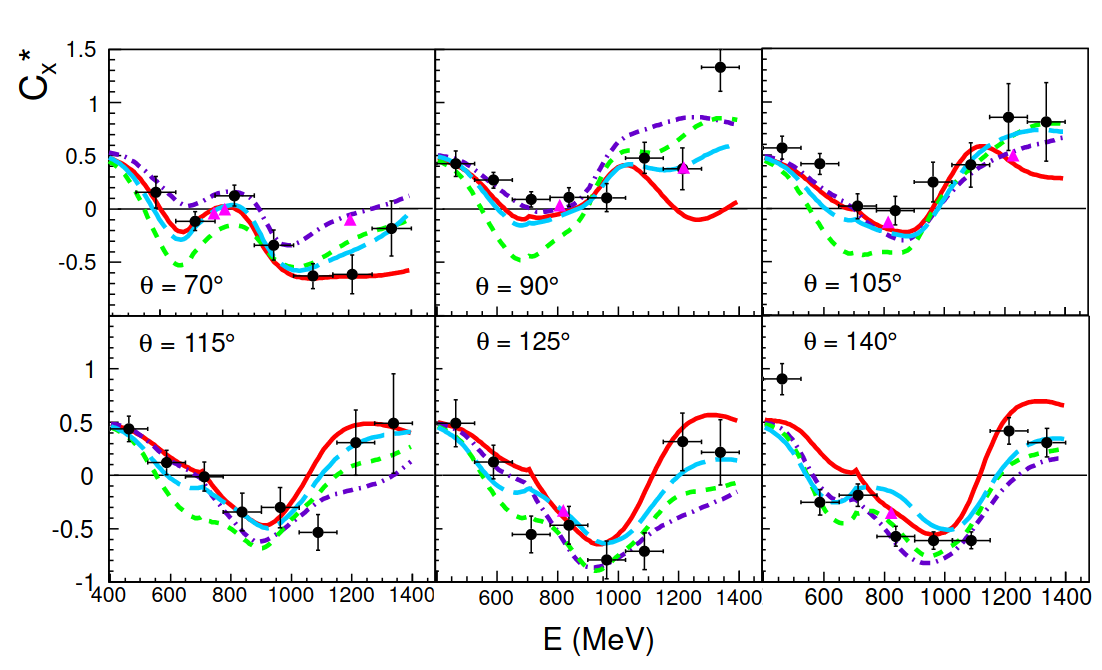}
	\caption{Left: The double polarization observable E for $\gamma p \to n \pi^+$  measured with a circularly polarized beam on a transversely polarized target by the CLAS experiment \cite{Strauch:2015zob}. Right: First measurement of the observable $C_x^*$ exploiting the polarization of the proton by a recoil polarimeter for the reaction $\gamma p \to p  \pi^0$ with the Crystal Ball/TAPS experiment \cite{Sikora:2013vfa}. \APS{\cite{Sikora:2013vfa}}{2013}}
	\label{fig:M:npi+_E_CLAS}	\label{fig:M:ppi0_Cx_MAMI}
\end{figure}

One of the well-measured double polarization observable is the observable E, which is closely related to the helicity asymmetry $\sigma_{3/2} - \sigma_{1/2}$, see for example \cite{Gottschall:2013uha,Gottschall:2019pwo}. This asymmetry has already been published by the GDH experiment at MAMI in 2002 and 2004 \cite{Ahrens:2002gu,Ahrens:2004pf,Ahrens:2006gp}, but a normalization, which leads to the observable E had been missing. This was first done by the CBELSA/TAPS collaboration for the $p\pi^0$ final state \cite{Gottschall:2013uha, Gottschall:2019pwo}. In the last years, the data base for this observable has been extend for all other final states by measurements from the CLAS collaboration for $n\pi^+$ \cite{Strauch:2015zob} and $p\pi^-$ \cite{Ho:2017kca} and by the Crystal Ball/TAPS experiment for $n\pi^0$ \cite{Dieterle:2017myg}. An example of one of these measurement is shown in Fig.~\ref{fig:M:npi+_E_CLAS}, left. Here, the observable E is shown in a measurement of the CLAS collaboration in comparison to predictions by different partial wave analyses (top row) as well as fits by the different models (bottom row). Due to several data sets, which have been included in the analyses at low energies, the predictions are very close to the data in the first energy bin. However, at the two higher bins, the predictions are vastly different and do not describe the data at all. By including the data points in the different analysis and performing a new fit, the data can be accurately described.


The measurement of recoil polarization observables has always been a challenge. There are two different methods used in the last 20 years. The first one was realized by using electroproduction with very low values of $Q^2$, resulting in final states indistinguishable from photoproduction \cite{Luo:2011uy}. Here, the High Momentum Spectrometer (HMS), located in Hall C at Jefferson Lab, were utilized to identify the decay products at selected angles and energies. The second method uses a recoil polarimeter, which can be inserted into an existing detector system. A first measurement utilizing this method at the Crystal Ball experiment at MAMI has been performed and published \cite{Sikora:2013vfa}. The data, which can be seen in Fig.~\ref{fig:M:ppi0_Cx_MAMI}, right, already allow first constraints for the different models. Still, the accuracy is not sufficient for any detailed studies, but this first experiment proves that this method is feasible and can be used for the determination of further recoil observables in the future.

\begin{figure}[tp]
	\includegraphics[width=0.5\textwidth]{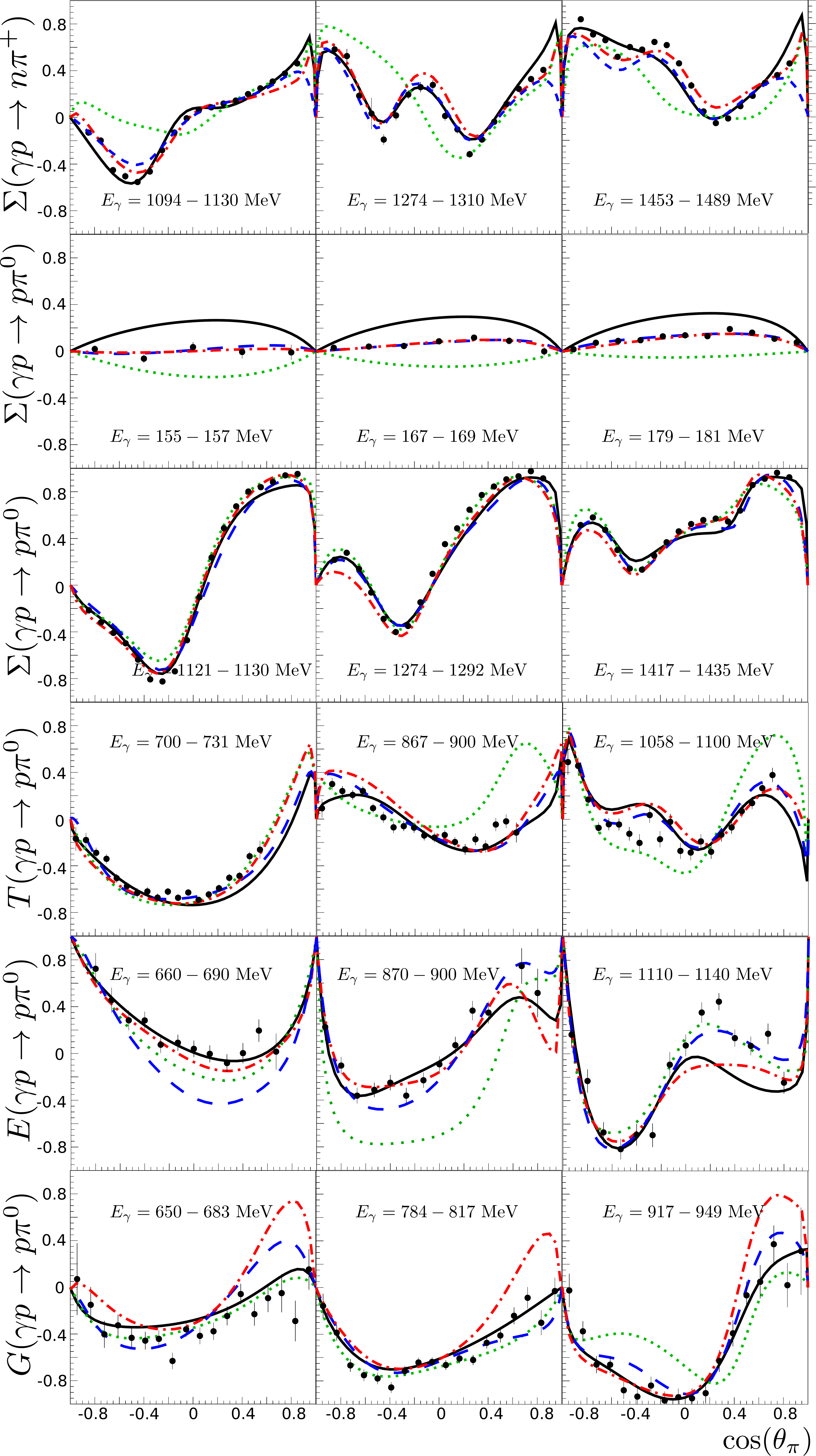}
	\includegraphics[width=0.5\textwidth]{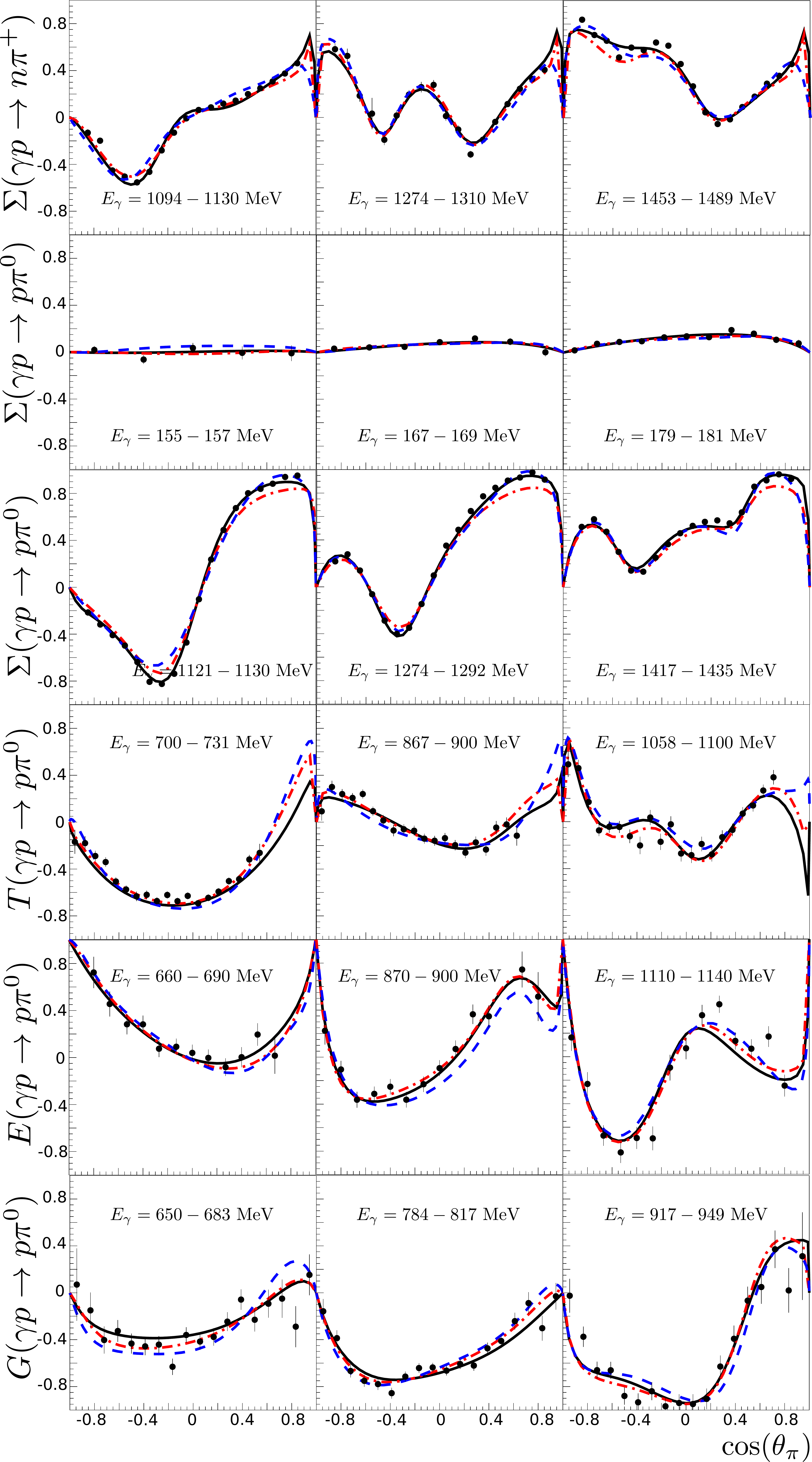}
	\caption{Selected data sets from \cite{Dugger:2013crn,Hornidge:2012ca,Hartmann:2015kpa,Gottschall:2013uha,Gottschall:2019pwo,Thiel:2012yj,Thiel:2016chx}. Left: Compared to predictions from the four different partial wave analyses: BnGa2011-02 (black solid line), J\"uBo2015B (blue dashed line), MAID2007 (green dotted line) and SAID CM12 (red dashed dotted line). Right: Compared to fits to the data:  BnGa (black  solid  line),  J\"uBo (blue  dashed line) and SAID (red dashed dotted line). Figures adapted from \cite{Beck:2016hcy}.}
	\label{fig:M:Npi_interpretation_anisovich}
\end{figure}

\begin{figure}[tbp]
	\centering
	\includegraphics[width=0.47\linewidth]{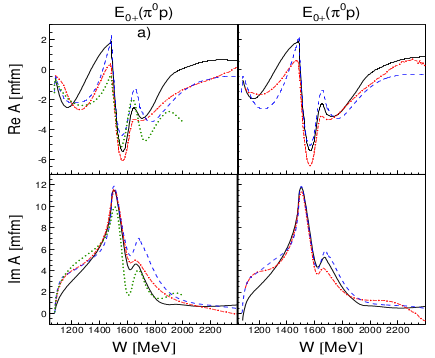}
	\includegraphics[width=0.47\linewidth]{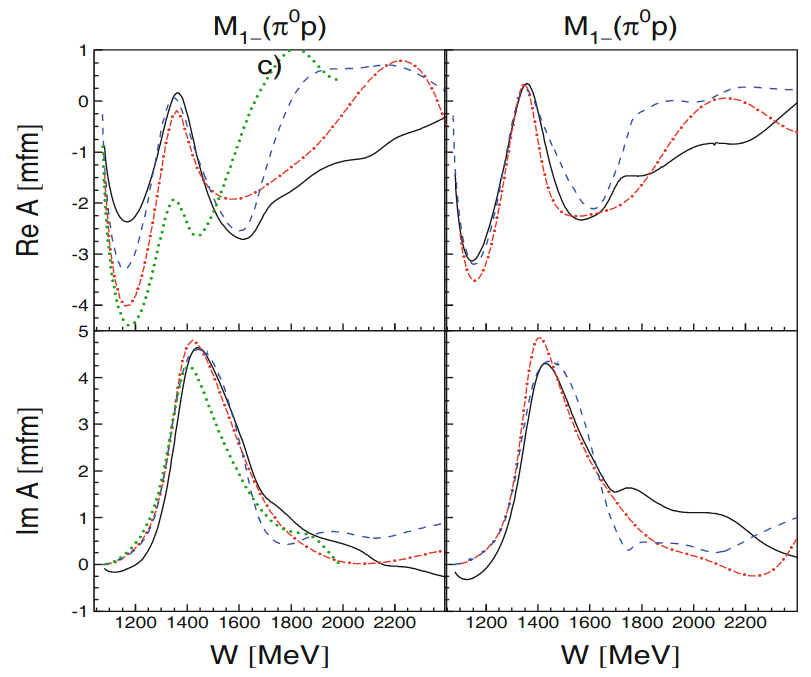}
	\caption{The real (top row) and imaginary (bottom row) part of the $E_{0+}$ and $M_{1-}$ multipoles are shown for the $p\pi^0$ final state before (left column) and after (right column) including the newly measured polarization observables in pion photoproduction. Figure adapted from \cite{Beck:2016hcy}.}
	\label{fig:M:Npi_interpretation_anisovich_multipoles}
\end{figure}

With this increasing amount of new data sets for pion photoproduction, discussion started how to properly interpret these results. In a combined effort between the different partial wave analysis groups, BnGa (see Sec.~\ref{sec:PWA:BnGaModel}), SAID (see Sec.~\ref{sec:PWA:SAIDModel}), MAID (see Sec.~\ref{sec:PWA:MAIDModel}) and J\"uBo (see Sec.~\ref{sec:PWA:JueBoModel}) the impact of the new measurements on the partial waves was investigated \cite{Anisovich:2015gia}. For this, all multipoles of the different models were compared without the new data sets. Afterwards, a selection of new data was chosen and incorporated into the models. The data sets, which were chosen for this, are shown in Fig.~\ref{fig:M:Npi_interpretation_anisovich}, on the left side compared to the predictions of the different models, on the right with the fits by the different PWA groups. As it can be expected, by fitting the new data sets, all three different groups were able to describe the data satisfactorily. However, the most interesting part of this exercise is the comparison of the multipoles to determine the impact of the new measurements. For this, a comparison of all multipoles before and after including the new data sets can be found in \cite{Beck:2016hcy}. An example of the $M_{1-}$ multipole can be found in Fig.~\ref{fig:M:Npi_interpretation_anisovich_multipoles}. It becomes obvious that the visible discrepancies in the real part of the $M_{1-}$ multipole are significantly reduced, once the new data sets are included. For further details see also Sec.~\ref{sec:interpretation}. 

\begin{figure}[tp]
	\centering
	\includegraphics[width=0.55\textwidth]{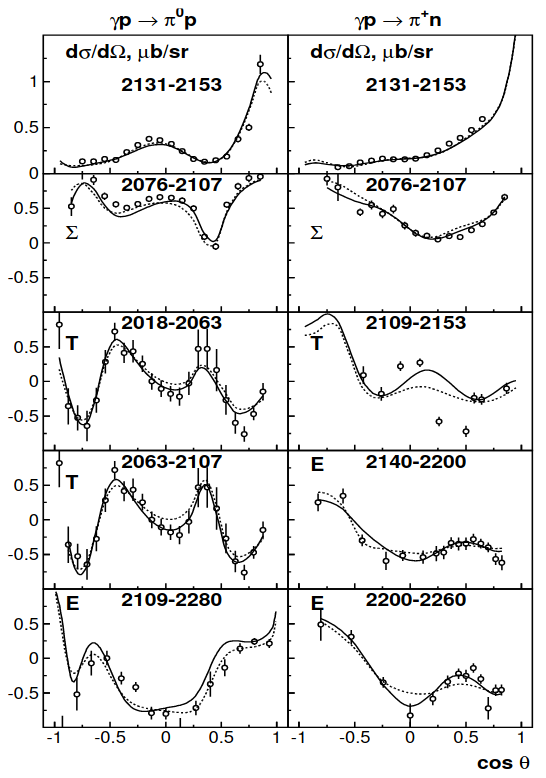}
	\includegraphics[width=0.42\textwidth]{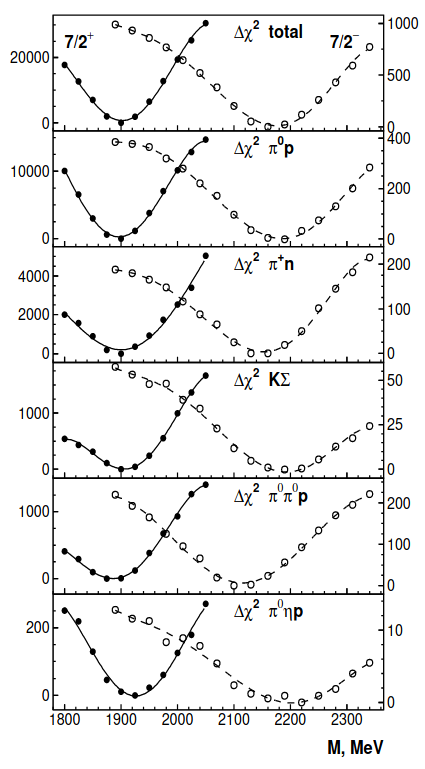}
	\caption{Left: Differential  cross  section \cite{Crede:2011dc,Dugger:2009pn}, target asymmetry T \cite{Hartmann:2014mya,Hartmann:2015kpa}, beam asymmetry \cite{Dugger:2013crn} and helicity asymmetry E \cite{Gottschall:2013uha,Strauch:2015zob} for the final states $p\pi^0$ and $n\pi^+$ compared to two fits from the BnGa group \cite{Anisovich:2015gia}: Best fit (solid curve) and fit without the $\Delta(2200)\frac{7}{2}^-$ (dotted curve). Right: Pseudo-$\chi^2$ distributions for mass scans in the $J^P = \frac{7}{2}^\pm$ partial waves. A clear signal for the $\Delta(1950)\frac{7}{2}^+$ could be found, but no parity partner at the same mass.  For further details see \cite{Anisovich:2015gia}.}
	\label{fig:M:delta2200_anisovich}
\end{figure}

In another interesting project, data sets from different experiments were combined in order to shed light on the question whether chiral symmetry is restored for the high-mass resonances. This would manifest itself in parity doublets, which are states of similar masses but with opposite parities. For further discussion see for example \cite{Glozman:1999tk,Cohen:2001gb}. In the past, several $\Delta$ resonances with different spins and parities were confirmed with masses around around 1.9~GeV, see \cite{Zyla:2020zbs}: $\Delta(1910) \frac{1}{2}^+$ and $\Delta(1900) \frac{1}{2}^-$, $\Delta(1920) \frac{3}{2}^+$ and $\Delta(1940) \frac{3}{2}^-$, $\Delta(1905) \frac{5}{2}^+$ and $\Delta(1930) \frac{5}{2}^-$, $\Delta(1950) \frac{7}{2}^+$ and $\Delta(2200) \frac{7}{2}^-$. The only state, which seems not to fit the pattern is the resonance $\Delta(2200) \frac{7}{2}^-$, which has a much higher mass. If chiral symmetry is indeed restored at this mass region, for each state, there would be a partner state with the same spin but parity of different sign around the same mass region. A combined search by the BnGa group was performed, utilizing several different observables from various experiments. For the reactions $\gamma p \to p \pi^0$ and $\gamma p \to n \pi^+$ the differential cross section and the observables $\Sigma$, $T$ and $E$ example distributions are shown in Fig.~\ref{fig:M:delta2200_anisovich}. Here, fits from the BnGa are shown with and without including the $\Delta(2200)\frac{7}{2}^-$. By including this resonance, the PWA group is able to describe the data points better, however, comparing the two solutions by the BnGa shows the sensitivity required to establish the resonance signal in multiple observables. To further investigate the probability of resonances in the $J^P = \frac{7}{2}^\pm$ partial waves, mass scans in these partial waves were performed using multichannel Breit-Wigner amplitudes \cite{Anisovich:2015gia}. In Fig.~\ref{fig:M:delta2200_anisovich}, right, the change of the resulting pseudo-$\chi^2$ of the total fit at different energies is shown for various final states. For the $J^P = \frac{7}{2}^+$ wave, a clear signature of the resonance is visible at the expected mass. When the same mass region is scanned for a $J^P = \frac{7}{2}^-$ resonance, the next indication of a resonance is visible around a mass of 2200~MeV. In total, no mass-degenerate  parity  partner  of the $\Delta(1950)\frac{7}{2}^+$ could be found in the data \cite{Anisovich:2015gia}.

\begin{table*}
\caption{A collection of polarization observable measurements for pion photoproduction. Hint: We do not distinguish between differential and total cross section data. Further details are given in the cited references. The symbols indicate which PWA (circle: BnGa-2019 \cite{Muller:2019qxg}, diamond: J\"uBo-2017 \cite{Ronchen:2018ury} and star: SAID-MA19  \cite{Briscoe:2019cyo}) use the data in their fits. \label{tab:measSinglePion1}}
\centering
\vspace*{5mm}
\begin{tabular}{clllc}
Observable & Energy range $E^{lab}_\gamma$ & Facility & Reference & Year\\
\hline
\multicolumn{5}{c}{$\gamma p \rightarrow p\pi^0$}\\
\hline
G & $\sim 650-2300$ MeV & CLAS & Zachariou et al. \cite{CLAS:2021udy} & 2021\\
E & $600-2310$ MeV & CBELSA/TAPS & Gottschall et al. \cite{Gottschall:2019pwo} $\circ$ $\star$ & 2021\\
$\sigma$, $\Sigma$ & $1300-2400$ MeV & LEPS2 BGOegg & Muramatsu et al. \cite{Muramatsu:2019sro} $\star$ & 2019 \\
$\sigma$ & $450-1400$ MeV & Crystal Ball/TAPS & Dieterle et al. \cite{Dieterle:2018adj}   & 2018 \\
$\sigma$ & $1275-5425$ MeV & CLAS & Kunkel et al. \cite{Kunkel:2017src} & 2018 \\
E, $\sigma_{1/2}$, $\sigma_{3/2}$ & $400-1450$ MeV & Crystal Ball/TAPS & Dieterle et al. \cite{Dieterle:2017myg} & 2017 \\
G & $617-1325$ MeV & CBELSA/TAPS & Thiel et al. \cite{Thiel:2016chx} $\circ$ $\diamond$ $\star$ & 2017 \\
$\Sigma$ & $320-650$ MeV & Crystal Ball/TAPS & Gardner et al. \cite{Gardner:2016irh} $\circ$ $\star$ & 2016 \\
T, F & $425-1445$ MeV &  Crystal Ball/TAPS & Annand et al. \cite{Annand:2016ppc} $\diamond$ $\star$ & 2016 \\
$\sigma$ & $218-1573$ MeV &  Crystal Ball/TAPS & Adlarson et al. \cite{Adlarson:2015byy} $\circ$ $\star$ & 2015 \\
T & $665-1900$ MeV & CBELSA/TAPS & Hartmann et al. \cite{Hartmann:2015kpa} $\circ$ $\diamond$ $\star$ & 2015 \\
P, H & $665-930$ MeV & CBELSA/TAPS & Hartmann et al. \cite{Hartmann:2015kpa} $\circ$ $\diamond$ $\star$ & 2015 \\
$\sigma$ & $\sim 140-190$ MeV & Crystal Ball/TAPS & Schumann et al. \cite{Schumann:2015ypa} $\star$ & 2015 \\
T, P, H & $\sim 665-930$ MeV & CBELSA/TAPS & Hartmann et al. \cite{Hartmann:2014mya} $\circ$  $\diamond$ $\star$ & 2014 \\
$C_x^\ast$ & $400-1400$ MeV & Crystal Ball/TAPS & Sikora et al. \cite{Sikora:2013vfa} $\diamond$ $\star$ & 2014 \\
E & $600-2300$ MeV & CBELSA/TAPS & Gottschall et al. \cite{Gottschall:2013uha} $\circ$ $\diamond$ $\star$ & 2014 \\
$\Sigma$ & $1102-1862$ MeV & CLAS & Dugger et al. \cite{Dugger:2013crn}$\circ$  $\diamond$ $\star$ & 2013 \\
$\Sigma$ & $144-180$ MeV & Crystal Ball/TAPS & Hornidge et al. \cite{Hornidge:2012ca} $\circ$ $\diamond$ $\star$ & 2013 \\
$\sigma$, $\Sigma$ & $163$ MeV & Crystal Ball/TAPS & Hornidge et al. \cite{Hornidge:2012ca}$\circ$  $\diamond$ $\star_\Sigma$ & 2013 \\
G & $620-1120$ MeV & CBELSA/TAPS & Thiel et al. \cite{Thiel:2012yj} $\circ$ $\diamond$ $\star$ & 2012 \\
P, $C_{x'}$, $C_{z'}$ & $1800-5600$ MeV &  GEp-III, GEp-2$\gamma$ & Luo et al. \cite{Luo:2011uy} $\diamond$ $\star$ & 2012 \\
$\sigma$ & $850-2500$ MeV & CBTAPS & Crede et al. \cite{Crede:2011dc} $\circ$ $\diamond$ $\star$ & 2011 \\
$\Sigma$ & $920-1680$ MeV & CBTAPS & Sparks et al. \cite{Sparks:2010vb} $\circ$ $\diamond$ $\star$ & 2010 \\
$\sigma$ & $144-800$ MeV & Crystal Ball/TAPS & Schumann et al. \cite{Schumann:2010js} $\diamond$ & 2010 \\
$\Sigma$ & $800-1400$ MeV & CBTAPS & Elsner at al. \cite{Elsner:2008sn} $\circ$ $\diamond$ $\star$ & 2009 \\
$\sigma$ & $675-2875$ MeV & CLAS & Dugger et al. \cite{Dugger:2007bt}$\circ$  $\diamond$ $\star$ & 2007 \\
$\sigma$ & $300-3000$ MeV & CBELSA & van Pee et al. \cite{vanPee:2007tw} $\circ$ & 2007 \\
$\sigma$, $\Sigma$ & $1500-2400$ MeV & LEPS & Sumihama et al. \cite{Sumihama:2007qa} $\diamond$ $\star$& 2007 \\
$\Sigma$ & $240-440$ MeV & TAPS & Beck et al. \cite{Beck:2006ye} $\circ$ $\diamond$ $\star$ & 2006 \\
$\sigma$ & $202-790$ MeV & TAPS & Beck et al. \cite{Beck:2006ye} $\diamond$ & 2006 \\
$\sigma$, $\Sigma$ & $550-1475$ MeV & GRAAL & Bartalini et al. \cite{Bartalini:2005wx} $\circ$ $\diamond$ $\star_\Sigma$ & 2005 \\
G & $340$ MeV & GDH & Ahrens et al. \cite{Ahrens:2005zq} $\circ$ $\diamond$ $\star$& 2005 \\
$\sigma$ & $300-3000$ MeV & CBELSA & Bartholomy et al. \cite{Bartholomy:2004uz} $\circ$  $\diamond$ $\star$ & 2005 \\
$\sigma$, $\sigma_{3/2} - \sigma_{1/2}$ & $\sim 145-450$ MeV & GDH & Ahrens et al. \cite{Ahrens:2004pf} $\circ$ $\diamond$ & 2004 \\
$\sigma$, $\sigma_{3/2} - \sigma_{1/2}$ & $550-790$ MeV & GDH & Ahrens et al. \cite{Ahrens:2002gu} $\circ$ $\diamond$ $\star_{\sigma_{3/2} - \sigma_{1/2}}$ & 2002 \\
P, $C_{x'}$, $C_{z'}$ & $2500-3100$ MeV & Hall A & Wijesooriya et al. \cite{Wijesooriya:2002uc} $\diamond$ $\star$ & 2002 \\
$\Sigma$ & $144-166$ MeV & TAPS & Schmidt et al. \cite{Schmidt:2001vg} $\diamond$ $\star$ & 2001 \\
$\sigma$, $\Sigma$ & $213-333$ MeV & LEGS & Blanpied et al. \cite{Blanpied:2001ae} $\circ$ $\diamond$ $\star_\Sigma$ & 2001 \\
$\Sigma$ & $500-1100$ MeV & Yerevan & Adamian et al. \cite{Adamian:2000yi} $\circ$ $\diamond$ $\star$ & 2001 \\
$\sigma$, $\Sigma$ & $270-420$ MeV & DAPHNE & Beck et al. \cite{Beck:1999ge}  &2000 \\
\hline
	\end{tabular}
\end{table*}

\begin{table*}
	\caption{Continuation of Tab.~\ref{tab:measSinglePion1}: A collection of polarization observable measurements for pion photoproduction. Hint: We do not distinguish between differential and total cross section data. Further details are given in the cited references. The symbols indicate which PWA (circle: BnGa-2019 \cite{Muller:2019qxg}, diamond: J\"uBo-2017 \cite{Ronchen:2018ury} and star: SAID-MA19  \cite{Briscoe:2019cyo}) use the data in their fits.\label{tab:measSinglePion2}}
	\centering
	\vspace*{5mm}
	\begin{tabular}{clllc}
		Observable & Energy range $E^{lab}_\gamma$ & Facility & Reference & Year\\
		\hline
		\multicolumn{5}{c}{$\gamma n \rightarrow n\pi^0$}\\
		\hline
		$\Sigma$ & $390-610$ MeV & Crystal Ball/TAPS & C. Mullen et al. \cite{Mullen:2021myn} $\star$ & 2021 \\
		$\sigma$ & $290-813$ MeV & Crystal Ball/TAPS & Briscoe et al. \cite{Briscoe:2019cyo} $\star$ & 2019\\
		$\sigma$ & $450-1400$ MeV & Crystal Ball/TAPS & Dieterle et al. \cite{Dieterle:2018adj} & 2018 \\
		E, $\sigma_{1/2}$, $\sigma_{3/2}$ & $450-1400$ MeV & Crystal Ball/TAPS & Dieterle et al. \cite{Dieterle:2017myg} & 2017 \\
		$\sigma$ & $450-1400$ MeV & Crystal Ball/TAPS & Dieterle et al. \cite{A2:2014fca} $\circ$ & 2014 \\
		$\Sigma$ & $600-1500$ MeV & GRAAL & DiSalvo et al. \cite{DiSalvo:2009zz} $\circ$ $\star$ & 2009 \\
		\hline
		\multicolumn{5}{c}{$\gamma p \rightarrow n\pi^+$}\\
		\hline
		G & $\sim 650-2300$ MeV & CLAS & Zachariou et al. \cite{CLAS:2021udy} & 2021\\
		$\sigma$, $\Sigma$ & $1500-2950$ MeV & LEPS & Kohri et al. \cite{Kohri:2017kto}  $\star$ & 2018 \\
		$E$ & $350-2370$ MeV & CLAS & Strauch et al. \cite{Strauch:2015zob} $\circ$  $\diamond$ $\star$ & 2015 \\
		$\Sigma$ & $1102-1862$ MeV & CLAS & Dugger et al. \cite{Dugger:2013crn} $\circ$ $\diamond$ $\star$ & 2013 \\
		$\sigma$ & $725-2875$ MeV & CLAS & Dugger et al. \cite{Dugger:2009pn} $\circ$ $\diamond$ $\star$ & 2009 \\
		$\sigma$, $\sigma_{3/2} - \sigma_{1/2}$ & $450-790$ MeV & GDH & Ahrens et al. \cite{Ahrens:2006gp} $\circ$ $\diamond$ $\star$ & 2006 \\
		G & $340$ MeV & GDH & Ahrens et al. \cite{Ahrens:2005zq} $\circ$ $\diamond$ $\star$& 2005 \\
		$\sigma$ & $1100-5535$ MeV & E94-104 (Hall A) & Zhu et al. \cite{Zhu:2004dy} $\diamond$ & 2005 \\
		$\sigma$, $\sigma_{3/2} - \sigma_{1/2}$ & $180-450$ MeV & GDH & Ahrens et al. \cite{Ahrens:2004pf} $\circ$ $\diamond$ $\star$ & 2004 \\
		$\Sigma$ & $800-1500$ MeV & GRAAL & Bartalini et al. \cite{Bartalini:2002cj} $\circ$ $\diamond$ $\star$ & 2002 \\
		$\sigma$, $\Sigma$ & $213-333$ MeV & LEGS & Blanpied et al. \cite{Blanpied:2001ae} $\circ$ $\diamond$ $\star_\Sigma$ & 2001 \\
		$\sigma$ & $300-2000$ MeV & ATHOS (Bonn) & Dannhausen et al. \cite{Dannhausen:2001yz} $\circ$ $\diamond$ $\star$ & 2001 \\
		$\Sigma$ & $550-1100$ MeV & GRAAL & Ajaka et al. \cite{Ajaka:2000rj} $\circ$  $\diamond$ $\star$ & 2000 \\
		$\sigma$ & $250-400$ MeV & EXP & Branford et al. \cite{Branford:1999cp} $\diamond$ $\star$ & 2000 \\
		$\sigma$, $\Sigma$ & $270-420$ MeV & DAPHNE & Beck et al. \cite{Beck:1999ge} $\circ$ $\diamond$ $\star$ &2000 \\
		\hline
		\multicolumn{5}{c}{$\gamma n \rightarrow p\pi^-$}\\
		\hline
		$\sigma$ & $149-162$ MeV & PIONS@MAX-lab  & Briscoe et al. \cite{Briscoe:2020qat} $\star$ & 2020 \\
		E & $700-2400$ MeV & CLAS & Ho et al. \cite{Ho:2017kca} $\star$ & 2017 \\
		$\sigma$ & $445-2510$ MeV & CLAS & Mattione et al. \cite{Mattione:2017fxc} $\star$ & 2017 \\
		$\sigma$ & $1050-3500$ MeV & CLAS & Chen et al. \cite{Chen:2012yv} $\circ$ & 2012 \\
		$\sigma$ & $301-455$ MeV & GDH & Briscoe at al. \cite{Briscoe:2012ni} $\star$ & 2012 \\
		$\Sigma$ & $700-1500$ MeV & GRAAL & Mandaglio et al. \cite{Mandaglio:2010sg} $\circ$ & 2010 \\
		$\sigma$ & $1000-3500$ MeV & CLAS & Chen et al. \cite{Chen:2009sda} $\circ$ & 2009 \\
		$\sigma$ & $1100-5535$ MeV & E94-104 & Zhu et al. \cite{Zhu:2004dy} & 2005 \\
		
		\hline
	\end{tabular}
\end{table*}

\subsubsection{Eta and Eta' Photoproduction}
\label{sec:measurements:EtaPhotoproduction}
The $\eta$ meson as well as the $\eta'$ meson have an isospin of $I=0$, which allows the exclusive study of $N^*$ resonances through the $N\eta$ and $N\eta'$ final states. A detailed review about these final states is given in \cite{Krusche:2014ava}.

\begin{figure}[p]
\begin{minipage}{0.46\textwidth}
\includegraphics[width=0.89\textwidth]{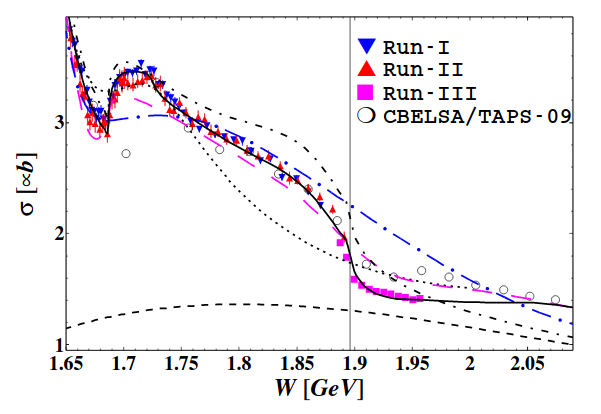}
\\[3mm]
\includegraphics[width=0.89\textwidth]{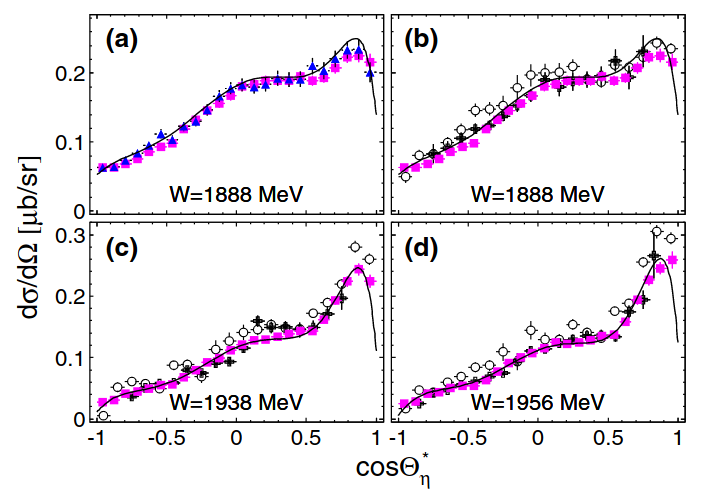}
\end{minipage}
\begin{minipage}{0.46\textwidth}
\includegraphics[width=\textwidth]{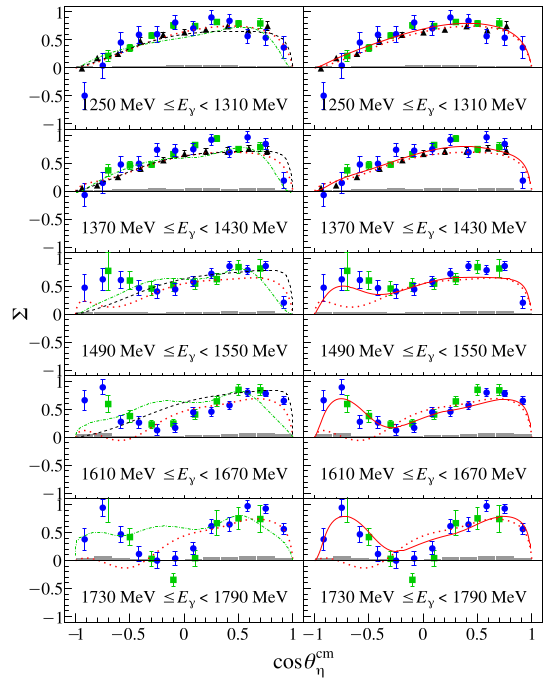}
\end{minipage}
	\begin{center}
	\includegraphics[width=0.8\textwidth]{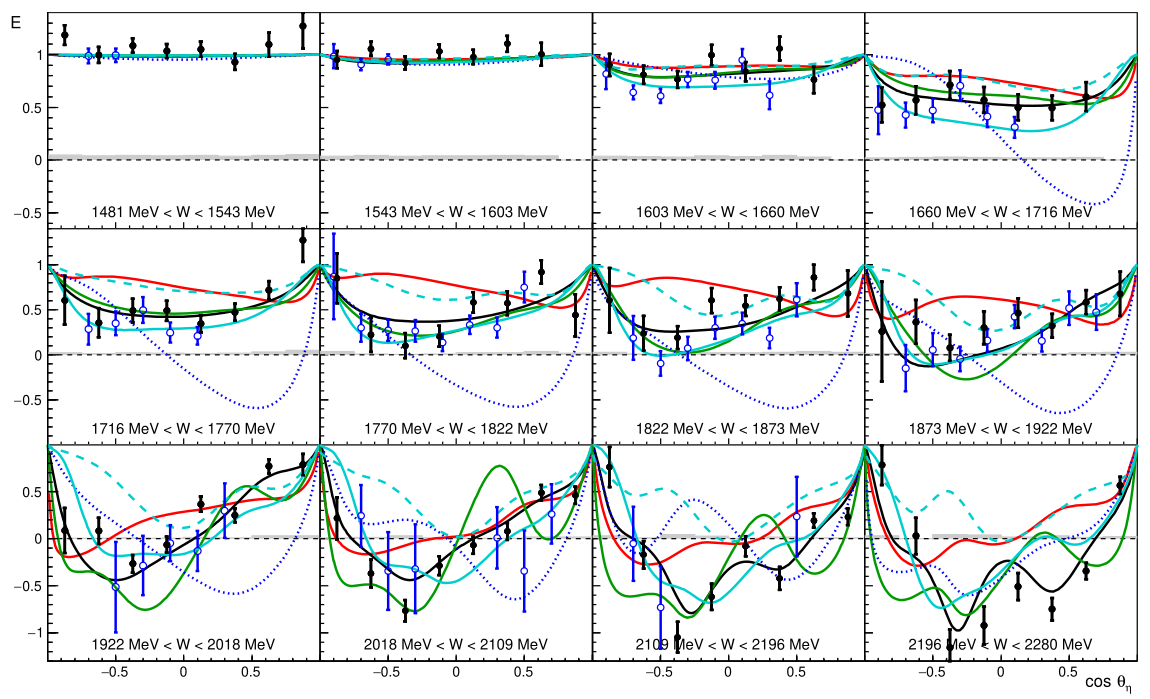}
	\end{center}
	\caption{Top left: The total ($\sigma$) and differential cross section ($\frac{d\sigma}{d\Omega}$) data are shown for CBELSA/TAPS (black open circles), CLAS (black crosses) and three different data sets (red triangles, blue triangles and magenta squares) of the Crystal Ball/TAPS experiment \cite{Kashevarov:2017kqb}. Top right: Four selected energy bins are shown for the beam asymmetry ($\Sigma$) for GRAAL (black triangles), CLAS (green squares) and CBELSA/TAPS data \cite{CBELSA/TAPS:2020yam}. Bottom: The double polarization observable E is shown for CLAS (blue, open circles) and CBELSA/TAPS (black points) data \cite{Muller:2019qxg}. All data are shown for the reaction $\gamma p \to p \eta$.}
	\label{fig:M:peta_CS_S_E}
\end{figure}

In $p\eta$ photoproduction, the unpolarized cross section is the most often and accurately measured observable. A full list of all measurements starting from year 2000 is given in Tab.~\ref{tab:measSingleEta}. The most precise measurement around the $N(1535)\frac{1}{2}^-(S_{11})$ resonance was performed by Krusche et al. \cite{Krusche:1995nv} and above it with a remarkable binning of up to 6.5~MeV width by the Crystal Ball/TAPS experiment (see Fig.~\ref{fig:M:peta_CS_S_E}) up to an incoming photon energy of 1.6 GeV \cite{Kashevarov:2017kqb}. Higher energies are covered by the CBELSA/TAPS \cite{Crede:2009zzb}, the BGOOD \cite{Alef:2019imq} and the CLAS \cite{Dugger:2002ft,Williams:2009yj,Hu:2020ecf} data. The LEPS experiment contributed to the database with precise data for the backward angles ($1<\cos\theta_{\eta}^{cms}<0.6$) for the photon energy range of 1.6 - 2.4 GeV \cite{Sumihama:2009gf}. Overall, there is a good consistency between all the data sets. This is remarkable since different $\eta$ decay modes were used to reconstruct the $\eta$ meson. While the neutral decay modes $\eta\to \gamma\gamma$ and $\eta\to \pi^0\pi^0\pi^0$ were measured with the GRAAL \cite{Bartalini:2007fg}, CBELSA/TAPS \cite{Crede:2009zzb} and Crystal Ball/TAPS \cite{McNicoll:2010qk,Kashevarov:2017kqb} experiments, the charged decay mode $\eta\to \pi^+\pi^-\pi^0$ was measured by the CLAS \cite{Dugger:2002ft,Williams:2009yj,Hu:2020ecf} and BGOOD \cite{Alef:2019imq} experiments. The LEPS experiment detected protons that scattered at forward angle and reconstructed the $\eta$ as a missing particle \cite{Sumihama:2009gf}. 

Fits of the differential cross section data revealed strong dominance of the two $S$-wave resonances $N(1535)\frac{1}{2}^-(S_{11})$ and $N(1650)\frac{1}{2}^-(S_{11})$ from the threshold region up to $W=1650$~MeV \cite{Anisovich:2005tf}, resulting in flat distributions in the differential cross sections. This is further supported by the measured $E$ observable of the CLAS \cite{Senderovich:2015lek} and CBELSA/TAPS \cite{Muller:2019qxg} collaborations, which show a value of around 1 in this region (see Fig.~\ref{fig:M:peta_CS_S_E}). The opening of the $K\Sigma$ decay channel of the $N(1650)\frac{1}{2}^-(S_{11})$ is well visible in the total cross section at around $W\approx 1.68$ GeV as a pronounced dip. At higher energies, additional strong t-channel contributions become visible in the differential cross section data as a forward peak (see Fig.~\ref{fig:M:peta_CS_S_E}). The aforementioned precise Crystal Ball/TAPS total cross section data \cite{Kashevarov:2017kqb} shows a drop in the cross section at exactly the $p\eta'$ photoproduction threshold ($W\approx 1.9$~GeV). This cusp-like structure gives evidence for a third $S$-wave resonance, the $N(1895)\frac{1}{2}^-(S_{11})$, as the cusp is caused by the opening of the $p\eta'$ decay channel of this resonance. This claim is further supported by the recently published beam asymmetry $\Sigma$ data of the CBELSA/TAPS collaboration \cite{CBELSA/TAPS:2020yam}. Here, the $p\eta'$ cusp was observed as well in a moment analysis of the beam asymmetry data. The large angular coverage of this data-set showed for the first time sensitivity up to $G$-waves in the $p\eta$ final state. The star-rating of the $N(1895)\frac{1}{2}^-(S_{11})$ has been upgraded from a 2-stars (PDG-2016 \cite{Patrignani:2016xqp}) to a 4-stars rating (PDG-2020 \cite{Zyla:2020zbs}). Very recently there are discussions on a possible two-pole structure of the $N(1895)\frac{1}{2}^-(S_{11})$ \cite{Khemchandani:2020exc}.

The precise determination of three $S$-wave resonances' parameters are key for understanding $\eta$ photoproduction. One of these parameters is the $N\eta$ branching ratio, which has been quite puzzling for the first two $S$-wave resonances. It was listed as 45-60\% for the $N(1535)\frac{1}{2}^-(S_{11})$ and 3-10\% for the $N(1650)\frac{1}{2}^-(S_{11})$ in the PDG-2010 \cite{Nakamura:2010zzi}. These values were mainly based on $\pi^-p\to \eta n$, and cross section and beam asymmetry data of $\eta$ photoproduction, which are not sufficient to unambiguously determine the contributing partial waves. The discrepancy between the two branching ratios was highly discussed and different explanations exist in literature. An overview is given in \cite{Muller:2019qxg}. The BnGa PWA performed a new fit including a set of the latest polarization observables $(T,P,H,G,E,\Sigma)$ measured by the CBELSA/TAPS collaboration as well as the $(\sigma,T,F)$ data of the A2 collaboration and $\Sigma$ and $E$ data of the CLAS collaboration. The new data helps to rule out solutions with two $J^P=\frac{5}{2}^+$ resonances and results in a $N\eta$ branching ratio of $(0.41\pm0.04)$ for the $N(1535)\frac{1}{2}^-(S_{11})$ and in $(0.33\pm0.04)$ for the $N(1650)\frac{1}{2}^-$ \cite{Muller:2019qxg}. The latest $\eta$MAID solution reports similar values with $(0.41\pm0.04)$ for the $N(1535)\frac{1}{2}^-(S_{11})$ and $(0.28\pm0.11)$ for the $N(1650)\frac{1}{2}^-(S_{11})$ \cite{Kashevarov:2017kqb}. Thus, the polarization observables helped to constrain the values of the branching ratios, reducing significantly the previously existing discrepancy. 

Apart from the three $S$-wave resonances, the PDG-2020 \cite{Zyla:2020zbs} lists the following resonances with significant branching ratios for the $p\eta$ final state: $N(1710)\frac{1}{2}^+ (P_{11})$ with $\Gamma_{N\eta}=(10-50)$\%, $N(1880)\frac{1}{2}^+(P_{11})$ with $\Gamma_{N\eta}=(5-55)$\%; $N(1720)\frac{3}{2}^+(P_{13})$ with $\Gamma_{N\eta}=(1-5)$\%, $N(1900)\frac{3}{2}^+(P_{13})$ with $\Gamma_{N\eta}=(2-14)$\%; $N(2060)\frac{5}{2}^-(D_{15})$ with $\Gamma_{N\eta}=(2-6)$\%; $N(1860)\frac{5}{2}^+(F_{15})$ with $\Gamma_{N\eta}=(2-6)$\%; $N(2190)\frac{7}{2}^-(G_{17})$ with $\Gamma_{N\eta}=(1-3)$\%. The resonances contributing to the $D_{13}$ wave have very small $\Gamma_{N\eta}$ values below 1\% as the beam asymmetry data \cite{Elsner:2007hm} showed for the $N(1520)\frac{3}{2}^-(D_{13})$ resonance in the threshold region through an interference term of the $D_{13}$ wave with the dominant $S_{11}$-wave. Comparisons of different PWA models still show large differences in the description of the $P,D,F,G\dots$ partial waves as the given large ranges of the branching ratios suggest. The existing database, especially for the polarization observables above $E_\gamma=$ 1 GeV, is not yet capable to constrain them sufficiently. Further experimental data is being measured and analyzed by the CBELSA/TAPS, Crystal Ball/TAPS and LEPS/BGOegg experiments.


The $n\eta$ final state is much less explored than the $p\eta$ final state since there are no free neutron targets. However, it is desirable to also investigate $\eta$ photoproduction off the neutron in order to gain information about the isospin structure of the electromagnetic excitation of the contributing resonances. Therefore, measurements were performed using light nuclei like deuterium or helium as target nuclei, e.g. with the TAPS experiment at MAMI \cite{Weiss:2002tn}, making it possible to compare quasi-free $\gamma n \to n\eta$ and $\gamma p \to p\eta$ reactions. According to the participant-spectator model, the $\eta$ mesons are produced quasi-free on one nucleon, which is referred to as the participant nucleon, while the other nucleon is considered to be a spectator nucleon. Weiss et al.~\cite{Weiss:2002tn} reported the total inclusive and exclusive cross sections for $\eta$ photoproduction off the deuteron with high statistics, covering the energy range of the first $S$-wave resonance $N(1535)\frac{1}{2}^-(S_{11})$. They report a neutron-to-proton cross section ratio of $\sigma_n/\sigma_p=0.66\pm0.01$ above the free nucleon production threshold, which suggests a dominant isovector-excitation for the $N(1535)\frac{1}{2}^-(S_{11})$ resonance. This was later also confirmed by the CBELSA/TAPS collaboration, reporting an isoscalar part of only $0.09\pm 0.01$ of the amplitude \cite{Jaegle:2011sw}. 

\begin{figure}[htb]
\includegraphics[width=0.66\textwidth]{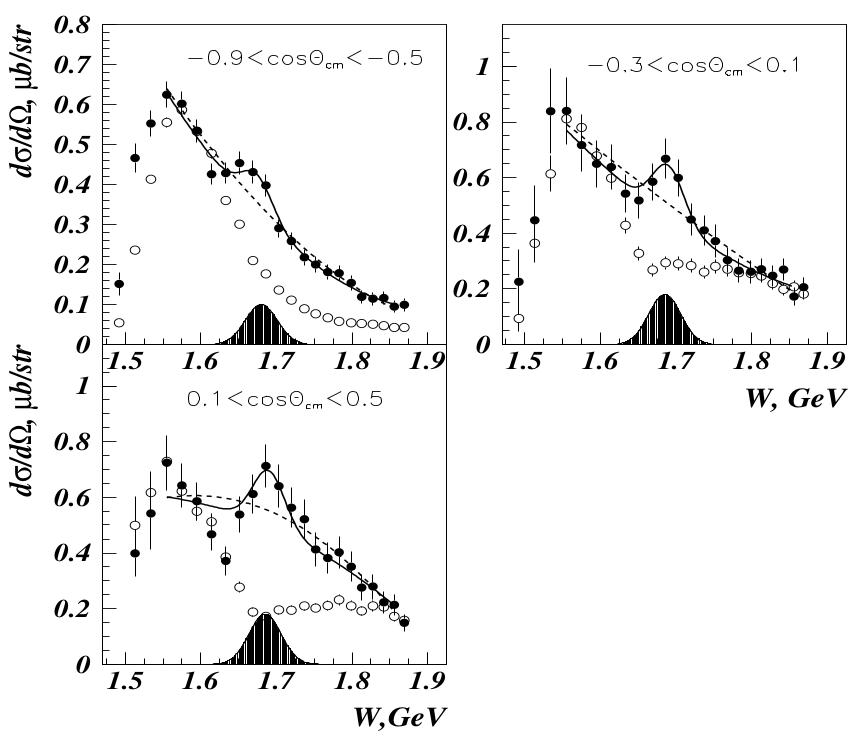}
\hspace*{0.3cm}
\includegraphics[width=0.28\textwidth]{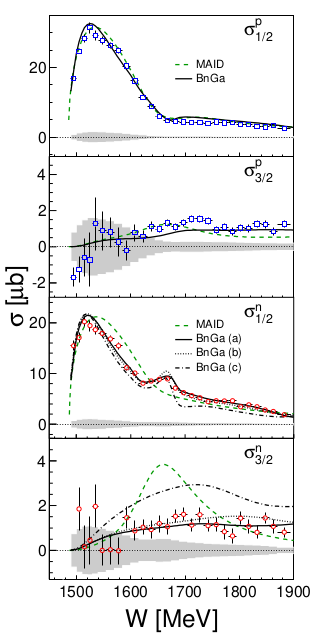}
	\caption{Left: The $\gamma n\to\eta n$ cross section data of the GRAAL experiment is shown as black circles, while the $\gamma p \to \eta p$ cross section, which has been normalized to the $N(1535)\frac{1}{2}^-$ resonance in the neutron cross section, is shown as open circles \cite{Kuznetsov:2007gr}. Right: The total helicity-dependent cross sections $\sigma_{1/2}$ and $\sigma_{3/2}$ are plotted for $\gamma n\to\eta n$ (red circles) and $\gamma p \to \eta p$ (blue squares) as measured with the Crystal Ball/TAPS experiment \cite{Witthauer:2017get}.}
	\label{fig:M:neta_E}
\end{figure}

To extend the covered energy range towards higher energies, several measurements were performed, e.g. by Kutznetsov et al.~\cite{Kuznetsov:2007gr} using the GRAAL experiment (see Fig.~\ref{fig:M:neta_E} on the right), Miyara et al.~\cite{Miyahara:2007zz} using the LNS (now ELPH) and Jaegle et al.~\cite{Jaegle:2008ux,Jaegle:2011sw} using the CBELSA/TAPS experiment utilizing a deuteron target, making it possible to compare quasi-free $\gamma n \to n\eta$ and $\gamma p \to p\eta$ reactions. Here, a surprising narrow structure was observed in the cross section data at $M\sim$ 1.68 GeV with a small width of $\Gamma \leq$ 30 MeV \cite{Kuznetsov:2007gr} in the $n\eta$ final state, which is not seen in the $p\eta$ final state (see Fig.~\ref{fig:M:neta_E} on the left). Quasi-free $\eta$ photoproduction was also studied at the Crystal Ball/TAPS experiment using a liquid deuterium and a liquid $^3$He as targets, measuring the angular dependence of the narrow structure \cite{Werthmuller:2013rba}. The cross section data confirms the existence of the narrow structure with a position of $W=(1670\pm5)$ MeV and an intrinsic width of $\Gamma=(30\pm15)$ MeV \cite{Werthmuller:2013rba}.

Different ansatzes exist to explain the nature of this narrow structure: it could be an intrinsic resonance with unusual properties \cite{Polyakov:2003dx,Arndt:2003ga,Choi:2005ki,Fix:2007st,Shrestha:2012ep}, a result of intermediate strangeness loop contributions \cite{Doring:2009qr}, a coupled-channel effect of already known resonances \cite{Shklyar:2006xw,Shyam:2008fr} or explained by interference effects of the $N(1535)\frac{1}{2}^-(S_{11})$ and $N(1650)\frac{1}{2}^-(S_{11})$ resonances \cite{Anisovich:2015tla}. Further experimental investigations were carried out by the Crystal Ball/TAPS \cite{Witthauer:2017get,Witthauer:2017wdb} and CBELSA/TAPS \cite{Witthauer:2017pcy} experiments using a circularly polarized photon beam in combination with a longitudinally polarized deuterated butanol target. Fig.~\ref{fig:M:neta_E} shows on the right the helicity dependent total cross sections $\sigma_{1/2}$ and $\sigma_{3/2}$ for both reactions $\gamma n \to n\eta$ and $\gamma p \to p\eta$ for the Crystal Ball/TAPS data \cite{Witthauer:2017get}. It is remarkable that the narrow structure is only visible in the helicity-$1/2$ cross section, indicating a relation to the helicity-1/2 amplitude and thus, to the $S_{11}$ and/or the $P_{11}$ partial waves. Additionally extracted Legendre coefficients and comparisons to different BnGa model predictions indicate a narrow resonance in the $P_{11}$ wave interfering with the $S_{11}$ wave, for the narrow structure \cite{Witthauer:2017get,Witthauer:2017wdb}. Following this publications, Tiator et al. \cite{Tiator:2018heh} fitted the cross section as well as the E data and could attribute this structure to an interference between different partial waves with a dominant contribution of the $N(1710)\frac{1}{2}^+$. In contrast to this, Suh et al. \cite{Suh:2018yiu} were able to describe the data by a narrow nucleon resonance $N(1685)\frac{1}{2}^+$. A final conclusion about this structure is still unavailable and can only be provided by additional high-statistics measurements. Apart from the narrow structure at $W=1685$~MeV, another narrow structure was reported at $W=1720$~MeV in the $n\eta$ final state \cite{Werthmuller:2015owc} as well as in the beam asymmetry $\Sigma$ for Compton scattering off protons \cite{Kuznetsov:2015nla}, which needs further investigation.


\begin{figure}[htb]
\begin{minipage}{0.36\textwidth}
\includegraphics[width=\textwidth]{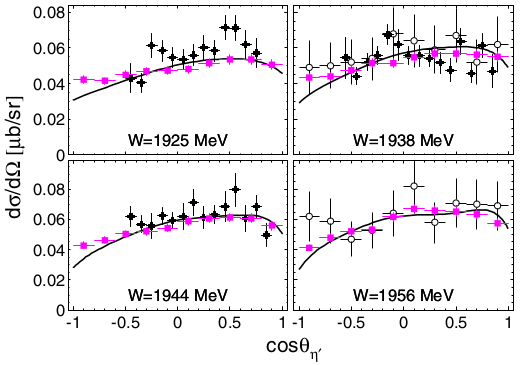}
\end{minipage}
\begin{minipage}{0.28\textwidth}
\includegraphics[width=\textwidth]{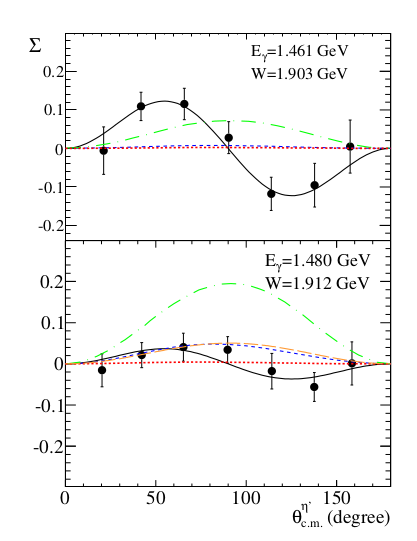}
\end{minipage}
\begin{minipage}{0.33\textwidth}
\includegraphics[width=\textwidth]{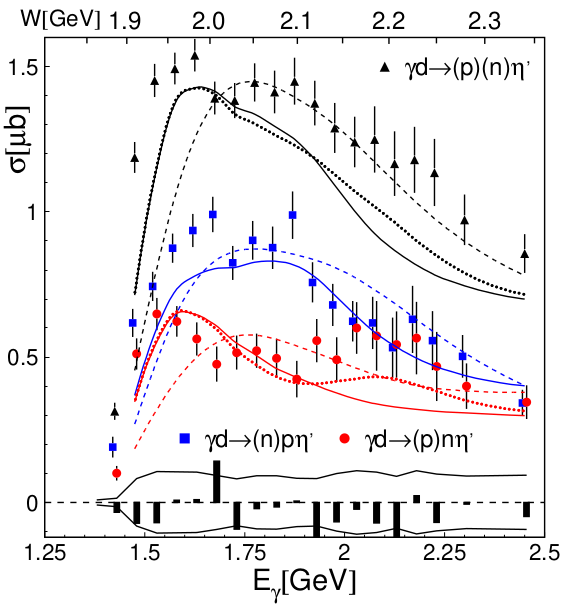}
\end{minipage}
	\caption{Left: The differential cross section \cite{Kashevarov:2017kqb} of the $p\eta'$ final state for four selected energy bins of the Crystal Ball/TAPS experiment (magenta squares) \cite{Kashevarov:2017kqb}, CLAS experiment (black points) \cite{Williams:2009aa} and CBELSA/TAPS (open circles) \cite{Crede:2009zzb} experiment. Center: The angular distributions of the beam asymmetry $\Sigma$ for two energy bins by the GRAAL experiment \cite{Sandri:2014nqz}. Right: The total inclusive (black triangles), quasi-free proton (blue squares) and neutron (red points) cross section data of the CBELSA/TAPS experiment \cite{Jaegle:2010jg}. \EPJA} 
	\label{fig:M:netaprime_sigmas}
\end{figure}

The database for the $N\eta'$ channel is very poor since its total cross section is only in the order of $1\mu$b. This makes it quite challenging to obtain high-precision data for the $N\eta'$ channel. Nevertheless, it is worth investigating this channel since the high mass of the $\eta'$ meson allows us to explore the high-mass region of the nucleon excitation spectrum, where a lot of missing resonances are predicted. The differential cross section was measured for the $p\eta'$ final state by the CLAS \cite{Dugger:2005my,Williams:2009yj}, the CBELSA/TAPS \cite{Crede:2009zzb} and in the threshold region by the Crystal Ball/TAPS \cite{Kashevarov:2017kqb} experiments. Figure \ref{fig:M:netaprime_sigmas} shows a comparison of the different data sets of the differential cross section for four different energy bins near the $p\eta'$ production threshold. The CBELSA/TAPS and CLAS data sets are in good agreement up to $W\approx 2$ GeV. Above it, there are some unresolved discrepancies between the two data sets as the CBELSA/TAPS data is systematically lying above the CLAS data over a large angular range. The LEPS experiment provides data for the very backward angles ($-1.0<\cos\theta_{C.M.}^{\eta'}<-0.8$) \cite{Morino:2013raa}.

The GRAAL collaboration published the beam asymmetry $\Sigma$ for two energy bins very close to the $\eta'$ photoproduction threshold (see Fig.~\ref{fig:M:netaprime_sigmas}, center) \cite{Sandri:2014nqz}. These results are quite unexpected since the angular distributions show a nodal structure with a $\sin$-like $\theta_{c.m.}$ dependence, indicating contributions from higher partial waves. In addition, the amplitude of $\Sigma$ shows a strong energy dependence. The GRAAL data is in direct contrast to the behavior of the beam asymmetry $\Sigma$ for the $p\eta$ final state which is rather flat or parabolic, showing clear $S$-wave dominance near threshold. The $\eta$MAID model \cite{Tiator:2018heh} introduces a very narrow $S_{11}$-wave resonance with a mass of $M\sim 1902$ MeV and a width of only $\Gamma\sim2$ MeV in order to describe the data through a $S$- and $F$-wave interference term. The BnGa model \cite{Anisovich:2018yoo} also needs a new narrow resonance in order to describe the data. However, here a $D_{13}$ resonance is assumed with a mass of $M\sim 1900$ MeV and a width of $\Gamma<3$ MeV. Both models can describe the data similarly well. While the Crystal Ball/TAPS cross section data also supports the existence of a new resonance \cite{Tiator:2018heh,Anisovich:2018yoo}, the CLAS beam asymmetry data \cite{Collins:2017sgu} do not have the precision in the threshold region to confirm the GRAAL data. However, the CLAS $\Sigma$ data set extends the covered energy range up to $E_\gamma=1836$ MeV. The measurement of more polarization observables e.g. $T,P,H$ could help to clarify the existence and nature of this potentially new resonance \cite{Tiator:2018heh}.  
\begin{table*}[p]
\caption{A collection of polarization observable measurements for $\eta$ photoproduction. Hint: We do not distinguish between differential and total cross section data. Further details are given in the cited references. The symbols indicate which PWA (circle: BnGa-2019 \cite{Muller:2019qxg}, diamond: J\"uBo-2017 \cite{Ronchen:2018ury} and star: SAID \cite{McNicoll:2010qk,SAIDURL}) use the data in their fits.} \label{tab:measSingleEta}
\centering
\vspace*{5mm}
\begin{tabular}{clllc}
Observable & Energy range $E^{lab}_\gamma$ & Facility & Reference & Year\\
\hline
\multicolumn{5}{c}{$\gamma p \rightarrow p\eta$}\\
\hline
$\sigma$ & $1200-4700$ MeV & CLAS & Hu et al. \cite{Hu:2020ecf} & 2020 \\
$\Sigma$ & $1130-1790$ MeV & CBELSA/TAPS & Afzal et al. \cite{CBELSA/TAPS:2020yam} $\circ$ & 2020 \\
$\sigma,\Sigma$ & $708-2500$ MeV & BGOOD & Alef et al. \cite{Alef:2019imq} & 2020 \\
E & $708-2300$ MeV & CBELSA/TAPS & M\"uller et al. \cite{Muller:2019qxg} $\circ$ $\star$ & 2020 \\
G & $708-1300$ MeV & CBELSA/TAPS & M\"uller et al. \cite{Muller:2019qxg} $\circ$ $\star$ & 2020 \\
H,P & $708-933$ MeV & CBELSA/TAPS & M\"uller et al. \cite{Muller:2019qxg} $\circ$ $\star$ & 2020 \\
T & $708-2620$ MeV & CBELSA/TAPS & M\"uller et al. \cite{Muller:2019qxg} $\circ$ $\star$ & 2020 \\
$\Sigma$ & $1070-1876$ MeV & CLAS & Collins et al. \cite{Collins:2017sgu} $\circ$ $\diamond$ $\star$ & 2017 \\
E, $\sigma_{1/2}$, $\sigma_{3/2}$ & $708-1400$ MeV & Crystal Ball/TAPS & Witthauer et al. \cite{Witthauer:2017wdb} & 2017\\
$\sigma$ & $708-1577$ MeV & Crystal Ball/TAPS & Kashevarov et al. \cite{Kashevarov:2017kqb} $\circ$ $\star$ & 2017\\
$E$ & $708-1994$ MeV & CLAS & Senderovich et al. \cite{Senderovich:2015lek} $\circ$ $\diamond$ $\star$ & 2016\\
$T,F$ & $708-1400$ MeV & Crystal Ball/TAPS & Akondi et al. \cite{Akondi:2014ttg} $\circ$ $\diamond$ $\star$ & 2014\\
$\sigma$ & $708-1400$ MeV & Crystal Ball/TAPS & McNicoll et al. \cite{McNicoll:2010qk} $\circ$ $\diamond$ $\star$ & 2010\\
$\sigma$ & $850-2550$ MeV & CBELSA/TAPS & Crede et al. \cite{Crede:2009zzb} $\circ$ $\diamond$ $\star$ & 2009\\
$\sigma$ & $845-3830$ MeV & CLAS & Williams et al. \cite{Williams:2009yj} $\diamond$ $\star$ & 2009\\
$\sigma$ & $1600-2400$ MeV & LEPS & Sumihama et al. \cite{Sumihama:2009gf} $\diamond$ $\star$ & 2009\\
$\sigma,\Sigma$ & $708-1500$ MeV & GRAAL & Bartalini et al. \cite{Bartalini:2007fg} $\circ$ $\diamond$ $\star$ & 2007\\
$\Sigma$ & $800-1400$ MeV & CBTAPS & Elsner et al. \cite{Elsner:2007hm} $\diamond$ $\star$ & 2007\\
$\sigma$ & $708-1150$ MeV & LNS & Nakabayashi et al. \cite{Nakabayashi:2006ut} $\diamond$ $\star$ & 2006\\
$\sigma$ & $750-3000$ MeV & CB-ELSA & Crede et al. \cite{Crede:2003ax} $\diamond$ $\star$ & 2005\\
$\sigma,\sigma_{1/2},\sigma_{3/2}$ & $780-790$ MeV & GDH & Ahrens et al. \cite{Ahrens:2003bp} $\diamond$ $\star$ & 2003\\
$\sigma$ & $750-1950$ MeV & CLAS & Dugger et al. \cite{Dugger:2002ft} $\diamond$ $\star$ & 2002\\

\hline
\multicolumn{5}{c}{$\gamma n \rightarrow n\eta$}\\
\hline
E, $\sigma_{1/2}$, $\sigma_{3/2}$ & $708-1400$ MeV & Crystal Ball/TAPS & Witthauer et al. \cite{Witthauer:2017wdb} $\circ$ & 2017\\
$\sigma$, E, $\sigma_{1/2}$, $\sigma_{3/2}$ & $708-1850$ MeV & CBELSA/TAPS & Witthauer et al. \cite{Witthauer:2017pcy} & 2017\\
$\sigma_{1/2}$, $\sigma_{3/2}$ & $708-1400$ MeV & Crystal Ball/TAPS & Witthauer et al. \cite{Witthauer:2017get} $\circ$ & 2016\\
$\sigma$ & $708-1400$ MeV & Crystal Ball/TAPS & Werthm\"uller et al. \cite{Werthmuller:2014thb} $\circ$ & 2014\\
$\sigma$ & $708-1400$ MeV & Crystal Ball/TAPS & Witthauer et al. \cite{Witthauer:2013tkm} & 2013\\
$\sigma$ & $708-1400$ MeV & Crystal Ball/TAPS & Werthm\"uller et al. \cite{Werthmuller:2013rba} $\circ$ & 2013\\
$\sigma$ & $708-2500$ MeV & CBELSA/TAPS & Jaegle et al. \cite{Jaegle:2011sw} & 2011\\
$\Sigma$ & $708-1500$ MeV & GRAAL & Fantini et al. \cite{Fantini:2008zz} & 2008\\
$\sigma$ & $708-2500$ MeV & CBELSA/TAPS & Jaegle et al. \cite{Jaegle:2008ux} $\circ$ & 2008\\
$\sigma$ & $600-1150$ MeV & LNS & Miyahara et al. \cite{Miyahara:2007zz} & 2007\\
$\sigma$ & $708-1400$ MeV & GRAAL & Kutznetsov et al. \cite{Kuznetsov:2007gr} & 2007\\
$\sigma$ & $630-750$ MeV & TAPS & Weiss et al. \cite{Weiss:2002tn} & 2003\\
\hline
\end{tabular}
\end{table*}

The CBELSA/TAPS collaboration measured the first data for the quasi-free $n\eta'$ final state, using a deuterium target \cite{Jaegle:2010jg}. Fig \ref{fig:M:netaprime_sigmas} shows on the right the measured total cross section for the inclusive ($\gamma d\to (p)(n)\eta$) and exclusive ($\gamma d\to (p)n\eta$ and $\gamma d\to p(n)\eta$) final states. Both proton $\sigma_p$ and neutron $\sigma_n$ cross section are very similar above $E_\gamma\approx1.9$ GeV which can be explained by the contribution of $t$-channel processes. However, significant differences are observed at lower energies. Here, it holds $\sigma_n<\sigma_p$. While a two bump structure is visible in $\sigma_p$, $\sigma_n$ only shows the first bump structure and the second one seems to be suppressed, indicating resonance contributions.

The $N\eta'$ data has been investigated by several models, out of which the latest four are discussed. Early models like Zhong and Zhao \cite{Zhong:2011ht} found evidence for $S$-wave resonances and also for the $N(2080)\frac{5}{2}^-(D_{15})$ resonance based on a chiral quark model. Using photon- and hadron-induced reactions cross section data, Huang et al.~\cite{Huang:2012xj} were able to describe the data with only $S$- and $P$-wave resonances ($N(1720)\frac{3}{2}^+(P_{13})$, $N(1925)\frac{1}{2}^-(S_{11})$, $N(2130)\frac{1}{2}^+(P_{11})$, $N(2050)\frac{1}{3}^+(P_{13})$. Fitting all the latest available data, the latest BnGa solution \cite{Anisovich:2018yoo} found important contributions from the following four resonances: $N(1895)\frac{1}{2}^-(S_{11})$, $N(1900)\frac{3}{2}^+(P_{13})$, $N(2100)\frac{1}{2}^+(P_{11})$ and $N(2120)\frac{3}{2}^-(D_{13})$ for $N\eta'$ final state, while $F$-waves have negligibly small contributions. In contrast to this, the latest $\eta$MAID model finds important contributions from $S$ and $F$ waves: $N(1895)\frac{1}{2}^-(S_{11})$, $N(1880)\frac{1}{2}^+(P_{11})$, $N(1860)\frac{5}{2}^+(F_{15})$ and $N(1990)\frac{7}{2}^-(F_{17})$ \cite{Tiator:2018heh}. To resolve these existing ambiguities a lot more data is needed for $\eta'$ photoproduction.

\begin{table*}[ht]
	\caption{A collection of polarization observable measurements for $\eta'$ photoproduction. Hint: We do not distinguish between differential and total cross section data. Further details are given in the cited references. The symbols indicate which PWA (circle: BnGa-2019 \cite{Muller:2019qxg}, diamond: J\"uBo-2017 \cite{Ronchen:2018ury} and star: SAID \cite{SAIDURL}) use the data in their fits.} \label{tab:measSingleEtaPrime}
	\centering
	\vspace*{5mm}
	\begin{tabular}{clllc}
		Observable & Energy range $E^{lab}_\gamma$ & Facility & Reference & Year\\
		\hline
\multicolumn{5}{c}{$\gamma p \rightarrow p\eta'$}\\
\hline
$\sigma$ & $1447-1577$ MeV & Crystal Ball/TAPS & Kashevarov et al. \cite{Kashevarov:2017kqb} $\circ$ $\star$ & 2017\\
$\Sigma$ & $1516-1836$ MeV & CLAS & Collins et al. \cite{Collins:2017sgu} $\circ$ $\star$ & 2017 \\
$\Sigma$ & $1461-1480$ MeV & GRAAL & Levi Sandri et al. \cite{Sandri:2014nqz} $\circ$ $\star$ & 2015 \\
$\sigma$ & $1500-3000$ MeV & LEPS & Morino et al. \cite{Morino:2013raa} & 2013\\
$\sigma$ & $1448-2500$ MeV & CBELSA/TAPS & Crede et al. \cite{Crede:2009zzb} $\circ$ $\star$ & 2009\\
$\sigma$ & $1506-3695$ MeV & CLAS & Williams et al. \cite{Williams:2009yj} $\circ$ $\star$ & 2009\\
$\sigma$ & $1527-2227$ MeV & CLAS & Dugger et al. \cite{Dugger:2005my} $\star$ & 2006\\
\hline
\multicolumn{5}{c}{$\gamma n \rightarrow n\eta'$}\\
\hline
$\sigma$ & $1448-2500$ MeV & CBELSA/TAPS & Jaegle et al. \cite{Jaegle:2010jg} & 2011\\
\hline
\end{tabular}
\end{table*}

\subsubsection{Vector Mesons}
\label{sec:measurements:VectorMesonPhotoproduction}
As already explained in Sec.~\ref{sec:formalisms:VectorMesonPhotoproduction}, similar polarization observables as for single meson photoproduction can also be extracted for vector mesons, alongside with Spin Density Matrix Elements (\textit{SDME}s). The topic of $\omega$ photoproduction became quite popular in the last years, since experiments were able to gather large amounts of statistics for this final state. The $\omega$ meson mainly decays into two decay modes: $\omega \to \pi^+ \pi^- \pi^0$ (BR 89.2\%) and $\omega \to \pi^0 \gamma$ (BR 8.28\%) \cite{Zyla:2020zbs}. Depending on the experiment, different decay modes were preferred. The CLAS experiment extracts the $\omega$ from its $\pi^+\pi^-\pi^0$ decay, which results in a large amount of data. However, here the background from other contributing final states is large and a proper treatment of the background is needed. The CLAS collaboration uses for this the   probabilistic  event-based  (Q factor) method \cite{Williams:2008ac,Roy:2017qwv}. The CBELSA/TAPS and the Crystal Ball/TAPS experiment both identify the $\omega$ via its $\pi^0 \gamma$ decay, which results in less statistics due to the reduced branching ratio. However, the background below the $\omega$ peak is here small and no special treatment becomes necessary.
 
First fits to cross section measurements of vector mesons were performed in 2000 by Oh et al.~\cite{Oh:2000zi}. Here, data from the 70s from SLAC \cite{Ballam:1972eq} was taken as well as the more recent SAPHIR measurements \cite{Klein:1998xy}. First indications of dominant contributions by a ``missing'' $N(1910)\frac{3}{2}^+$ and a $N(1960)\frac{3}{2}^-$, later identified as the two-star $N(2080)\frac{3}{2}^-$, were found in their fits, but they were able to show that a detailed analysis of the states and the reaction mechanism can only be achieved by the measurement of polarization observables. Shortly afterwards, Titov and Lee \cite{Titov:2002iv} used an effective Lagrangian approach and found dominant contribution from a $N(1680)\frac{5}{2}^+$ and a $N(1520)\frac{3}{2}^-$ in the near threshold region.

Quark Model calculations by Zhao \cite{Zhao:2000tb} found dominant contributions of a $N(1720)\frac{3}{2}^+$ and a $N(1680)\frac{5}{2}^+$. Later, a coupled channel approach by Penner and Mosel \cite{Penner:2002md}, using the published cross section data, saw significant contributions by the $N(1710)\frac{1}{2}^+$ and $N(1900)\frac{3}{2}^+$. To solve these discrepancies by different calculations, the measurement of \textit{SDME}s by the SAPHIR collaboration \cite{Barth:2003kv} were included in this model \cite{Shklyar:2004ba}. In this unitary effective Lagrangian coupled-channel formalism, Shklyar et al.~\cite{Shklyar:2004ba} found dominant contributions of the $N(1675)\frac{5}{2}^-$ and the $N(1680)\frac{5}{2}^+$. However, they saw a strong interference pattern between the resonances and a $\pi^0$ exchange, which makes it difficult to determine further resonance contributions in this channel. This problem can be solved by further high statistics measurements including polarization observables.

\begin{figure}[p]
 	\centering
 	\includegraphics[width=0.45\textwidth]{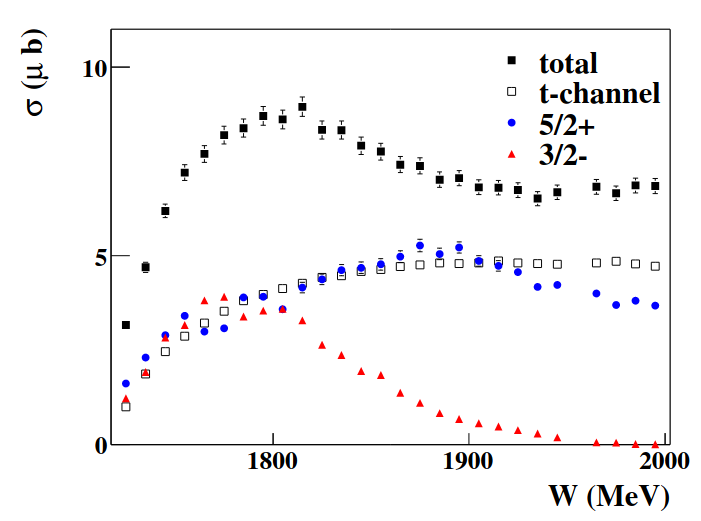}
	\includegraphics[width=0.45\textwidth]{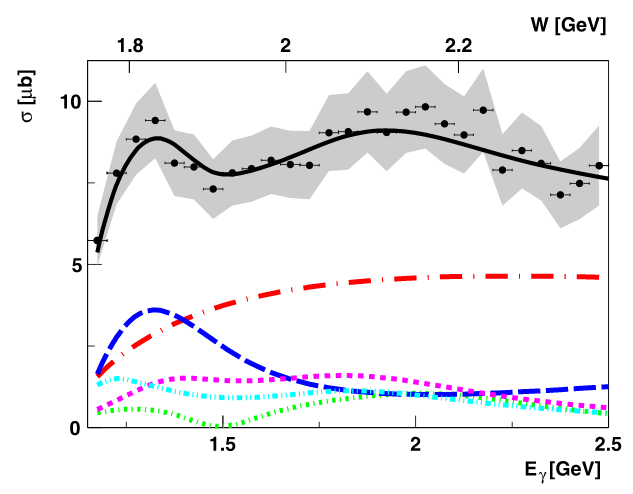}
	\includegraphics[width=0.75\textwidth]{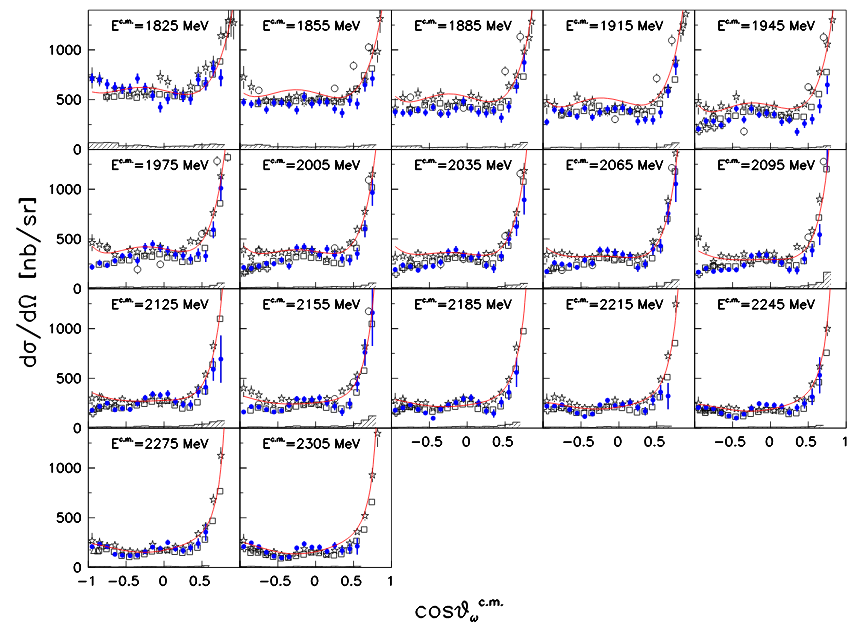}
	\caption{Top Left: Total cross section for the $p\omega$ final state by the CLAS collaboration \cite{Williams:2009ab}. A partial wave analysis based on this unpolarized data was performed to determine the contributing amplitudes \cite{Williams:2009aa}. \APS{\cite{Williams:2009aa}}{2009}
	Top Right: Total cross section for the $p\omega$ final state by the CBTAPS experiment \cite{Wilson:2015uoa} with a fit including polarization measurements. The largest contributions to the fit by the BnGa group are the Pomeron exchange (red dashed-dotted), resonant $J^P=\frac{3}{2}^+$ wave (blue dashed), $\frac{3}{2}^-$ wave (pink short dashed), $\frac{5}{2}^+$ (cyan double dotted-dashed) and $\frac{1}{2}^-$ (green short dashed-dotted). 
	Bottom: Differential cross section measured by the BGOegg collaboration \cite{Muramatsu:2020xbx}, compared to the measurements by CLAS \cite{Williams:2009ab} (open squares), LEPS \cite{Sumihama:2009gf} (open crosses)  and \cite{Morino:2013raa} (open triangles), and two measurements by CBTAPS \cite{Dietz:2015ppa} (open circles) and \cite{Wilson:2015uoa} (open stars). \APS{\cite{Muramatsu:2020xbx}}{2020}
	}
	\label{fig:M:pomega_crossection_wilson} 	\label{fig:M:pomega_crossection_muramatsu}
\end{figure}

One of the first high statistics measurement of the cross section as well as three unpolarized \textit{SDME}s were published in 2009 by the CLAS collaboration \cite{Williams:2009ab}. At they same time, an event-based mass-independent partial wave analysis of the data was published as well \cite{Williams:2009aa}. The result of this fit can be found in Fig.~\ref{fig:M:pomega_crossection_wilson} top left. Here, evidence for the nucleon resonances $N(1680)\frac{5}{2}^+$ and $N(1700)\frac{3}{2}^1$ near threshold was found. In addition, the resonance $N(2190)\frac{7}{2}^-$ can be seen in the data as well as a $\frac{5}{2}^+$ state around 2~GeV.

Two examples of more recent measurements of the total and differential cross section for $\omega$ photoproduction are presented in Fig.~\ref{fig:M:pomega_crossection_wilson}, top right and bottom. The measurement of the total cross section up to $E_\gamma = 2.5$~GeV was performed by the CBTAPS collaboration and was in detail analyzed by the BnGa group \cite{Wilson:2015uoa}. The data set was fitted in combination with the observables  $\Sigma_{(\pi)}$, $G_{(\pi)}$ and E, measured by the CBELSA/TAPS collaboration \cite{Eberhardt:2015lwa}. Here, the observables $G_{(\pi)}$ and $E$ were the first double polarization observables extracted for vector meson photoproduction. In the fits by the BnGa group \cite{Denisenko:2016ugz} it became evident that at threshold a significant contribution by the $J^P = \frac{3}{2}^+$ partial wave is visible, presumably due to the $N(1720) \frac{3}{2}^+$ resonance below threshold. This analysis saw contributions from 12 resonances and determined new branching ratios for them. They were also able to find evidence for the poorly known states like the $N(2000) \frac{5}{2}^+$ or the $N(2120) \frac{3}{2}^-$. 
In the differential cross section, see for example the measurement of the LEPS2/BGOegg collaboration \cite{Muramatsu:2020xbx} in Fig.~\ref{fig:M:pomega_crossection_wilson} (right), a strong forward peaking behavior becomes visible, which is consistent with an increased t-channel exchange. Different exchange particles, Pomeron (natural) or $\pi$ (unnatural), were fitted to the observables by the BnGa group. Here, the polarization observables, for example the beam asymmetry $\Sigma$, are of major importance, since they allow a distinction between natural and unnatural exchange \cite{Titov:2002zy}. The results of the BnGa fits indicate a large contribution from Pomeron exchange in contrast to the unnatural exchange by $\pi$ mesons.

\begin{figure}[thb]
	\includegraphics[width=0.5\textwidth]{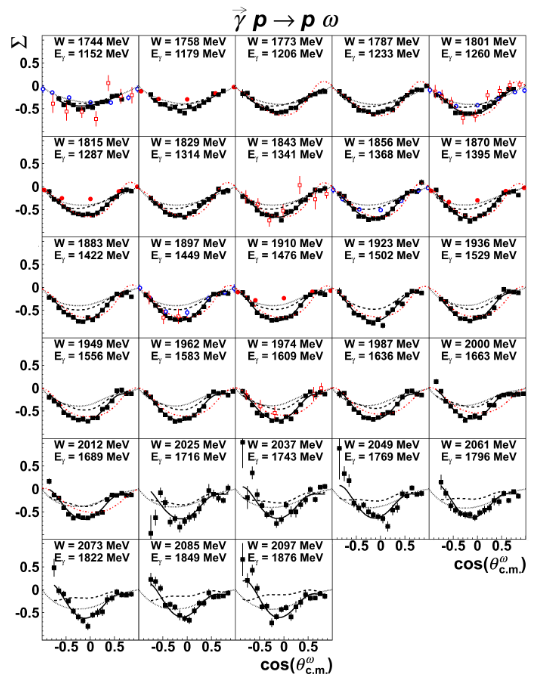}
	\includegraphics[width=0.5\textwidth]{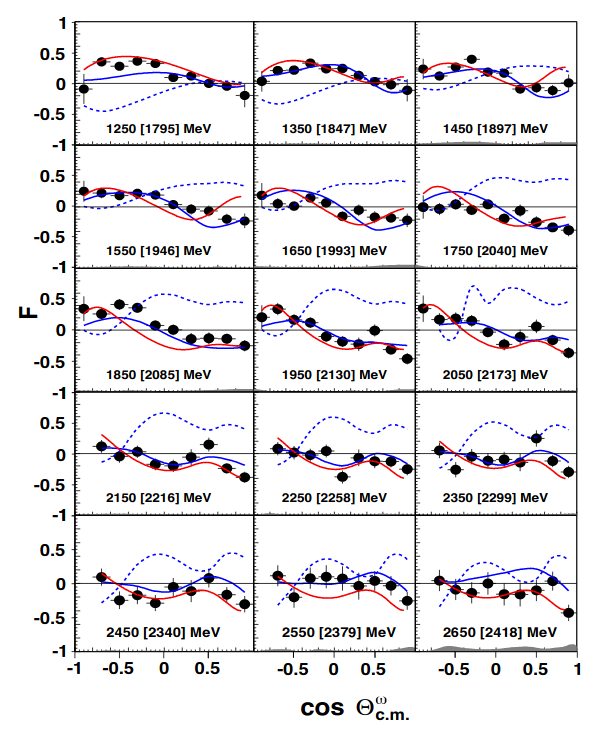}
	\caption{Left: Measurement of the beam asymmetry $\Sigma$  for $\omega$ photoproduction by the CLAS collaboration \cite{Collins:2017vev}, compared to previous data from CBTAPS  (open red squares \cite{Klein:2008aa}) and GRAAL (blue open circles \cite{Ajaka:2006bn} and red solid circles \cite{Vegna:2013ccx}) as well as to calculation by the BnGa groups without the data set (black dashed line) and including the data set (black solid line). For further details see \cite{Collins:2017vev}.
	Right: Double polarization observable F for $\omega$ photoproduction measured by the CLAS collaboration \cite{Roy:2018tjs} compared to curves by the BnGa group (prediction: blue dashed \cite{Denisenko:2016ugz}, fit: blue \cite{Roy:2018tjs}) and Wei et al. (fit: red \cite{Wei:2019imo}) \APS{\cite{Roy:2018tjs}}{2019}.}
	\label{fig:M:pomega_Sigma_collins}
	\label{fig:M:pomega_F_roy}
\end{figure}

Since 2015, the data base for $\omega$ photoproduction has been extensively increased due to several new measurements on polarization observables by different collaborations. One measurement of the beam asymmetry $\Sigma$ by the CLAS collaboration \cite{Collins:2017vev} is shown in Fig.~\ref{fig:M:pomega_Sigma_collins} (left). As mentioned earlier, the t-channel exchange is visible in forward direction, which makes the data points in the medium angular range sensitive to reactions in the s- and u-channel. This allows the extraction of information about intermediate resonant states. The beam asymmetry is compared to the calculations by the BnGa group mentioned earlier. In this calculation \cite{Denisenko:2016ugz} the beam asymmetry measurement of the CBELSA/TAPS experiment \cite{Eberhardt:2015lwa}, which covers just a small region of the CLAS measurement, is already included. 

Another example of the double polarization measurements performed by the CLAS collaboration is the observable F \cite{Roy:2018tjs}, which can be extracted using a circularly polarized beam impinging on transversely polarized target. This observable, which is shown in Fig.~\ref{fig:M:pomega_F_roy} (right), is part of a collection of double polarization observables published by the CLAS collaboration, see Tab.~\ref{tab:measSingleVector}. All these new polarization measurements by the CLAS collaboration were also included in the fit by the BnGa group, which results in a satisfying description of the data, see for example Fig.~\ref{fig:M:pomega_Sigma_collins} left and right. The new solution confirms the dominant partial waves, which were already found in \cite{Denisenko:2016ugz}. A significant contribution of the $N(2120) \frac{3}{2}^-$ could be found, in addition the contribution to the $N(2000) \frac{5}{2}^+$ increased by about 50\% \cite{Roy:2018tjs}. This result is consistent with the single channel PWA, which was performed by the CLAS collaboration in 2009 \cite{Williams:2009aa}.

The recent polarized and unpolarized measurements were analyzed in an effective Lagrangian approach by Wei et al.~\cite{Wei:2019imo}. Here, the reaction amplitude include $\pi$ and $\eta$ exchange in the t-channel, nucleon resonances in the s-channel as well as in the u-channel and a contact current. The fit could describe the data satisfactorily and found evidence for the nucleon resonances  $N(1520)\frac{3}{2}^-$, $N(1700)\frac{3}{2}^-$, $N(1720)\frac{3}{2}^+$, $N(1860)\frac{5}{2}^+$, $N(1875)\frac{3}{2}^-$, $N(1895)\frac{1}{2}^-$ and $N(2060)\frac{5}{2}^-$ \cite{Wei:2019imo}. The t-channel $\pi$ exchange was found to dominate the background of the nucleon resonances.

\begin{figure}[th]
	\centering
	\includegraphics[width=0.4\textwidth]{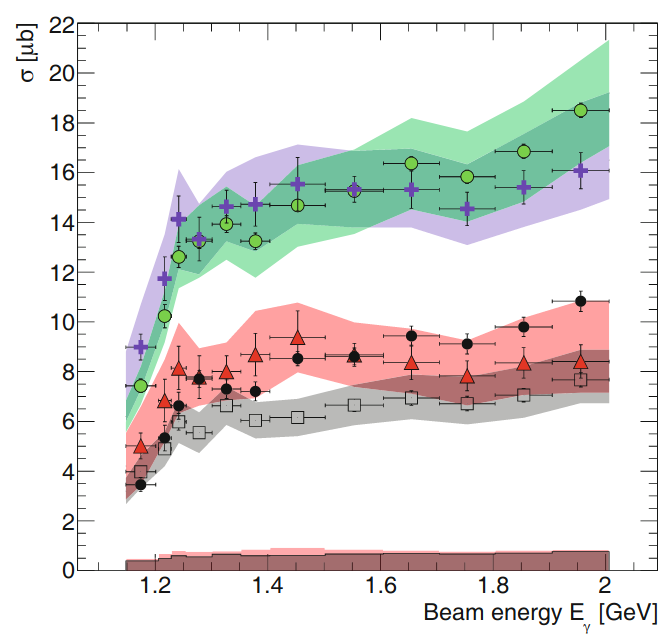}
	\hspace{2cm}
	\includegraphics[width=0.23\textwidth]{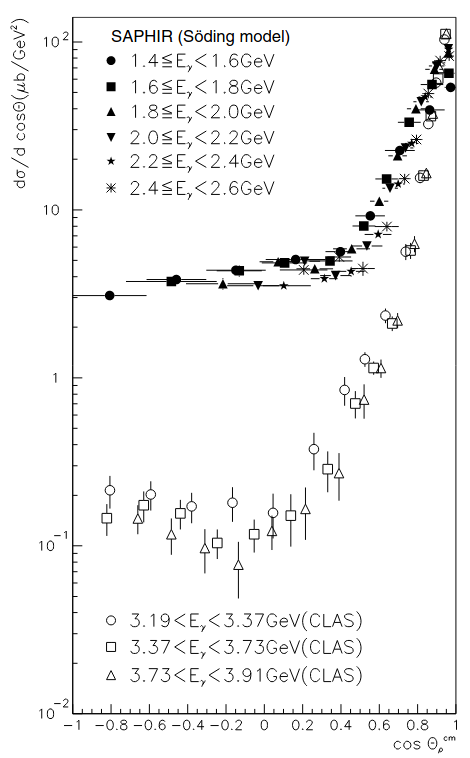}
	\caption{Left: Cross section for $\omega$ photoproduction \cite{Dietz:2015ppa} off the bound proton (open squares), bound neutron (red triangles), sum of both cross sections (purple crosses), quasi-free inclusive cross-section $\sigma_{incl}$ (green circles) and the cross section off the neutron calculated by $\sigma_{incl} - \sigma^{bound}_p$ (black circles). 
	Right: Differential Cross section for photoproduction of the $\rho$ meson by the SAPHIR experiment \cite{Wu:2005wf} (filled marker) compared to the data of the CLAS collaboration \cite{Battaglieri:2001xv} (open marker) at higher photon energies. \EPJA}
	\label{fig:M:nomega_crosssection_dietz}
	\label{fig:M:prho_crosssection_wu}
\end{figure}

The data on $\omega$ photoproduction off the neutron is sparse. Up to now, only one publication is available, where the cross section of the free and of the bound proton was determined in combination with the cross section of the bound neutron \cite{Dietz:2015ppa}. The most interesting topic here is the comparison between the cross section of proton and neutron. The measurement of the $\eta$ cross section off the neutron showed substantial differences, as explained in Sec.~\ref{sec:measurements:EtaPhotoproduction}. Interestingly, for the $\omega$ the measured cross section for the reactions off the neutron is larger than the cross section off the proton, see Fig.~\ref{fig:M:nomega_crosssection_dietz} (left). The interpretation of this fact is still unclear and further measurements, providing a smaller statistical error, are needed. 

Photoproduction of $\phi$ mesons gained already interest decades ago, since it is sensitive to the strangeness content of the nucleon. Earlier models reported that an enhancement in the cross section could be attributed to a sizable strange quark admixture, see for example \cite{Henley:1991ge}. In the last years, a few measurements of the cross section as well as \textit{SDME}s by the CLAS collaboration, by LEPS and by SAPHIR became available. The first measurements of the cross section by the CLAS collaboration \cite{Anciant:2000az,McCormick:2003bb} observed a slight rise in the cross section in backward direction, which could be attributed to nucleon exchange in the u-channel. This is effect can be attributed to a appreciable strangeness contribution in the nucleon \cite{Dey:2014tfa}. 

In the cross section measurement by LEPS \cite{Mibe:2005er} in 2005, a structure around $E_\gamma \approx 2$~GeV was observed, which did not comply with the expected rise in the cross section due to t-channel exchange. Further measurements with high statistics by the CLAS collaboration \cite{Dey:2014tfa} confirmed this structure and could show that it arises from the t-channel exchange in forward direction and is dominantly observed in the charged decay channel. Several different ideas like rescattering or interference effects were discussed, but no final conclusion could be drawn from the data. Further investigations with high statistics data sets including polarization are needed to answer this question.

Data of photoproduction of $\rho$ mesons is quite sparse. There were only two measurements of the cross section in the last years, one from the CLAS collaboration \cite{Battaglieri:2001xv} and one from the SAPHIR experiment \cite{Wu:2005wf}. A substantial difficulty  for the measurement of this reactions comes from the fact that the $\rho$ meson decays via the $\pi^+\pi^-$ final state \cite{Zyla:2020zbs}. Here, a separation of the signal of the $\rho$ meson from background contributions for example of $\Delta$ production needs to be done carefully. An example of the differential cross section measured by the SAPHIR collaboration \cite{Wu:2005wf} and the CLAS measurement \cite{Battaglieri:2001xv} can be seen in Fig.~\ref{fig:M:prho_crosssection_wu} (right). In the differential cross section, a strong forward peaking behavior is visible, which is expected from the t-channel exchange. Contributing resonances for this final states have been extracted by the model by Hunt et al.~\cite{Hunt:2018wqz}. Contributions were for example observed by the $N(1880)\frac{1}{2}^+$ and the $N(2060)\frac{5}{2}^-$, although further validation would be desirable.

\begin{table*}
\caption{A collection of polarization observable measurements for vector meson photoproduction. Hint: We do not distinguish between differential and total cross section data. Further details are given in the cited references. }\label{tab:measSingleVector}
\centering
\vspace*{5mm}
\begin{tabular}{clllc}
Observable & Energy range $E^{lab}_\gamma$ & Facility & Reference & Year\\
\hline
\multicolumn{5}{c}{$\gamma p \rightarrow p\omega$}\\
\hline
$\sigma$, $\Sigma$, SDME & $1300-2400$ MeV & LEPS2 & Muramatsu et al. \cite{Muramatsu:2020xbx} & 2020\\
P, H & $1200-2000$ MeV & CLAS & Roy et al. \cite{Roy:2018tjs} & 2019\\
F & $1200-2700$ MeV & CLAS & Roy et al. \cite{Roy:2018tjs} & 2019\\
$\Sigma$ & $1108-2100$ MeV & CLAS & Roy et al. \cite{Roy:2017qwv} & 2018\\
T & $1108-2800$ MeV & CLAS & Roy et al. \cite{Roy:2017qwv} & 2018\\
$\Sigma$ & $1152-1876$ MeV & CLAS & Collins et al. \cite{Collins:2017vev} & 2017\\
E & $1108-2300$ MeV & CLAS & Akbar et al. \cite{Akbar:2017uuk} & 2017\\
E & $1108 - 2300$ MeV & CBELSA/TAPS & Eberhardt et al. \cite{Eberhardt:2015lwa} & 2015\\
G & $1108 - 1300$ MeV & CBELSA/TAPS & Eberhardt et al. \cite{Eberhardt:2015lwa} & 2015\\
$\sigma$, SDME & $1108-2500$ MeV & CBTAPS & Wilson et al. \cite{Wilson:2015uoa} & 2015\\
$\sigma$ & $1500-3000$ MeV & LEPS & Morino et al. + \cite{Morino:2013raa} & 2015\\
$\sigma$ & $1200-1600$ MeV & CBTAPS & Dietz et al. \cite{Dietz:2015ppa} & 2015\\
$\sigma$ & $1108-1400$ MeV & Crystal Ball/TAPS & Strakovsky et al. \cite{Strakovsky:2014wja} & 2015\\
$\Sigma$ & $1100-1500$ MeV & GRAAL & Vegna et al. \cite{Vegna:2013ccx} & 2015\\
$\sigma$, SDME & $1108-3829$ MeV & CLAS & Williams et al. \cite{Williams:2009ab} & 2009\\
$\Sigma$ & $1108-1700$ MeV & CBTAPS & Klein et al. \cite{Klein:2008aa} & 2008\\
$\sigma$, $\Sigma$ & $1108-1500$ MeV & GRAAL & Ajaka et al. \cite{Ajaka:2006bn} & 2006\\
$\sigma$ & $1108-2600$ MeV & SAPHIR & Barth et al. \cite{Barth:2003kv} & 2003\\
$\sigma$ & $3190-3910$ MeV & CLAS & Battaglieri et al. \cite{Battaglieri:2002pr} & 2003\\
\hline
\multicolumn{5}{c}{$\gamma n \rightarrow n\omega$}\\
\hline
$\sigma$ & $1400-3400$ MeV & CLAS & Chetry et al. \cite{Chetry:2018kxz} & 2018\\
$\Sigma$ & $1100-1500$ MeV & GRAAL & Vegna et al. \cite{Vegna:2013ccx} & 2015\\
$\sigma$ & $1200-1600$ MeV & CBTAPS & Dietz et al. \cite{Dietz:2015ppa} & 2015\\
\hline
\multicolumn{5}{c}{$\gamma p \rightarrow p\phi$}\\
\hline
$\sigma$, SDME & $1500-2900$ MeV & LEPS & Mizutani et al. \cite{Mizutani:2017wpg} & 2017\\
$\sigma$, SDME & $1600-3830$ MeV & CLAS & Dey et al. \cite{Dey:2014tfa} & 2014\\
$\sigma$ & $1600-3600$ MeV & CLAS & Seraydaryan et al. \cite{Seraydaryan:2013ija} & 2014\\
$\sigma$, SDME & $1650-3590$ MeV & CLAS & Qian et al. \cite{Qian:2009ab} & 2009\\
$\sigma$, SDME & $1573-2370$ MeV & LEPS & Mibe et al. \cite{Mibe:2005er} & 2005\\
SDME & $3300-3900$ MeV & CLAS & McCormick et al. \cite{McCormick:2003bb} & 2004\\
$\sigma$ & $1573-2650$ MeV & SAPHIR & Barth et al. \cite{Barth:2003bq} & 2003\\
$\sigma$ & $3300-3900$ MeV & CLAS & Anciant et al. \cite{Anciant:2000az} & 2000\\
\hline
\multicolumn{5}{c}{$\gamma p \rightarrow p\rho$}\\
\hline
$\sigma$, SDME & $1086-2600$ MeV & SAPHIR & Wu et al. \cite{Wu:2005wf} & 2005\\
$\sigma$ & $3190-3910$ MeV & CLAS & Battaglieri et al. \cite{Battaglieri:2001xv} & 2001\\
\hline
\end{tabular}
\end{table*}

\subsubsection{Strangeness Photoproduction}
\label{sec:measurements:StrangenessPhotoproduction}
Strangeness photoproduction offers a different insight to the non-strange excited states ($N^*,\Delta^*$) as some resonances are expected to couple more strongly to $K\Lambda$ and $K\Sigma$ final states than to $\pi N$ or to $\pi\pi N$ final states \cite{Capstick:1993kb,Capstick:1998uh}. The most studied reaction channels are $\gamma p \to K^+ \Lambda $ and $\gamma p \to K^+ \Sigma^0 $ (see Tab.~\ref{tab:measStrangeness}). In addition, strangeness photoproduction has the advantage that the hyperon polarization degree can be determined through the parity violating weak decay and by measuring the angular distribution of the decay products ($\Lambda \to p \pi^-$ and for $\Sigma^0$ via $\Sigma^0 \to \Lambda \gamma; \Lambda \to p \pi^-$) \cite{Glander:2003jw}. This means recoil polarization observables can be easily accessed in these reactions, which are a lot more challenging to measure compared to the already discussed photoproduction reactions. Therefore, these kaon-hyperon $KY$ final states offer the best chance of obtaining a rich database and thus, a complete experiment (see Sec.~\ref{sec:formalisms:CompleteExperiments}) for single meson photoproduction.

The isospin of $\Lambda$ is $0$, while the isospin of $\Sigma$ is $1$. While the $K\Lambda$ final states are only sensitive to $N^*$ resonances in the intermediate state like the $p\eta$ and $p\eta'$ final states, the $K\Sigma$ final states have intermediate states with possible isospin of $1/2$ ($N^*$) or $3/2$ ($\Delta^*$).

\begin{figure}[tbhp]
	\begin{center}
		\includegraphics[width=0.7\textwidth]{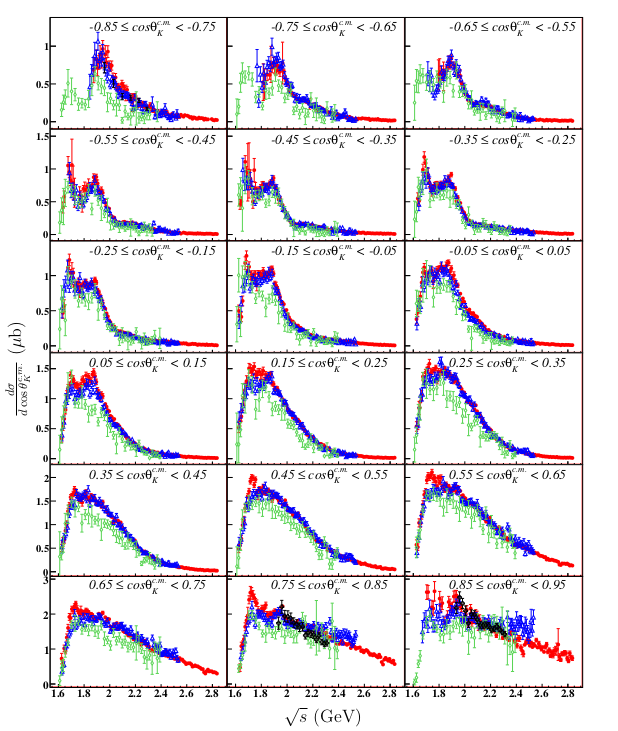}
	\end{center}
	\caption{The differential cross section as a function of the center of mass energy $\sqrt{s}$ is shown for $K^+\Lambda$ for different angles. Data were measured at CLAS (closed red circles \cite{McCracken:2009ra}, open blue triangles \cite{Bradford:2005pt}), SAPHIR (open green diamonds) \cite{Glander:2003jw} , LEPS (open black crosses) \cite{Sumihama:2007qa,Hicks:2007aa} . \cite{McCracken:2009ra}. \APS{\cite{McCracken:2009ra}}{2010} }
	\label{fig:M:K+Lambda_crossection_Glander}
\end{figure}

The LEPS, SAPHIR, BGOOD and CLAS spectrometers were well suited for the detection of the charged particles like the charged kaon, the proton or charged pion in the final state. Therefore, these experimental facilities contributed mainly to the database of the $K^+\Lambda$ and $K^+\Sigma^0$ final states. The GRAAL collaboration performed $dE/dx$ and TOF measurements to reconstruct the charged particles and also contributed to the existing database. It is worth mentioning that Jude et al.~\cite{Jude:2013jzs} were able to obtain differential cross section data for both final states as well using the Crystal Ball/TAPS setup (see Fig.~\ref{fig:a2setup}). Here, a novel technique to identify the $K^+$ was developed, studying the energy and timing characteristics of the energy deposits of the $K^+$ weak decay products in the calorimeter.

\begin{figure}[tb]
	\includegraphics[width=0.48\textwidth]{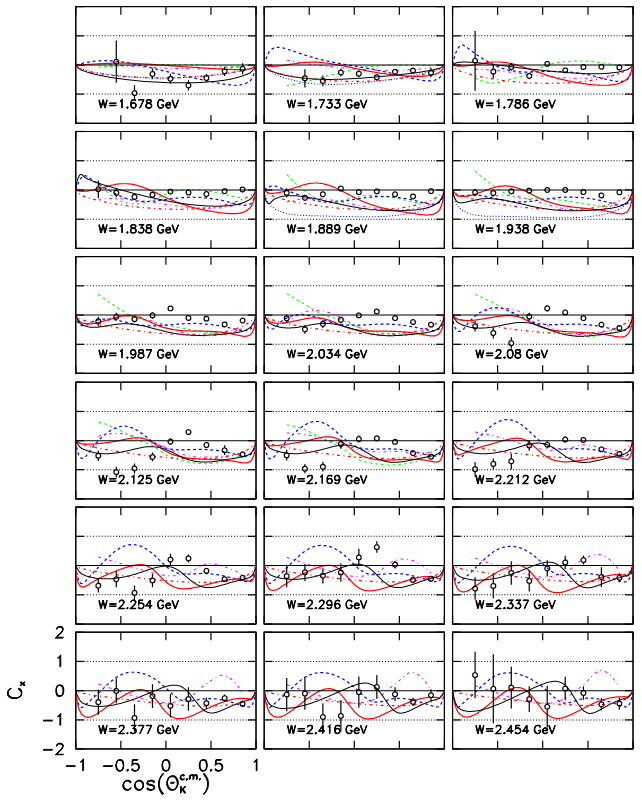}
	\includegraphics[width=0.47\textwidth]{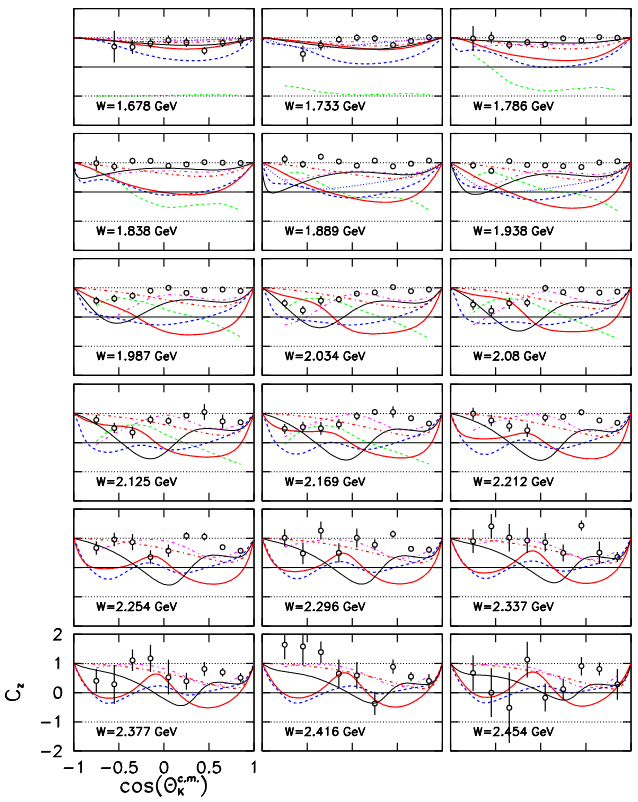}
	\caption{The beam-recoil polarization observables $C_x$ (left) and $C_z$ (right) for $K^+\Lambda$ measured at the CLAS experiment \cite{Bradford:2006ba}. \APS{\cite{Bradford:2006ba}}{2007}}
	\label{fig:M:K+Lambda_CxCz_Bradford}
\end{figure}

Differential cross section data of the $K^+\Lambda$ channel from SAPHIR \cite{Glander:2003jw}, LEPS \cite{Sumihama:2007qa} and CLAS \cite{McNabb:2003nf} show an enhancement at $E_\gamma\approx 1450$ MeV ($W\approx 1900$ MeV) for backward $K^+$ angles (see Fig.~\ref{fig:M:K+Lambda_CxCz_Bradford}). According to the single-channel isobar Kaon-MAID model of Mart and Bennhold \cite{Mart:1999ed} this enhancement is caused by the $N(1895)\frac{3}{2}^- (D_{13})$ resonance, which was also predicted by a relativistic quark model calculation \cite{Glander:2003jw}. At forward angles, the differential cross section for $K^+\Lambda$ shows a forward peak and increases due to a large contribution of $K^+$, $K^*(892)$,and $K_1(1270)$ meson exchanges in the $t$-channel \cite{Sumihama:2007qa}. In contrast to this, the differential cross section data for the $K^+\Sigma^0$ channel decreases at forward angles and $K$ exchange in the $t$-channel does not seem to play a dominant role \cite{Sumihama:2007qa}. It is noteworthy to mention that the CLAS collaboration measured the differential cross section and the recoil polarization observable $P_\Lambda$ with high precision with a 10 MeV-wide energy binning \cite{McCracken:2009ra}. However, different resonance contributions have been suggested based on the included data set in the PWA due to the existing discrepancies between the CLAS \cite{McCracken:2009ra} and the SAPHIR \cite{Glander:2003jw} data sets \cite{Mart:2006dk}. Here, the latest high-precision data from BGOOD \cite{Alef:2020yul} was able to resolve this issue for the very forward angles $\cos\theta^K_{CM}>0.9$, where they report a good agreement to the CLAS data. In addition, this data helps to constrain the $t$-channel $K$ and $K^*$ exchange at forward angles and low momentum transfer. More data regarding strangeness photoproduction can be expected from the BGOOD facility in the future  \cite{Alef:2019imq,Jude:2019qqd,Jude:2020byj,Jude:2020bwn}.

\begin{figure}[th]
\begin{center}
\includegraphics[width=0.87\textwidth]{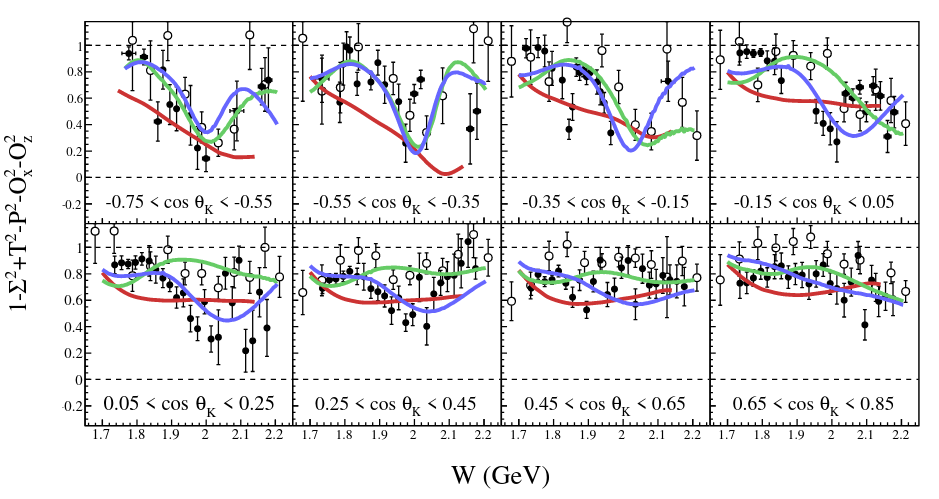}
\end{center}
\caption{Consistency check using the Fierz relation and all measured polarization observables for the reaction $\gamma p \to K^+ \Lambda$ \cite{GlasgowPrivComm} at CLAS: $C_x^2+C_z^2$ (open circles) \cite{Bradford:2006ba} has to equal $1- O_x^2-O_z^2-\Sigma^2+T^2-P^2$ (filled circles) \cite{Paterson:2016vmc}. The colored lines give the evaluated values from the three models: ANL-Osaka \cite{Kamano:2013iva} (red line), BnGa-2014 \cite{Gutz:2014wit} (green line) and a BnGa refit \cite{Paterson:2016vmc} (blue line).}
\label{fig:M:K+Lambda_FierzRelation_Paterson}
\end{figure}

The CLAS collaboration measured for the first time the beam-recoil double polarization observables $C_x$ and $C_z$ for the $K^+\Lambda$ (see Fig.~\ref{fig:M:K+Lambda_CxCz_Bradford}) and $K^+\Sigma^0$ final states \cite{Bradford:2006ba}. Furthermore, the polarization observables $\Sigma, T, P_\Lambda, O_x, O_z$ were all measured at CLAS as well \cite{McNabb:2003nf,McCracken:2009ra,Paterson:2016vmc}. This allowed for a consistency check between the measured data-sets using the Fierz identity, which holds \begin{equation}
\label{eq:M:FierzRelation}
O_x^2+O_z^2+C_x^2+C_z^2 + \Sigma^2-T^2+P^2=1.
\end{equation}
Fig.~\ref{fig:M:K+Lambda_FierzRelation_Paterson} shows the comparison between $C_x^2+C_z^2$ and $1-O_x^2-O_z^2 - \Sigma^2 +T^2-P^2$, which is fairly good. Ireland et al.~\cite{Ireland:2019uja} used the recent CLAS polarization data to estimate a value for the decay parameter $\alpha_-$ of the $\Lambda \to p\pi^-$ decay. It was determined as $0.721(6)(5)$, which is in agreement with the new $\alpha_-$ value reported by the BES III collaboration \cite{Ablikim:2018zay}. However, both latest values are significantly higher than the PDG value of $\alpha_-=0.642(13)$. As a consequence, all recoil polarization observables, that rely on the $\alpha_-$ value for the extraction of the polarization observables, will have to be re-considered in the future.

The J\"uBo group fitted all the latest CLAS data and studied the impact of the $\gamma p \to K^+\Lambda$ channel on the nucleon excitation spectrum \cite{Ronchen:2018ury}. They report significant changes in the resonance parameters of the $N(1710)\frac{1}{2}^+(P_{11})$ and $N(1720)\frac{3}{2}^+ (P_{13})$ resonances. In addition, the existence of the $N(1900)\frac{3}{2}^+(P_{13})$ and possibly the $N(2060)\frac{5}{2}^-(D_{15})$ was confirmed and further indications for possible new dynamically generated poles in the $D_{13}$ and $P_{31}$ waves were found. The BnGa group     \cite{Anisovich:2017bsk} extracted the $E_{0+}$, $M_{1-}$, $E_{1+}$, and $M_{1+}$ multipoles of the reaction $\gamma p \to K^+\Lambda$ using their energy-dependent multichannel fit and using an energy-independent multichannel L$+$P formalism \ref{sec:PWA:LPlusFormalism}. They find evidence for the following three resonances: 
$N(1895)\frac{1}{2}^-(S_{11})$, $N(1880)\frac{1}{2}^+(P_{11})$ and $N(1900)\frac{3}{2}^+(P_{13})$.


To achieve an isospin separation, data is needed not only for $\gamma p \to K^+\Sigma^0$, but also for the reaction $\gamma p \to K^0\Sigma^+$. Here, the decay modes $K^0_S \to \pi^+\pi^-$ and $\Sigma^+ \to \pi^0 p$ or $\Sigma^+\to n\pi^+$ were used by the SAPHIR \cite{Lawall:2005np} and the CLAS facilities \cite{Nepali:2013bp}, while the CBTAPS \cite{Castelijns:2007qt,Ewald:2011gw,Ewald:2014ozu} and the Crystal Ball/TAPS \cite{Aguar-Bartolome:2013fqw,Akondi:2018shh} facilities used the neutral decay mode $K^0_S \to \pi^0\pi^0$. 

\begin{figure}[th]
\includegraphics[width=0.38\textwidth]{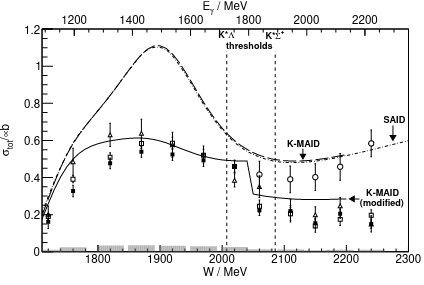}
\includegraphics[width=0.57\textwidth]{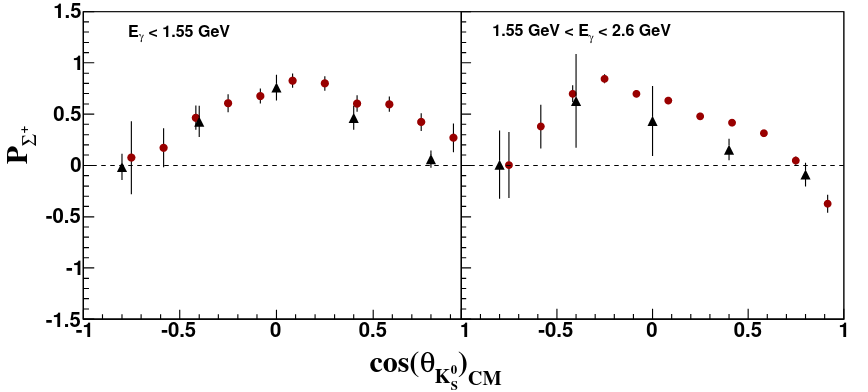}
\caption{Left: The total cross section for the reaction $\gamma p \to K^0\Sigma^+$ of CBTAPS (filled \cite{Ewald:2011gw} and open \cite{Castelijns:2007qt} squares) and SAPHIR (triangles) \cite{Lawall:2005np}. The circles show the sum of $K^0\Sigma^+$ \cite{Ewald:2011gw} and $K^{0*}\Sigma^+$ \cite{Nanova:2008kr} cross section data above the $K^*$ threshold. \cite{Ewald:2011gw} Right: The hyperon recoil polarization $P_{\Sigma^+}$ of CLAS (red circles) \cite{Nepali:2013bp} and SAPHIR (black triangles) \cite{Lawall:2005np}. \APS{\cite{Nepali:2013bp}}{2013}}
\label{fig:M:K0Sigma+_P_Nepali}
\end{figure}

So far, only the cross section, the recoil polarization observable and the beam asymmetry have been published for this reaction. The total cross section is shown in Fig.~\ref{fig:M:K0Sigma+_P_Nepali} on the left. The data sets from Ewald et al.~\cite{Ewald:2011gw}, Castelijns et al.~\cite{Castelijns:2007qt} and Lawall et al.~\cite{Lawall:2005np} show a drop in the total cross section right between the thresholds of $K^*\Lambda$ and $K^*\Sigma$ photoproduction at $E_\gamma \approx 1800$~MeV as indicated by the dashed lines. According to Ewald et al.~\cite{Ewald:2011gw}, the angular dependence of the differential cross section changes remarkably in this energy region. The differential cross section rises with higher incident photon energy and shows an increasing forward peaking, which can be attributed to an increasing $t$-channel contribution associated with $K^{0*}$ exchange. However, at $E_\gamma \approx 1800$ this behavior changes to a flat angular distribution and the cross section drops significantly, suggesting a change from $t$-channel mechanism to $s$-channel production mechanism of $K^0\Sigma^+$ around the $K^*$ threshold region. The Kaon-MAID model was able to reproduce the data significantly better by manually setting the $K^{0*}$ exchange to zero above the $K^{*}\Sigma^+$ threshold. In addition, if the $K^*$ vector meson is produced as a free particle above the $K^*$ threshold, it is expected that the vanishing strength from the $K^0\Sigma^+$ channel should contribute to the $K^{0*}\Sigma^+$ channel. This was tested by adding up the cross sections from the $K^0\Sigma^+$ \cite{Ewald:2011gw} and $K^{0*}\Sigma^+$ \cite{Nanova:2008kr} above the $K^*$ threshold, which leads to a smooth transition at the threshold region (see Fig.~\ref{fig:M:K0Sigma+_P_Nepali}). However, more polarization observables are needed to confirm this cusp-like behavior.


Based on SU(6) symmetry and considering the spin-flavor configurations of the hyperons, it follows for the hyperon polarization $P_{\Sigma^0}\approx -P_{\Lambda}$ and based on isospin symmetry $P_{\Sigma^0}\approx P_{\Sigma^+}$. Comparing the recoil polarization observable $P$ for the reactions $\gamma p \to K^+\Lambda$ and $\gamma p \to K^+\Sigma^0$, Dey et al.~\cite{Dey:2010hh} showed that the SU(6) symmetry is explicitly broken for mid- and backward-angles of the $K^+$ in the center-of-mass system. However, for large $\sqrt{s}$ and forward angles, the symmetry seems to hold \cite{Dey:2010hh}. Fig.~\ref{fig:M:K0Sigma+_P_Nepali} shows on the right the hyperon polarization $P_{\Sigma^+}$ for the reaction $\gamma p \to K^0\Sigma^+$. The polarization is highest in the mid-angular range. Comparing the data to $K^+\Sigma^0$ data, shows a similar trend for the hyperon polarization for forward kaon angles and high $\sqrt{s}$. However, large differences are observed at $\sqrt{s}\approx2$ GeV \cite{Nepali:2013bp}.    It is assumed that the differences may be related to different production mechanisms in the two energy ranges.

\begin{figure}[th]
\begin{center}
\includegraphics[width=0.5\textwidth]{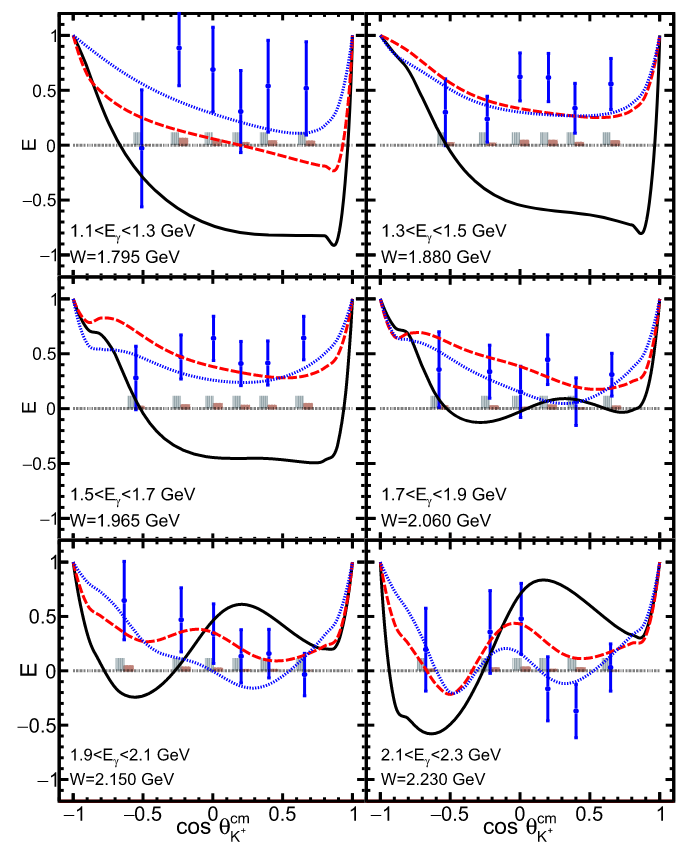}
\end{center}
\caption{The double polarization observable $E$ of the CLAS facility for the reaction $\gamma n \to K^+\Sigma^-$ \cite{Zachariou:2020kkb}. The black line represents the BnGa-2017 solution \cite{Anisovich:2017pox}, the dashed red line the solution including the new helicity asymmetry data and the dotted blue line the solution with an additional $D_{13}$ resonance.}
\label{fig:M:K+Sigma-_E_Zachariou}
\end{figure}

Only a sparse databasis, comprised mainly of the cross section data, exists for the reactions on the neutron, e.g. for $\gamma n \to K^0\Lambda$ \cite{Compton:2017xkt,Akondi:2018shh}, $\gamma n \to K^0\Sigma^0$ \cite{Akondi:2018shh} and $\gamma n \to K^+\Sigma^-$ \cite{Kohri:2006yx,AnefalosPereira:2009zw}. First data for the $E$ observable was published as well for $\gamma n \to K^0\Lambda$ and $\gamma n \to K^0\Sigma^0$ \cite{Ho:2018riy}, albeit with poor statistics. Fig.~\ref{fig:M:K+Sigma-_E_Zachariou} shows the results of $E$ for the reaction $\gamma n \to K^+\Sigma^-$, which were measured at the CLAS facility \cite{Zachariou:2020kkb}. This new data is sensitive to the interference of the $S_{11}$ and $P_{13}$ partial waves, changing significantly the photocouplings of the $N(1895)\frac{1}{2}^-(S_{11})$, $N(1720)\frac{3}{2}^+(P_{13})$ and $N(1900)\frac{3}{2}^+(P_{13})$ resonances. In addition, indications for the \textit{missing} $N(2120)\frac{3}{2}^-(D_{13})$ resonance are given within the Bonn-Gatchina PWA \cite{Muller:2019qxg}. Further discussion can be found in \cite{Zachariou:2020kkb,}.

Further implications for potential \textit{missing} resonances above $E_\gamma=2$ GeV is expected to result from photoproduction data of vector mesons like $K^{*+}(892)$ in the reaction $\gamma p \to K^{*+}\Lambda$, which has a high production threshold of $W=2007$ MeV. More details are given in \cite{Anisovich:2017rpe,Kim:2011rm,Kim:2014hha,Wang:2017tpe,Wang:2018vlv,Wang:2019mid}. 
  
  
\begin{table*}[p]
	\caption{A collection of polarization observable measurements for strangeness photoproduction. Hint: We do not distinguish between differential and total cross section data. Further details are given in the cited references. The symbols indicate which PWA (circle: BnGa-2019 \cite{Muller:2019qxg}, diamond: J\"uBo-2017 \cite{Ronchen:2018ury} and star: SAID \cite{SAIDURL}) use the data in their fits.}\label{tab:measStrangeness}
	\centering
	\vspace*{5mm}
	\begin{tabular}{clllc}
		Observable & Energy range $E^{lab}_\gamma$ & Facility & Reference & Year\\
		\hline
		\multicolumn{5}{c}{$\gamma p \rightarrow K^+ \Lambda$}\\
		\hline
		$\sigma, P_{\Lambda}$ & $911-1394$ MeV & BGOOD & Alef et al. \cite{Alef:2020yul} & 2021\\
		$\sigma, \Sigma$ & $1500-3000$ MeV & LEPS & Shiu et al. \cite{Shiu:2017wiw} & 2018\\
		$\Sigma$, T, P, $O_x$, $O_z$ & $1710-2190$ MeV & CLAS & Paterson et al. \cite{Paterson:2016vmc} $\circ$ $\diamond$ $\star$ & 2016\\
		$\sigma$ & $911-1450$ MeV & Crystal Ball/TAPS & Jude at al. \cite{Jude:2013jzs} $\circ$ $\star$ & 2014\\
		$\sigma$, $P_\Lambda$ & $1620-2840$ MeV & CLAS & McCracken et al. \cite{McCracken:2009ra} $\circ$ $\diamond$ $\star$ & 2010\\
		$O_x$, $O_z$, T & $911-1500$ MeV & GRAAL & Lleres et al. \cite{Lleres:2008em} $\circ$ $\diamond$ $\star$ & 2009\\
		$\Sigma$, $P_\Lambda$ & $911-1500$ MeV & GRAAL & Lleres et al. \cite{Lleres:2007tx} $\circ$ $\diamond$ $\star$ & 2007\\
		$\sigma$, $\Sigma_{K\Lambda}$ & $1500-2370$ MeV & LEPS & Hicks et al. \cite{Hicks:2007aa} $\diamond$ $\star_\Sigma$ & 2007\\
		$C_x$, $C_z$ & $895-2942$ MeV & CLAS & Bradford et al. \cite{Bradford:2006ba} $\circ$ $\diamond$ $\star$ & 2007\\
		$\sigma$ & $895-2942$ MeV & CLAS & Bradford et al. \cite{Bradford:2005pt} $\diamond$ $\star$ & 2006\\
		$\sigma$, $\Sigma$ & $1500-2400$ MeV & LEPS & Sumihama et al. \cite{Sumihama:2005er} $\diamond$ $\star$ & 2006\\
		$\sigma$,$P_\Lambda$ & $911-2600$ MeV & SAPHIR & Glander et al. \cite{Glander:2003jw} $\diamond$ $\star$ & 2004\\
		$\sigma$, $P_\Lambda$ & $911-2325$ MeV & CLAS & McNabb et al. \cite{McNabb:2003nf} $\diamond$ $\star$ & 2004\\
		$\Sigma$ & $1500-2400$ MeV & LEPS & Zegers et al. \cite{Zegers:2003ux} $\diamond$ $\star$ & 2003\\	
		\hline
		\multicolumn{5}{c}{$\gamma p \rightarrow K^+ \Sigma^0$}\\
		\hline
		$\sigma$ & $1688-1973$~MeV & BGOOD & Jude et al. \cite{Jude:2020byj} & 2021 \\
		$\sigma, \Sigma$ & $1500-3000$ MeV & LEPS & Shiu et al. \cite{Shiu:2017wiw} & 2018\\
		$\Sigma$, T, P, $O_x$, $O_z$ & $1710-2190$ MeV & CLAS & Paterson et al. \cite{Paterson:2016vmc} $\circ$ $\star$ & 2016\\
		$\sigma$ & $1046-1450$ MeV & Crystal Ball/TAPS & Jude at al. \cite{Jude:2013jzs} $\circ$ $\star$ & 2014\\
		$\sigma$, $P_\Sigma$ & $1053-3829$ MeV & CLAS & Dey et al. \cite{Dey:2010hh} $\circ$ $\star$ & 2010\\
		$\Sigma$, $P_\Sigma$ & $1046-1500$ MeV & GRAAL & Lleres et al. \cite{Lleres:2007tx} $\circ$ $\star$ & 2007\\
		$C_x$, $C_z$ & $895-2942$ MeV & CLAS & Bradford et al. \cite{Bradford:2006ba} $\circ$ $\star$ & 2007\\
		$\sigma$ & $895-2942$ MeV & CLAS & Bradford et al. \cite{Bradford:2005pt} $\star$ & 2006\\
		$\sigma$, $\Sigma$ & $1500-2400$ MeV & LEPS & Kohri et al. \cite{Kohri:2006yx} $\star$ & 2006\\	
		$\sigma$, $\Sigma$ & $1500-2400$ MeV & LEPS & Sumihama et al. \cite{Sumihama:2005er} $\star$ & 2006\\
		$\sigma$,$P_\Sigma$ & $1046-2600$ MeV & SAPHIR & Glander et al. \cite{Glander:2003jw} $\star$ & 2004\\
		$\sigma$, $P_\Sigma$ & $1046-2325$ MeV & CLAS & McNabb et al. \cite{McNabb:2003nf} $\star$ & 2004\\
		$\Sigma$ & $1500-2400$ MeV & LEPS & Zegers et al. \cite{Zegers:2003ux} $\circ$ $\star$ & 2003\\
		\hline
		\multicolumn{5}{c}{$\gamma p \rightarrow K^0 \Sigma^+$}\\
		\hline
		$\sigma$ & $1047-1855$ MeV & Crystal Ball/TAPS & Akondi et al. \cite{Akondi:2018shh} & 2019\\
		$\Sigma$, $P_\Sigma$ & $1047-1650$ MeV & CBTAPS & Ewald et al. \cite{Ewald:2014ozu} & 2014\\
		$P_\Sigma$ & $1000-3500$ MeV & CLAS & Nepali et al. \cite{Nepali:2013bp} $\circ$ & 2013\\
		$\sigma$, $P_\Sigma$ & $1047-1450$ MeV & Crystal Ball/TAPS & Aguar-Bartolome et al. \cite{Aguar-Bartolome:2013fqw} $\circ$ $\star$ & 2013\\
		$\sigma$ & $1047-2250$ MeV & CBTAPS & Ewald et al. \cite{Ewald:2011gw} $\circ$ & 2012\\
		$\sigma$, $P_\Sigma$ & $1047-2300$ MeV & CBTAPS & Castelijns et al. \cite{Castelijns:2007qt} $\circ$ & 2008\\
		$\sigma$, $P_\Sigma$ & $1047-2600$ MeV & SAPHIR & Lawall et al. \cite{Lawall:2005np} $\circ$ $\star$ & 2005\\
		\hline
	\end{tabular}
\end{table*}

\begin{table*}[ht]
	\caption{A collection of polarization observable measurements for strangeness photoproduction. Hint: We do not distinguish between differential and total cross section data. Further details are given in the cited references. The symbols indicate which PWA (circle: BnGa-2019 \cite{Muller:2019qxg}, diamond: J\"uBo-2017 \cite{Ronchen:2018ury} and star: SAID \cite{SAIDURL}) use the data in their fits.}\label{tab:measStrangeness2}
	\centering
	\vspace*{5mm}
	\begin{tabular}{clllc}
		Observable & Energy range $E^{lab}_\gamma$ & Facility & Reference & Year\\
		\hline
		\multicolumn{5}{c}{$\gamma n \rightarrow K^0 \Lambda$}\\
		\hline
		$\sigma$ & $915-1365$ MeV & Crystal Ball/TAPS & Akondi et al. \cite{Akondi:2018shh} & 2018\\
		E & $1071-2449$ MeV & CLAS & Ho et al. \cite{Ho:2018riy} & 2018\\
		$\sigma$ & $500-3600$ MeV & CLAS & Compton et al. \cite{Compton:2017xkt} & 2017\\
		\hline
		\multicolumn{5}{c}{$\gamma n \rightarrow K^+ \Sigma^-$}\\
		\hline
		E & $1100-2300$ MeV & CLAS & Zachariou et al. \cite{Zachariou:2020kkb} & 2020\\
		$\sigma$ & $1100-3600$ MeV & CLAS & Anefalos Pereira et al. \cite{AnefalosPereira:2009zw} & 2010\\
		$\sigma$, $\Sigma$ & $1500-2400$ MeV & LEPS & Kohri et al. \cite{Kohri:2006yx} $\circ$ $\star$ & 2006\\	
		\hline
		\multicolumn{5}{c}{$\gamma n \rightarrow K^0 \Sigma^0$}\\
		\hline
		$\sigma$ & $1051-1365$ MeV & Crystal Ball/TAPS & Akondi et al. \cite{Akondi:2018shh} & 2019\\
		E & $1071-2449$ MeV & CLAS & Ho et al. \cite{Ho:2018riy} & 2018\\	
		\hline
	\end{tabular}
\end{table*}

\subsection{Multi Meson Photoproduction}
\label{sec:measurements:MultiMesonPhotoproduction}

\begin{figure}[th]
	\raisebox{0.1\height}{\includegraphics[width=0.25\textwidth]{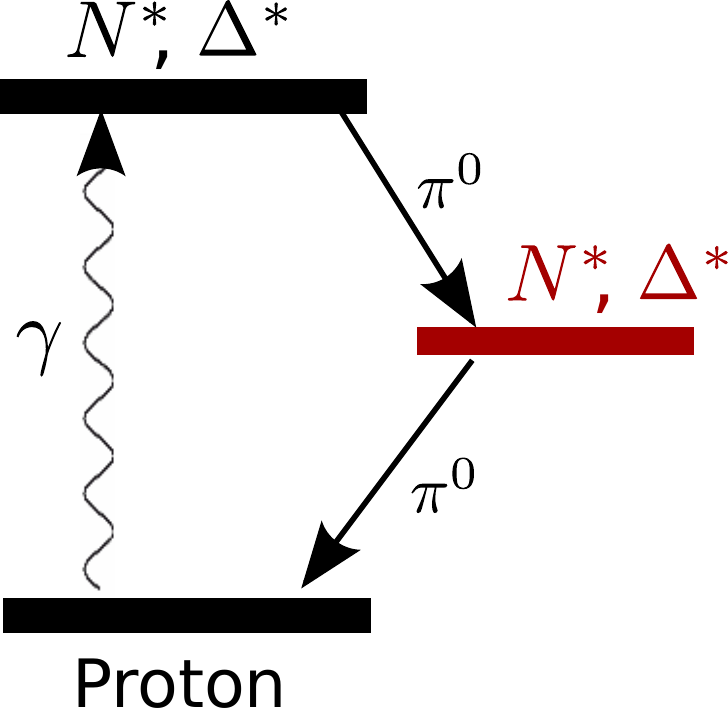}}
	\includegraphics[width=0.7\textwidth]{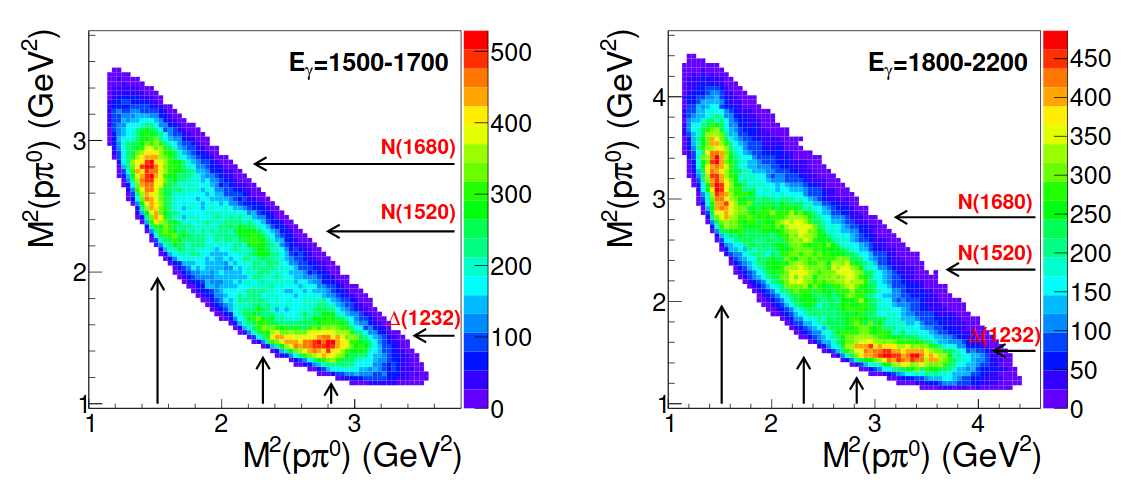}
	\caption{Left: Sketch of the cascading decay of a resonance via a second resonant state. Right: Dalitz plot of the reaction $p\pi^0\pi^0$ for two different energies \cite{Sokhoyan:2015fra}. The intermediate states are visible as horizontal and vertical bands and marked by arrows. \EPJA}
	\label{fig:M:ppi0pi0_dalitz}
\end{figure}

In the comparison of cross sections for different final states, like it is shown in Fig.~\ref{fig:Intro:crosssections}, it becomes obvious that at high beam energies, nature seems to invest more into the production of final states including multiple particles than in single meson photoproduction. This makes the investigation of final states like $\pi\pi$ or $\pi\eta$ especially interesting in the search for heavy resonances. In these reactions, the excited state will decay in most cases by emittance of a meson into an intermediate resonance, which will then decay into the ground state via a second meson, see Fig.~\ref{fig:M:ppi0pi0_dalitz}, left. To visualize the intermediate states, Dalitz plots can be analyzed. Two example plots for the reaction $\gamma p \to p \pi^0\pi^0$ can be seen in Fig.~\ref{fig:M:ppi0pi0_dalitz}, right. Here, the squared invariant mass of one $p\pi^0$ pair is plotted against the invariant mass squared of the second pion with the proton. The intermediate resonances are visible as horizontal and vertical bands. In the example for $p\pi^0\pi^0$, which is shown in Fig.~\ref{fig:M:ppi0pi0_dalitz}, clear signatures from the $\Delta(1232)\frac{3}{2}^+$, the $N(1520)\frac{3}{2}^-$ and the $N(1680)\frac{5}{2}^+$ as intermediate states are visible.

These processes make the multi-meson final states interesting for the partial wave analyses. If there are states, which cannot directly decay into the ground state via emittance of a single meson, multi meson photoproduction might provide a different method to access these states.

\begin{figure}[tp]
	\centering
	\includegraphics[width=0.6\textwidth]{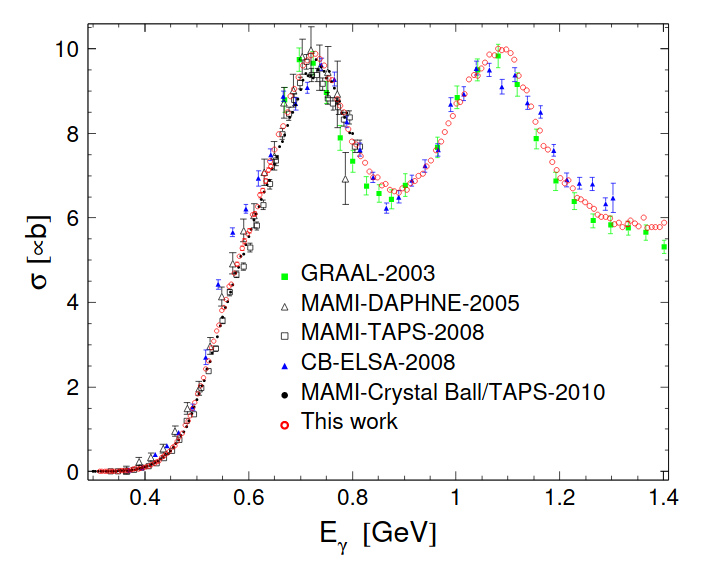}\\
	\includegraphics[width=0.8\textwidth]{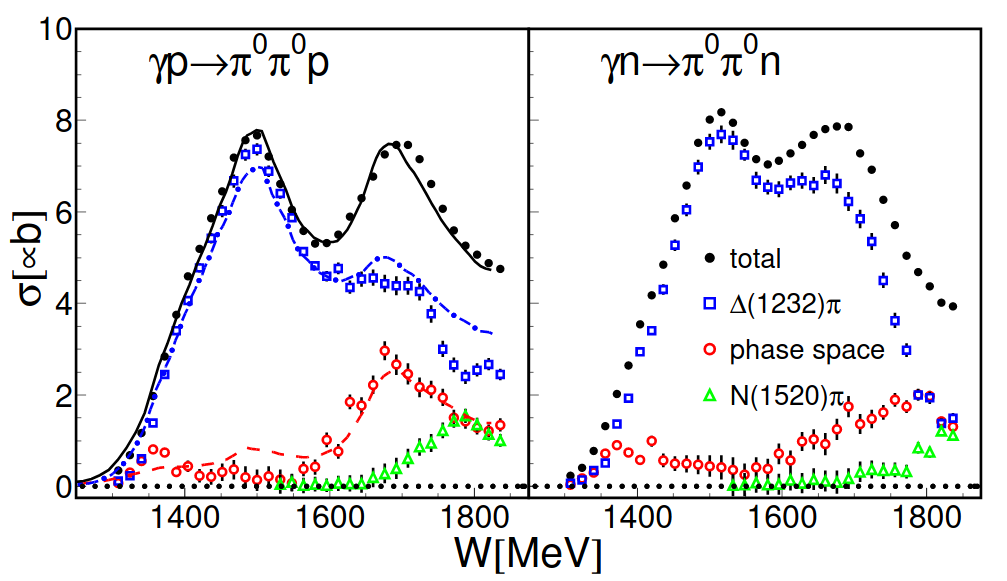}
	\caption{Top: The cross section for $p\pi^0\pi^0$ measured by the Crystal Ball/TAPS experiment (red circles) \cite{Kashevarov:2012wy}, compared to previous measurements: GRAAL (green) \cite{Assafiri:2003mv}, CBELSA (blue triangles) \cite{Thoma:2007bm},\cite{Sarantsev:2007aa}, DAPHNE (open triangles) \cite{Ahrens:2005ia}, TAPS (open squares) \cite{Sarantsev:2007aa} and Crystal Ball/TAPS (filled circles) \cite{Schumann:2010js}. \APS{\cite{Kashevarov:2012wy}}{2012}
		Bottom: The total cross section off the proton and off the neutron extracted by the Crystal Ball/TAPS experiment \cite{Dieterle:2015kos} with contributions from different sequential decays or phase space. The results are compared to the PWA for free protons from \cite{Thoma:2007bm}. For further details see \cite{Dieterle:2015kos}. \EPJA
	}
	\label{fig:M:ppi0pi0_crosssection}
	\label{fig:M:npi0pi0_crosssection}
\end{figure}

For a detailed analysis of the final states comprising multiple mesons, not only the total and differential cross sections need to be investigated, but also the distributions of the cross sections for different kinematic variables like the invariant masses or angles of a particle subsystem. One example of the total cross section for the $p\pi^0\pi^0$ case is shown in Fig.~\ref{fig:M:ppi0pi0_crosssection}, top. Here, the data is compared to older measurements and the different data sets agree well within their errors. Two structures are visible, which need to be further identified by a partial wave analysis. From the angular distributions, moments $W_{LM}$ could be extracted and used to give further insight into the composition of the structures \cite{Kashevarov:2012wy}. Here, large contributions of the $J=\frac{3}{2}$ waves could be found also in the region below the $N(1520)\frac{3}{2}^-$, a result which is in agreement to the previous findings from the helicity difference of the cross section in \cite{Ahrens:2005ia}.  

In another interesting measurement, the Crystal Ball/TAPS experiment determined the cross section off the neutron \cite{Dieterle:2015kos}, see Fig.~\ref{fig:M:npi0pi0_crosssection}, bottom. Here, two similar structures are visible, although the structure do not appear as pronounced as for the reaction off the proton. Interestingly, the value of the cross section appears to be of similar order for both peaks. This result is in contradiction to the previous measurement by the GRAAL collaboration, where they saw a much more pronounced second structure for the reaction off the neutron \cite{Ajaka:2007zz}, compare also \cite{Dieterle:2015kos}.

For a more detailed analysis of the contributions to the $N\pi^0\pi^0$ final states, Dieterle et al.~\cite{Dieterle:2015kos} fitted simulated line shapes of a phasespace distribution and of the sequential decays $\Delta(1232)\pi$ and $N(1520)\pi$ to the invariant mass distributions. Angular distributions, which can give access to the quantum numbers of the states, as well as the beam helicity  asymmetries were ignored in this analysis. As a result, the first structure of the cross section can be in both cases - production off the proton and off the neutron - described by a sequential decay of $\Delta(1232)\pi$, see Fig.~\ref{fig:M:npi0pi0_crosssection}. For the second structure, only slight contributions are visible for the sequential decay of $N(1520)\pi$, most could be attributed either to phase space or to the decay via the $\Delta$ resonance. The distributions for the reaction off the proton can be compared to results from previous partial wave analyses from \cite{Thoma:2007bm} and they agree well within their uncertainties. However, for a further detailed investigation of the contributions the study of angular distributions as well as polarization measurements is necessary.

\begin{figure}[ht]
	\includegraphics[width=0.45\textwidth]{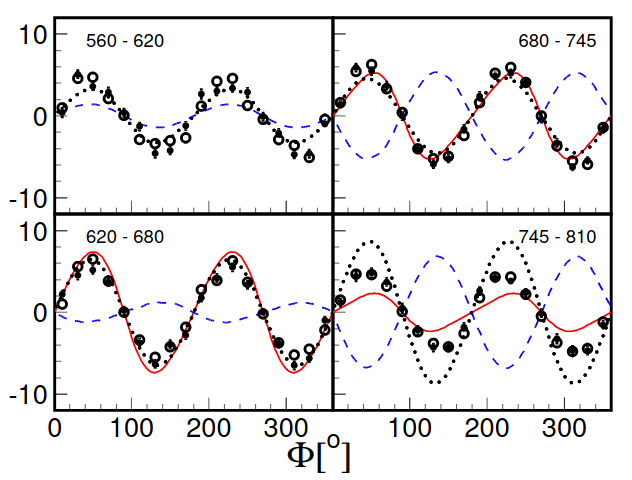}
	\includegraphics[width=0.45\textwidth]{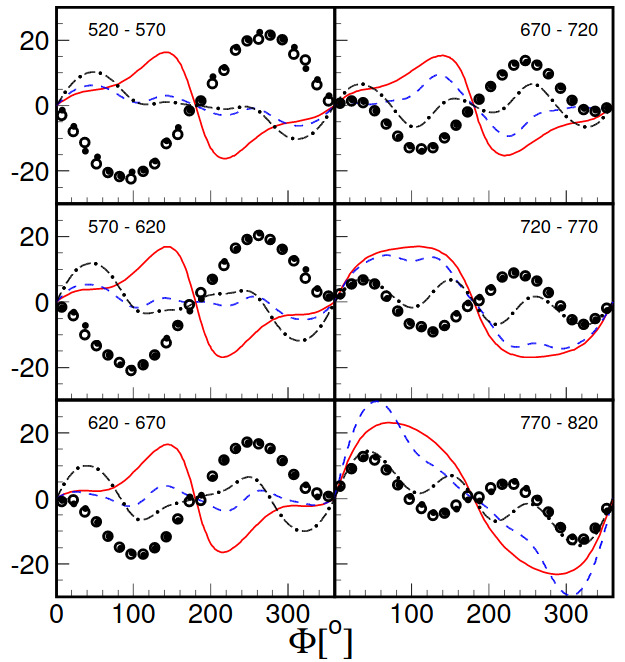}
	\caption{Left: The beam helicity asymmetry $I^\odot(\Phi)$ (filled symbols) and $-I^\odot(2\pi-\Phi)$ (open symbols) for $p\pi^0\pi^0$ for four different energy bins \cite{Zehr:2012tj} compared to different model calculations: Two-Pion-MAID (Red solid curves) \cite{Fix:2005if}, Valencia model (blue dashed curves) \cite{Roca:2002vd}, BnGa fit (blackdotted curves) \cite{Sarantsev:2007aa}. Right: Beam helicity asymmetry $I^\odot(\Phi)$ (filled symbols) and $-I^\odot(2\pi-\Phi)$ (open symbols) for $n\pi^0\pi^+$ \cite{Zehr:2012tj} compared to Two-Pion-MAID (Red solid curves) \cite{Fix:2005if}, Valencia model with (blue dashed curves) and without $\rho$ contributions (black dash-dotted curves) \cite{Fix:2005if}. \EPJA}
	\label{fig:M:npi0pi+_beamhelicity}
	\label{fig:M:ppi0pi0_beamhelicity}
\end{figure}

The easiest way to analyze polarization experiments for two meson photoproduction can be achieved by the so-called \textit{quasi two-body kinematics}. Here, two of the three final state particles are regarded as a single pseudo particle. By this simplification, the same observables can be extracted than for a single pseudoscalar meson (see Sec.~\ref{sec:formalisms:SingleMesonPhotoproduction}). Examples of publications, where observables using this simplified ansatz have been extracted, can be seen in Tab.~\ref{tab:measMultiMeson1} and \ref{tab:measMultiMeson2}. However, to extract the complete information from the measured final state, exploiting the full \textit{three-body kinematics}, leading to the observables mentioned in Sec.~\ref{sec:formalisms:MultiMesonPhotoproduction}, is crucial. Therefore we will in this chapter only discuss the latter observables.

A first step to shed further light on the reaction mechanisms is the measurement of the beam helicity asymmetry $I^\odot$, which can be extracted by using a circularly polarized photon beam. This observable is given by the difference between the two helicity states of the polarized beam. First measurements of this polarization observable were done by the CLAS experiment for the $p\pi^+\pi^-$ final state \cite{Strauch:2005cs}. Here, the data was compared to two different models, one by Mokeev et al.~\cite{Mokeev:2001qg,Mokeev:2004fcd,Burkert:2004rh} and another by Fix and Arenh\"ovel \cite{Fix:2005if}. Although both models were able to describe the previously measured data well, large discrepancies were observed by comparison with the newly measured beam helicity asymmetry. This hints to the fact that for a proper understanding of multi-meson final states, models are needed which are able to describe unpolarized as well as polarized data simultaneously.

In 2012, measurements were published by the Crystal Ball/TAPS collaboration for $p\pi^0\pi^0$ and $n\pi^0\pi^+$ \cite{Zehr:2012tj}, which can be seen in Fig.~\ref{fig:M:ppi0pi0_beamhelicity}. Here, not only the observable $I^\odot$ is plotted, but it is also compared to the modified observable under the intrinsic symmetry conditions. No major deviations between the two modes are visible, which indicates that all systematic uncertainties are under control. For the $p\pi^0\pi^0$ final state, the results from the MAID model \cite{Fix:2005if} and the BnGa analysis \cite{Sarantsev:2007aa} are in agreement with the data, in contrast to the Valencia model \cite{Roca:2002vd}, which is in most energy bins out of phase to the data. For the final state $n\pi^0\pi^+$, the results look different. Here, none of the models was able to describe the data at all, in most cases even the phase is completely different. At the time of this publication, no data of the final state $n\pi^0\pi^+$  was available for the BnGa model. As a test, the data was also compared to the Valencia model without the $\rho$ contributions. For the invariant masses and the total cross section, including the $\rho$ meson contributions improved the descriptions of the data a lot, see \cite{Zehr:2012tj}. Surprisingly, this is not the case for the beam helicity asymmetry $I^\odot$. This indicates that the reaction mechanisms are not understood in detail and further observables are necessary.

\begin{figure}
	\includegraphics[width=\textwidth]{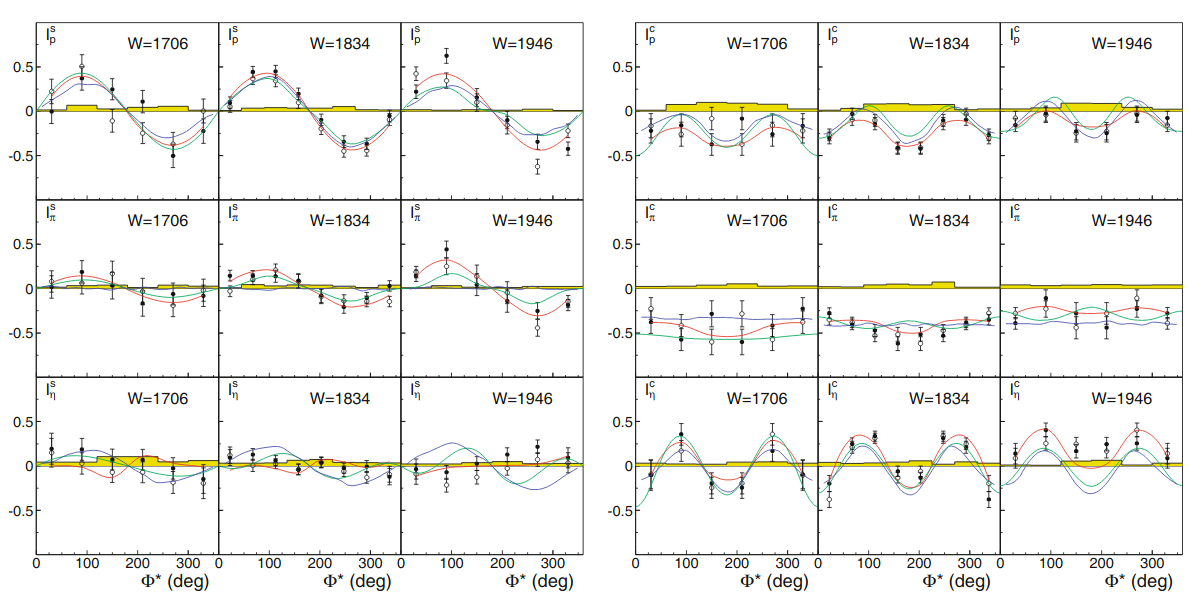}
	\caption{Three-body beam asymmetries $I_s$ (left) and $I_c$ (right) \cite{Gutz:2014wit}. Closed symbols: $I_s(\phi^\ast),I_c(\phi^\ast)$ , open symbols: $-I_s(2\pi-\phi^\ast),I_c(2\pi-\phi^\ast)$. Compare to different PWA curves: BnGa (red) \cite{Gutz:2014wit}, Fix et al. (green) \cite{Fix:2010nv}, D\"oring et al. (blue) \cite{Doring:2010fw}. \EPJA}
	\label{fig:M:ppi0eta_IsIc}
\end{figure}

In the last years, the amount of data for polarization observables exploiting the full three-body kinematics increased substantially. One example is the measurement of the observables $I_c$ and $I_s$, which have been measured for the final states $p\pi^0\pi^0$ and $p\pi^0\eta$ by the CBELSA/TAPS collaboration \cite{Gutz:2009zh,Gutz:2014wit,Sokhoyan:2015fra}. These observables can be measured by impinging a linearly polarized photon beam on an unpolarized target. For single meson photoproduction, this combination allows the measurement of the beam asymmetry $\Sigma$. For two meson photoproduction, the equivalent observable can be accessed from the full three-body kinematics by regarding the two mesons as a pseudo particle. For more information see the detailed description in \cite{Gutz:2014wit}.

An example of the observables $I_s$ and $I_c$ for $\pi^0\eta$ photoproduction are shown in Fig.~\ref{fig:M:ppi0eta_IsIc}. Here, both observables are plotted against the angle $\phi^\ast$, which is the angle between the decay and the reaction plane, for different energy bins. In addition, the recoiling particle, $p$, $\pi^0$ or $\eta$ is given as a small index at each observable. The data has been fitted by the BnGa partial wave analysis group.  At threshold, they found dominant contributions by the $\Delta(1700) \frac{3}{2}^-$, which was also seen by the Valencia group \cite{Doring:2005bx}. This fact is in agreement with the prediction that the resonance is dynamically  generated  by meson-baryon interactions \cite{Doring:2010fw}. For higher energies, the dominance of the $J^P = \frac{3}{2}^-$ partial wave continues, where the $\Delta(1940)\frac{3}{2}^-$ contributes. This was first claimed by the BnGa group \cite{Horn:2008qv} and later confirmed by the Fix model \cite{Fix:2010nv}.

\begin{figure}
	\includegraphics[width=0.5\textwidth]{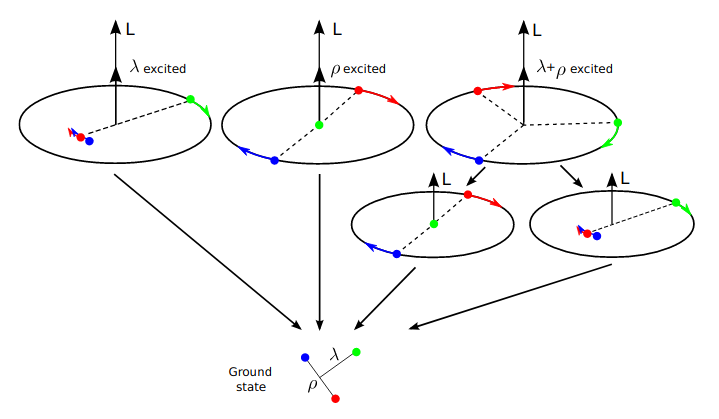}
	\includegraphics[width=0.5\textwidth]{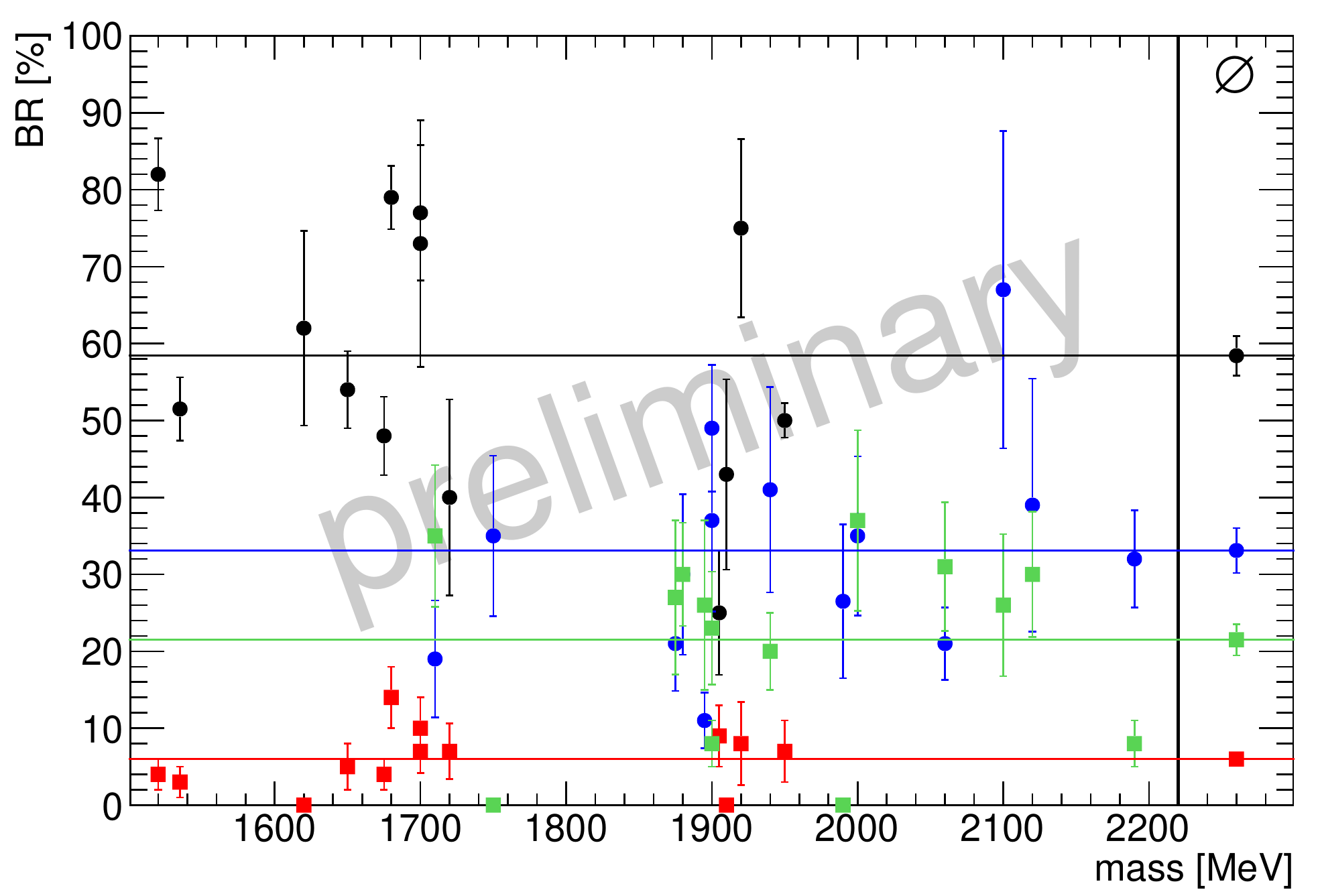}
	\caption{Left: For $L=2$ excitation, the interaction of the quarks can be described by two oscillators, $\rho$ and $\lambda$. When just one of these oscillators is excited, a direct decay into the ground state can occur. When both oscillators are excited, a sequential de-excitation is needed to reach the ground state \cite{Thiel:2015kuc}. Right: Branching ratios of the resonances, which are included in the BnGa PWA, plotted for different decay modes. Resonances with one oscillator excited: decaying directly into the ground states $N\pi$ or $\Delta\pi$ (black dots) or decaying into excited states like $N(1520)\pi$ (red squares). Mixed-oscillator states decaying into the ground states (blue  dots) and excited  states (green squares). For further details see \cite{Seifen2020}.}
	\label{fig:M:twomeson_oscillatorexcitation_thiel}
\end{figure}

As mentioned before, multi-meson photoproduction can give information about cascading decays. This can be exploited to determine the excitation mode inside the nucleon. For resonances with an orbital angular momentum of $L=2$, one can describe the excitation of the three quarks by an oscillator model with two oscillators $\rho$ and $\lambda$. There are, for the case of $L=2$, three different options, how the oscillators can be excited: either $\rho$ or $\lambda$ can be excited, or both oscillators can be excited at the same time, compare for example \cite{Thiel:2015kuc}. These cases are illustrated in Fig.~\ref{fig:M:twomeson_oscillatorexcitation_thiel}, left. In the case were both oscillators are excited, a direct decay into the ground state is suppressed and a decay via an intermediate state favored. 

Regarding the resonances, three different cases can be realized: resonances with just one of the oscillators excited, resonances with both oscillators excited and the coherent sum of these two modes (mixed oscillator states). In reality, no resonances with both oscillators excited have been found so far.

In Fig.~\ref{fig:M:twomeson_oscillatorexcitation_thiel}, right, different branching ratios from the BnGa partial wave analysis are plotted. Resonances, which have just one oscillator excited dominantly decay directly into the ground state and only with a small branching ratio into intermediate excited states.  This means that if just one of the oscillators is excited, a direct decay into the ground state is highly likely.
For mixed oscillator resonant states, both decay modes occur with almost similar probability \cite{Seifen2020}. They are therefore more likely to exhibit a cascading decay.

\begin{table*}[tp]
\caption{A collection of polarization observable measurements for two pseudoscalar meson photoproduction. Hint: We do not distinguish between differential and total cross section data. Further details are given in the cited references. The symbols indicate which PWA (circle: BnGa-2019 \cite{Muller:2019qxg}, diamond: J\"uBo-2017 \cite{Ronchen:2018ury} and star: SAID \cite{SAIDURL}) use the data in their fits.}\label{tab:measMultiMeson1}
\centering
\vspace*{5mm}
\begin{tabular}{clllc}
Observable & Energy range $E^{lab}_\gamma$ & Facility & Reference & Year\\
\hline
\multicolumn{5}{c}{$\gamma p \rightarrow p \pi^0 \pi^0$}\\
\hline
$\text{P}_\text{x}, \text{P}_\text{y}, T, H, P$ & $650 - 2600$ MeV & CBELSA/TAPS & Seifen et al. \cite{Seifen2020} $\circ$ & 2020\\
$E, \sigma_{1/2}, \sigma_{3/2}$ & $\sim 431-1455$ MeV & Crystal Ball/TAPS & Dieterle et al. \cite{Dieterle:2020vug}& 2020\\
$\text{I}_0$ & $432-1374$ MeV & Crystal Ball/TAPS & Dieterle et al. \cite{Dieterle:2015kos} & 2015\\
$\text{I}_0, \Sigma$ & $600-2500$ MeV & CB/TAPS & Sokhoyan et al. \cite{Sokhoyan:2015fra} $\circ$ & 2015\\
$\text{I}_0$ & $\sim 750-2500$ MeV & CBELSA/TAPS & Thiel et al. \cite{Thiel:2015kuc}& 2015\\
$\text{I}^\text{c}, \text{I}^\text{s}$ & $970-1650$ MeV & CB/TAPS & Sokhoyan et al. \cite{Sokhoyan:2015fra} $\circ$ & 2015\\
$\text{I}^\odot$ & $\sim 600-1400$ MeV & Crystal Ball/TAPS & Oberle et al. \cite{Oberle:2014rap}& 2014\\
$\text{I}_0$ & $309-1400$ MeV & Crystal Ball/TAPS & Kashevarov et al. \cite{Kashevarov:2012wy} $\circ$ & 2012\\
$\text{I}_0$ & $309-800$ MeV & Crystal Ball/TAPS &Zehr et al. \cite{Zehr:2012tj} & 2012\\
$\text{I}^\odot$ & $550-820$ MeV & Crystal Ball/TAPS & Zehr et al. \cite{Zehr:2012tj}& 2012\\
$\text{I}^\odot$ & $560-810$ MeV & Crystal Ball/TAPS & Krambrich et al. \cite{Krambrich:2009te}& 2009\\
$\text{I}_0$ & $309-820$ MeV & TAPS, CBELSA & Sarantsev et al. \cite{Sarantsev:2007aa}& 2008\\
$\text{I}_0$ & $400-1300$ MeV & CBELSA & Thoma et al. \cite{Thoma:2007bm} $\circ$& 2008\\
$\Sigma$ & $650-1450$ MeV & CBELSA  & Thoma et al. \cite{Thoma:2007bm} $\circ$& 2008\\
$\text{I}_0$, $\sigma_{1/2}$, $\sigma_{3/2}$ & $400-800$ MeV & DAPHNE & Ahrens et al. \cite{Ahrens:2005ia}& 2005\\
$\text{I}_0$ & $300-425$ MeV & TAPS & Kotulla et al. \cite{Kotulla:2003cx} & 2004\\
$\text{I}_0, \Sigma$ & $650-1450$ MeV & GRAAL & Assafiri et al. \cite{Assafiri:2003mv} $\circ$ & 2003\\
$\text{I}_0$ & $309-820$ MeV & TAPS & Wolf et al. \cite{Wolf:2000qt} & 2000\\
$\text{I}_0$ & $200-820$ MeV & TAPS & Kleber et al. \cite{Kleber:2000qs} & 2000\\
\hline
\multicolumn{5}{c}{$\gamma p \rightarrow p \pi^+ \pi^-$}\\
\hline
$\text{I}_0$ & $\sim 895-1663$ MeV & CLAS & Golovatch et al. \cite{Golovatch:2018hjk} $\circ$ &2019\\
$\text{I}^\odot$ & $1100-5400$ MeV & CLAS & Badui et al. \cite{Badui:2016bay}& 2016\\
$\text{I}_0$ & $800-1100$ MeV & NKS2 & Hirose et al. \cite{Hirose:2009zz}& 2009\\
$\text{I}^\odot$ & $575 - 815$ MeV & Crystal Ball/TAPS & Krambrich et al. \cite{Krambrich:2009te}& 2009\\
$\text{I}_0$ & $400-800$ MeV & DAPHNE & Ahrens et al. \cite{Ahrens:2007zzj}& 2007\\
$\text{I}_0$ & $\sim 560 - 2560$ MeV & SAPHIR & Wu et al. \cite{Wu:2005wf}& 2005\\
$\text{I}^\odot$ & $502-2350$ MeV & CLAS & Strauch et al. \cite{Strauch:2005cs}& 2005\\
\end{tabular}
\end{table*}
\begin{table*}[tp]
\caption{(Continuation of Tab.~\ref{tab:measMultiMeson1}) A collection of polarization observable measurements for two pseudoscalar meson photoproduction. Hint: We do not distinguish between differential and total cross section data. Further details are given in the cited references. The symbols indicate which PWA (circle: BnGa-2019 \cite{Muller:2019qxg}, diamond: J\"uBo-2017 \cite{Ronchen:2018ury} and star: SAID-MA19  \cite{Briscoe:2019cyo}) use the data in their fits.}\label{tab:measMultiMeson2} 
\centering
\vspace*{5mm}
\begin{tabular}{clllc}
Observable & Energy range $E^{lab}_\gamma$ & Facility & Reference & Year\\
\hline
\multicolumn{5}{c}{$\gamma p \rightarrow p \pi^0 \eta$}\\
\hline
$\sigma_{1/2}$, $\sigma_{3/2}$, E & $\sim 895 - 1450$~MeV & Crystal Ball/TAPS & K\"aser et al. \cite{Kaser:2018bba} & 2018 \\
$\text{I}_0$&$\sim 930 - 1450$~MeV&Crystal Ball/TAPS& V. Sokhoyan et al. \cite{A2:2018vbv} & 2018 \\
$\text{I}_0$ & $\sim 930-2500$ MeV & CB/TAPS & Gutz et al. \cite{Gutz:2014wit} $\circ$ & 2014\\
$\Sigma$ & $\sim1070-1550$ MeV & CB/TAPS & Gutz et al. \cite{Gutz:2014wit} $\circ$ & 2014\\
$\text{I}^\text{c}, \text{I}^\text{s}$ & $\sim1081-1550$ MeV & CB/TAPS & Gutz et al. \cite{Gutz:2014wit} $\circ$ & 2014\\
$\text{I}^\text{c}, \text{I}^\text{s}$ & $970-1650$ MeV & CBELSA/TAPS & Gutz et al. \cite{Gutz:2009zh} $\circ$ & 2010\\
$\text{I}_0$ & $950-1400$ MeV & Crystal Ball/TAPS & Kashevarov et al. \cite{Kashevarov:2009ww} & 2009\\
$\text{I}_0$ & $\sim 1070-2860$ MeV & CB & Horn et al. \cite{Horn:2008qv} $\circ$ & 2008\\
$\text{I}_0$ & $\sim 1070-2860$ MeV & CB & Horn et al. \cite{Horn:2007pp} $\circ$ & 2008\\
$\text{I}_0, \Sigma$ & $\sim 930-1500$ MeV & GRAAL & Ajaka et al. \cite{Ajaka:2008zz} & 2008\\
$\Sigma$ & $970-1650$ MeV & CBELSA/TAPS & Gutz et al. \cite{Gutz:2008zz} & 2008\\
$\text{I}_0$ & $1000-1150$ MeV & GeV-$\gamma$ at LNS? & Nakabayashi et al. \cite{Nakabayashi:2006ut} & 2006\\
\hline
\multicolumn{5}{c}{$\gamma n \rightarrow n \pi^0 \eta$}\\
\hline
$\sigma_{1/2}$, $\sigma_{3/2}$, E & $\sim 895 - 1450$~MeV & Crystal Ball/TAPS & K\"aser et al. \cite{Kaser:2018bba} & 2018 \\
\hline
\multicolumn{5}{c}{$\gamma p \rightarrow n \pi^+ \pi^0$}\\
\hline
$\text{I}^\odot$ & $\sim 550-820$ MeV & Crystal Ball/TAPS & Zehr et al. \cite{Zehr:2012tj} & 2012\\
$\text{I}_0$ & $\sim 325-800$ MeV & Crystal Ball/TAPS & Zehr et al. \cite{Zehr:2012tj} & 2012\\
$\text{I}^\odot$ & $520 - 820$ MeV & Crystal Ball/TAPS & Krambrich et al. \cite{Krambrich:2009te} & 2009\\
$\text{I}_0$ & $400-800$ MeV & DAPHNE & Ahrens et al. \cite{Ahrens:2003na} & 2003\\
$\text{I}_0$ & $300-820$ MeV & TAPS & Langgärtner et al. \cite{Langgartner:2001sg} & 2001\\
\hline
\multicolumn{5}{c}{$\gamma n \rightarrow n \pi^0 \pi^0$}\\
\hline
$\text{I}_0$ & $\sim 430-1371$ MeV & Crystal Ball/TAPS & Dieterle et al. \cite{Dieterle:2015kos} & 2015\\
$\text{I}^\odot$ & $\sim 600-1400$ MeV & Crystal Ball/TAPS & Oberle et al. \cite{Oberle:2014rap} &2013\\
$\text{I}_0, \Sigma$ & $\sim 600-1500$ MeV & GRAAL & Ajaka et al. \cite{Ajaka:2007zz} & 2007\\
\hline
\multicolumn{5}{c}{$\gamma p \rightarrow p K^+ K^-$}\\
\hline
$\text{I}_0$ & $3000-3800$ MeV & CLAS & Lombardo et al. \cite{Lombardo:2018gog} & 2018\\
$\text{I}^\odot$ & $1100-5400$ MeV & CLAS & Badui et al. \cite{Badui:2016bay} & 2016\\
\hline
\multicolumn{5}{c}{$\gamma p \rightarrow p \pi^0 \omega$}\\
\hline
$\text{I}_0$ & $1383 -2970$ MeV & CBELSA & Junkersfeld et al. \cite{CB-ELSA:2007bzd} & 2007
\end{tabular}
\end{table*}

%% file: PWA_PhenomenologicalAna.tex
\label{sec:PWA}

In order to extract resonance parameters (masses, widths, branching fractions) from scattering-data, the amplitude has to be parameterized as a function of energy in some suitable way. Then, this parameterization has to be fitted to the data. To be more precise, the thus fitted model-function has to be {\it analytically continued}~\cite{Behnke:1962,Cartan:1966} to complex energies, in order to determine the {\it pole-parameters} of the resonances (pole-positions and residues) in the complex energy plane. Only pole parameters coming from analytic functions can comprise a model-independent characterization of resonances~(See the article on resonances in the PDG~\cite{Zyla:2020zbs}, as well as the publications~\cite{Dalitz:1970ga,Hadzimehmedovic:2011ua}.). This chapter shall serve as a brief introduction to the most important aspects of this method. Analyses of this kind are quite broadly denoted as {\it energy-dependent} (ED) fits. 

In contrast, a complementary approach to such ED fits consists of the procedures of the CEA (Complete Experiment Analysis) and TPWA (Truncated Partial Wave Analysis) introduced in chapter~\ref{sec:formalisms}. Here, one tries to extract the amplitudes themselves as parameters defined at discrete bins in phase-space\footnote{Or, for the TPWA one extracts partial waves at discrete points in energy (or, for more than $3$ final-state particles, at discrete points in a higher-dimensional space of multiple energy-variables, respectively), since angular distributions have already been analyzed in this case.} without any model-assumptions concerning the form of these amplitudes as functions of energy. Thus, neither a CEA nor a TPWA itself can yield an analytic continuation to resonance poles. The approaches of CEA and TPWA are less sophisticated than the methods to be described in this chapter, but still they have the advantage of, ideally, total model-independence. They can be solved uniquely in the theoretically idealized case of a {\it complete experiment}~(Sec.~\ref{sec:formalisms:CompleteExperiments}).

In Light Baryon Spectroscopy, one uses a hierarchy of reactions such as those described in chapters~\ref{sec:formalisms} and~\ref{sec:measurements} to study the spectrum. Since many strongly interacting particles are known to exist~\cite{Zyla:2020zbs}, reactions are manifold and their kinematic thresholds are usually quite closely located in energy. In most cases considered in Light Baryon Spectroscopy, a few thresholds exist over so-called {\it resonance-regions} in energy. Moreover, the same resonance is in many cases allowed to couple to many possible channels. For these reasons, usually a so-called {\it coupled-channels} approach is used in ED fits.

\subsection{$S$-Matrix principles} \label{sec:PWA:SMatrixPrinciples}

Selected principles and methods from the so-called analytic {\it $S$-Matrix theory}~\cite{AnalyticSMatrix} are often employed in ED fits. The $S$-Matrix, or {\it scattering matrix} is an abstract operator in the Hilbert-space of asymptotic scattering-states, which defines the probability for transitioning from an initial state $\left| a \right>$ to a final state $\left| b \right>$ by its matrix-element $\left< b \right| \hat{S} \left| a \right>$. It is more correct to state that the probability is given by the modulus-squared of the matrix element. It should be mentioned that the $S$-Matrix can act in the space of discrete and continuous 'quantum numbers' (spin, isospin, four-momenta, $\ldots$), as well as in channel-space.

The part of the $S$-Matrix which contains non-interactions is usually split off additively, which can be done via the introduction of the $T$-Matrix $\hat{T}$\footnote{It is a convention to introduce the factors of '$2$' and '$i$' in this equation. We keep the equations in section~\ref{sec:PWA:SMatrixPrinciples} consistent with this particular convention. For the models discussed in section~\ref{sec:PWA:CompendiumModels}, the convention can be different.}~\cite{AnalyticSMatrix,PeskinSchroeder:1995,Chung:1995dx}:
\begin{equation}
 \hat{S} = \mathbbm{1} + 2 i \hat{T} \mathrm{.} \label{eq:PWA:SAndTMatrix}
\end{equation}
Matrix-elements of the $\hat{T}$-operator are then the amplitudes to be modeled (cf. section~\ref{sec:formalisms:SpinAmplitudes}). $S$-Matrix theory sets the requirements~\cite{AnalyticSMatrix} of {\it Unitarity}, {\it Analyticity} and {\it Crossing}, about which we now provide some more detail. 

\subsubsection{Unitarity} \label{sec:PWA:Unitarity}

The conservation of probabilities is intimately linked to the mathematical property of unitarity. Fixing a given initial state $\left| i \right>$, then the total probability for transitioning from this particular initial state to any allowed final state {\it has to be} $1$. Defining a general superposition for the initial state as $\left| i \right> := \sum_{n} a_{n} \left| n \right>$ with $\sum_{n} \left| a_{n} \right|^{2} = 1$, one can write the following for the probabilities of the system to arrive in some other state~\cite{AnalyticSMatrix}:
\begin{equation}
 1 = \sum_{m}  \left| \left< m \right| \hat{S} \left| i \right> \right|^{2} = \sum_{m} \left< i \right| \hat{S}^{\dagger} \left| m \right> \left< m \right| \hat{S} \left| i \right> = \left< i \right| \hat{S}^{\dagger}  \hat{S}  \left| i \right> = \sum_{n'} a_{n'}^{\ast} a_{n} \left< n' \right| \hat{S}^{\dagger}  \hat{S} \left| n \right>  . \label{eq:ConservationProbUnitarityProof}
\end{equation}
This can only be fulfilled for any $\hat{S}$ and any superposition $\left| i \right>$ in case the $S$-Matrix is {\it unitary}: $\hat{S}^{\dagger} \hat{S} \equiv \mathbbm{1}$. 

It is interesting to investigate the kinds of constraints this simple unitarity equation implies for the transition operator $\hat{T}$. Combining the general decomposition~\eqref{eq:PWA:SAndTMatrix} with the unitarity-relation for $\hat{S}$, one obtains the so-called {\it optical theorem} in operator-notation \cite{AnalyticSMatrix,PeskinSchroeder:1995}:
\begin{equation}
 \frac{1}{2i} \left( \hat{T} - \hat{T}^{\dagger} \right) = \hat{T}^{\dagger} \hat{T} \mathrm{.} \label{eq:PWA:UnitarityTMatrix} 
\end{equation}
Evaluating matrix-elements of this equation between the asymptotic scattering states $\left| a \right>$ and $\left| b \right>$ results in the following for the process $a \rightarrow b$
\begin{equation}
 \frac{1}{2i} \left( \left< b \right| \hat{T} \left| a \right> - \left< b \right| \hat{T}^{\dagger} \left| a \right> \right) = \frac{1}{2i} \left( \left< b \right| \hat{T} \left| a \right> - \left< a \right| \hat{T} \left| b \right>^{\ast} \right) \equiv \left< b \right| \hat{T}^{\dagger} \hat{T} \left| a \right> \mathrm{.} \label{eq:PWA:UnitarityTMatrixStep2}
\end{equation}
 To obtain the desired form of the unitarity-equation, one has to insert a resolution of the identity $\mathbbm{1} \equiv \sum_{n} \left| n \right> \left< n \right|$ into the operator-product on the right-hand-side. The sum here only runs over intermediate states $\left| n \right>$ which are allowed by kinematics (as well as, if present, further internal symmetries). Thus, the further one ascends in the total CMS-energy, the more terms occur in the sum~\cite{AnalyticSMatrix}. Defining $\mathcal{T}_{ba} := \left< b \right| \hat{T} \left| a \right>$, we get:
\begin{equation}
  \frac{1}{2i} \left( \mathcal{T}_{ba} - \mathcal{T}_{ab}^{\ast} \right) \equiv \left< b \right| \hat{T}^{\dagger} \hat{T} \left| a \right> = \sum_{n} \left< b \right| \hat{T}^{\dagger} \left| n \right> \left< n \right| \hat{T} \left| a \right> = \sum_{n} \mathcal{T}_{nb}^{\ast} \mathcal{T}_{na} \mathrm{.} \label{eq:PWA:UnitarityTMatrixStep3ShortNotation}
\end{equation}
The symbol '$\sum_{n}$' in this equation encompasses the sum over all kinematically allowed intermediate states and discrete and continuum quantum numbers, which actually turns it into an integration-summation. Therefore, unitarity equations such as~\eqref{eq:PWA:UnitarityTMatrixStep3ShortNotation} are generally integral-equations and thus are a lot more complicated than this simple notation may suggest.

In case the asymptotic- and the intermediate scattering states in equation~\eqref{eq:PWA:UnitarityTMatrixStep3ShortNotation} contain at most $2$-, $3$-, $\ldots$ particles, one talks about the constraints of {\it two-, three-, $\ldots$ body unitarity}. It is well known that in case two-body unitarity equations are evaluated in the partial-wave basis, the resulting constraints for the partial-wave amplitudes are purely algebraic and none of the intermediate-state integrals survive~\cite{MartinSpearman,Gribov:2009zz}. The equations of three-, four-, $\ldots$ body unitarity on the other hand are necessarily integral-constraints, even when evaluated in the partial-wave basis (see reference~\cite{Fleming:1964zz} and derivations in the appendix of~\cite{Mikhasenko:2019vhk}). Thus, unitarity-constraints for more than two-body systems are generally harder to implement. Three-body unitarity in particular has been a topic of great current research-interest, for the analysis of scattering in the infinite volume~\cite{Mai:2017vot,Jackura:2018xnx,Mikhasenko:2019vhk} as well as of lattice-QCD data in a finite volume~\cite{Mai:2017bge,Mai:2018djl,Briceno:2017tce,Hansen:2020otl}.

Keeping unitarity-constraints intact in parametrizations of amplitudes is very important. In case of pure (coupled channels-) two-body unitarity, the so-called {\it $K$-Matrix approach} \cite{Aitchison:1972ay,Chung:1995dx} has turned out useful for this task. One observes that the operator-relation~\eqref{eq:PWA:UnitarityTMatrix} can be rewritten as
\begin{equation}
 \left( \hat{T}^{-1} + i \mathbbm{1} \right)^{\dagger} = \hat{T}^{-1} + i \mathbbm{1} \mathrm{.} \label{eq:PWA:OptTheoremRewritten}
\end{equation}
This expression motivates the definition of the $K$-Matrix operator \cite{Chung:1995dx}
\begin{equation}
 \hat{K}^{-1} := \hat{T}^{-1} + i \mathbbm{1} \mathrm{,} \label{eq:PWA:KMatrixDef}
\end{equation}
which can be solved for the $T$-Matrix as follows \cite{Chung:1995dx}:
\begin{equation}
 \hat{T} = \hat{K} \left( \mathbbm{1} - i \hat{K} \right)^{-1} \equiv \left( \mathbbm{1} - i \hat{K} \right)^{-1} \hat{K} \mathrm{.} \label{eq:PWA:TMatrixInTermsOfKMatrix}
\end{equation}
Furthermore, equation~\eqref{eq:PWA:OptTheoremRewritten} states that in order to obtain a unitary $T$-Matrix, the $K$-operator has to be {\it hermitean}: $\hat{K}^{\dagger} = \hat{K}$.

One can also reverse this argument and proceed to construct a model for an ED fit by first specifying the functional form of the $K$-Matrix in a suitable basis of states. In case the thus constructed $K$-Matrix is hermitean, unitary $S$- and $T$-Matrices result automatically. However, apart from that a lot of freedom still exists for the specific parameterization of the $K$-matrix.

Finally, it is worth mentioning that unitarity-constraints can be very useful for {\it resolving the discrete ambiguities} which are known to plague the fully model-independent TPWAs discussed in chapter~\ref{sec:formalisms}. This has been shown rigorously by Crichton, Martin, Atkinson and others~\cite{Crichton:1966,Martin:1969xs,Atkinson:1972hr,Atkinson:1973wt,Atkinson:1973je,Berends:1973mg,Cornille:1974zp,Atkinson:1977ia,Martin:2020jlu}, but only in the context of elastic two-body unitarity for scattering of spinless particles (for a comprehensive review of these issues, see~\cite{Bowcock:1976ax}). Still, we also expect a similar resolution of discrete ambiguities to carry over to more complicated cases.



\subsubsection{Analyticity} \label{sec:PWA:Analyticity}

The transition amplitudes $\mathcal{T}_{ba} := \left< b \right| \hat{T} \left| a \right>$ are functions of external Lorentz-invariants for the described process. For an on-shell $2\rightarrow 2$ reaction, these invariants would be just the Mandelstam-variables $s$ and $t$, while for a $2 \rightarrow n$ reaction with $n \geq 3$, one would have a set of $3 (2 + n) - 10$ well-selected external invariants (cf. appendix~\ref{sec:appendix:Kinematics}). 

The principle of {\it Analyticity} postulates the physical transition amplitudes, i.e. those defined for real physical values of the external invariants, to be boundary values of multivariate {\it analytic functions} (i.e. {\it complex dif\-fe\-ren\-ti\-ab\-le functions}) \cite{Behnke:1962,Cartan:1966}, of the external invariants. For the simple example of a $2 \rightarrow 2$ reaction, Mandelstam formulated this analyticity requirement as a double-dispersion relation \cite{Mandelstam:1958xc}. In practice however, one typically encounters either discussions of $2$-body channels with the variable $t$ held fixed, or of cases where the $2$-body amplitude has been expanded into partial waves. In these cases, the relevant (partial-wave-) amplitudes are ordinary single-variable analytic function on the complex energy-plane. Analyticity is known to be linked to the fundamental principle of {\it microcausality}. This has been demonstrated in the past for specific examples \cite{GellMann:1954db}, but a general proof of the connection is still lacking \cite{AnalyticSMatrix}.

In a bit more detail, amplitudes are not only required to be purely analytic functions, but rather {\it meromorphic} functions \cite{Behnke:1962}. This means that isolated singularities (poles) are allowed to exist~\cite{AnalyticSMatrix}. Even extended singularities (branch-points and cuts) are not excluded. Furthermore, where and of which kind these singularities have to be is dictated by physics. First-order poles are either stable bound states or resonances. As already mentioned above, {\it only poles} are accepted as model-independent signals for resonances \cite{Zyla:2020zbs}. Therefore, pole-parameters (positions and residues) should be extracted once a model has been fitted. Branch-points on the other hand correspond, roughly speaking, to {\it thresholds}. The analytic amplitudes are actually multivalued functions defined on multiple {\it Riemann sheets}~\cite{Behnke:1962}.

The opening of every new channel introduces a branch-point on the real energy axis which is located at the threshold energy $s = \left( \sum_{i} m_{i} \right)^{2}$ (cf. appendix Tab.~\ref{tab:appendix:thresholds})~\cite{AnalyticSMatrix}. This type  of real branch-point caused by stable asymptotic states, can lead to a pronounced 'cusp'-effect in partial-wave amplitudes. An example of such a cusp effect can be found for example in Sec.~\ref{sec:measurements:EtaPhotoproduction}. Furthermore, complex branch-points exist as well but they are caused by intermediate states where at least one particle is unstable~\cite{Ceci:2011ae}.

The branching-behavior of the amplitude is essentially dictated by unitarity-equations. To see this, consider the left-hand side of equation~\eqref{eq:PWA:UnitarityTMatrixStep3ShortNotation}, which for the simple case of a $2 \rightarrow 2$-reaction with scalar particles and time-reversal invariant interactions, reduces to the imaginary part $\mathrm{Im} \left[ \mathcal{T}_{ba}  \right]$. In more general cases this statement no longer holds, but what remains true is that this left-hand-side defines a {\it discontinuity}~\cite{Fleming:1964zz,Mikhasenko:2019vhk} of the amplitude across an appropriately defined point-region in the space of external invariants. One can write this symbolically as follows:
\begin{equation}
 \mathrm{Disc} \left[ \mathcal{T}_{ba} \right] = \sum_{n} \mathcal{T}_{nb}^{\ast} \mathcal{T}_{na} \mathrm{.} \label{eq:PWA:UnitarityTMatrixStep3}
\end{equation}
Once the energy is raised, additional terms are kinematically allowed on the right-hand-side of equation~\eqref{eq:PWA:UnitarityTMatrixStep3}, which leads to additional singularities. This is a particular example of physics implying singularities of the amplitude. 

\begin{figure}[htb]
\centering
 \hspace*{10pt}  \begin{overpic}[width=0.625\textwidth]%
      {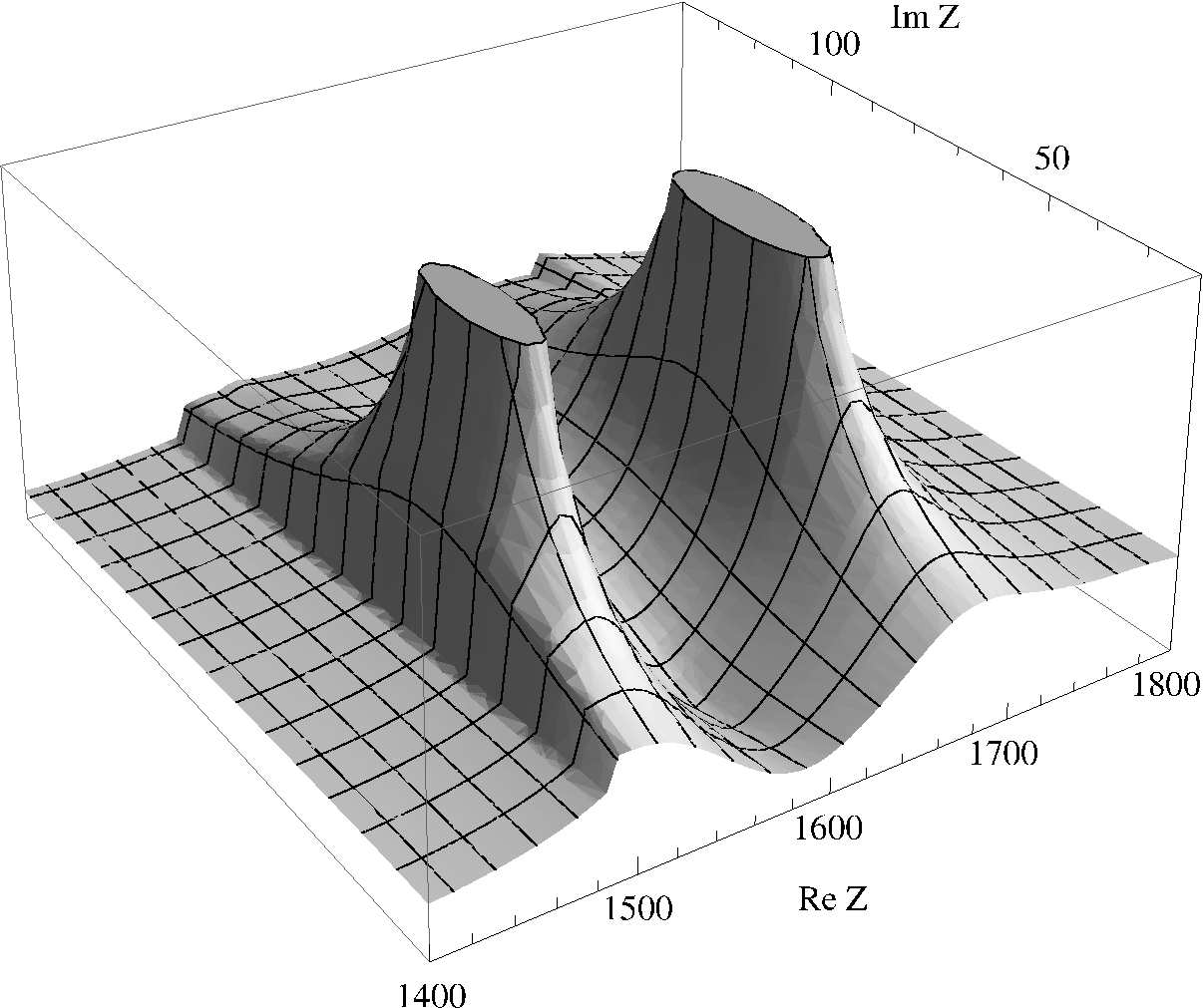}
\end{overpic}
\vspace*{10pt} 
 \caption{The plot shows the modulus $\left| T \right|$ of the partial wave $S_{11}$ (i.e. orbital angular momentum $\ell = 0$, isospin $I = \frac{1}{2}$ and total angular momentum $J=\frac{1}{2}$) of the meson-exchange model by the J\"{u}lich-Bonn group (cf.~Sec.~\ref{sec:PWA:JueBoModel}), as a function of complex energy $Z$ $\left[\mathrm{MeV}\right]$ \cite{Doring:2009yv}. The poles of the nucleon resonances $N^{\ast} (1535)$ and $N^{\ast} (1650)$ can be clearly seen. Slightly below the $N^{\ast} (1535)$-pole, the discontinuity of the $\eta N$-branch cut is visible as a sharp step. In the same way, one can still see the $\rho N$-cut with complex branch-point, behind the $N^{\ast} (1650)$-pole.}
 \label{fig:PWA:DoeringEtAlPoles}
\end{figure}

\begin{figure}[htb]
\centering
 \hspace*{10pt}  \begin{overpic}[width=0.55\textwidth]%
      {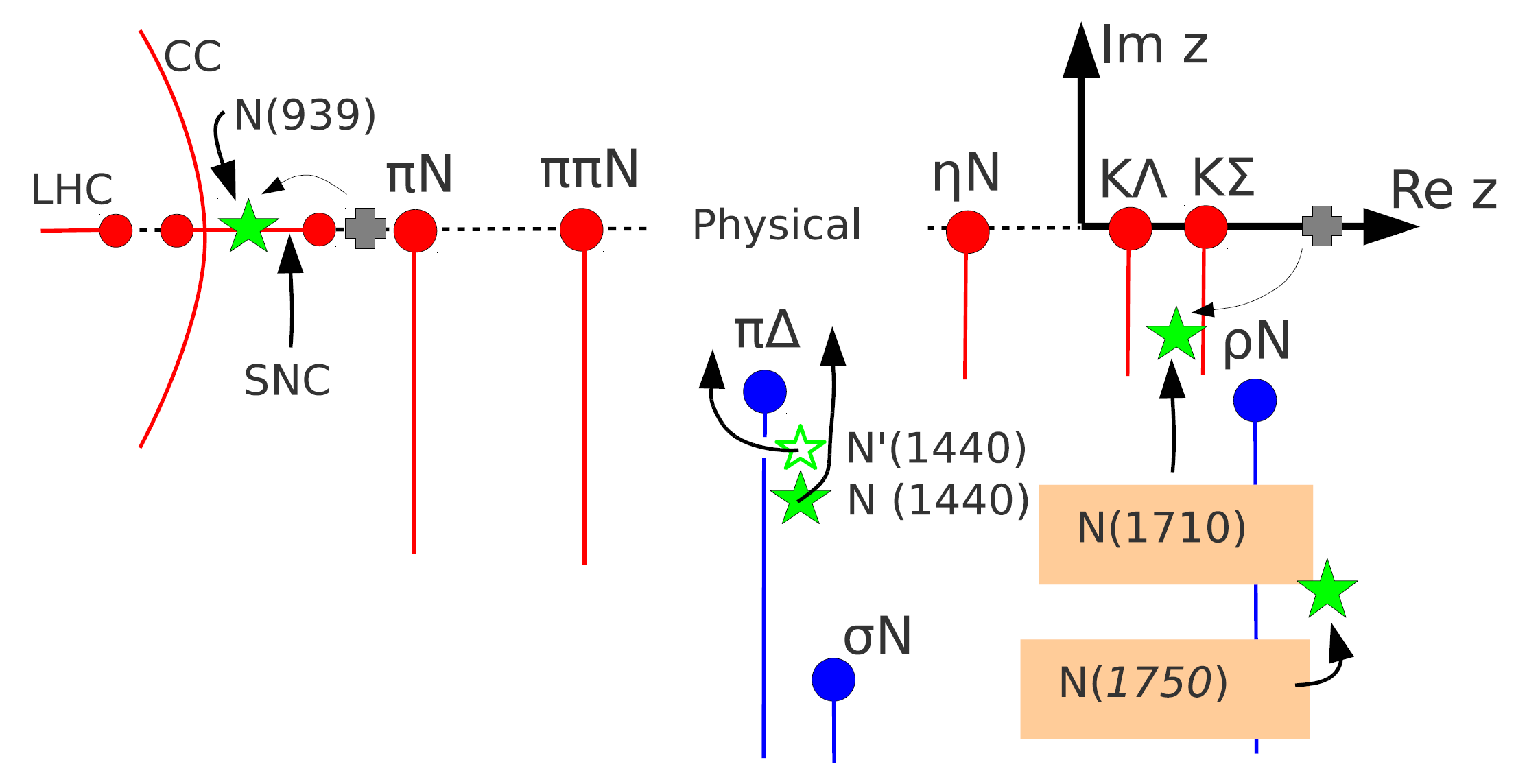}
\end{overpic}
\vspace*{10pt} 
 \caption{As an example, the analytic structure of the $P_{11}$-partial wave in $\pi N \rightarrow \pi N$ (i.e. orbital angular momentum $\ell = 0$, isospin $I = \frac{1}{2}$ and total angular momentum $J=\frac{1}{2}$) of the dynamical coupled-channels model by the J\"{u}lich-Bonn group (cf.~Sec.~\ref{sec:PWA:JueBoModel}) is shown in the plane of complex energies $z$ \cite{Ronchen:2012eg}. Shown are branch-points of thresholds on the real energy-axis (red dots) as well as of the effective intermediate $\pi \Delta$-, $\sigma N$- and $\rho N$-states, which are located in the complex plane (blue dots). The branch-cuts (red and blue lines) have been rotated downwards. Poles belonging to nucleon resonances and the ground-state nucleon are shown as green stars. In addition, the left-hand-cut (LHC), circular cut (CC) and short-nucleon-cut (SNC) are indicated (see discussions in~\cite{Doring:2009yv,Ronchen:2012eg}). \EPJA}
 \label{fig:PWA:JuBoAnalyticStructure}
\end{figure}

The above-mentioned branch-points due to the opening of new channels in direct ($s$-channel) processes fall under the collective term of {\it right-hand-singularities} (i.e., they appear in the right half of the complex energy plane of the considered amplitude). Moreover, there exist {\it left-hand-singularities} as well, which come to pass due to thresholds in so-called {\it crossed channels} (i.e. $t$-, $u$-channels) (see section~\ref{sec:PWA:Crossing} for a brief discussion of crossing). Furthermore, particle-exchanges in the crossed channels can produce a very complicated analytic structure in the partial-wave amplitudes, e.g. so-called circular cuts can occur~\cite{MartinSpearman,Doring:2009yv}, which is generated by the angular partial-wave projection integrals. In many practical models, complicated left-hand-singularities are either ignored or modelled in an alternative and simpler manner (see section~\ref{sec:PWA:CompendiumModels}). Illustrations for the analytic structure of a partial wave in a realistic model, and in particular for the occurring resonance poles, are shown in Figures~\ref{fig:PWA:DoeringEtAlPoles} and~\ref{fig:PWA:JuBoAnalyticStructure}. 

In the individual reactions ubiquitous in Light Baryon Spectroscopy, one generally has to model the amplitude for a (possible) dense population of multiple resonances and also including the effects of non-resonant background-processes. These complications lead to the necessity for intricate models such as those described in section~\ref{sec:PWA:CompendiumModels}.


\subsubsection{Crossing} \label{sec:PWA:Crossing}

From the Feynman-rules in quantum field theoretical perturbation theory \cite{PeskinSchroeder:1995}, one can deduce another property of the $S$-Matrix. It is implied that the {\it same} analytic functions have to describe different physical processes, which are related to each other by so-called {\it crossing}-transformations \cite{AnalyticSMatrix}. A crossing-transformation moves one particle from the initial/final state to the final/initial state, replaces the particle's four-momentum $p$ by $-p$ and conjugates it into its own antiparticle. 
%
%
In most references (cf. \cite{AnalyticSMatrix}), $2 \rightarrow 2$-reactions related by crossing are discussed. For example, the processes $\pi^{+} p \longrightarrow \pi^{+} p$ ($s$-channel), $\pi^{+}  \pi^{-} \longrightarrow p \bar{p}$ ($t$-channel) and $\pi^{+} \bar{p} \longrightarrow \pi^{+} \bar{p}$ ($u$-channel) are crossing-related. It is however also allowed to relate $2$-body scattering processes to decays via crossing-transformations, such as in the derivation of the so-called Khuri-Treiman equations~\cite{Aitchison:1966lpz,Pasquier:1968zz,Mikhasenko:2019vhk}. 

For the different models discussed in section~\ref{sec:PWA:CompendiumModels}, a precise implementation of the crossing-property is not given. As an example, recent dispersive precision-studies of $\pi N$-scattering using the Roy-Steiner equations~\cite{Hoferichter:2015dsa,Hoferichter:2015hva} have featured a precise implementation of crossing.

%


\subsection{A compendium of models} \label{sec:PWA:CompendiumModels}
Seven of the most prominent recent models used for energy-dependent (ED) fits in Light Baryon Spectroscopy are outlined in the following. From now on, we resort back to the notations and symbols used by the respective authors in the original publications. 
The discussion of the models will by no means be complete. However, we hope that this section can provide a useful guideline to the literature. 

The models currently on the market fulfill the $S$-Matrix principles discussed in section~\ref{sec:PWA:SMatrixPrinciples} with a varying degree of rigor. In practice, one has to make approximations and the exact fulfillment of unitarity, analyticity and crossing is an idealization which in most cases cannot be fully satisfied. The approximations should keep the models tractable, while also maintaining the quality of the physics results extracted from the data. 

The list of models described in the following is not complete. Further recently developed ED fit approaches are for instance given by the analyses performed by the Giessen-group \cite{Shklyar:2012zz, Shklyar:2012js}. The historic analyses of $\pi N$-scattering by H\"{o}hler \cite{Hoehler84}, Cutkosky and collaborators \cite{Cutkosky:1979fy} should of course also be mentioned here.

\subsubsection{The SAID-analysis} \label{sec:PWA:SAIDModel} 
The \underline{S}cattering \underline{A}nalysis \underline{I}nteractive \underline{D}ial-in~\cite{Arndt:1995bj,Arndt:2006bf,Workman:2012hx,Workman:2012jf,Beck:2016hcy,}, or SAID, provided by the George Washington University (GWU) is one of the earliest ED analyses whose results were published on the internet~\cite{SAIDURL}. In the most recent fits, the SAID-group employs phenomenological $K$-matrix type approaches for the analysis of pion- and photon-induced reactions \cite{Workman:2012hx, Workman:2012jf}. The $T$-Matrix for pion-nucleon elastic scattering is given by
\begin{equation}
 T_{\alpha \beta} = \left[ 1 - \bar{K} C \right]^{-1}_{\alpha \sigma} \bar{K}_{\sigma \beta} \mathrm{,} \label{eq:PWA:SAIDPiNElasticParametrization}
\end{equation}
where $C$ is a Chew-Mandelstam function \cite{Workman:2012hx, Workman:2012jf} and the greek indices~$(\alpha,\beta,\sigma)$ run over the included hadronic channels: $\pi N$, $\pi \Delta$, $\rho N$ and $\eta N$. A sum over the repeated index $\sigma$ is implicit. The imaginary part of the Chew-Mandelstam function is generally given by the phase-space, i.e. $C = \mathrm{Re} \left[C \right] + i \rho$. Dropping the real parts would correspond a pure '$K$-matrix approximation'. However, the SAID-group goes beyond this approximation by retaining the real part~\cite{Workman:2012hx}. Furthermore, the imaginary part is continued below channel thresholds. It has to be mentioned that the pion-nucleon analyses published by SAID have since a long time served as input to other ED PWA-approaches~(for instance the J\"{u}lich-Bonn-, ANL-Osaka-, MAID- and KSU-analyses, sections~\ref{sec:PWA:JueBoModel},~\ref{sec:PWA:ANLOsakaModel},~\ref{sec:PWA:MAIDModel} and~\ref{sec:PWA:KentStateApproach}). Therefore, the SAID model defines an important reference point in the research community for these reactions.

The photon-induced reactions are implemented in an approach that closely resembles the so-called $P$-vector ('production-vector') formalism~\cite{Aitchison:1972ay,Chung:1995dx} (see also section~\ref{sec:PWA:BnGaModel}):
\begin{equation}
 T_{\alpha \gamma} = \left[ 1 - \bar{K} C \right]^{-1}_{\alpha \sigma} \bar{K}_{\sigma \gamma} \mathrm{.} \label{eq:PWA:SAIDPiNElasticParametrization2}
\end{equation}
In this expression, the $K$-Matrix elements $\bar{K}_{\sigma \gamma}$ quantify the coupling strengths of the $\gamma N$-system to the purely hadronic channels listed above. The energy-dependence of the $K$-Matrix elements in the SAID-formalism is defined purely in terms of phenomenological polynomials. The only resonance which is put in as a genuine $K$-Matrix pole is the $\Delta (1232) \frac{3}{2}^{+}$. All other $T$-Matrix poles are generated dynamically, i.e. by the factor $\left[ 1 - \bar{K} C \right]^{-1}$ which models final-state interactions. For this reason, the $T$-Matrices for the elastic pion-nucleon- and the photoproduction reactions can only have the same pole-structure in this formalism. 

Different partial wave analyses of SAID are available for the comparison with photoproduction data at the SAID website \cite{SAIDURL}: the recent SAID-MA19 solution \cite{Briscoe:2019cyo} for all pion reactions ($\gamma p \to \pi^0 p$, $\gamma p \to \pi^+ n$, $\gamma n\to \pi^- p$, giving for the first time $\gamma n$ couplings for selected resonances and $\gamma n \to \pi^0 n$) as well as independent solutions for the $p\eta$ \cite{McNicoll:2010qk}, $p\eta'$, $K\Lambda$ and $K\Sigma$ final states. It has to be mentioned here that the $p \eta$- and $p \eta'$ final states have been analyzed using methods other than the above-described Chew-Mandelstam $K$-matrix.

\subsubsection{The Bonn-Gatchina model} \label{sec:PWA:BnGaModel} 
The Bonn-Gatchina partial-wave analysis~\cite{Anisovich:PWABook,Anisovich:2011fc,Sarantsev:2016fac,Beck:2016hcy} is a flexible tool for the analysis of many different reactions in an elaborate coupled-channels formalism. The fundamentals of the formalism have been published in many works over the years, see for instance \cite{Anisovich:PWABook,Anisovich:2011fc}. At the present moment, the model has been around for almost $20$~years. Regularly updated results are made accessible to the public on the Bonn-Gatchina webpage~\cite{BoGaURL}. In its most recent version, the Bonn-Gatchina model utilizes a so-called '$N/D$-inspired approach', which is reviewed in reference~\cite{Beck:2016hcy}. The following description relies mainly on this review. 

The model is a coupled-channel $K$-Matrix analysis with a multitude of channels. The transition-amplitude is written as\footnote{The Bonn-Gatchina group consistently uses the letter '$A$' for the $T$-Matrix.}
$\hat{\bm{A}}(s)$ with matrix-elements $A_{ab}(s)$ for channels $a$ and $b$, where the channel-indices include $2$-body systems such as $\pi N$, $\eta N$, $K \Lambda$, $\pi \Delta$ and the electromagnetic channel $\gamma N$. Furthermore, $3$-body states such as $\pi \pi N$ and $\pi \eta N$ are included into the analysis as well. 

The BnGa-group uses the so-called covariant tensor-formalism \cite{Anisovich:PWABook} for their definition of the partial-wave decompositions of amplitudes. The dependence on 'good' quantum-numbers (total angular momentum $J$, parity $P$, isospin $I$) is only made implicit in the following expressions. In the employed Ansatz, the transition-amplitude in matrix-form reads as follows
\begin{equation}
 \hat{\bm{A}}(s) = \hat{\bm{K}} \left( \hat{\mathbbm{1}} - \hat{\bm{B}} \hat{\bm{K}} \right)^{-1} \mathrm{.} \label{eq:PWA:BnGaTransitionAmplitude}
\end{equation}
The diagonal matrix $\hat{\bm{B}}$ contains rescattering loop-diagrams and is in principle the same as the Chew-Mandelstam function mentioned in section~\ref{sec:PWA:SAIDModel}. The $K$-matrix is parametrized in a 'pole$+$background'-Ansatz, i.e.
\begin{equation}
 K_{ab} (s) := \sum_{\alpha} \frac{g_{a}^{(\alpha)} g_{b}^{(\alpha)}}{M_{\alpha}^{2} - s} + f_{ab} (s) \mathrm{.} \label{eq:PWA:BonnGatchinaKMatrix}
\end{equation}
In this expression, $M_{\alpha}$ is the mass of the resonance '$\alpha$' and $g_{a}^{(\alpha)}$ is its coupling to the channel '$a$'. The energy-dependent functions $f_{ab}(s)$ describe non-resonant transitions $a \rightarrow b$. However, in most partial waves the Bonn-Gatchina group assumes constant~$f_{ab}$. Notable exceptions are the partial waves $S_{11}$ (relative orbital angular momentum $\ell = 0$; total angular momentum $J=1/2$; isospin $I=1/2$) and $S_{31}$ ($\ell = 0$; $J=3/2$; $I=1/2$), where the Bonn-Gatchina group uses the following parametrization:
\begin{equation}
 f_{ab} (s) = \frac{f^{(1)}_{ab} + f^{(2)}_{ab} \sqrt{s}}{s - s_{0}^{ab}} \mathrm{.} \label{eq:PWA:fabBackgroundSpecialCase}
\end{equation}
The constants $f_{ab}^{(i)}$ and $s_{0}^{ab}$ have to be fitted to the data. The reasoning behind these parametrizations is the following: the functions~$f_{ab}(s)$ inserted as 'background' at the $K$-matrix level~\eqref{eq:PWA:BonnGatchinaKMatrix} have to satisfy the following two basic requirements
\begin{itemize}
 \item[(i)] the chosen parametrizations have to introduce as few new parameters as possible. The number of channels introduced into the Bonn-Gatchina analysis is anyways very large, i.e. the channel-indices ‘$a$’ and ‘$b$’ run over many values (cf. lists given above).
 \item[(ii)] the parametrizations have to be analytic in energy, while introducing no new right-hand singularities in the complex energy-plane (cf. section~\ref{sec:PWA:Analyticity}). Further left-hand singularities, which are situated outside of the physical region, are however in principle allowed.
\end{itemize}
The simplest possibility which satisfies requirements~(i) and~(ii) is simply a constant term $f_{ab}$. In some specific waves however (i.e. the $S_{11}$ and $S_{31}$ waves), the constant is not sufficient, due to crossed-channel (i.e. $t$- and $u$-channel) exchanges which have significant projections into these waves (cf. the discussion on left-hand singularities in section~\ref{sec:PWA:Analyticity}). In order to model these partial-wave projected crossed-channel exchanges with as few parameters as possible, the Bonn-Gatchina group utilizes the left-hand pole parametrization~\eqref{eq:PWA:fabBackgroundSpecialCase}.

The $\gamma N$-channel is introduced in the so-called $P$-vector approach~\cite{Aitchison:1972ay,Chung:1995dx} (see also section~\ref{sec:PWA:SAIDModel}). The photoproduction-amplitude for the process $a \rightarrow b$ reads
\begin{equation}
 A_{a} = \hat{P}_{b} \left( \hat{\mathbbm{1}} - \hat{B} \hat{K} \right)^{-1}_{ba} \mathrm{.} \label{eq:PWA:AmplitudeInPVectorApproach}
\end{equation}
The $P$-vector is again parametrized in a pole-background form (cf.  equation~\eqref{eq:PWA:BonnGatchinaKMatrix}):
\begin{equation} 
  P_{b} (s) := \sum_{\alpha} \frac{g_{\gamma N}^{(\alpha)} \hspace*{1pt} g_{b}^{(\alpha)}}{M_{\alpha}^{2} - s} + \tilde{f}_{\left( \gamma N \right)b} (s) \mathrm{.} \label{eq:PWA:BonnGatchinaPVector}
\end{equation}
The photo-couplings $g_{\gamma N}^{(\alpha)}$ have to be fitted to the data and the functions $\tilde{f}_{\left( \gamma N \right)b} (s)$ describe non-resonant transitions. In practical analyses, the $\tilde{f}_{\left( \gamma N \right)b}$ are again assumed as constant. Furthermore, for reactions where only the decay is weakly coupled, like e.g. those with $3$-body final states, the BnGa-group uses a so-called $D$-vector approach. For reactions where both the production and decay are weakly coupled, the so-called $PD$-vector method is utilized (see reference~\cite{Beck:2016hcy} for more details on both approaches). 

The Bonn-Gatchina group uses so-called {\it reggeized} amplitudes~\cite{AnalyticSMatrix} in order to describe the forward-peaks in the data present at high energies, which originate from $t$-channel exchanges. The Regge amplitudes are projected to partial waves and all projections to lower waves (up to the $D$-waves) are taken out of the amplitudes~\cite{AndreyPrivComm}. Then, the Bonn-Gatchina group introduces to the $K$-matrix part the contribution~\eqref{eq:PWA:fabBackgroundSpecialCase}, which provides a good first-order description of the $t$- (and $u$-) channel contributions. Therefore, the reggeized amplitudes have only high partial waves in them. This procedure is necessary to avoid unitarity-violations~\cite{AndreyPrivComm}. For more details on and explicit expressions for the Regge amplitudes, see references~\cite{Anisovich:PWABook,Beck:2016hcy}.


The Bonn-Gatchina partial-wave analysis has a reputation for being well-optimized and thus it allows for fits of the above-described model in short runtimes. This facilitates tests of and comparisons among many different scenarios for resonance contents. At the present moment, a lot of the newly-included resonances in the summary tables of the PDG~\cite{Zyla:2020zbs} stem from the Bonn-Gatchina analysis.

The most recent BnGa PWA is the BnGa-2019 solution \cite{Muller:2019qxg} for the $\gamma p$ data and is available at their website \cite{BoGaURL}. The $\gamma n$ data is used to determine only the resonance $\gamma n$-couplings, which will be published in the future. Further recent developments include extensive partial-wave analyses for the determination of the properties of hyperons~\cite{Matveev:2019igl,Sarantsev:2019xxm}.

  
\subsubsection{The J\"{u}lich-Bonn model} \label{sec:PWA:JueBoModel} 
The J\"{u}lich-Bonn model~\cite{Doring:2009yv,Doring:2009bi,Doring:2010ap,Ceci:2011ae,Ronchen:2012eg,Ronchen:2014cna,RonchenDiss:2014,Ronchen:2015vfa,Ronchen:2016sjn,Doring:2016snk,Beck:2016hcy,Ronchen:2018ury} is a dynamical coupled-channels (DCC) model for a combined analysis of pion- and photon-induced reactions. Recently, the model has been extended to an approach for single-meson electroproduction~\cite{Mai:2021vsw}. The electroproduction model is called the J\"{u}lich-Bonn-Washington model. Results can be inspected at a recently created website~\cite{JuBoURL}.

Recursive Lippmann-Schwinger equations are at the core of this analysis. For the pion-induced reactions~\cite{Ronchen:2012eg}, the following equation is solved for the partial-wave matrix elements of $T$:
\begin{equation}
 T_{\mu \nu} (q,p^{\prime};W) = V_{\mu \nu} (q,p^{\prime};W) + \sum_{\kappa} \int_{0}^{\infty} dp \hspace*{2pt} p^{2} \hspace*{2pt} V_{\mu \kappa} (q,p;W) G_{\kappa} (p;W) T_{\kappa \nu} (p,p^{\prime};W) \mathrm{.} \label{eq:PWA:JuelichModelPiNLippmannSchwinger}
\end{equation}
The indices $\mu$, $\nu$ and $\kappa$ denote the final, initial and intermediate channels. They run over the genuine $2$-body channels $\pi N$, $\eta N$, $K \Sigma$ and $K \Lambda$ as well as the 'effective' $2$-body channels consisting of a stable particle and an unstable isobar: $\pi \Delta$, $\sigma N$ and $\rho N$. The effective $2$-body channels are used to model the $3$-particle intermediate state $\pi \pi N$. For this reason, $2$-body unitarity is satisfied exactly in the J\"{u}lich approach and $3$-body unitarity is fulfilled approximately \cite{Ronchen:2012eg,Doring:2009yv}. The $G_{\kappa}$ denote $2$- and $3$-body propagators. Kinematical variables in equation~\eqref{eq:PWA:JuelichModelPiNLippmannSchwinger} are the total center-of-mass energy $W$ and the outgoing and incoming $3$-momenta $q = \left| \vec{q} \right|$ and $p^{\prime} = \left| \vec{p}^{\prime} \right|$. The momenta can violate the on-mass-shell condition in this approach~\cite{Beck:2016hcy}. The scattering-potential $V_{\mu \nu}$ in equation~\eqref{eq:PWA:JuelichModelPiNLippmannSchwinger} is derived from effective lagrangians \cite{Ronchen:2012eg}. It contains bare $s$-channel poles as genuine resonances. Furthermore, $t$- and $u$-channel exchanges of a large collection of known light mesons and baryons are included as well. The potential is described in more detail in reference~\cite{Ronchen:2012eg}. 

Apart from the genuine resonances which are built in 'by hand', the Lippmann-Schwinger equation~\eqref{eq:PWA:JuelichModelPiNLippmannSchwinger} can produce so-called {\it dynamically generated} resonances. The latter are $T$-Matrix poles which arise as a by-product of the solution of equation~\eqref{eq:PWA:JuelichModelPiNLippmannSchwinger} and thereby implicitly from an iteration of the initially parameterized potential to all orders. The ability to systematically examine such dynamically generated resonances is one of the most prominent features of the model.

The J\"{u}lich-Bonn model of single-meson photoproduction~\cite{Ronchen:2014cna,Ronchen:2015vfa} is a phenomenological extension of the had\-ro\-nic meson-baryon amplitudes. The photoproduction amplitude consists of a pre-defined photoproduction kernel~$V_{\mu \gamma}$ plus a rescattering term, which in its form resembles the second term on the right hand side of equation~\eqref{eq:PWA:JuelichModelPiNLippmannSchwinger}. This results in the following equation for the photoproduction multipoles $M_{\mu \gamma}$ (cf. section~\ref{sec:formalisms:SingleMesonPhotoproduction}):
\begin{equation}
 M_{\mu \gamma} (q;W) = V_{\mu \gamma} (q;W) + \sum_{\kappa} \int_{0}^{\infty} dp \hspace*{2pt} p^{2} \hspace*{2pt} T_{\mu \kappa} (q,p;W) G_{\kappa} (p;W) V_{\kappa \gamma} (p;W) \mathrm{.} \label{eq:PWA:JuelichModelPhotoProdLippmannSchwinger}
\end{equation}
The $\gamma N$-state is labeled by the index '$\gamma$' and the $T$-matrix $T_{\mu \kappa}$ encompasses the full meson-baryon model, which is a solution of equation~\eqref{eq:PWA:JuelichModelPiNLippmannSchwinger}.
The kernel $V_{\mu \gamma}$ can again be decomposed into pole- and non-pole parts and the J\"{u}lich-Bonn group parameterized the non-pole terms and the creation-vertex functions for the resonances as energy-dependent polynomials \cite{Ronchen:2014cna}. On the contrary, hadronic annihilation-vertex functions have been imported from the meson-baryon model. Explicit expressions for the photoproduction-kernel can be found in \cite{Ronchen:2014cna,RonchenDiss:2014}.

The J\"{u}lich-Bonn group is well-known for implementing unitarity- and analyticity-constraints (cf. sections~\ref{sec:PWA:Unitarity} and~\ref{sec:PWA:Analyticity}) in a way that is as theoretically rigorous as possible. The group has described the analytic continuation of the meson-baryon model over a staggering amount of $256$ Riemann-sheets in detail in reference~\cite{Doring:2009yv}. This number of sheets holds for the channel-space described in reference~\cite{Doring:2009yv}, but the analytic structure of the model has probably further complicated since then.

\subsubsection{The ANL-Osaka model} \label{sec:PWA:ANLOsakaModel} 

The ANL-Osaka model~\cite{Sato:1996gk,Matsuyama:2006rp,Julia-Diaz:2007qtz,Kamano:2013iva,Kamano:2013ssa,Sato:2014paa,Kamano:2015hxa,Kamano:2016bgm,Kamano:2019gtm} is a DCC-approach for the combined analysis of pion- and photon-induced reactions. Apart from the J\"{u}lich-Bonn analysis (section~\ref{sec:PWA:JueBoModel}), it is the complementary major DCC-model for Light Baryon Spectroscopy currently on the market.

The ANL-Osaka group defines the reaction amplitudes~$T_{\alpha, \beta} (p_{\alpha},p_{\beta};E)$ in the partial-wave basis (total angular momentum and total isospin quantum numbers are implicit in the following) as a solution of the following set of coupled-channel integral equations~\cite{Kamano:2013iva,Kamano:2013ssa,Sato:2014paa}, which are formally the same as equation~\eqref{eq:PWA:JuelichModelPiNLippmannSchwinger}:
\begin{align}
 &T_{\alpha, \beta} (p_{\alpha},p_{\beta};E) = V_{\alpha \beta} (p_{\alpha},p_{\beta}) + \sum_{\delta} \int p^{2} \hspace*{1pt} dp \hspace*{2pt}  V_{\alpha \delta} (p_{\alpha},p) G_{\delta} (p,E) T_{\delta \beta} (p,p_{\beta};E) \mathrm{,} \label{eq:PWA:ANLOsakaModelLippmannSchwinger} \\
 & \text{where } V_{\alpha \beta} (p_{\alpha},p_{\beta}) =  v_{\alpha \beta} (p_{\alpha},p_{\beta})  + \sum_{N^{\ast}}  \frac{\Gamma^{\dagger}_{N^{\ast},\alpha} (p_{\alpha}) \Gamma_{N^{\ast},\beta} (p_{\beta}) }{E - M_{N^{\ast}}^{0}}  \mathrm{.} \label{eq:PWA:ANLOsakaModelPotential}
\end{align}
Here, the greek indices encompass the reaction channels~$\alpha, \beta, \delta = \pi N, \gamma N, \eta N , K \Sigma, K \Lambda, \pi \Delta, \rho N, \sigma N$, where the last three channels are effective $2$-body channels used to model the $3$-body system~$\pi \pi N$ (cf. section~\ref{sec:PWA:JueBoModel}) and~$G_{\delta} (p,E)$ is the Green's function (i.e. propagator) of channel~$\delta$. The potential~\eqref{eq:PWA:ANLOsakaModelPotential} is constructed from non-resonant processes~$v_{\alpha \beta} (p_{\alpha},p_{\beta})$, which are given by meson-exchange mechanisms calculated from effective lagrangians via the method of unitary transformations~\cite{Kamano:2013iva}. The second contribution to the potential~\eqref{eq:PWA:ANLOsakaModelPotential} is given by the $s$-channel exchange of 'bare' $N^{\ast}$-states with bare masses~$M_{N^{\ast}}^{0}$ and the vertex functions~$\Gamma_{N^{\ast},\beta} (p_{\beta})$ describe the strength of the decay~$N^{\ast} \rightarrow \beta$. The ANL-Osaka model satisfies exact $2$-body unitarity and approximate $3$-body unitarity, similarly to the J\"{u}lich-Bonn approach (section~\ref{sec:PWA:JueBoModel}). However, the authors of reference~\cite{Kamano:2013iva} state that within the numerical precision of their analyses, unitarity is practically exact. The description provided in equations~\eqref{eq:PWA:ANLOsakaModelLippmannSchwinger} and~\eqref{eq:PWA:ANLOsakaModelPotential} above is somewhat simplified. A more detailed decomposition of the whole ANL-Osaka model-function can be found for instance in reference~\cite{Kamano:2013iva}.

Up to this point, the formalisms of ANL-Osaka and J\"{u}lich-Bonn (section~\ref{sec:PWA:JueBoModel}) look very similar. However, differences lie in the details. For instance, regarding the treatment of $t$- and $u$-channel exchanges and their relation via crossing transformations (cf. section~\ref{sec:PWA:Crossing}), the J\"{u}lich-Bonn model contains some constraints~\cite{Krehl:1999ak} which ANL-Osaka does not have. Furthermore, although the channel space of both models is exactly the same, details on the construction of the scattering potentials differ. Firstly, the exchanged mesons and baryons for the nonresonant potentials differ~\cite{Ronchen:2014cna,Kamano:2013iva}. Secondly, the ANL-Osaka group uses the above-mentioned method of unitary transformations for the construction of their potential, which J\"{u}lich-Bonn does not.

The fitting strategy of the ANL-Osaka group consists of first analyzing the pion-induced reactions, i.e. data on~$\pi N \rightarrow \pi N$, $\pi N \rightarrow \eta N$, $\pi N \rightarrow K \Lambda$ and $\pi N \rightarrow K \Sigma$, in order to obtain a first stable solution for the meson-baryon $T$-matrix. The photon-induced reactions are then analyzed in a second step. In a fit published in $2013$~\cite{Kamano:2013iva}, the ANL-Osaka group was able to describe a large collection of data with $10$ isospin-$\frac{1}{2}$ resonances and~$10$ isospin~$\frac{3}{2}$-resonances below $2$~GeV. More recent developments include the determination of the isospin-decomposition for the $\gamma N \rightarrow N^{\ast}$ transition amplitudes~\cite{Kamano:2016bgm}.

\subsubsection{The MAID-analysis} \label{sec:PWA:MAIDModel} 
The \underline{MA}inz unitary \underline{I}sobar mo\underline{D}el~\cite{Drechsel:1998hk,Drechsel:2007if,Beck:2016hcy,Tiator:2016jzo,Kashevarov:2017vyl,Tiator:2018pjq,Tiator:2019dfj}, or MAID, is a flexible tool for ED analyses of photon- and electron-induced reactions. The results can be accessed at the url~\cite{MAIDURL}. MAID is generally not a coupled-channels model, although for pion photoproduction specifically, some sort of coupling to pion-nucleon scattering is implemented. At the core of the model lies the definition of the $T$-Matrix as a 'background$+$resonance'-Ansatz:
\begin{equation}
 t_{\gamma \pi} (W,Q^{2}) \equiv t_{\gamma \pi}^{B} \left(W,Q^{2}\right) + t_{\gamma \pi}^{R} \left(W,Q^{2}\right) \mathrm{,} \label{eq:PWA:MAIDTMatrixAnsatz}
\end{equation}
where $W$ is the total center-of-mass energy and $Q^{2}$ the photon virtuality. The 'good' quantum numbers are collected in a multi-index $\alpha = \left\{ j,\ell,\ldots \right\}$, which then uniquely specifies each partial wave. The background-part is defined via the following expression:
\begin{equation}
 t_{\gamma \pi}^{B,\alpha} (W,Q^{2}) \equiv v_{\gamma \pi}^{B,\alpha} \left(W,Q^{2}\right) \left[ 1 + i t^{\alpha}_{\pi N} \left( W \right) \right]  \mathrm{.} \label{eq:PWA:MAIDPhenBackground}
\end{equation}
The potential $v_{\gamma \pi}^{B,\alpha}$ is defined by Born-Terms and the amplitude $t^{\alpha}_{\pi N}$ models the pion-nucleon rescattering. The amplitudes of elastic pion-nucleon scattering are implemented as: $t^{\alpha}_{\pi N} = \left[ \eta_{\alpha} \exp \left( 2 i \delta_{\alpha} \right) - 1 \right] /2i$. The phase-shifts~$\delta_{\alpha}$ and inelasticities~$\eta_{\alpha}$ for the pion-nucleon reaction are imported from the SAID-analysis (cf. section~\ref{sec:PWA:SAIDModel}). 

The resonant part of the $T$-Matrix~\eqref{eq:PWA:MAIDTMatrixAnsatz} is defined by (a sum over) the following Breit-Wigner type amplitudes:
\begin{equation}
 t_{\gamma \pi}^{R} \left(W,Q^{2}\right) := \bar{\mathcal{A}}^{R}_{\alpha} \left(W,Q^{2}\right) \frac{f_{\gamma N} (W) \Gamma_{\mathrm{tot.}} (W) M_{\mathrm{R}} f_{\pi N} (W) }{M_{\mathrm{R}}^{2} - W^{2} - i M_{\mathrm{R}} \Gamma_{\mathrm{tot.}} (W)} \exp \left[ i \phi_{R} \left(W,Q^{2}\right) \right] \mathrm{.} \label{eq:PWA:MAIDResonanceAnsatz}
\end{equation}
Here, $M_{\mathrm{R}}$ and $\Gamma_{\mathrm{tot.}} (W)$ are mass and width of the resonance, $f_{\gamma N}$ and $f_{\pi N}$ are Breit-Wigner factors relating total and partial decay-widths and $\phi_{R}$ is an additional phase needed for the unitarization. The electromagnetic coupling $\bar{\mathcal{A}}^{R}_{\alpha}$ is for most resonances parameterized as
\begin{equation}
 \bar{\mathcal{A}}^{R}_{\alpha} \left(W,Q^{2}\right) \equiv \bar{\mathcal{A}}^{R}_{\alpha} \left(Q^{2}\right) = \mathcal{A}_{\alpha} (0) \left( 1 + a_{1} Q^{2} + a_{2} Q^{4} + \ldots \right) \exp \left[ - b_{1} Q^{2} \right] \mathrm{,} \label{eq:PWA:MAIDResonancecouplingQ2Dependence}
\end{equation}
except for the $\Delta(1232) \frac{3}{2}^{+}$-resonance, where~$\bar{\mathcal{A}}^{R}_{\alpha}$ depends on the $3$-momentum $k \left(W,Q^{2}\right)$ of the photon. 

Recently, the MAID-group analyzed $\eta$- and $\eta^{\prime}$-photoproduction~\cite{Tiator:2016jzo,Tiator:2019dfj} and started using Regge-amplitudes for some select reactions~\cite{Kashevarov:2017vyl}. An easy-to-read and concise summary on the model can be found in reference~\cite{Tiator:2018pjq}.

\subsubsection{The Kent State University (KSU-) model} \label{sec:PWA:KentStateApproach}

The ED-fit model from Kent State University (KSU) was first proposed by Manley~\cite{Manley:2003fi} and has in recent years been applied in a variety of analyses for pion- and photon-induced reactions by Manley, Shrestha and Hunt~\cite{Shrestha:2012ep,Hunt:2018tvt,Hunt:2018mrt,Hunt:2018wqz}. The KSU-group typically analyzes single-energy partial waves from multiple reactions, which are fitted at once using their flexible ED coupled-channels model. These single-energy partial waves are taken in part from PWAs performed by other groups (for instance SAID, section~\ref{sec:PWA:SAIDModel}), or from TPWA-fits performed by the KSU-group itself. The method of fitting single-energy partial-waves has the merit of avoiding to introduce any model-bias a priori, compared to the method of fitting the ED-model directly to the data.

The ED KSU-model introduces the following multiplicative decomposition of the unitary and symmetric $S$-Matrix $\text{{\bf S}}$ in the partial-wave basis~\cite{Shrestha:2012ep,Hunt:2018wqz} (matrices also act in channel-space):
\begin{equation}
 \text{{\bf S}} = \text{{\bf B}}^{\bm{T}} \text{{\bf R}} \text{{\bf B}}   .  \label{eq:PWA:KSUDecompSMatrix}
\end{equation}
Here, $\text{{\bf R}}$ is the resonant part of the amplitude and purely describes $s$-channel processes. The matrix $\text{{\bf B}}$ is a product of background-matrices, which are individually unitary and symmetric: 
\begin{equation}
   \text{{\bf B}} = \text{{\bf B}}_{\text{{\bf 1}}} \text{{\bf B}}_{\text{{\bf 2}}} \ldots \text{{\bf B}}_{\text{{\bf n}}}  . \label{eq:PWA:KSUBMatrixDecomposition}
\end{equation}
For parameterizations of the individual background-contributions $\text{{\bf B}}_{\text{{\bf i}}}$, see \cite{Manley:2003fi}. The resonant part $\text{{\bf R}}$ can be written either in terms of the resonant $T$-Matrix $\text{{\bf T}}_{\text{{\bf R}}}$ or in terms of a $K$-Matrix $\text{{\bf K}}$~\cite{Hunt:2018wqz}:
\begin{equation}
  \text{{\bf R}} = \text{{\bf I}} + 2 i \text{{\bf T}}_{\text{{\bf R}}} =  \left( \text{{\bf I}} + i \text{{\bf K}} \right)  \left( \text{{\bf I}} - i \text{{\bf K}} \right)^{-1}  , \label{eq:PWA:KSUResAmpDecomposition}
\end{equation}
where $\text{{\bf K}}$ has to be hermitean (cf.~Sec.~\ref{sec:PWA:Unitarity}). In addition, $\text{{\bf K}}$ is also symmetric in order to satisfy time-reversal symmetry. Resonances are included at the level of the $K$-matrix in the following form~\cite{Shrestha:2012ep}:
\begin{equation}
 \text{{\bf K}}_{ij} = \sum_{\alpha = 1}^{N} X_{i \alpha} X_{j \alpha} \tan \delta_{\alpha}  . \label{eq:PWA:KSUResonantKMatrix}
\end{equation}
In this expression, $i,j$ are channel-indices, '$\alpha$' denotes an individual resonance, $N$ is the number of resonances present within the energy-range of the fit and $\delta_{\alpha}$ is an energy-dependent scattering-phase. Each factor~$X_{i \alpha}$ is related to the branching ratio for resonance $\alpha$ to decay into the channel '$i$'. Thus for each resonance, one has to have:~$\sum_{i} \left( X_{i \alpha} \right)^{2} = 1$.

For the simple special case of only one single resonance, the phase~$\delta$ has an energy-dependent motion according to the Breit-Wigner form
\begin{equation}
 \tan \delta(W)  = \frac{\Gamma / 2}{M - W}   , \label{eq:PWA:KSUOneResPhaseMotion}
\end{equation}
with $X_{i}^{2} = \Gamma_{i} / \Gamma$, where~$\Gamma_{i}$ is the partial width for the $i$-th channel and~$\Gamma$ is the total width of the resonance. In this example, the single resonance leads to an isolated pole in the resonant $T$-matrix:
\begin{equation}
  \left[ \text{{\bf T}}_{\text{{\bf R}}} \right]_{ij} = X_{i} \cdot X_{j} \frac{\Gamma / 2}{M - W - i \Gamma / 2} . \label{eq:PWA:KSUSpecialCaseOneResonance}
\end{equation}
For the general case with $N$ resonances, the resonances still contribute in the simple additive way~\eqref{eq:PWA:KSUResonantKMatrix} at the level of the $K$-matrix. However, at the level of the resonant $T$-matrix, contributions from individual resonance-poles are not separable any more, but rather a re-coupling of poles occurs defined in terms of calculable energy-dependent coefficients~\cite{Shrestha:2012ep}. This non-separability of the resonant $T$-matrix is necessary to maintain unitarity.

In earlier fits involving Shrestha~\cite{Shrestha:2012ep}, the KSU-group performed a combined fit of the reactions~$\pi N \rightarrow \pi N$, $\pi N \rightarrow \pi \pi N$, $\pi N \rightarrow \eta N$, $\pi N \rightarrow K \Lambda$ and $\gamma N \rightarrow \pi N$. In the later works involving Hunt~\cite{Hunt:2018tvt,Hunt:2018mrt,Hunt:2018wqz}, the photoproduction reactions~$\gamma p \rightarrow \eta p$, $\gamma n \rightarrow \eta n$ and $\gamma p \rightarrow K^{+} \Lambda$ were analyzed in addition. In the most recently published solution~\cite{Hunt:2018wqz}, the KSU-group needed~$22$ isospin-$\frac{1}{2}$ resonances, as well as~$12$ states with isospin~$\frac{3}{2}$, in order to describe the data.

\subsubsection{The L+P-formalism} \label{sec:PWA:LPlusFormalism}

The Laurent+Pietarinen- (L+P-) formalism has been originally introduced as a single-channel pole-extraction method by {\v{S}}varc and collaborators~\cite{Svarc:2013laa,Svarc:2014zja,Svarc:2014sqa,Svarc:2015usk,Svarc:2020waq,Svarc:2021kql}. It has been designed to extract pole-parameters from data for single-energy partial waves on the real energy-axis, which could either stem from a TPWA performed on real data, or from a discretization of the solution of some ED-model. As a mathematical construction, the L+P-formalism is simpler and thus easier to implement than the other models discussed in this chapter. However, the method has already found success. For instance, pole-parameters for nucleon resonances extracted from Karlsruhe-Helsinki partial waves using this method~\cite{Svarc:2014zja} have been included into the summary tables of the particle data group~\cite{Zyla:2020zbs}. The particle data group also included the L+P-formalism as a possible pole-extraction method into their reviews.

The basic idea of the method is to construct the simplest analytic function which can model a partial-wave amplitude with a well-defined analytic structure in terms of poles and branch-cuts. The $T$-Matrix element in the partial-wave basis $T(W)$ is thus written as a function of complex energy $W$ via a Laurent-expansion\footnote{In the situation with more than one pole, it would be more precise to call the expansion~\eqref{eq:PWA:basicLPlusPAnsatz} a {\it Mittag-Leffler expansion}. Still, we will stick with the name Laurent-expansion in this review.} for the general situation with $N_{\text{pole}}$ poles, plus a background-term~\cite{Svarc:2013laa}: 
\begin{equation}
  T (W) = \sum_{i = 1}^{N_{\text{pole}}} \frac{a_{-1}^{(i)}}{W - W_{i}} + B(W)    .  \label{eq:PWA:basicLPlusPAnsatz}
\end{equation}
Here, $W_{i}$ are the pole-positions in the complex energy-plane, the $a_{-1}^{(i)}$ are the complex residues for each pole $i$ and $B(W)$ is the background-function which in particular has to model all the relevant branch-cuts.


The background-function $B(W)$ is unknown. The only useful initial data one possesses on this function is given by its well-defined set of branch-points and -cuts (cf.~Sec.~\ref{sec:PWA:Analyticity}). One can now use analyticity and find a general power-series expansion for a function $F(W)$ that has a branch-point at some $x_{P} \in \mathbb{C}$. The parameters defining this expansion then have to be fitted out of the data on the single-energy partial waves. Such an expansion has been found by Ciulli and Fischer using conformal-mapping techniques~\cite{CIULLI1961465} and it takes the form of a power-series in terms of the so-called conformal variable $Z(W)$: 
\begin{equation}
 F(W) = \sum_{n = 0}^{N_{\text{Piet.}}} c_{n} Z^{n} (W) \text{, with } Z(W) = \frac{\alpha - \sqrt{x_{P} - W}}{\alpha + \sqrt{x_{P} - W}} . \label{eq:PWA:BasicPietarinenExp} 
\end{equation}
Here, $\alpha$ is a real shape- (i.e. tuning-) parameter and $c_{n}$ are real expansion-coefficients. Expansions such as~\eqref{eq:PWA:BasicPietarinenExp} have been used prominently by Pietarinen in analyses of $\pi N$-scattering~\cite{Pietarinen:1972nk}. Therefore, they have been termed {\it Pietarinen expansions}~\cite{Hoehler84}.

All this defines the L+P-formalism in terms of a certain number $N_{\text{pole}}$ of poles and a certain number of branch-cuts. For multiple branch cuts, one can just define $B(W)$ as the sum of multiple Pietarinen-expansions of the form~\eqref{eq:PWA:BasicPietarinenExp}. 

The fit-strategy proposed by the authors of the method~\cite{Svarc:2013laa,Svarc:2014zja,Svarc:2020waq} starts from a philosophy of {\it minimality}, i.e. one starts with the minimal possible number of poles and raises this number until a satisfactory fit is obtained. The number of branch-points is in principle known due to the physical context of the analyzed partial wave. However, due to the large number of parameters introduced with each Pietarinen-expansion, the authors generally propose as a best strategy to use just three Pietarinen expansions. Using this approach, one can write the L+P-equations in a more general 'coupled-channels-form' as follows (the residues are here explicitly decomposed into real- and imaginary parts)~\cite{Svarc:2015usk,Svarc:2020waq}:
{\allowdisplaybreaks
\begin{align}
  T^{a} (W) &= \sum_{j = 1}^{N_{\text{pole}}} \frac{x_{j}^{a} + i y_{j}^{a}}{W_{j} - W} + \sum_{k = 0}^{K^{a}} c_{k}^{a} X^{a} (W)^{k} + \sum_{l = 0}^{L^{a}} d_{l}^{a} Y^{a} (W)^{l} + \sum_{m = 0}^{M^{a}} e_{m}^{a} Z^{a} (W)^{m}  , \label{eq:PWA:LPlusPDef} \\
  X^{a} (W) &= \frac{\alpha^{a} - \sqrt{x_{P}^{a} - W}}{\alpha^{a} + \sqrt{x_{P}^{a} - W}}, \hspace*{2.5pt} Y^{a} (W) = \frac{\beta^{a} - \sqrt{x_{Q}^{a} - W}}{\beta^{a} + \sqrt{x_{Q}^{a} - W}}, \hspace*{2.5pt} Z^{a} (W) = \frac{\gamma^{a} - \sqrt{x_{R}^{a} - W}}{\gamma^{a} + \sqrt{x_{R}^{a} - W}}, \label{eq:PWA:LPlusPPietDef} \\
  W, W_{j} &\in \mathbb{C} \text{; } N_{\text{pole}} \text{ : number of poles} \text{; } x_{P}^{a}, x_{Q}^{a}, x_{R}^{a} \in \mathbb{R} \text{ (or } \in \mathbb{C} \text{)} \text{ : branch-points}; \nonumber \\
  x_{i}^{a}, y_{i}^{a} &, c_{k}^{a}, d_{l}^{a}, e_{m}^{a}, \alpha^{a}, \beta^{a}, \gamma^{a} \in \mathbb{R} \text{; } K^{a}, L^{a}, M^{a} \in \mathbb{N} \text{ : number of Pietarinen-coeff.'s in 'channel' } a , 
  \label{eq:PWA:LPlusPSymbols}
\end{align}
}%
where '$a$' is an index labeling either multiple partial waves belonging to one particular reaction, or even multiple partial-waves for a multitude of reactions. Further details on the fitting-procedure, including information on smoothness penalty-functions not shown here, can be found in~\cite{Svarc:2013laa,Svarc:2020waq}.

The roles of the three Pietarinen-expansions in the Ansatz~\eqref{eq:PWA:LPlusPDef} are as follows: The first Pietarinen-expansion is supposed to model an effective branch-cut which is supposed to take account of all left-hand-singularities (cf.~Sec.~\ref{sec:PWA:Analyticity}), the second one represents the elastic threshold of the analyzed partial-wave and thus has its branch-point fixed at the correct energy and the third one is again an effective branch-cut, which represents contributions from all higher inelastic thresholds. The branch-points of the first and third Pietarinen-functions are thus allowed to vary freely in the fit.

The advantages of the L+P-method consist of its mathematical simplicity and its flexibility to model general analytic functions, which has resulted in successful fits of partial waves with very complicated analytic structures in the past~\cite{Svarc:2013laa,Svarc:2014zja,Svarc:2014sqa}. However, the method also has its drawbacks, the first one being that it does not correctly reproduce the analytic structure of more complicated models such as the J\"{u}lich-Bonn model (cf.~Sec.~\ref{sec:PWA:JueBoModel}), on all Riemann-sheets. Instead, the L+P-parametrization is generally only a representation of such more complicated model-amplitudes in a limited region of the complex energy-plane on the physical Riemann-sheet, close to the real axis where the input-data are located. This is in most cases enough to extract good estimates for resonance pole-parameters. Furthermore, the formalism as defined in equations~\eqref{eq:PWA:basicLPlusPAnsatz} and~\eqref{eq:PWA:BasicPietarinenExp} generally violates unitarity. Methods to implement elastic unitarity close to the elastic threshold exist, which use certain penalty-techniques~\cite{Hoehler84}. Regarding the implementation of coupled-channel unitarity into the L+P-formalism, nothing has been published at the time of this writing.



\subsubsection{Further Analyses}
Besides the models described above, several different analyses exist, which were only focused on certain final states. These analyses are mentioned here for reference and not described in detail.

Since the data sets of measurements off neutrons are only sparse in comparison the the measurements off protons, neutron data is generally not included in the overall fits of the partial wave analyses. Data of neutrons can be used to extract the isospin multipoles by exploiting the Watson's theorem. This approach can be used for low-lying resonances and has been demonstrated in \cite{Workman:2010xc}. There exists also a separate, recent analysis by the BnGa group \cite{Anisovich:2017afs}, were the data sets have been used for consistency checks as well as to determine the helicity couplings of the $N^*$ resonances. Here, the couplings are given as residues of the pole positions in the complex plane as well as the Breit-Wigner parameters.

For the measurement of the vector meson $\omega$, several different models have been developed in the last decades. The first fits were performed by Oh et al.~\cite{Oh:2000zi} and used unpolarized data sets. Later, coupled channel fits included newly measured cross section data \cite{Penner:2002md}. The most sophisticated analysis was published in combination with the measurements by the CLAS collaboration, see \cite{Williams:2009aa}. Further details can be found in Sec.~\ref{sec:measurements:VectorMesonPhotoproduction}.

Reactions containing two mesons in the final states are often included into the partial wave analyses, like it is done for example for the BnGa partial wave analysis and the MAID model. There exist also several separate analyses, which focus only on (specific) multi meson final states. One model, which need to mentioned here is the Valencia model \cite{Roca:2002vd}, which focused on the low energetic $\pi^0\pi^0$ data in order to describe the $\sigma$ ($f_0(500)$) meson. Worth mentioning here is also the model from Fix and Arenh\"ovel \cite{Fix:2005if,Fix:2010nv}, which includes $\pi^0\pi^0$ and $\pi^0\eta$ photoproduction of nucleons and deuterons. The phenomenological model developed by Mokeev et al.~\cite{Mokeev:2001qg,Mokeev:2004fcd,Mokeev:2020hhu} focuses on the description of final states containing two charged pions in order to describe the recent measurements by the CLAS collaboration. A more recent approach is a field-theoretical description of the electromagnetic production of two pions off nucleons \cite{Haberzettl:2018hcl}.  This model was applied to photoproduction as well as electroproduction. Further details and examples can be found in Sec.~\ref{sec:measurements:MultiMesonPhotoproduction}.

In the strangeness sector, the single-channel isobar Kaon-MAID model of Mart and Bennhold \cite{Mart:1999ed} was one of the earliest available models, along with the model by Guidal \cite{Guidal:1999qi}, a dynamical coupled channel approach by Diaz et al. \cite{Julia-Diaz:2006rvy} and a unitary coupled-channel effective Lagrangian model by Shklyar et al. \cite{Shklyar:2005xg}. Additionally, the BnGa \cite{Gutz:2014wit}, J\"uBo \cite{Ronchen:2018ury} and ANL-Osaka \cite{Kamano:2013iva} groups have started to include some of the strangeness final states in their coupled channel analyses as well. 


%% file: Interpretation.tex
\label{sec:interpretation}

\subsection{Impact on the Partial Waves }

The different partial wave analysis groups (see Sec.~\ref{sec:pwa}) expanded their own respective database bit by bit over the years, which they use to extract the resonance parameters through fits to the measured observables. It is marked in Tables \ref{tab:measSinglePion1} - \ref{tab:measMultiMeson2} which data sets are included in the latest BnGa \cite{Muller:2019qxg}, J\"uBo \cite{Ronchen:2018ury} and SAID  \cite{Briscoe:2019cyo,SAIDURL} solutions, respectively.

\begin{figure}[ht]
\centering
\includegraphics[width=0.5\textwidth]{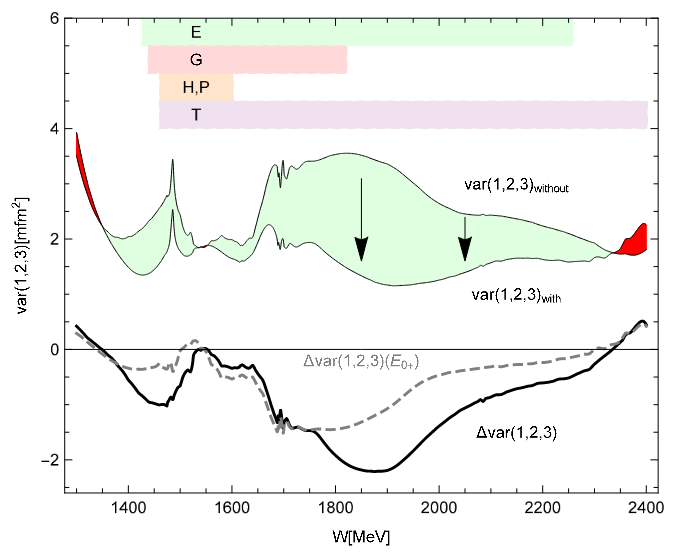}
\includegraphics[width=0.45\textwidth]{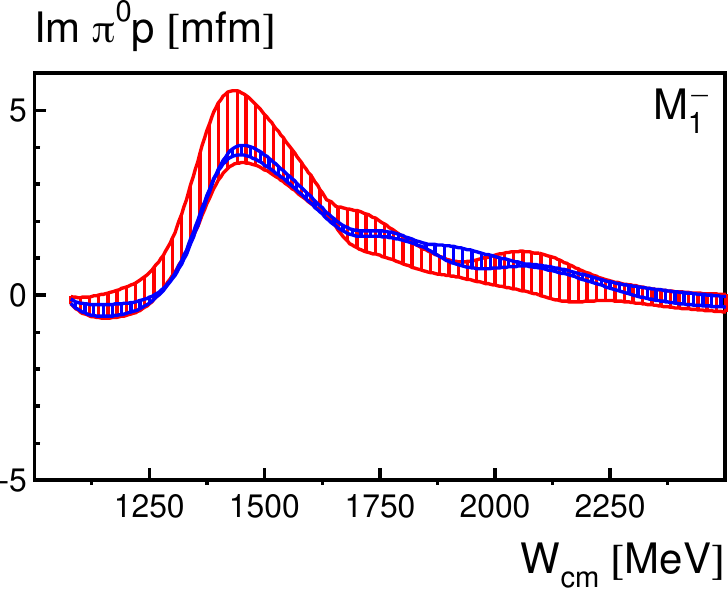}
\caption{Left: The variance of all three PWA groups (BnGa, J\"uBo and SAID) summed over all multipoles up to $L=4$ of the reaction  $\gamma p \to p\pi^0$ is shown once without the polarization observables $E,G,H,P$ and $T$, and once with them. The green area shows the reduction of the variance over the mass range. The difference of the variances $\Delta\text{var}(1,2,3)$ is displayed as well. The dashed line shows the difference only for the $E_{0+}$ multipole. Taken from \cite{Beck:2016hcy}. \EPJA Right:  The red shaded areas give the spread of different fits that were derived from the older BnGa2011-01 and BnGa2011-02 solutions, while the blue shaded area gives the range of different fits of the BnGa2014 solution which included the latest polarization data $E,G,H,P$ and $T$. Modified from \cite{Hartmann:2015kpa}.}
\label{fig:interpretation:pwadiffpi0}
\end{figure}

\begin{figure}[htp]
\includegraphics[width=0.24\textwidth]{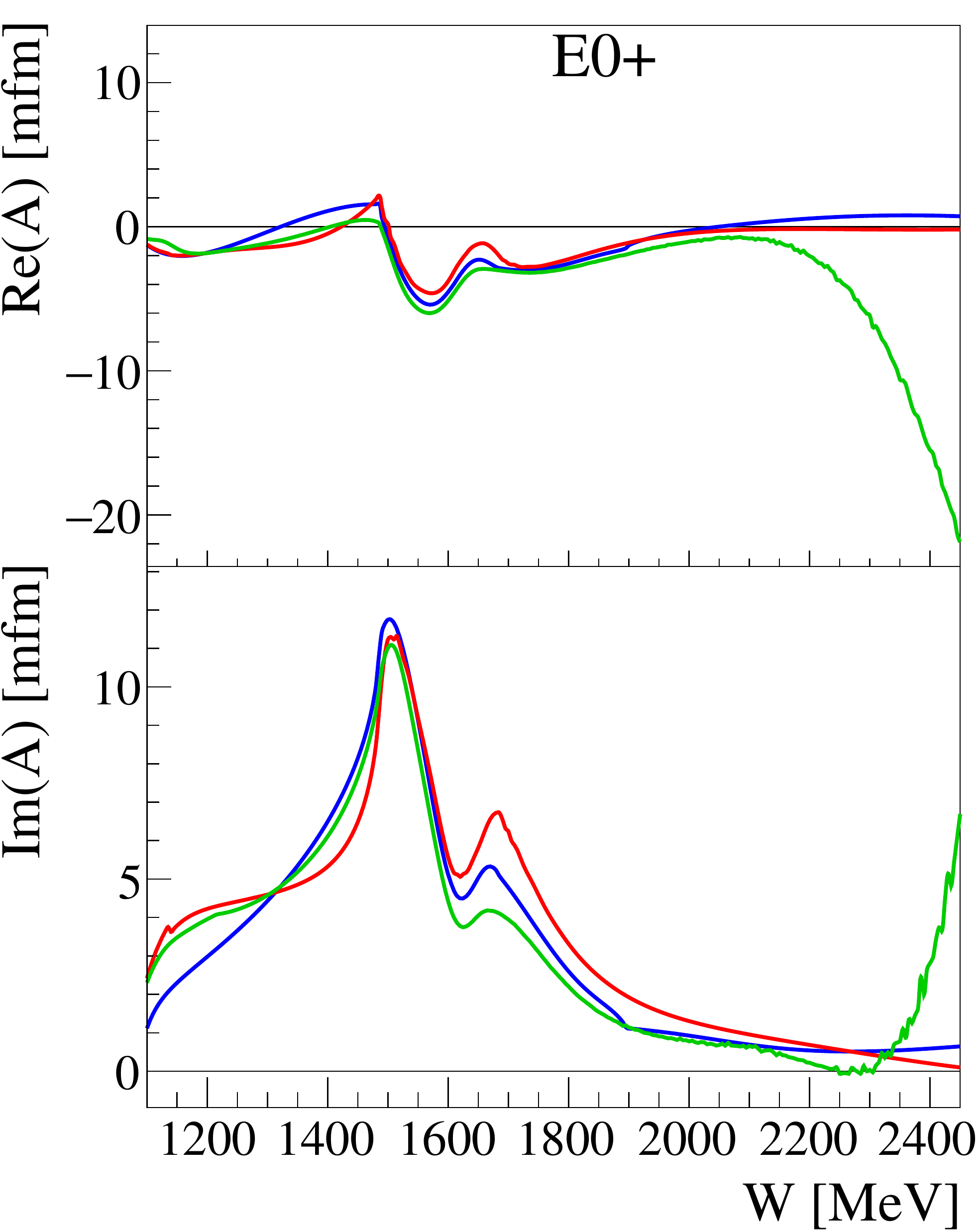}
\includegraphics[width=0.24\textwidth]{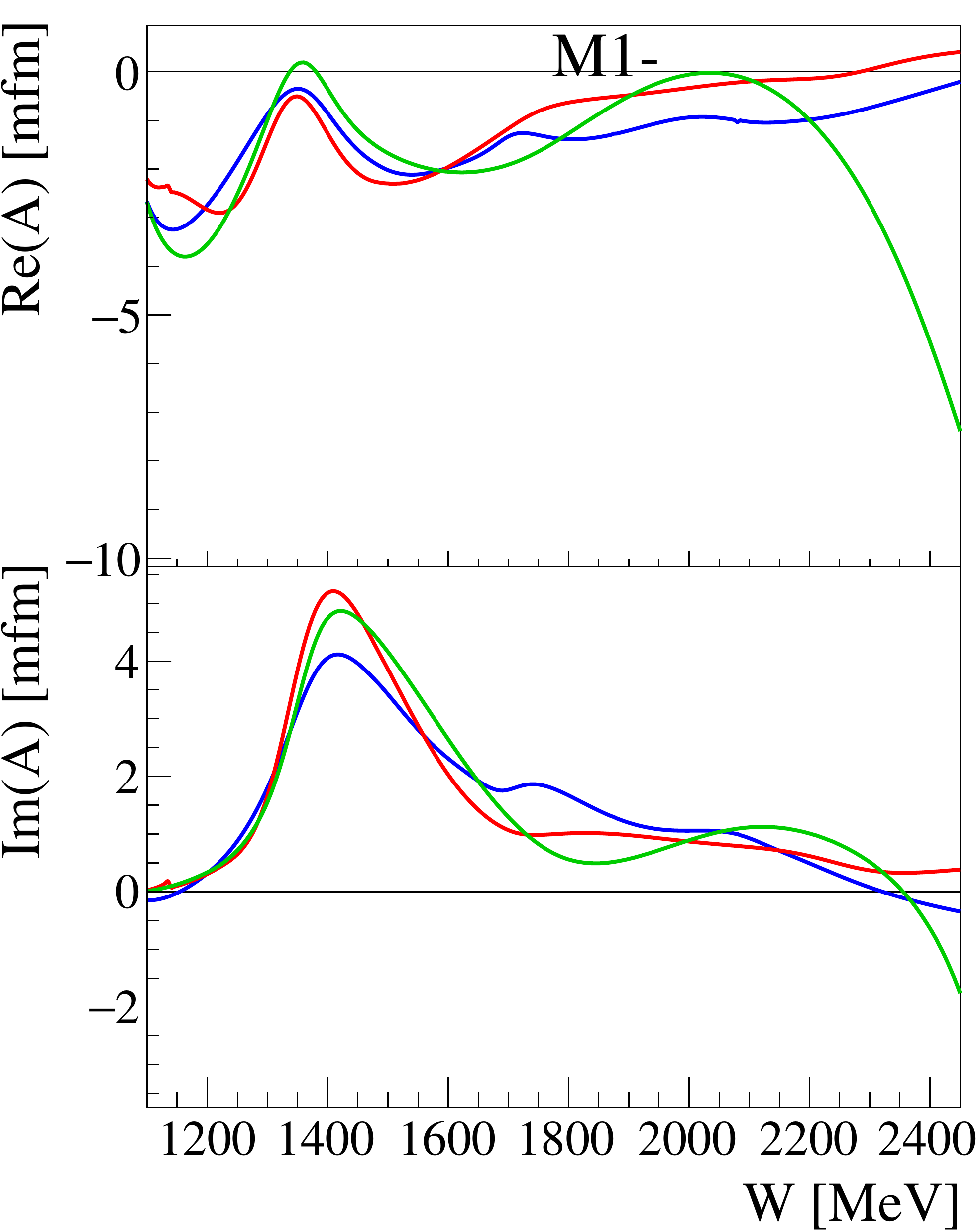}
\includegraphics[width=0.24\textwidth]{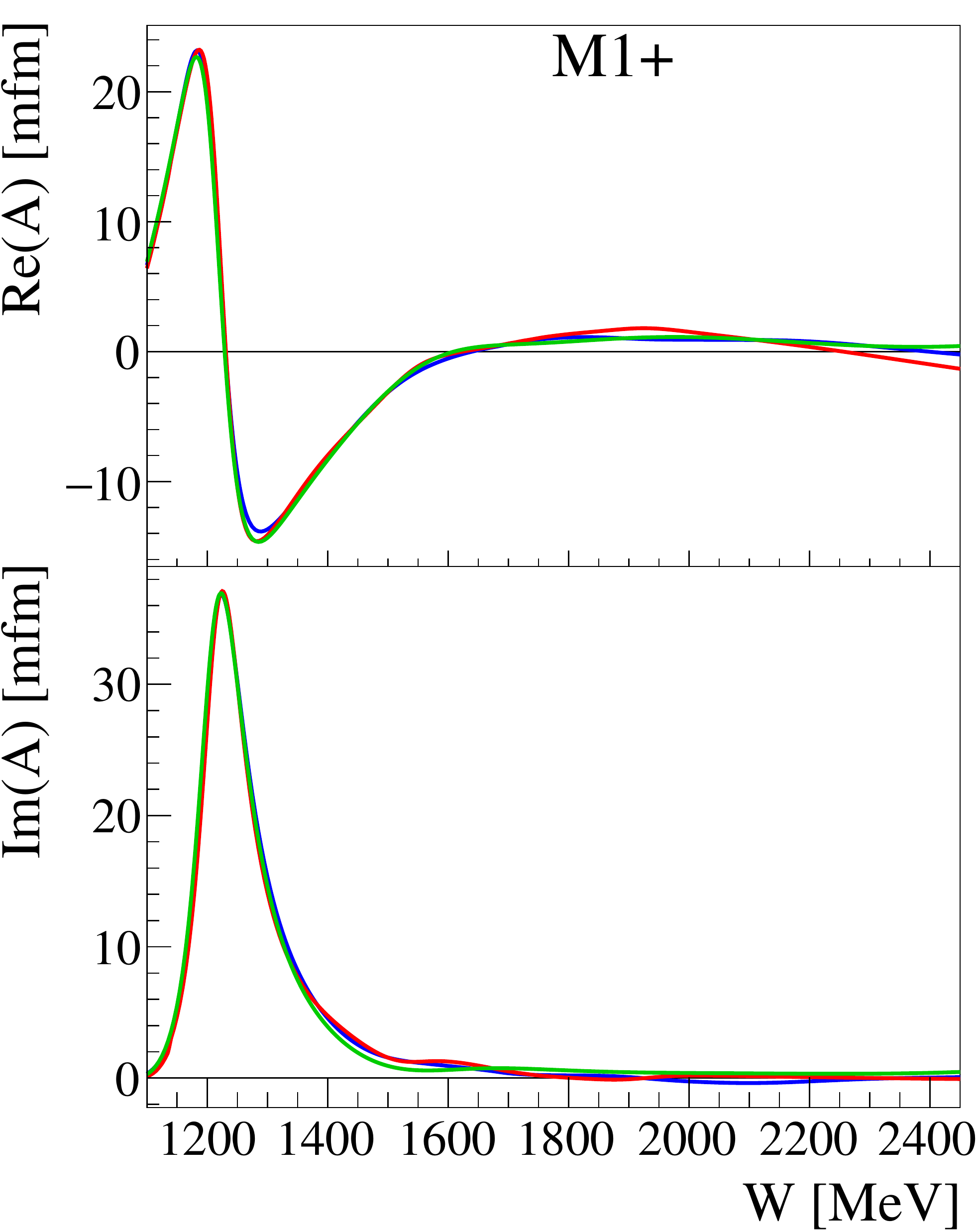}
\includegraphics[width=0.24\textwidth]{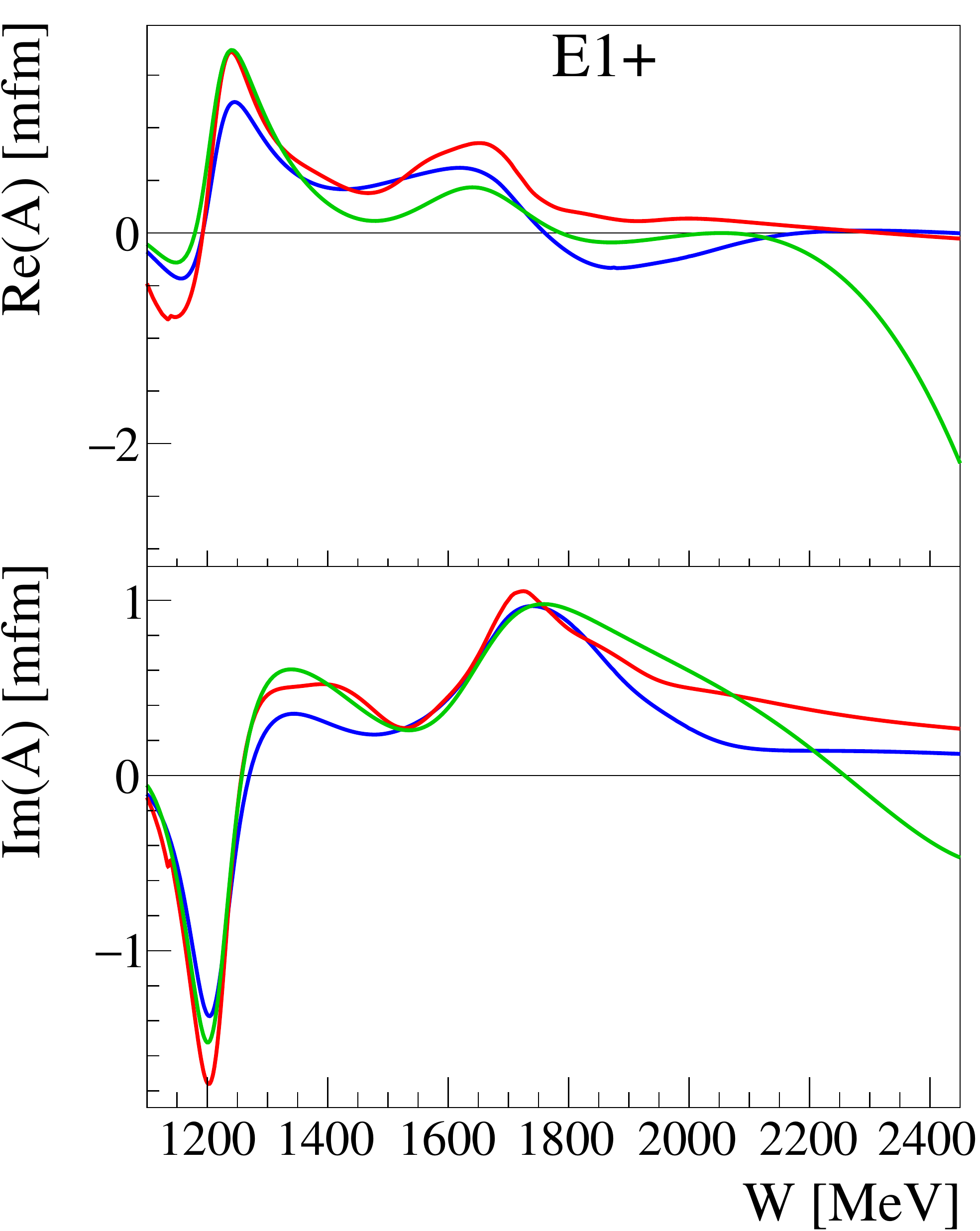}

\includegraphics[width=0.24\textwidth]{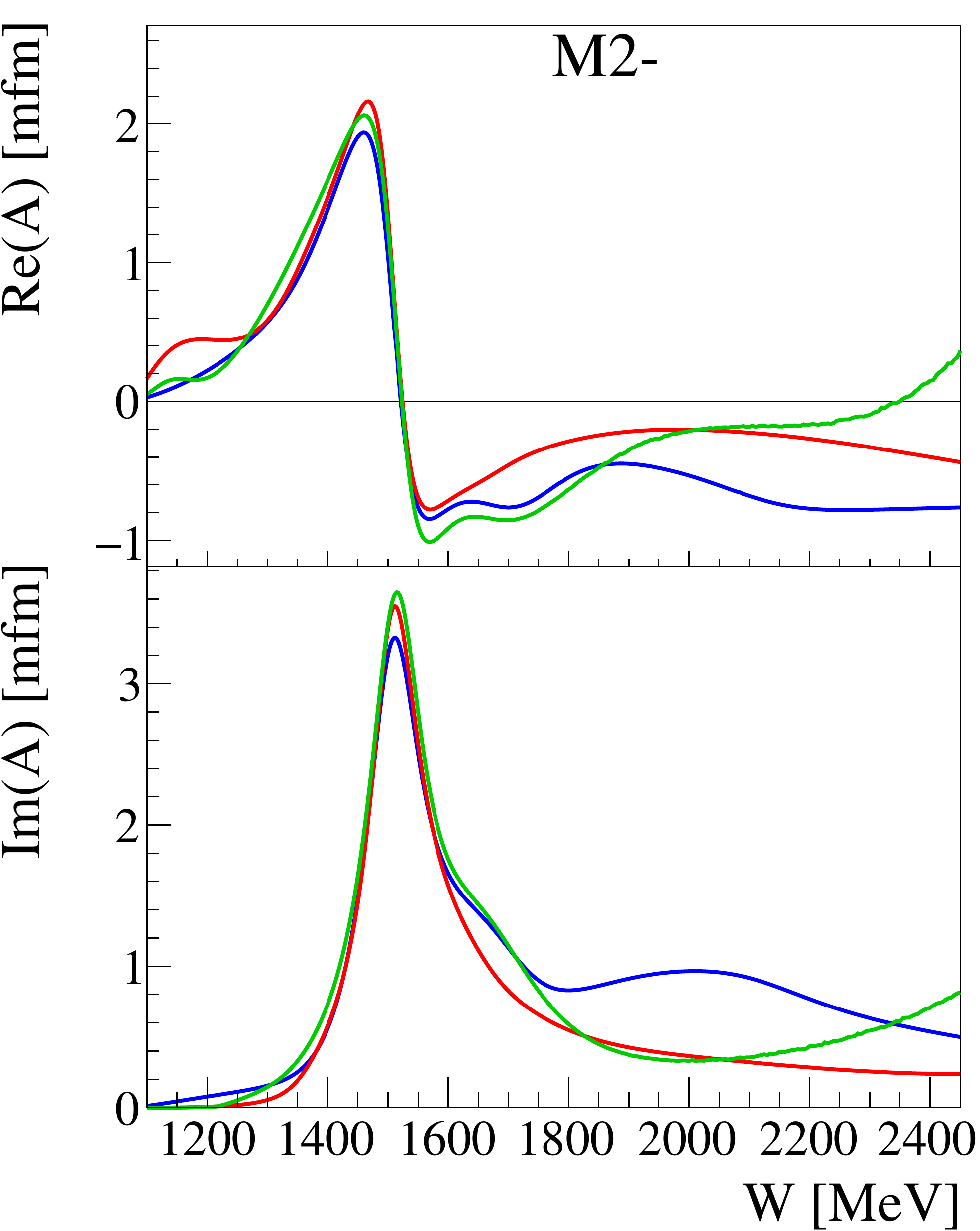}
\includegraphics[width=0.24\textwidth]{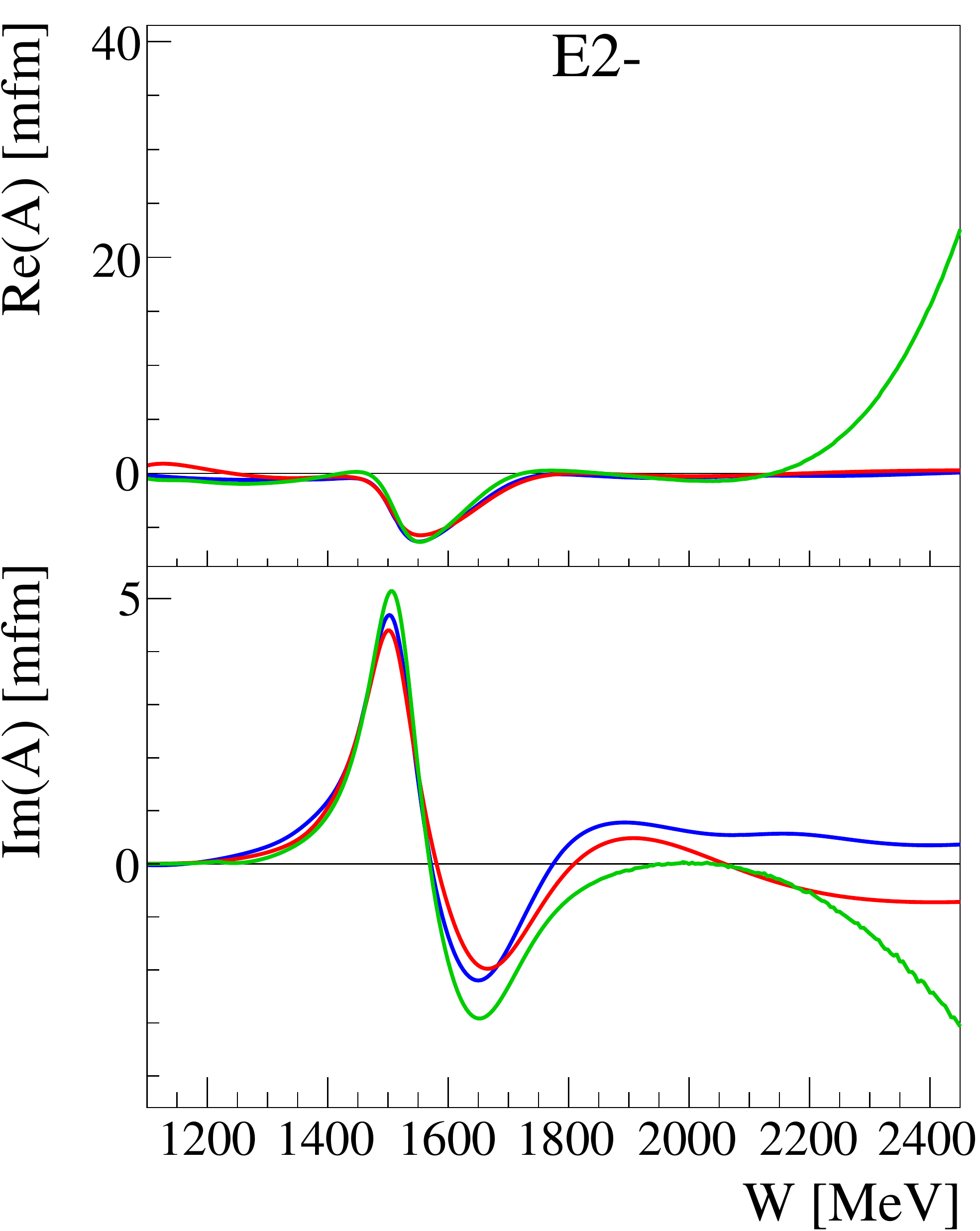}
\includegraphics[width=0.24\textwidth]{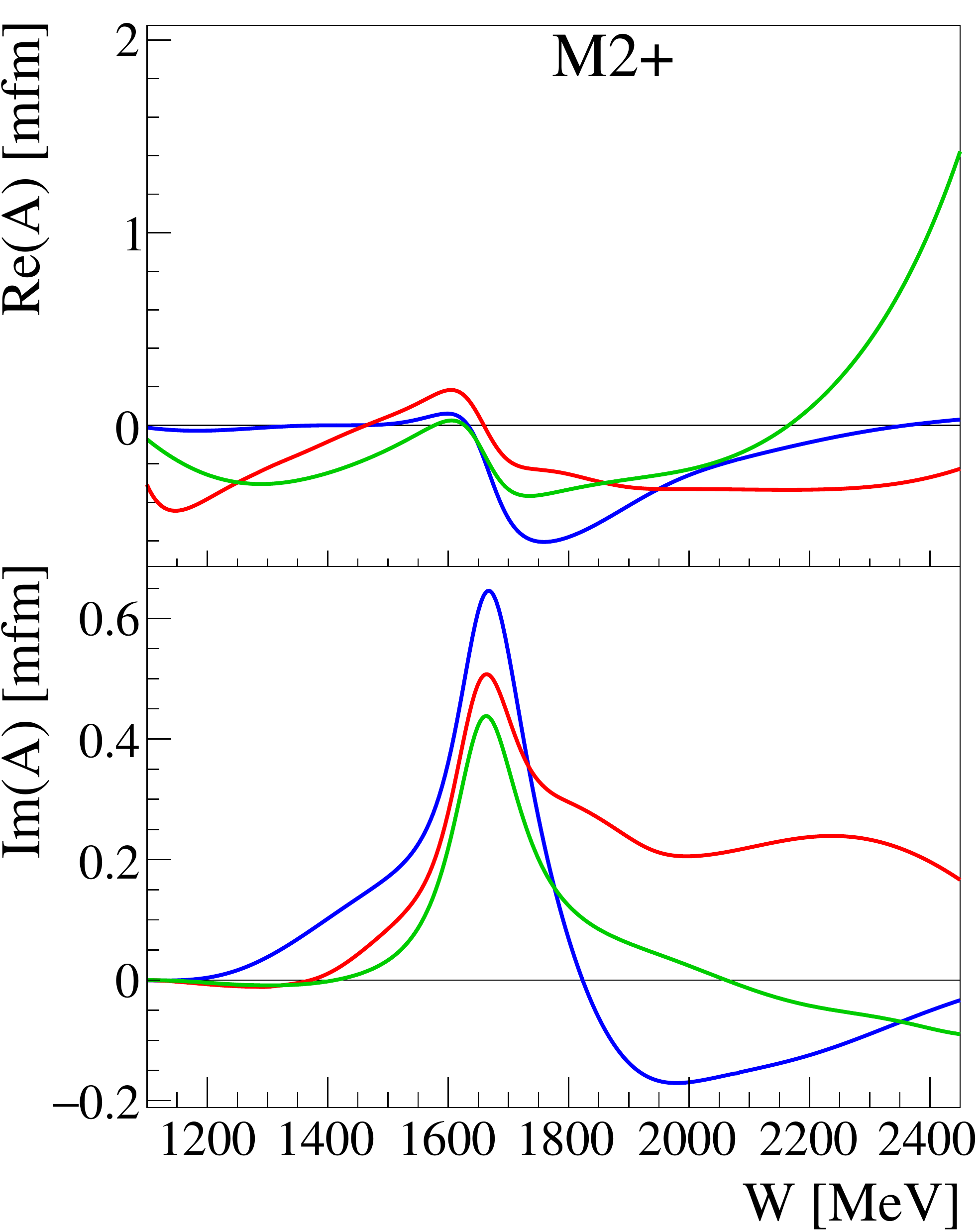}
\includegraphics[width=0.24\textwidth]{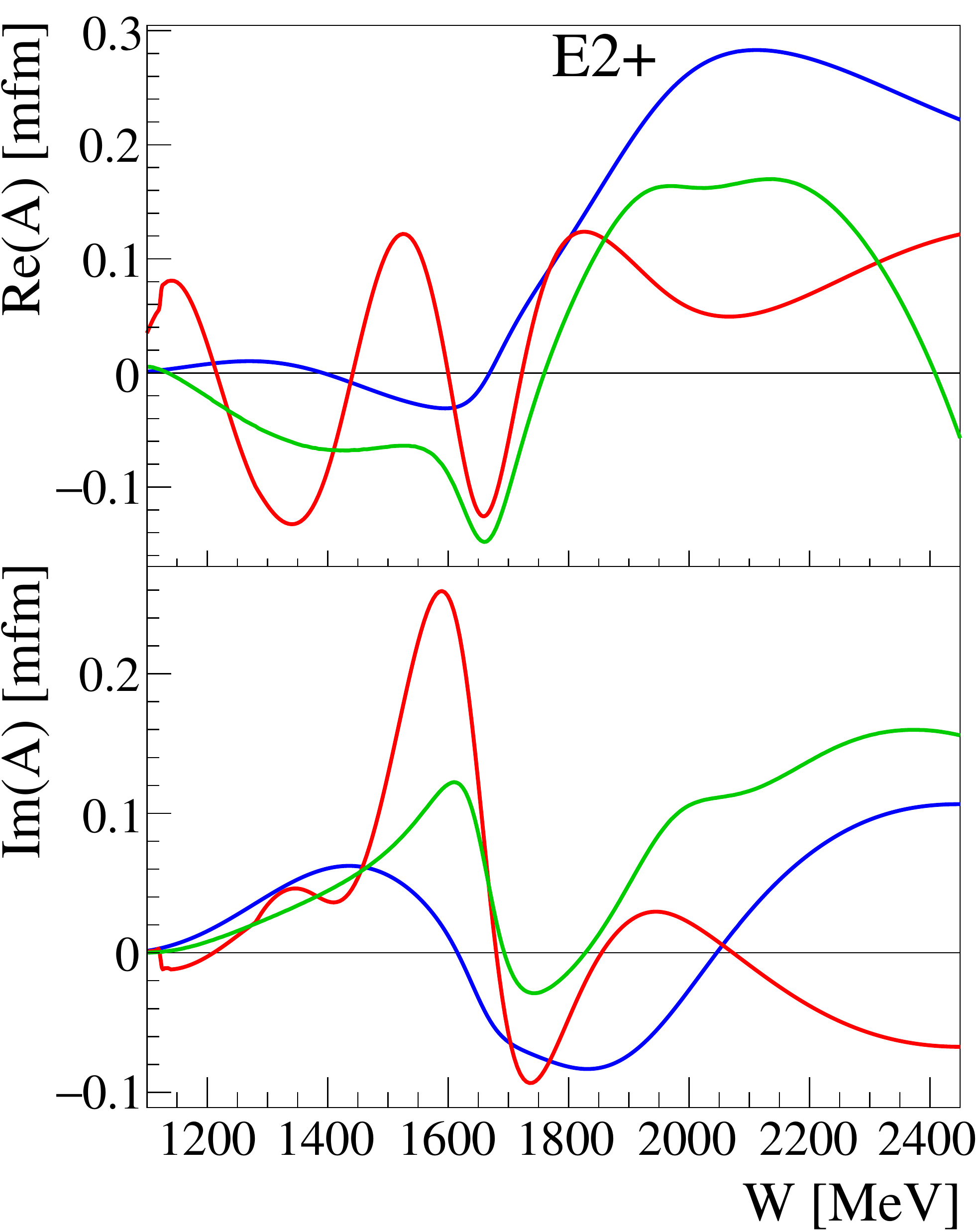}

\includegraphics[width=0.24\textwidth]{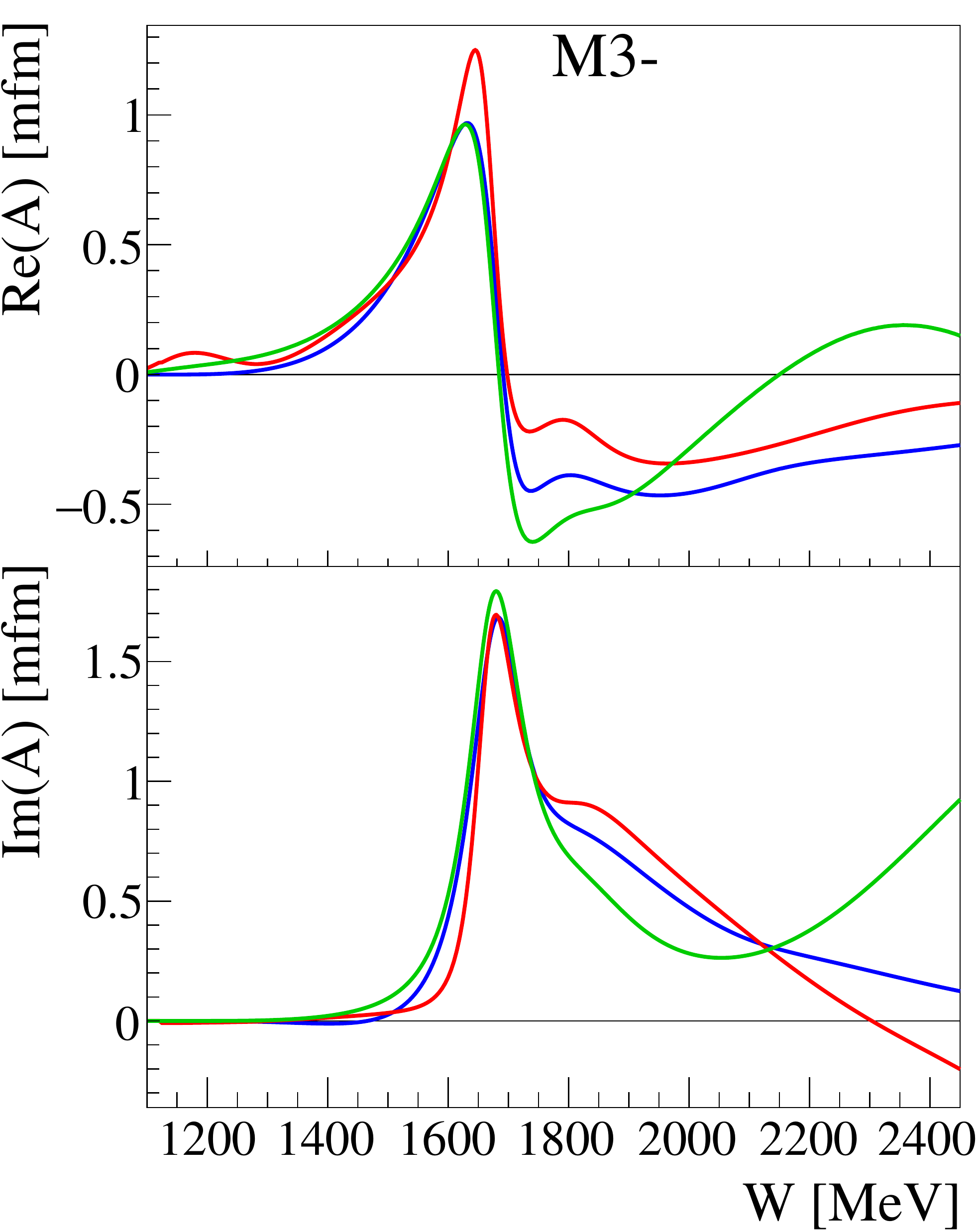}
\includegraphics[width=0.24\textwidth]{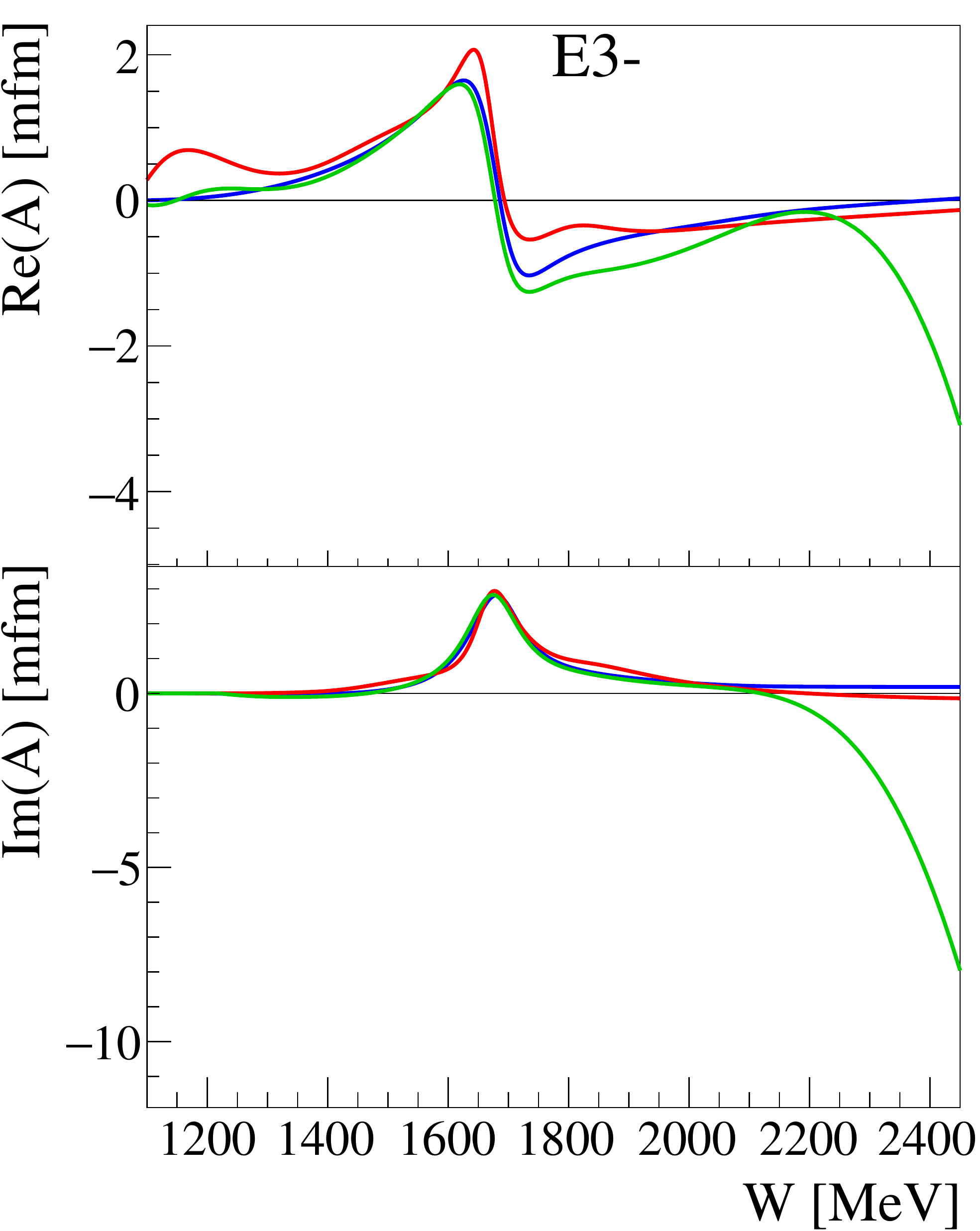}
\includegraphics[width=0.24\textwidth]{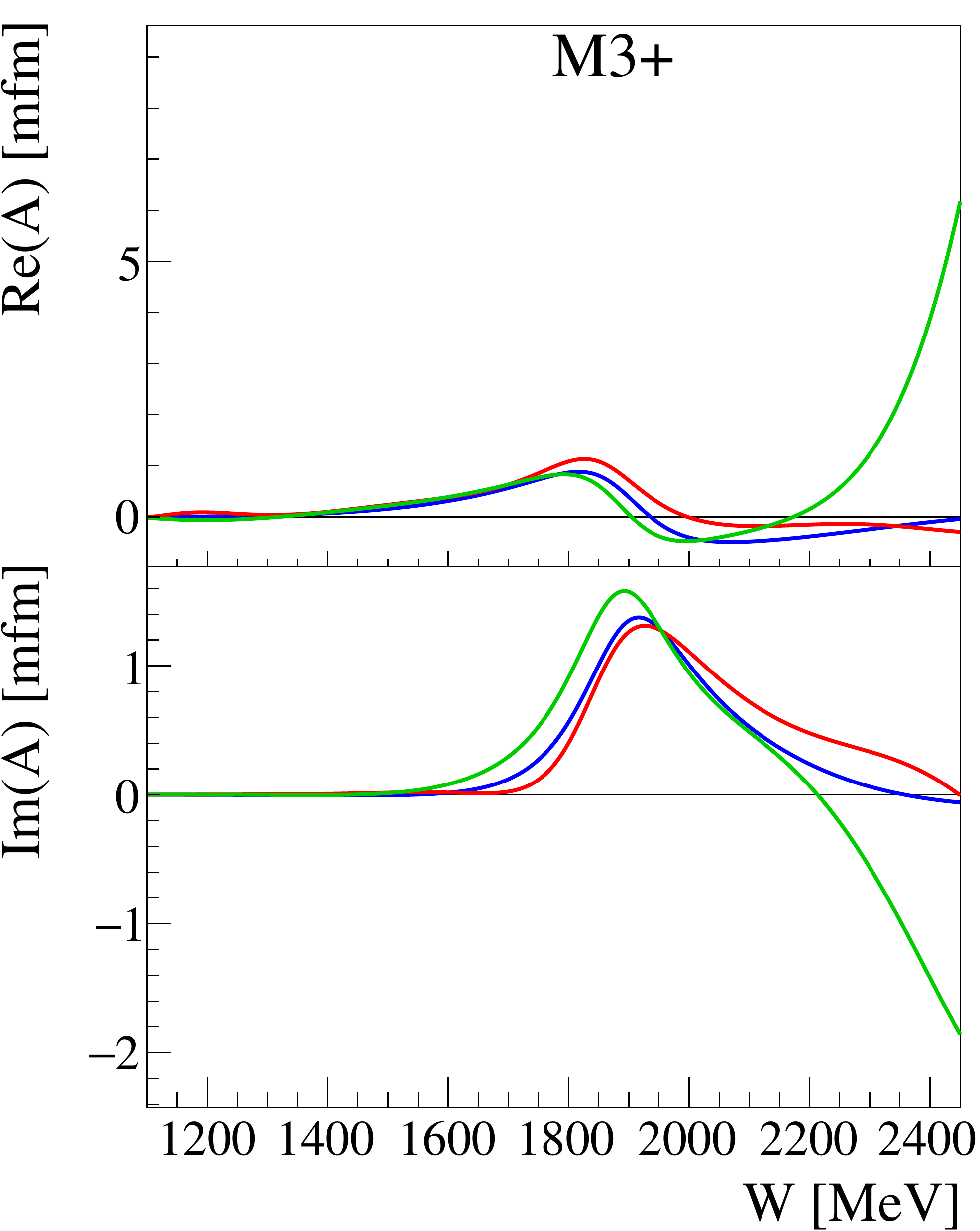}
\includegraphics[width=0.24\textwidth]{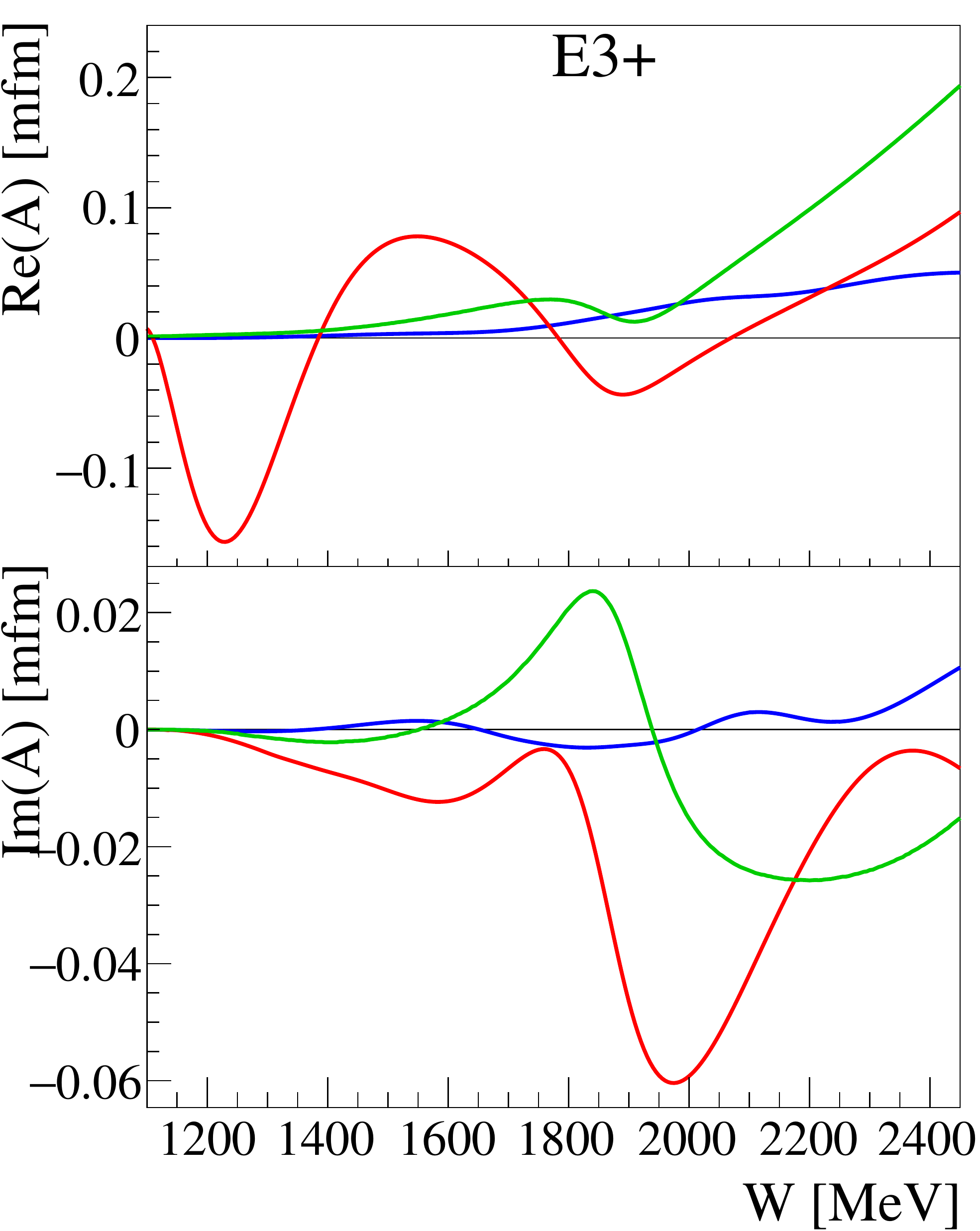}

\includegraphics[width=0.24\textwidth]{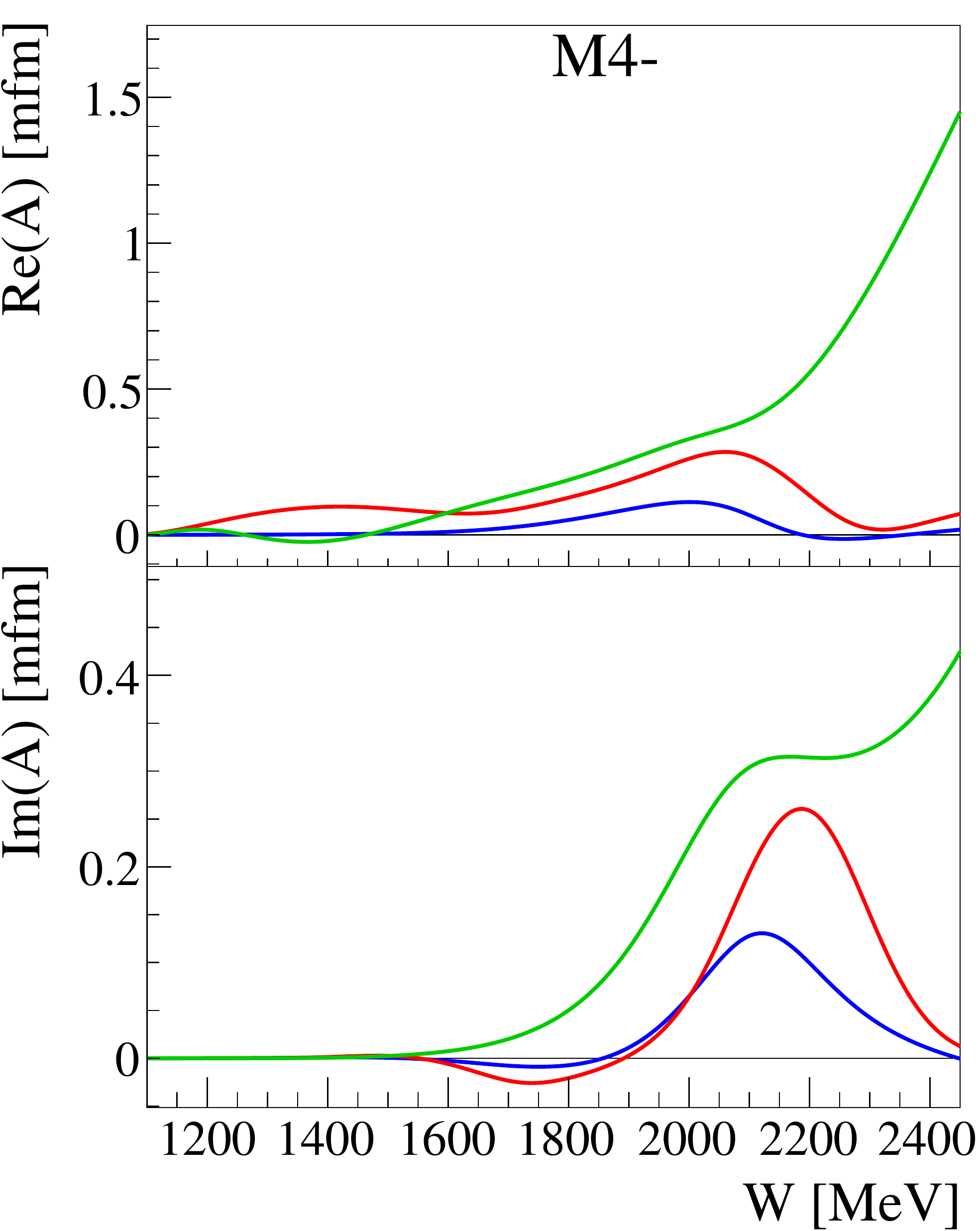}
\includegraphics[width=0.24\textwidth]{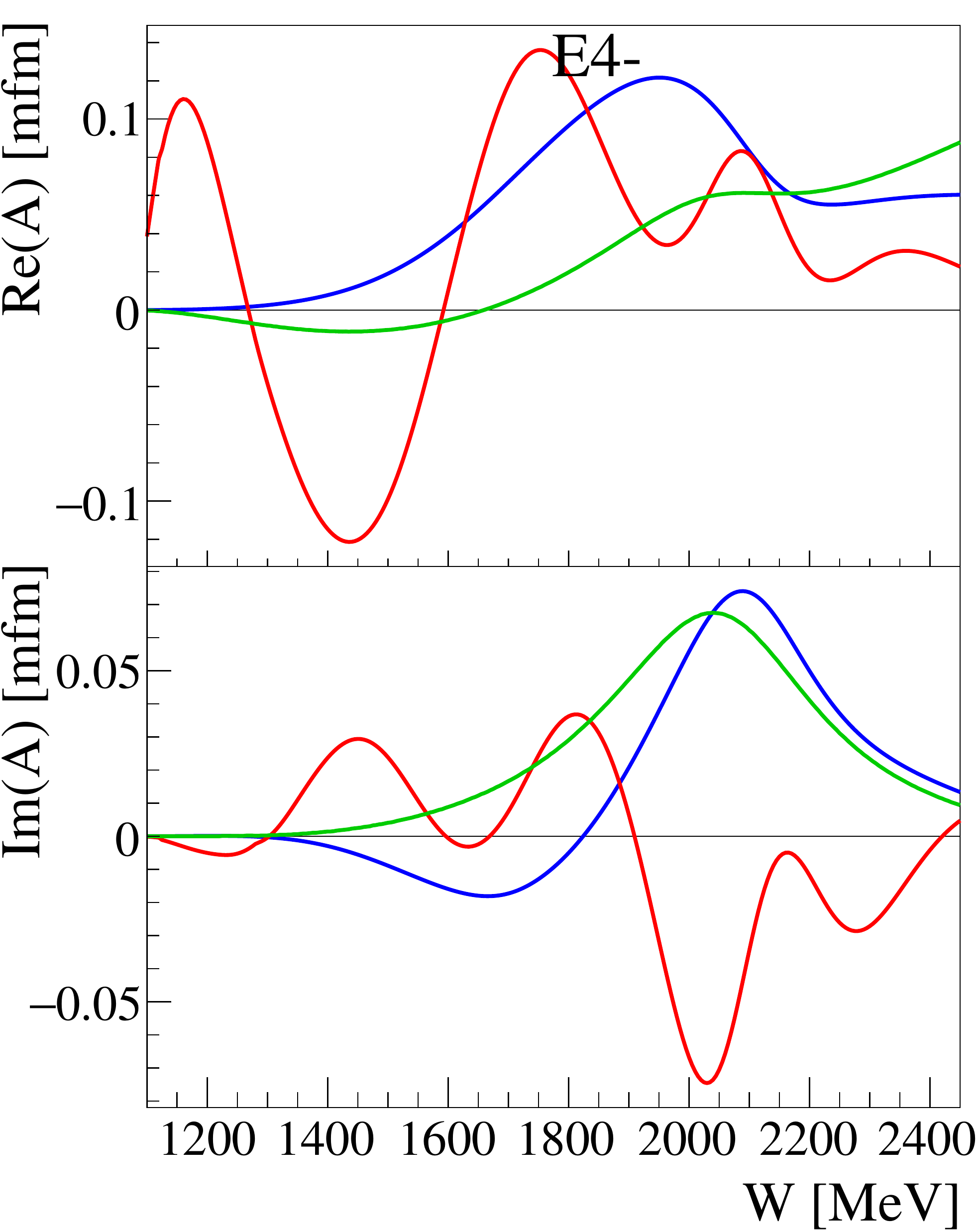}
\includegraphics[width=0.24\textwidth]{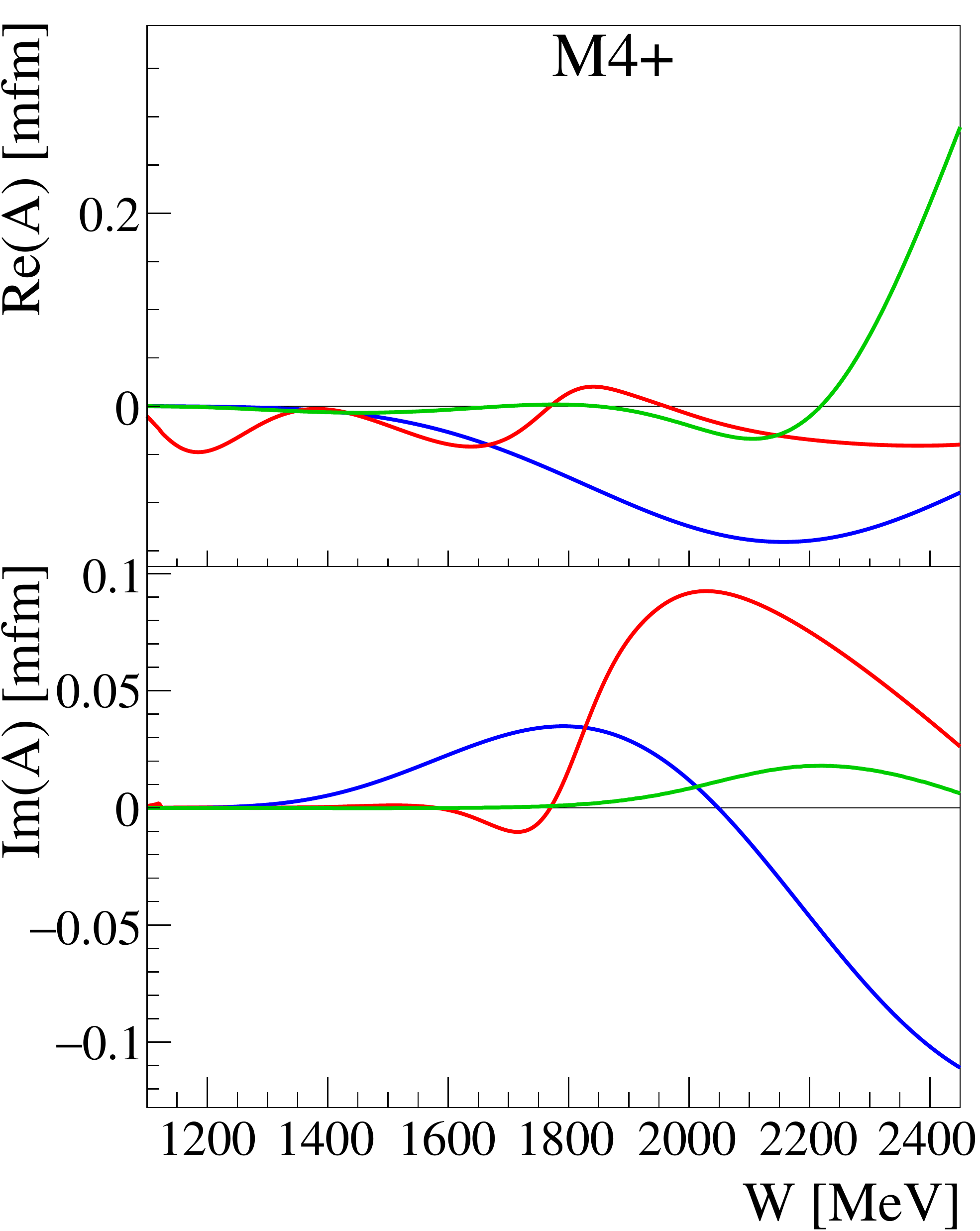}
\includegraphics[width=0.24\textwidth]{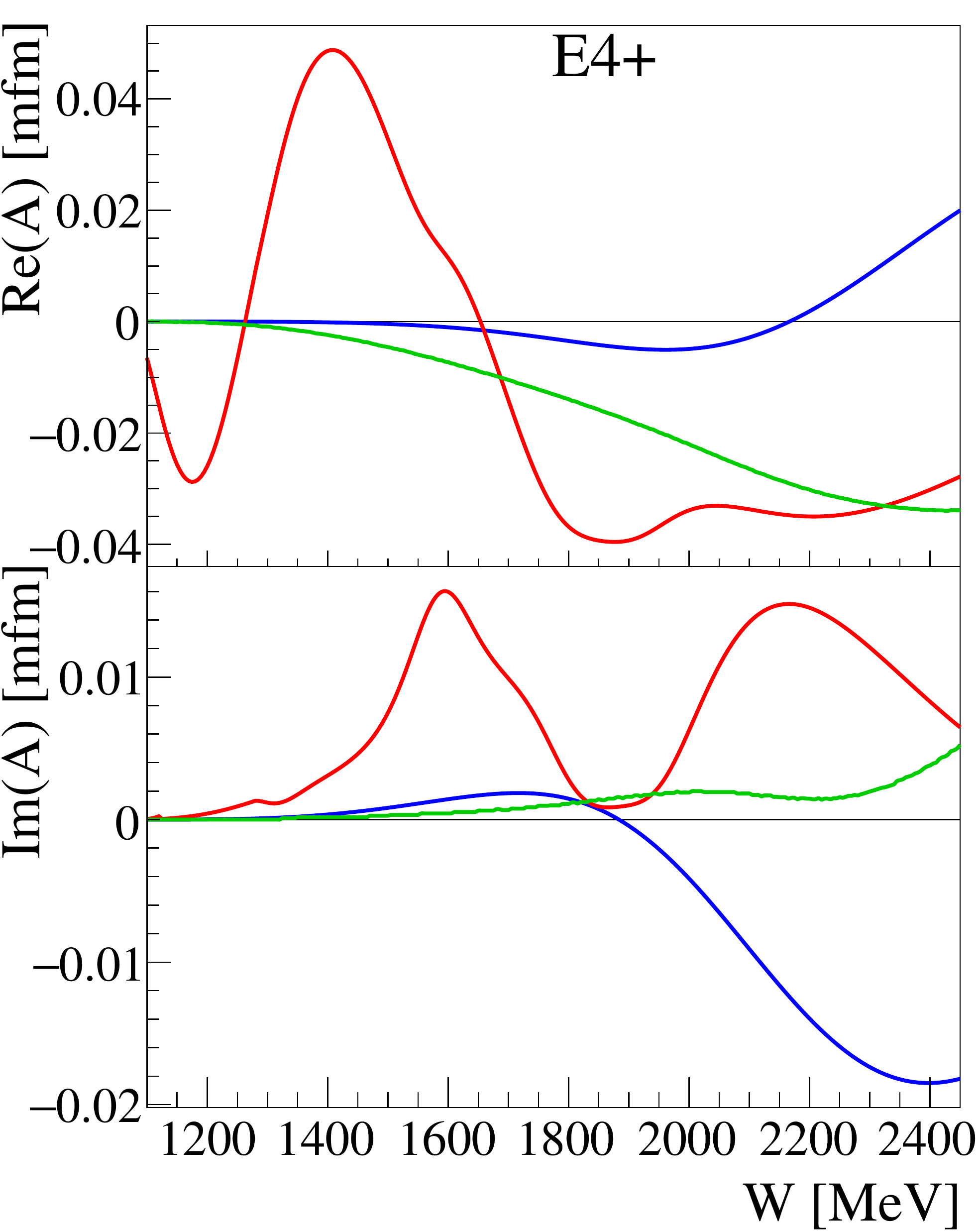}
\caption{The real and imaginary parts of the $p\pi^0$ multipoles. The colors are as follows: BnGa-2019 (blue) \cite{Muller:2019qxg}, J\"uBo17 (red) \cite{Ronchen:2018ury} and SAID-MA19 (green) \cite{Briscoe:2019cyo}.}
\label{fig:interpretation:pi0multipoles}
\end{figure}

The impact of the increased database on the partial waves is discussed in the following for the reaction $\gamma p \to p\pi^0$ as it has the largest existing database.
Fig.~\ref{fig:interpretation:pi0multipoles} shows the comparison of the latest PWA solutions (BnGa, J\"uBo and SAID) for the $\gamma p \to p\pi^0$ multipoles for up to $L=4$. Below $W=1800$ MeV, there is a good agreement between the three PWA, while the higher mass range shows larger discrepancies. Here, only the extracted multipoles are shown without their errors, since most analyses do not publish their error bands. \newline 
\noindent Anisovich et al.~\cite{Beck:2016hcy} investigated the impact of a chosen set of measured polarization observables $E,G,H,P$ and $T$ for the consistency between the different PWA groups (BnGa, J\"uBo and SAID) regarding the extracted $\gamma p \to p\pi^0$ partial waves. A comparison of the multipoles before and after including the chosen polarization observables is shown  for the $E_{0+}$ and $M_{1-}$ multipoles as an example in Fig.~\ref{fig:M:Npi_interpretation_anisovich_multipoles}. The variance of all three PWAs summed over all multipoles up to $L=4$ is visualized in Fig.~\ref{fig:interpretation:pwadiffpi0} on the left. It is evident that the variance reduced significantly when taking the set of polarization observables into account. Furthermore, it was found that large parts of the inconsistency between the PWA group solutions can be attributed to the $E_{0+}$ multipole, as can be seen in Fig.~\ref{fig:M:Npi_interpretation_anisovich_multipoles} and Fig.~\ref{fig:interpretation:pwadiffpi0}. 
Looking at the different PWA group solutions individually, Hartmann et al.~\cite{Hartmann:2015kpa} also demonstrated that the latest measured polarization data in photoproduction reactions is crucial for the reduction of the spread of different possible fit solutions within one PWA and thus, help to constrain the partial waves. This is shown in Fig.~\ref{fig:interpretation:pwadiffpi0} on the right exemplary for the BnGa PWA and for the imaginary part of the $M_{1-}$ multipole \cite{Hartmann:2015kpa}.

It is worth analyzing the extent to which a \textit{complete experiment}~(sections~\ref{sec:formalisms:CompleteExperiments}) has been obtained in the many new photoproduction measurements over the years. Most promising are here surely the single-meson photoproduction measurements~(section~\ref{sec:measurements:SingleMesonPhotoproduction}). The point of view of the complete experiment is useful due to the fact that it is completely model-independent. Therefore, one can make a heuristic argument that once a truly model-independent complete experiment is obtained, all the energy-dependent PWA-models discussed in chapter~\ref{sec:PWA} should converge to the same partial-wave solution as well.

The complete-experiment problem can be analyzed both for the problem of the extraction of the full spin-amplitudes (section~\ref{sec:formalisms:SpinAmplitudes}), but also for the TPWA (or single-energy fits, cf. section~\ref{sec:formalisms:PWExpansion}). The extraction-problem for the full amplitudes cannot be solved uniquely with the datasets measured for any reaction so far. Even for the final states with the most measured polarization observables, i.e. the $\pi^{0} p$, $\eta p$ and $K^{+} \Lambda$ final states~(cf. Tables~\ref{tab:measSinglePion1},~\ref{tab:measSingleEta} and~\ref{tab:measStrangeness}), the amplitude-extraction cannot be unique since it requires polarization observables to be measured from at least three different classes, cf.~section~\ref{sec:formalisms:CompleteExperiments}. 

For the TPWA, the situation is different: here it has shown rigorously that for the idealized situation of a mathematically 'exact' TPWA, five carefully selected observables can constitute a complete experiment and that they have to be selected from at least two different classes (section~\ref{sec:formalisms:CompleteExperiments}). This criterion is indeed fulfilled by measurements in several single-meson photoproduction reactions: for the $\pi^{0} p$ and $\eta p$ final states (Tables~\ref{tab:measSinglePion1} and~\ref{tab:measSingleEta}), the group~$\mathcal{S}$ and $\mathcal{BT}$-observables have all been measured, with good kinematical overlap between most datasets. For the $K^{+} \Lambda$ final state (Table~\ref{tab:measStrangeness}), the group $\mathcal{S}$ and $\mathcal{BR}$ observables have been measured and published. Several further observables from the $\mathcal{BT}$ class are expected to be published by the CLAS collaboration in the next few years. The benchmark reaction for the complete experiment in the TPWA is surely given by the $\pi^{0} p$ final state~(Table~\ref{tab:measSinglePion1}): the kinematical overlap between datasets, but even more importantly the statistical precision of individual measurements, is best here. For the $\pi^{0} p$-reaction, a fully model-independent TPWA was performed with all group $\mathcal{S}$ plus three $\mathcal{BT}$ observables, using $\ell_{\mathrm{max}} = 3$, in reference~\cite{YannickPhD}. The relatively narrow energy region where all relevant datasets overlap can also be seen on the left in Figure~\ref{fig:interpretation:pwadiffpi0}, in the range $W = 1450 \ldots 1550$ MeV. Here, the fully model-independent TPWA with $\ell_{\mathrm{max}} = 3$ allowed for a well-separated global minimum in $2$ from $8$ energy bins. In all remaining bins, the solution pool was polluted by local minima. This problem was remedied by fixing the $F$-waves to a PWA-model, which resulted in a unique solution. The reason for this behavior is given by statistical as well as systematic measurement uncertainties. Similar (but not exactly equal) results were found in an investigation of $K^{+} \Lambda$ photoproduction by Sandorfi and collaborators~\cite{Sandorfi:2010uv}. 

For the remaining reactions of vector- and multi-meson photoproduction, a complete experiment is not yet even close neither for the full amplitude-extraction problem, nor for the TPWA.

While the idealized situation of a complete experiment is thus still not fulfilled, the community is moving in the right direction with further polarization measurements.

\subsection{Impact on the Listing of Baryon Resonances}

To investigate the influence of the strongly increased data basis since 2000, as discussed in Sec.~\ref{sec:measurements}, a comparison of the states listed in the \textit{Listing of Baryon Resonances} by the Particle Data Group\footnote{https://pdglive.lbl.gov} (PDG)  can be useful. In Tabs.~\ref{tab:int:rev-n-resonances} and \ref{tab:int:rev-delta-resonances}, the recent listing of the $N^*$ and $\Delta$ baryons from \cite{Zyla:2020zbs} are compared to the values as in 2002 \cite{ParticleDataGroup:2002ivw}. A comparison of the full table of baryon resonances, including hybrids, heavy quark states and pentaquarks can be found in the appendix, Tab.~\ref{tab:int:baryon_table_pdg}.



For the $N^*$ and $\Delta$ baryons composed of light quarks, Tabs.~\ref{tab:int:rev-n-resonances} and \ref{tab:int:rev-delta-resonances} show a significant increase in the amount of $N^*$ resonances: here 9 new states were added within the last years. Additionally, an increase in the star rating can be observed for several states, like for example the $N(2100)$ and the $\Delta(2200)$, whose significances were increased from * (``Evidence is poor'') to *** (``Evidence ranges from likely to certain''). To investigate the responsible measurements, a survey of the contributing final states can be found in Tab.~\ref{tab:int:rev-n-resonances} for the $N^*$ resonances and in Tab.~\ref{tab:int:rev-delta-resonances} for the $\Delta$ resonances. As already discussed in Sec.~\ref{sec:measurements}, the amount of photoproduction data has strongly increased in the last years and is now dominating the partial wave analysis fits. As a result, most of the newly found resonant states have a significant evidence from photoproduction and are mostly more dominant there than for pion scattering. For example the four-star resonance $N(1895)$, which has only been rated with one star (``Evidence is poor'') from pion scattering. In contrast, this resonance has strong contributions from the $\eta$ and $\eta'$ final states, where it is rated with four stars (``Evidence is certain''). Further insight into Tab.~\ref{tab:int:rev-n-resonances} and Tab.~\ref{tab:int:rev-delta-resonances} show the addition of new final states. Since 2002, data of the final states $N\sigma$, $N\omega$ and $N\eta'$ were added, leading to an increased influence on the significance of the resonances. For these reactions, the data basis still is insufficient compared to other final states, but new measurements at different facilities tackle this issue.

\begin{table}[tb]	
	\resizebox{\textwidth}{!}{%
\begin{tabular}{l|l|l|l||llllllllll}
	Particle & $J^P$ & overall& PWA & $N\gamma$ & $N\pi$ & $\Delta \pi$ & $N\sigma$ & $N\eta$ & $\Lambda K$ & $\Sigma K$ & $N\rho$ & $N\omega$ & $N\eta'$\\\hline
	$N$ & $1/2^+$ & ****& & & & & & & & & & & \\
	$N(1440)$ & $1/2^+$ & ****& $\circ$ $\diamond_{g}$ $\star$ $\triangleright$ & ***{\color{red}*} & **** & ***{\color{red}*} & {\color{red}***} & {\color{red}-} & & & {\color{red}-} & & \\
	$N(1520)$ & $3/2^-$ & ****& $\circ$ $\diamond$ $\star$ $\triangleright$ & **** & **** & **** & {\color{red}**} & *{\color{red}***} & & & {\color{red}- - - -}& & \\
	$N(1535)$ & $1/2^-$ & ****& $\circ$ $\diamond$ $\star$ $\triangleright$ & ***{\color{red}*} & **** & *{\color{red}**} & {\color{red}*} & **** & & & {\color{red}- -}& & \\
	$N(1650)$ & $1/2^-$ & ****& $\circ$ $\diamond$ $\star$ $\triangleright$ & ***{\color{red}*} & **** & *** & {\color{red}*} & *{\color{red}***} & *{\color{red}- -} & {\color{red}- -}&{\color{red}- -} & & \\
	$N(1675)$ & $5/2^-$ & ****& $\circ$ $\diamond$ $\star$ $\triangleright$ & **** & **** & **** & {\color{red}***} & * & * & {\color{red}*} &{\color{red}-} & & \\
	$N(1680)$ & $5/2^+$ & ****& $\circ$ $\diamond$ $\star$ $\triangleright$ & **** & **** & **** & {\color{red}***} & {\color{red}*} & {\color{red}*} & {\color{red}*} &{\color{red}- - - - } & & \\
	$N(1700)$ & $3/2^-$ & ***& $\circ$ $\triangleright$ & ** & *** & **{\color{red}*} & {\color{red}*} & * &{\color{red}- -} &{\color{red}-} &{\color{red}-} & & \\
	$N(1710)$ & $1/2^+$ & ***{\color{red}*}&$\circ$ $\diamond$ $\triangleright$& ***{\color{red}*} & ***{\color{red}*} & *{\color{red}-} & & **{\color{red}*} & ** & * & * & {\color{red}*} & \\
	$N(1720)$ & $3/2^+$ & ****& $\circ$ $\diamond$ $\star$ $\triangleright$ & **{\color{red}**} & ****  & *{\color{red}**} & {\color{red}*} & * & **{\color{red}**} & * & *{\color{red}-} &{\color{red} *} & \\
	{\color{red}$N(1860)$} & {\color{red}$5/2^+$} & {\color{red}**}& $\triangleright$& {\color{red}*} & {\color{red}**} & & {\color{red}*} & {\color{red}*} & & & & & \\
	{\color{red}$N(1875)$} & {\color{red}$3/2^-$} & {\color{red}***}& $\circ$ $\triangleright$& {\color{red}**} & {\color{red}**} & {\color{red}*} & {\color{red}**} & {\color{red}*} & {\color{red}*} & {\color{red}*} & {\color{red}*} & {\color{red}*} & \\
	{\color{red}$N(1880)$} & {\color{red}$1/2^+$} & {\color{red}***}&$\circ$ $\triangleright$ & {\color{red}**} & {\color{red}*} & {\color{red}**} & {\color{red}*} & {\color{red}*} & {\color{red}**} & {\color{red}**} & & {\color{red}**} & \\
	{\color{red}$N(1895)$} & {\color{red}$1/2^-$} & {\color{red}****}& $\circ$ $\triangleright$& {\color{red}****} & {\color{red}*} & {\color{red}*} & {\color{red}*} & {\color{red}****} & {\color{red}**} & {\color{red}**} & {\color{red}*} & {\color{red}*} &{\color{red} ****} \\
	$N(1900)$ & $3/2^+$ & **{\color{red}**}&$\circ$ $\diamond$ $\triangleright$ & {\color{red}****} & ** & {\color{red}**} & {\color{red}*} & {\color{red}*} & {\color{red}**} & {\color{red}**} & {\color{red}-} & {\color{red}*} &{\color{red} **} \\
	$N(1990)$ & $7/2^+$ & **& $\circ$ $\diamond$ $\triangleright$ & *{\color{red}*} & ** & & & * & * & * & & & \\
	$N(2000)$ & $5/2^+$ & **& $\circ$ $\star$ & {\color{red}**}& *{\color{red}-} & *{\color{red}*} & {\color{red}*} & * & {\color{red}-} &{\color{red}-} &{\color{red}- -} & {\color{red}*} & \\
	{\color{red}$N(2040)$} & {\color{red}$3/2^+$} & {\color{red}*}& $\triangleright$& & {\color{red}*} & & & & & & & & \\
	{\color{red}$N(2060)$} & {\color{red}$5/2^-$} & {\color{red}***}& $\circ$ $\diamond_{g}$ $\triangleright$ & {\color{red}***} & {\color{red}**} & {\color{red}*} & {\color{red}*} & {\color{red}*} & {\color{red}*} & {\color{red}*} & {\color{red}*} & {\color{red}*} & \\
	$N(2100)$ & $1/2^+$ & *{\color{red}**}& $\circ$ $\triangleright$ & {\color{red}**} & *{\color{red}**} & {\color{red}**} & {\color{red}**} & * & {\color{red}*} & & {\color{red}*} & {\color{red}*} & {\color{red}**} \\
	{\color{red}$N(2120)$} & {\color{red}$3/2^-$} & {\color{red}***}& $\circ$ $\triangleright$ & {\color{red}***} & {\color{red}**} & {\color{red}**} & {\color{red}**} & & {\color{red}**} & {\color{red}*} & & {\color{red}*} & {\color{red}*}\\
	$N(2190)$ & $7/2^-$ & ****& $\circ$ $\diamond$ $\star$ $\triangleright$ & *{\color{red}***} & **** & {\color{red}****} & {\color{red}**} & * & *{\color{red}*} & * & * & {\color{red}*} & \\
	$N(2220)$ & $9/2^+$ & ****& $\circ$ $\diamond$ $\star$ & {\color{red}**} & **** & & & * & {\color{red}*} & {\color{red}*} & & & \\
	$N(2250)$ & $9/2^-$ & ****& $\circ$ $\diamond$ $\star$ $\triangleright$ & {\color{red}**} & **** & & & * & {\color{red}*} & {\color{red}*} & & & \\
	{\color{red}$N(2300)$} & {\color{red}$1/2^+$} & {\color{red}**}& & & {\color{red}**} & & & & & & & & \\
	{\color{red}$N(2570)$} & {\color{red}$5/2^-$} & {\color{red}**}& & & {\color{red}**} & & & & & & & & \\
	$N(2600)$ & $11/2^-$ & *** & $\star$ & & *** & & & & & & & & \\
	$N(2700)$ & $13/2^+$ & ** & & & ** & & & & & & & & 
\end{tabular}
}
\caption{Overview of the $N^\ast$ resonances and their decay channels as of 2020 \cite{Zyla:2020zbs}. In red are the changes to the edition of 2002 highlighted \cite{ParticleDataGroup:2002ivw}. The symbols in the PWA column indicate which resonance is seen by the different PWA (circle: BnGa-2019 \cite{Muller:2019qxg}, diamond: J\"uBo-2017 \cite{Ronchen:2018ury}, star: SAID-MA19 \cite{Briscoe:2019cyo} and right triangle: KSU model \cite{Hunt:2018wqz}). The subscript $g$ indicates dynamically generated resonances as reported by the according PWA group. However, we do not list resonances here, that are seen by any PWA group, but are not identified with resonance states listed in the PDG.}
\label{tab:int:rev-n-resonances}
\end{table}

\begin{table}[ht]
\begin{tabular}{l|l|l|l||llllll}
	Particle & $J^P$ & overall& PWA & $N\gamma$ & $N\pi$ & $\Delta \pi$ & $\Sigma K$ & $N\rho$ & $\Delta \eta$\\\hline
	$\Delta(1232)$ & $3/2^+$ & **** & $\circ$ $\diamond$ $\star$ $\triangleright$ & **** & **** & & & & \\
	$\Delta(1600)$ & $3/2^+$ & ***{\color{red}*}& $\circ$ $\diamond_{g}$ $\triangleright$ & **{\color{red}**} & *** & ***{\color{red}*} & & {\color{red}-} & \\
	$\Delta(1620)$ & $1/2^-$ & ****& $\circ$ $\diamond$ $\star$ $\triangleright$ & ***{\color{red}*} & **** & **** & & {\color{red}- - - -} & \\
	$\Delta(1700)$ & $3/2^-$ & ****& $\circ$ $\diamond$ $\star$ $\triangleright$ & ***{\color{red}*} & **** & ***{\color{red}*} & * & *{\color{red}-} & \\
	$\Delta(1750)$ & $1/2^+$ & * & & {\color{red}*} & * & & {\color{red}*} & & \\
	$\Delta(1900)$ & $1/2^-$ & **{\color{red}*}& $\circ$ $\triangleright$ & *{\color{red}**} & **{\color{red}*} & * & *{\color{red}*} & *{\color{red}-} & \\
	$\Delta(1905)$ & $5/2^+$ & ****& $\circ$ $\diamond$ $\star$ $\triangleright$ & ***{\color{red}*} & **** & ** & * & *{\color{red}-} & {\color{red}**} \\
	$\Delta(1910)$ & $1/2^+$ & ****& $\circ$ $\diamond$ $\star$ $\triangleright$ & *{\color{red}**} & **** & *{\color{red}*} & *{\color{red}*}  &{\color{red}-} & {\color{red}*} \\
	$\Delta(1920)$ & $3/2^+$ & ***& $\circ$ $\diamond$ $\triangleright$ & *{\color{red}**} & *** & **{\color{red}*} & *{\color{red}*} & & {\color{red}**} \\
	$\Delta(1930)$ & $5/2^-$ & ***& $\diamond$ $\star$ $\triangleright$ & *{\color{red}-} & *** & {\color{red}*} & * & & \\
	$\Delta(1940)$ & $3/2^-$ & *{\color{red}*}& $\circ$ $\triangleright$ & {\color{red}*} & *{\color{red}*} & {\color{red}*} & & & {\color{red}*} \\
	$\Delta(1950)$ & $7/2^+$ & ****& $\circ$ $\diamond$ $\star$ $\triangleright$ & **** & **** & **{\color{red}- -} & *{\color{red}**} &{\color{red}-} & \\
	$\Delta(2000)$ & $5/2^+$ & **& & {\color{red}*} & {\color{red}**} & {\color{red}*} & & *{\color{red}-} & \\
	$\Delta(2150)$ & $1/2^-$ & *& & & * & & & & \\
	$\Delta(2200)$ & $7/2^-$ & *{\color{red}**}& $\circ$ $\diamond$ & {\color{red}***} & *{\color{red}*} & {\color{red}***} & {\color{red}**} & & \\
	$\Delta(2300)$ & $9/2^+$ & ** & & & ** & & & & \\
	$\Delta(2350)$ & $5/2^-$ & *& & & * & & & & \\
	$\Delta(2390)$ & $7/2^+$ & * & & & * & & & & \\
	$\Delta(2400)$ & $9/2^-$ & **& $\diamond$ $\star$ & {\color{red}**} & ** & & & & \\
	$\Delta(2420)$ & $11/2^+$ & ****& $\star$ & * & **** & & & & \\
	$\Delta(2750)$ & $13/2^-$ & **& & & ** & & & & \\
	$\Delta(2950)$ & $15/2^+$ & **& & & ** & & & & 	
\end{tabular}
\caption{Overview of the $\Delta^\ast$ resonances and their decay channels as of 2020 \cite{Zyla:2020zbs}. In red are the changes to the edition of 2002 highlighted \cite{ParticleDataGroup:2002ivw}. The symbols in the PWA column indicate which resonance is seen by the different PWA (circle: BnGa-2019 \cite{Muller:2019qxg}, diamond: J\"uBo-2017 \cite{Ronchen:2018ury}, star: SAID-MA19 \cite{Briscoe:2019cyo} and right triangle: KSU model \cite{Hunt:2018wqz}). The subscript $g$ indicates dynamically generated resonances as reported by the according PWA group. However, we do not list resonances here, that are seen by any PWA group, but are not identified with resonance states listed in the PDG.}
\label{tab:int:rev-delta-resonances}
\end{table}

One of the largest impacts stems from the accessibility of high precision measurements in final states, which allow a selection of the isospin. Final states like $\eta$ or $\eta'$ photoproduction as well as final states containing a $\Lambda$ only allow the occurrence of intermediate nucleon resonances $N^*$, since they are isospin 0 particles. The high precision measurements of these final states in the last years played a substantial role in the increase of number and star rating for the $N^*$ resonances. Details about the exact measurements can be found in Sec.~\ref{sec:measurements:EtaPhotoproduction} and Sec.~\ref{sec:measurements:StrangenessPhotoproduction}.

At high energies, a remarkable pattern of parity doublets was observed in the mapping of the excitation spectrum in the last years \cite{Anisovich:2011ye}. As proposed for example by \cite{Glozman:1999tk,Cohen:2001gb}, this effect might be due to the restoration of chiral symmetry at high energies, which should manifest itself by the emergence of parity doublets. Recently, further attention was given to a set of $\Delta$ resonances around 2~GeV: $\Delta(1910) \frac{1}{2}^+$ and $\Delta(1900) \frac{1}{2}^-$, $\Delta(1920) \frac{3}{2}^+$ and $\Delta(1940) \frac{3}{2}^-$, $\Delta(1905) \frac{5}{2}^+$ and $\Delta(1930) \frac{5}{2}^-$ and $\Delta(1950) \frac{7}{2}^+$. As it was already described in Sec.~\ref{sec:measurements:PionPhotoproduction}, the only missing state to complete this pattern of parity doublets was a resonance with $J^P =\frac{7}{2}^-$ with a mass around 2000~MeV. For this, a detailed study of different observables and various final states was performed, as it is described in \cite{Anisovich:2015gia}. The fits to the data sets as well as the $\chi^2$ for different final states can be seen in Fig.~\ref{fig:M:delta2200_anisovich}. This combined high precision study revealed no indication for a resonance with the desired quantum numbers and a mass of $\sim 2$~GeV. The state closest in mass with quantum numbers $J^P =\frac{7}{2}^-$ could be identified as the $\Delta(2200) \frac{7}{2}^-$, which is about $200$~MeV off. In this combined analysis, the high precision of the data sets is exploited for a detailed search for new states. This points the way to future searches for missing resonances, which can only be achieved by combined investigations of high precision data sets and shows that light baryon spectroscopy has reached the region of high precision physics.

Besides of polarization experiments, two other completely different approaches provided information about baryon resonances in the last years. First of all, the HADES experiment  used $\pi N$ scattering to shed light on the role of the $N(1520)\frac{3}{2}^-$ resonance in $N\rho$ production \cite{Adamczewski-Musch:2020tfm}. This experiment reveals that even 20 years after the end of the last $N\pi$ scattering experiments focused on baryon spectroscopy, this initial state can provide valuable and new information.

The second approach, which was exploited by the CLEO-c \cite{CLEO:2010fre} and the BES experiment \cite{BESIII:2012ssm}, is the analysis of the decay of the $\psi(2S)$ state. This state, which was produced by $e^+e^-$ collisions, has a large branching ratio ($\sim 15\%$) for its decay into light hadrons \cite{CLEO:2008kwj,Zyla:2020zbs}. These decay products can be analyzed in terms of baryon resonances and two new high mass states, $N(2300)\frac{1}{2}^+$ and $N(2570)\frac{5}{2}^-$, could be identified \cite{BESIII:2012ssm}. These analyses demonstrate a novel ansatz in the field of baryon spectroscopy, which can lead to valuable insight into the excitation spectrum, especially in the high mass region.

Recently, discussions started about the observation of two-pole structures in the complex plane and their treatment in the \textit{Listing of Baryon Resonances}. One example of this is the $\Lambda(1405)$, which is in detail discussed by Mei{\ss}ner in \cite{Meissner:2020khl}. The $\Lambda(1405)$ exhibits a lighter pole close to the $\pi\Sigma$ threshold and a second, higher pole close to $K^-p$ threshold.  Another example is the $N(1895)\frac{1}{2}^-$, which decay pattern can also be described by a two-pole structure, see the analysis of Khemchandani et al.~\cite{Khemchandani:2020exc}. These structures are currently not reflected in the \textit{Listing of Baryon Resonances} and discussions started about how to properly treat observations like this.

Similarly to the two-pole structures, discussions also started about the occurrence of triangle singularities in baryon spectroscopy. Triangle singularities are a further example on how certain physical processes dictate the analytic structure of the amplitude (cf. section~\ref{sec:PWA:Analyticity}). Such singularities have been first found to generate mesonic states, as shown by Mikhasenko et al.~\cite{Mikhasenko:2015oxp}. They were able to demonstrate that the signal of the $a_1(1420)$ could in fact be attributed to a triangle singularity. Investigations have started about the impact of this result for baryon spectroscopy. First publications have now been submitted, were signatures of this process could also be observed in measurements of baryonic states \cite{Feijoo:2021jtr,CBELSATAPS:2021fvv}. Further discussions about this topic can be expected in the future. 

\begin{figure}
	\centering
	\includegraphics[width=0.8\textwidth]{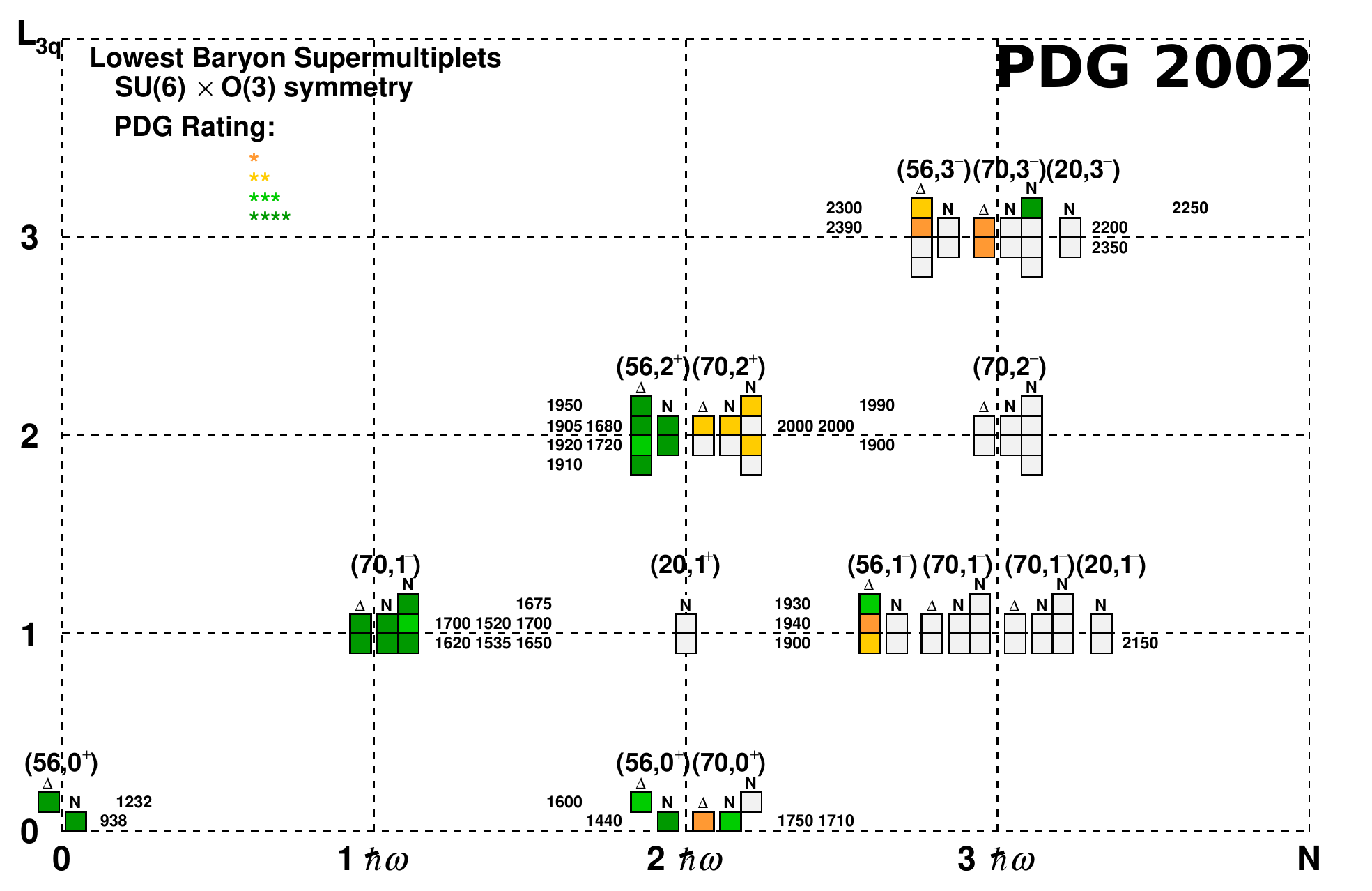}
	\caption{The $N^*$ and $\Delta^*$ resonances - as listed in the PDG 2002 \cite{ParticleDataGroup:2002ivw} - assigned to the $SU(6)$ multiplets. Note that this assignment is model-dependent. Courtesy of T. Seifen.}
	\label{fig:interpretation:multiplets2000}
\end{figure}
\begin{figure}
	\centering
	\includegraphics[width=0.8\textwidth]{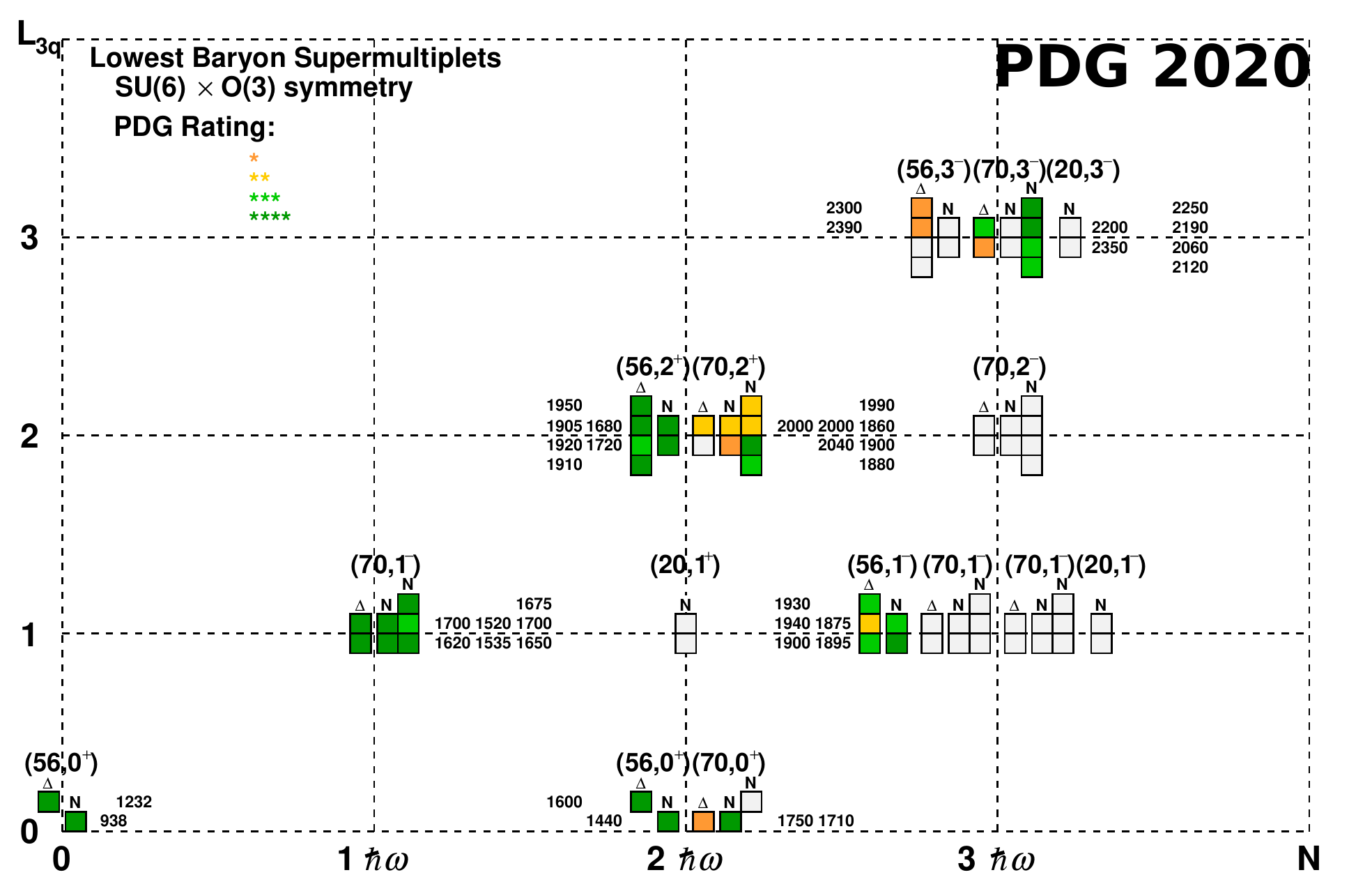}
	\caption{The recent $N^*$ and $\Delta^*$ resonances from the PDG 2020 \cite{Zyla:2020zbs} assigned to the $SU(6)$ multiplets. Note that this assignment is model-dependent. Courtesy of T. Seifen.}
	\label{fig:interpretation:multiplets2020}
\end{figure}

To further shed light onto the spectrum of the resonances with the newly found states, we inspect the patterns of the three-quark-states using the harmonic oscillator quark-model, as it was explained by Hey and Kelly in 1982 \cite{Hey:1982aj}. Considering only light quarks (\textit{up}, \textit{down} and \textit{strange}), the spin-flavor wave function for baryons can be classified according to the following decomposition of the direct product of three fundamental representations of the group $SU(2)\otimes SU(3) = SU(6)$ into a direct sum of irreducible representations \cite{Klempt:2012fy}:
\begin{align}
	6 \otimes 6 \otimes 6 = 56_S \oplus 70_M \oplus 70_M \oplus 20_A .
\end{align}
The 56-plet is symmetric as denoted by the $S$, the 70-plets are mixed-symmetric ($M$) and the 20-plet is anti-symmetric ($A$) with respect to the exchange of any two quarks. These representations can themselves be decomposed according to $SU(2)\otimes SU(3)$ as follows:
\begin{align}
 56 &= ~^4 10\oplus~^2 8\\
 70 &= ~^2 10 \oplus ~^4 8 \oplus ~^2 8 \oplus ~^2 1\\
 20&= ~^2 8 \oplus ~^4 1.
\end{align}
Here, the spectroscopic notation $~^{2s+1}n$ is used, where $s$ is the total quark spin and $n $ denotes the dimension of the $SU(3)$ representation. The measured resonances can be assigned to the states in the $SU(6)$ multiplets \cite{Menze:1988zv}, as explained in detail for example in \cite{Klempt:2012fy}. The status of resonances as of 2002 \cite{ParticleDataGroup:2002ivw} can be found in Fig.~\ref{fig:interpretation:multiplets2000} compared to the recent situation of 2020 \cite{Zyla:2020zbs} in Fig.~\ref{fig:interpretation:multiplets2020}. Here, the multiplets are shown for the total angular momentum of the three quarks $L_{3q}$ versus the main quantum number $N = 2n+L$. Multiplets are denoted in the plots according to the notation~$\left( D, L_{3q}^{P} \right)$, where~$D$ is the dimension of the irreducible representation of~$SU(6)$, $L_{3q}$ the three-quark angular momentum and~$P$ the parity of the resonance. Higher-mass states can be assigned to the higher lying excitation bands ($N > 3\hbar\omega$) and are omitted in the plots. Further details can be found in \cite{Klempt:2012fy}.

A comparison between both figures reveals that the newly identified resonances can be nicely assigned to predicted states by the harmonic oscillator model. Several multiplets, like the $(70,2^+)$ multiplet in the second excitation band, have been filled by the new states. However, it can be observed that the new resonances dominantly fill up existing multiplets instead of providing first states for previously empty multiplets. Interestingly, several multiplets, like the $(20,1^-)$ multiplet in the second excitation band and the $(70,2^-)$ multiplet in the third band, remain completely empty. The reason of this is currently unknown, but different theories exist. As for the $(20,1^-)$ multiplet, the states need to have a totally anti-symmetric spin-flavor wave function. In quark models, both oscillators of the spatial wave function need to be excited simultaneously \cite{Klempt:2012fy,Klempt:2017lwq}. Whether this configuration makes the formation of states unlikely is under discussion. In other publications, resonances are also assigned to the $(20,1^-)$ multiplet, like it was done in \cite{Crede:2013kia}, however, the basis for this claim is weak. Here, stronger evidence is required in order to draw any conclusions. It might also be possible that states in these multiplets can only be reached via cascading decays, which can be accessed my multi-meson final states. The (predicted) mass ranges of the empty multiplets lie well within the energy region covered by the active experiments. Therefore, additional high-statistics measurements can provide the desired answers in the future.



%% file: Conclusion.tex
During the last two decades, major changes happened in the field of light baryon spectroscopy. Improved accelerator facilities together with new high rate detector system allowed the measurement of an increasing amount of data. New targets providing polarized nucleons and beams, which were composed of polarized photons provided access to high precision measurements of large sets of polarization observables for various final states. This vast quantity of new data posed additional challenges for the different partial wave analysis groups.

The future of baryon spectroscopy can lie in different directions. First experiments proved that even well-measured reactions like pion nucleon scattering can lead to new information \cite{Adamczewski-Musch:2020tfm} and other experiments provided insight by using different methods like the analysis of $\psi(2S)$ decays \cite{CLEO:2010fre,BESIII:2012ssm}. 

Further contributions can be expected by the new CLAS12 experiment \cite{Burkert:2020akg} at Jefferson Lab, which utilizes electroproduction. Electroproduction allows to investigate the $Q^2$ evolution of the excited states and therefore can shed light on possible hybrid states \cite{Dudek:2012ag}. First results of a combined analysis for photo- and electroproduction have been published recently \cite{Mokeev:2020hhu}.

Also located at the Jefferson Lab is the GlueX experiment \cite{GlueX:2020idb}, which uses high energetic photoproduction to shed light on exotic mesonic states. Baryon resonances are certainly also being produced in their data, the possibility to extract information about these states from their data is under investigation.

Currently, new experiments are built up, which also plan to provide further input on hyperon spectroscopy like the PANDA experiment \cite{PANDA:2021ozp}, which will investigate $p\bar{p}$ collisions at the FAIR facility at GSI, Germany. 

In a more distant future, the planned KLF facility \cite{KLF:2020gai} at JLab can provide an impact for strange hadron spectroscopy. They plan to utilize a $K_L$ beam impinging on a hydrogen or deuteron target to generate a large amount of hyperons, which can then be measured with the GlueX detector. Further insight into the structure of the nucleons is expected by the future EIC collider \cite{AbdulKhalek:2021gbh}. Here, collisions between high energetic electrons and ions or protons are planned to shed light on different research topics, including hyperon spectroscopy \cite{Briscoe:2021cay}. 

In total, light baryon spectroscopy has reached the region of high precision physics. There are striking future opportunities in this field and a vast amount of new and diverse data can be expected in the near future. This will provide a fruitful environment for researchers and will lead to an outstanding impact in the understanding of QCD.

%% file: appendix.tex
\subsection{Important kinematic relations} \label{sec:appendix:Kinematics}

This appendix collects some important kinematic laws which can help the understanding of the main text, as well as sketches of their derivations. Many general results are then specified for the important case of single meson photoproduction. For more details, we refer to the references~\cite{BycklingK,Zyla:2020zbs}. 

We begin with $2$-body reactions (or $2 \rightarrow 2$-reactions)
\begin{equation}
 1 + 2 \longrightarrow 3 + 4 , \label{eq:appendix:2BodyReaction}
\end{equation}
among particles with $4$-momenta $p_{1},\ldots,p_{4}$. These have to fulfill $4$-momentum conservation
\begin{equation}
 p_{1} + p_{2} = p_{3} + p_{4} \mathrm{.} \label{eq:appendix:4MomentumCons}
\end{equation}
It is customary to define the Lorentz-invariant {\it Mandelstam variables}
\begin{equation}
 s = \left( p_{1} + p_{2} \right)^{2} \mathrm{,} \hspace*{5pt} t = \left( p_{1} - p_{3} \right)^{2} \mathrm{,} \hspace*{5pt} u = \left( p_{1} - p_{4} \right)^{2} \mathrm{.} \label{eq:appendix:MandelstamVariables}
\end{equation}
In case all the initial and final particles are on their mass-shell, the Mandelstam variables are not independent, but fulfill the constraint $s+t+u = \sum_{j} m_{j}^{2}$. Thus, for a $2 \rightarrow 2$-process of on-shell particles, only two kinematic variables are fully independent in order to fully characterize the process. 

Important examples for $2$-body reactions discussed in this review are pion-nucleon-scattering, i.e. $\pi + N \to \pi + N$, as well as single-meson photoproduction $\gamma + N \to \varphi + B$, with a pseudoscalar meson $\varphi$ and recoil-baryon $B$ in the final state.

\begin{figure}[h]
\includegraphics[width=0.92\textwidth,trim=0 80 0 55,clip]{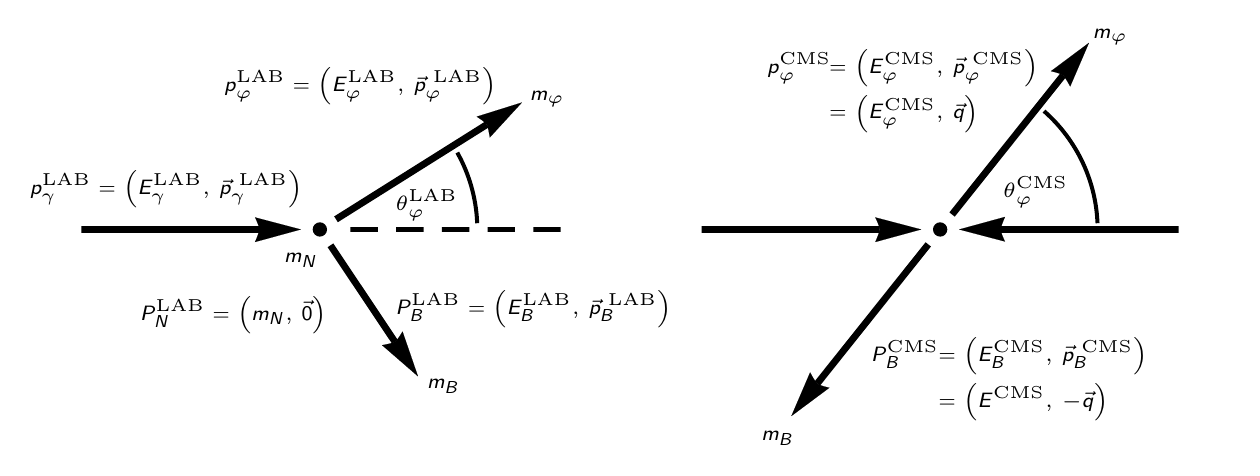}
\caption{The LAB- and CMS-frames are illustrated for the example-reaction of single pseudoscalar meson photoproduction $\gamma + N \to \varphi + B$.}
\label{fig:appendix:LABVsCMSVectorPlot}
\end{figure}

The two pertinent reference-frames used to specify kinematics for $2$-body reactions are given by the laboratory (LAB-) frame and the center-of-mass (CMS-) frame. Both frames are illustrated in Figure~\ref{fig:appendix:LABVsCMSVectorPlot} for the example of single-meson photoproduction. As mentioned above, the kinematics of the $2 \rightarrow 2$-reaction is fully specified by two independent kinematical invariants, for instance $(s,t)$. It is often customary to switch to a different pair of variables in order to specify kinematics, the only requirement being that the new pair is in one-to-one correspondence to $(s,t)$. For this, one often chooses the pair $(W,\theta)$ composed of the CMS-scattering angle $\theta$ and the total energy $W$ in the CMS, with $W = E_{1}^{\text{CMS}} + E_{2}^{\text{CMS}} = E_{3}^{\text{CMS}} + E_{4}^{\text{CMS}}$. The definition in equation~\eqref{eq:appendix:MandelstamVariables} and the specification of the kineamtics shown in Figure~\ref{fig:appendix:LABVsCMSVectorPlot} imply the following relation among $s$ and $W$: $W = \sqrt{s}$.


\clearpage

Further important points to be discussed regarding $2 \rightarrow 2$-reactions are the following:
\begin{itemize}
 \item[(i)] Calculation of threshold-energies: Evaluating the quantity $s = W^{2}$ in the LAB-frame of a fixed-target experiment (fixing the particle with mass $m_{2}$ to be at rest), one can quickly establish an expression for $E_{1}^{\text{LAB}}$ in terms of $W$, $m_{1}$ and $m_{2}$:
\begin{equation}
  E_{1}^{\mathrm{LAB}} = \frac{W^{2} - m_{1}^{2} - m_{2}^{2}}{2 m_{2}} , \label{eq:appendix:E1LABGeneralExpression}
\end{equation}
The minimal amount of CMS-energy necessary to produce the final state of the generic reaction $1 + 2 \rightarrow 3 + 4$ is $W_{\text{thr.}} = m_{3} + m_{4}$. Thus, we immediately obtain the minimal LAB-energy for the beam-particle '1' at threshold:
\begin{equation}
  E_{1,\text{thr.}}^{\mathrm{LAB}} = \frac{(m_{3} + m_{4})^{2} - m_{1}^{2} - m_{2}^{2}}{2 m_{2}} . \label{eq:appendix:E1LABGeneral2BodyThreshold}
\end{equation}
We specify these expressions further to the process of single-meson photoproduction, i.e. $\gamma + N \rightarrow \varphi + B$. The general expression for the LAB-energy of the incident photon becomes:
\begin{equation}
  E_{\gamma}^{\mathrm{LAB}} = \frac{W^{2} - m_{N}^{2}}{2 m_{N}} . \label{eq:appendix:EgammaLABPhotoprodExpression}
\end{equation}
The evaluation of the LAB-energy at threshold then proceeds via the following general expression:
\begin{equation}
\label{eq:appendix:GeneralThreshold}
E_{\gamma,\text{thr.}} = \frac{m_B^2+ m_\varphi^2+2m_B m_\varphi - m_N^2}{2m_N}
\end{equation}
For the important special case of pion photoproduction, the nucleon-masses cancel and one is left with:
\begin{equation}
\label{eq:appendix:PiProductionthreshold}
E_{\gamma,\text{thr.}} = \frac{m_\pi^2+2m_N m_\pi}{2m_N} .
\end{equation}
A tabular of the most important thresholds for photoproduction reactions can be found in Tab.~\ref{tab:appendix:thresholds}.

Next, we turn our attention to the CMS-frame. The definition of $W$, when using the relativistic energy-momentum relation for particles '1' and '2', reads:
\begin{equation}
  W = E_{1}^{\text{CMS}} + E_{2}^{\text{CMS}} = \sqrt{k^{2} + m_{1}^{2}} + \sqrt{k^{2} + m_{2}^{2}} . \label{eq:appendix:WCMSDefinitionInTermsOfK}
\end{equation}

\begin{table}[ht]
	\centering
	\begin{minipage}[t]{0.4\textwidth}
\begin{tabular}{l|l|l}
	Reaction & $E_{\gamma,thr.}$~[MeV] & $W_{thr.}$~[MeV] \\\hline
	$\gamma p \to \pi^0 p$ & 144.7 & 1073.2\\
	$\gamma n \to \pi^0 n$ & 144.7 & 1074.5\\
	$\gamma n \to \pi^- p$ & 148.5 & 1077.8\\
	$\gamma p \to \pi^+ n$ & 151.4 & 1079.1\\
	$\gamma p \to \eta p$ & 707.8 & 1486.1\\
	$\gamma n \to \eta n$ & 707.6 & 1487.4\\
	$\gamma p \to \eta' p$ & 1446.6 & 1896.1\\
	$\gamma n \to \eta' n$ & 1446.0 & 1897.3\\
	$\gamma p \to \omega p$ & 1109.1 & 1720.9\\
	$\gamma n \to \omega n$ & 1108.6 & 1722.2\\
	$\gamma p \to \rho p$ & 1095.5 & 1713.5\\
	$\gamma p \to \phi p$ & 1573.3 & 1957.7\\

\end{tabular}
	\end{minipage}
	\hspace*{0.8cm}
	\begin{minipage}[t]{0.4\textwidth}
\begin{tabular}{l|l|l}
	Reaction & $E_{\gamma,thr.}$~[MeV] & $W_{thr.}$~[MeV]\\\hline
	$\gamma p \to K^+\Lambda$ & 911.1 & 1609.4	 \\
	$\gamma n \to K^0 \Lambda$ & 915.3 & 1613.3\\
	$\gamma p \to K^+\Sigma^0$ & 1046.2 & 1686.3\\
	$\gamma p \to K^0 \Sigma^+$ & 1047.4 & 1687.0\\
	$\gamma n \to K^+ \Sigma^-$ & 1052.1 & 1691.1\\
	$\gamma n \to K^0 \Sigma^0$ & 1050.6 & 1690.3\\
	$\gamma p \to \pi^0\pi^0 p$ & 308.8 & 1208.2\\
	$\gamma n \to \pi^0\pi^0 n$ & 308.7 & 1209.5\\
	$\gamma p \to \pi^+\pi^0 n$ & 316.4 & 1214.1\\
	$\gamma p \to \pi^+\pi^- p$ & 320.7 & 1217.4\\
	$\gamma p \to \pi^0\eta p$ & 931.3 & 1621.1\\
	$\gamma p \to K^+K^- p$ & 1506.9 & 1925.6\\
\end{tabular}
\end{minipage}
\caption{Different photoproduction reactions and their threshold energies. }
\label{tab:appendix:thresholds}
\end{table}

This equation has the following well-known solution for the modulus of the CMS three-momentum $k$:
\begin{equation}
  k = \frac{1}{2 W} \sqrt{\left\{ W^{2} - \left( m_{1} - m_{2} \right)^{2} \right\} \left\{ W^{2} - \left( m_{1} + m_{2} \right)^{2} \right\}}  . \label{eq:appendix:GeneralSolutionBreakupMomentum}
\end{equation}
An analogous expression holds for the final-state three-momentum $q$ in terms of $W$, $m_{3}$ and $m_{4}$. Upon specifiying to single-meson photoproduction, we get the following formula
\begin{equation}
  k = E_{\gamma}^{\text{CMS}} = \frac{W^{2} - m_{N}^{2}}{2 W}  . \label{eq:appendix:InitialCMSMomentumPhotoproduction}
\end{equation}
This expression can be used to evaluate threshold-energies in the CMS, again by setting $W = W_{\text{thr.}} = m_{3} + m_{4}$.

\item[(ii)] Relations among scattering angles and $t$: The Mandelstam-variable $t$ becomes important in connection to angles instead of energies. The CMS-angle $\theta_{3}^{\mathrm{CMS}} \equiv \theta$ is fixed by Lorentz-invariants via the following expression
\begin{equation}
 \cos \theta = \frac{ t - m_{3}^{2} + 2 k \sqrt{m_{3}^{2} + q^{2}} }{ 2 q k } \mathrm{.} \label{eq:appendix:CosThetaCMSInvariants}
\end{equation}
The CMS- and LAB-angles $\theta$ and $\theta_{3}^{\text{LAB}}$ can be related using the equation
\begin{equation}
\tan \left( \theta_{3}^{\mathrm{LAB}} \right) = \frac{\sin \left( \theta \right)}{\gamma \left( \cos \left( \theta \right) + \frac{\beta}{\beta_{3}^{\mathrm{CMS}}} \right)} \mathrm{.} \label{eq:appendix:thetaMLAB}
\end{equation}
Here, $\gamma = 1/\sqrt{1 - \beta^{2}}$, $\beta$ is the LAB-velocity of the center of mass and $\beta_{3}^{\mathrm{CMS}}$ is the velocity of particle '3' in the CMS frame.

\item[(iii)] Lorentz-invariant phase-space volume: for the $2 \rightarrow 2$-reaction, the Lorentz-invariant phase-space volume $\rho = \int d \Phi_{2}$ can be evaluated analytically and it is an important quantity which appears in the definitions of (polarization) observables, unitarity relations, as well as (Breit-Wigner) resonance formulas, for instance. The result is simply $\rho = \frac{q}{k}$, with CMS three-momenta $q$ and $k$. For single-meson photoproduction, we have: 
\begin{equation}
 \rho = \frac{q}{k} = \frac{\sqrt{ \left\{ W^{2} - \left( m_{\varphi} + m_{B} \right)^{2} \right\} \left\{ W^{2} - \left( m_{\varphi} - m_{B} \right)^{2} \right\}}}{W^{2} - m_{N}^{2}}   . \label{eq:appendix:PhaseSpaceQOverK}
\end{equation}
\end{itemize}
When considering more general reactions with $n$-particles in the final state ('$2 \rightarrow n$-reactions'):
\begin{equation}
 1 + 2 \longrightarrow 1' + 2' + ... + n' , \label{eq:appendix:nBodyReaction}
\end{equation}
the general kinematic description gets more complicated drastically.
The phase-space is now generally described by $3 (2 + n) - 10$ independent kinematical variables\cite{AnalyticSMatrix}. For instance, the photoproduction of two pions
\begin{equation}
 \gamma + N \longrightarrow \pi + \pi + N , \label{eq:appendix:2PionProductionProcess}
\end{equation}
requires a general description using $5$ kinematical variables. Which combination of variables to choose is highly dependent on the employed convention and switching between different phase-space parametrizations is often far from trivial. Furthermore, the volume of the Lorentz-invariant phase-space cannot be evaluated analytically any more, once one goes beyond $2 \rightarrow 2$-processes. 

However, since overall $4$-momentum conservation still holds, one can find a simple generalization for the Mandelstam-variable $s$:
\begin{equation}
 s = W^{2} = \left( p_{1} + p_{2} \right)^{2}  = \left( p_{1'} + \ldots + p_{n'} \right)^{2}  .  \label{eq:appendix:nBodyReactionMandelstamS}
\end{equation}
This generalized variable can be used to calculate threshold-energies.

In case we assume the $n$-body final state to originate from photoproduction, i.e. from $\gamma + N$ in the initial state, we immediately obtain the following formulas for the threshold-energies in the CMS- and LAB-frames:
\begin{equation}
 k_{\text{thr.}} =  \frac{\left( m_{1'} + m_{2'} + \ldots + m_{n'} \right)^{2} - m_{N}^{2}}{2 \left( m_{1'} + m_{2'} + \ldots + m_{n'} \right)} , \hspace*{2.5pt}  E_{\gamma \text{,thr.}} = \frac{\left( m_{1'} + m_{2'} + \ldots + m_{n'} \right)^{2} - m_{N}^{2}}{2 m_{N}}  .  \label{eq:appendix:ThresholdEnergiesGeneralNBodyFinalState}
\end{equation}
\newpage
\subsection{Full Table of Baryon Resonances}
\begin{table}[h]
	\resizebox{\textwidth}{!}{%
		\begin{tabular}{|lll|lll|lll|lll|lll|}
			\hline
			Particle & $J^P$ & Rating & Particle & $J^P$ & Rating & Particle & $J^P$ & Rating & Particle & $J^P$ & Rating & Particle & $J^P$ & Rating \\
			\hline	
			p        & $1/2^+$  & **** & $\Delta$ (1232)  & $3/2^+$  & **** & $\Sigma^+$      & $1/2^+$  & **** & $\Xi^0$            &   $1/2^+$& ****        &{\color{red} $\Xi_{cc}^{++}$  }    &  &{\color{red} ***} \\
			n        & $1/2^+$  & **** & $\Delta$ (1600)  & $3/2^+$  & ***{\color{red}*} & $\Sigma^0 $     & $1/2^+$  & **** & $\Xi^-$            &   $1/2^+$& ****        & $\Lambda^0_b$        & {\color{red}$1/2^+$} & *** \\
			N (1440) & $1/2^+$  & **** & $\Delta$ (1620)  & $1/2^-$ & **** & $\Sigma^-$      & $1/2^+$  & **** & $\Xi$ (1530)       &   $3/2^+$& ****        & {\color{red}$\Lambda_b$ (5912)$^0$} & {\color{red}$1/2^-$} & {\color{red}***} \\
			N (1520) & $3/2^-$ & **** & $\Delta$ (1700)  & $3/2^-$ & **** & $\Sigma$ (1385) & $3/2^+$  & **** & $\Xi$ (1620)       &  & *                & {\color{red}$\Lambda_b$ (5920)$^0$} & {\color{red}$3/2^-$} & {\color{red}***} \\
			N (1535) & $1/2^-$ & **** & $\Delta$ (1750)  & $1/2^+$  & *    & $\Sigma$ (1580) & $3/2^-$ & *    & $\Xi$ (1690)       &  & ***              & {\color{red}$\Lambda_b$ (6146)$^0$} & {\color{red}$3/2^+$} & {\color{red}***} \\
			N (1650) & $1/2^-$ & **** & $\Delta$ (1900)  & $1/2^-$ & **{\color{red}*}  & $\Sigma$ (1620) & $1/2^-$ & *    & $\Xi$ (1820)        &   $3/2^-$ &***        & {\color{red}$\Lambda_b$ (6152)$^0$} & {\color{red}$5/2^+$} & {\color{red}***} \\
			N (1675) & $5/2^-$ & **** & $\Delta$ (1905)  & $5/2^+$  & **** & $\Sigma$ (1660) & $1/2^+$  & ***  & $\Xi$ (1950)        &  & ***              & {\color{red}$\Sigma_b$    }       & {\color{red}$1/2^+$} & {\color{red}***} \\
			N (1680) & $5/2^+$  & **** & $\Delta$ (1910)  & $1/2^+$  & **** & $\Sigma$ (1670) & $3/2^-$ & **** & $\Xi$ (2030)        &   {\color{red}$\geq$ 5/2 ?}& *** & {\color{red}$\Sigma^*_b$ }        & {\color{red}$3/2^+$} &{\color{red} *** }\\
			N (1700) & $3/2^-$ & ***  & $\Delta$ (1920)  & $3/2^+$  & ***  & $\Sigma$ (1750) & $1/2^-$ & ***  & $\Xi$ (2120)        &  & *                & {\color{red}$\Sigma_b$ (6097)$^+$ } &  & {\color{red}***} \\
			N (1710) & $1/2^+$  & ***{\color{red}*} & $\Delta$ (1930)  & $5/2^-$ & ***  & $\Sigma$ (1775) & $5/2^-$ & **** & $\Xi$ (2250)        &  & **               &{\color{red} $\Sigma_b$ (6097)$^-$}  &  & {\color{red}***} \\
			N (1720) & $3/2^+$  & **** & $\Delta$ (1940)  & $3/2^-$ & *{\color{red}*}   &\color{red} $\Sigma$ (1780) & \color{red}$3/2^+$  & \color{red}*    & $\Xi$ (2370)        &  & **               & $\Xi^-_b$            & {\color{red}$1/2^+$} & *{\color{red}**} \\
			\color{red}N (1860) & \color{red} $5/2^+$  &  \color{red}**   & $\Delta$ (1950)  & $7/2^+$  & **** & $\Sigma$ (1880) & $1/2^+$  & **   & $\Xi$ (2500)        &  & *                & $\Xi^0_b$            & {\color{red}$1/2^+$} & *{\color{red}**} \\
			\color{red}N (1875) &  \color{red}$3/2^-$ &  \color{red}***  & $\Delta$ (2000)  & $5/2^+$  & **   & \color{red}$\Sigma$ (1900) & \color{red}$1/2^-$ &\color{red} **   & $\Omega^-$          &  {\color{red} $3/2^+$} &****        & {\color{red}$\Xi'_b$ (5935)$^-$}   &{\color{red} $1/2^+$} &{\color{red} ***} \\
			\color{red}N (1880) &  \color{red}$1/2^+$  & \color{red}***  & $\Delta$ (2150)  & $1/2^-$ & *    & \color{red}$\Sigma$ (1910) &\color{red}$3/2^-$ &\color{red} ***  &{\color{red} $\Omega$ (2012)$^-$ }  &  {\color{red} ?$^-$ } &{\color{red}***    }     & {\color{red}$\Xi_ b$ (5945)$^0$  }   & {\color{red}$3/2^+$} & {\color{red}***} \\
			\color{red}N (1895) &  \color{red}$1/2^-$ &  \color{red}**** & $\Delta$ (2200)  & $7/2^-$ & *{\color{red}**}  & $\Sigma$ (1915) & $5/2^+$  & **** & $\Omega$ (2250)$^-$   &  & ***              & {\color{red}$\Xi_b$ (5955)$^-$  }   & {\color{red}$3/2^+$} & {\color{red}***} \\
			N (1900) & $3/2^+$  & **{\color{red}**} & $\Delta$ (2300)  & $9/2^+$  & **   & $\Sigma$ (1940) & $3/2^+$  & *    & $\Omega$ (2380)$^-$   &  & **               & {\color{red}$\Xi_b$ (6227)$^-$ }    &  & {\color{red}***} \\
			N (1990) & $7/2^+$  & **   & $\Delta$ (2350)  & $5/2^-$ & *    & \color{red}$\Sigma$ (2010) & \color{red}$3/2^-$ & \color{red}*    & $\Omega$ (2470)$^-$   &  & **               & {\color{red}$\Xi_b$ (6227)$^0$  }   &  &{\color{red} ***} \\
			N (2000) & $5/2^+$  & **   & $\Delta$ (2390)  & $7/2^+$  & *    & $\Sigma$ (2030) & $7/2^+$  & **** & $\Lambda^+_c$        &  {\color{red} $1/2^+$} &****        & {\color{red}$\Omega_b^-$   }         & {\color{red}$1/2^+$ }& {\color{red}***} \\
			\color{red}N (2040) &\color{red} $3/2^+$  &\color{red} *    & $\Delta$ (2400)  & $9/2^-$ & **   & $\Sigma$ (2070) & $5/2^+$  & *    & $\Lambda_c$ (2595)$^+$ &  {\color{red} $1/2^-$}& ***        & {\color{red}$\Omega_b$ (6316)$^-$  }           &  & {\color{red}* }\\
			\color{red}N (2060) &\color{red} $5/2^-$ & \color{red}***  & $\Delta$ (2420)  & $11/2^+$ & **** & $\Sigma$ (2080) & $3/2^+$  & *    & $\Lambda_c$ (2625)$^+$ &  {\color{red} $3/2^-$}& ***        & {\color{red}$\Omega_b$ (6330)$^-$   }          &  & {\color{red}*} \\
			N (2100) & $1/2^+$  & *{\color{red}**}  & $\Delta$ (2750)  & $13/2^-$ & **   & $\Sigma$ (2100) & $7/2^-$ & *    & $\Lambda_c$ (2765)$^+$ &  & *                & {\color{red}$\Omega_b$ (6340)$^-$ }    &  &{\color{red} *} \\
			\color{red} N (2120) & \color{red}$3/2^-$ &\color{red} ***  & $\Delta$ (2950)  & $15/2^+$ & **   & \color{red}$\Sigma$ (2160) & \color{red}$1/2^-$ & \color{red}*    & {\color{red}$\Lambda_c$ (2860)$^+$} &  {\color{red} $3/2^+$}&{\color{red} *** }        & {\color{red}$\Omega_b$ (6350)$^-$  }   &  &{\color{red} *} \\
			N (2190) & $7/2^-$ & **** & $\Lambda$        & $1/2^+$  & **** & \color{red}$\Sigma$ (2230) &\color{red} $3/2^+$  & \color{red}*    & $\Lambda_c$ (2880)$^+$ &  {\color{red} $5/2^+$}& **{\color{red}* }        & {\color{red}$P_c$ (4312)$^+$   }       &  & {\color{red}*} \\
			N (2220) & $9/2^+$  & **** & \color{red}$\Lambda$        & \color{red}$1/2^-$ & \color{red}**   & $\Sigma$ (2250) &       & ***  &{\color{red} $\Lambda_c$ (2940)$^+$} & {\color{red}  $3/2^-$}&{\color{red} ***  }      & {\color{red}$P_c$ (4380)$^+$  }  &  & {\color{red}*} \\
			N (2250) & $9/2^-$ & **** & $\Lambda$ (1405) & $1/2^-$ & **** & $\Sigma$ (2455) &       & **   & $\Sigma_c$ (2455)    &  {\color{red} $1/2^+$}& ****        & {\color{red}$P_c$ (4440)$^+$ }               &  & {\color{red}*} \\
			\color{red}N (2300) & \color{red}$1/2^+$  & \color{red}**   & $\Lambda$ (1520) & $3/2^-$ & **** & $\Sigma$ (2620) &       & **   & $\Sigma_c$ (2520)    & {\color{red}  $3/2^+$}& ***         & {\color{red}$P_c$ (4457)$^+$   }              &  & {\color{red}*} \\
			\color{red}N (2570) & \color{red}$5/2^-$ & \color{red}**   & $\Lambda$ (1600) & $1/2^+$  & ***{\color{red}*} & $\Sigma$ (3000) &       & *    &{\color{red} $\Sigma_c$ (2800) }   &  & {\color{red}***       }       &           &  &  \\
			N (2600) & 1$1/2^-$& ***  & $\Lambda$ (1670) & $1/2^-$ & **** & $\Sigma$ (3170) &       & *    & $\Xi^+_c$            &  {\color{red} $1/2^+$}& ***         &           &  &  \\
			N (2700) & 1$3/2^+$ & **   & $\Lambda$ (1690) & $3/2^-$ & **** &                 &  &          & $\Xi^0_c$            &  {\color{red} $1/2^+$}& ***{\color{red}*  }      &      &  &  \\
			&  &     & \color{red}$\Lambda$ (1710) & \color{red}$1/2^+$  & \color{red}*    &                 &  &          & $\Xi'^+_c$          &  {\color{red} $1/2^+$}& ***         &      &  &  \\
			&  &     & $\Lambda$ (1800) & $1/2^-$ & ***  &                 &  &          & $\Xi'^0_c$           &  {\color{red} $1/2^+$}& ***         &      &  &  \\
			&  &     & $\Lambda$ (1810) & $1/2^+$  & ***  &                 &  &          & $\Xi_c$ (2645)       & {\color{red}  $3/2^+$}& ***         &  &  &  \\
			&  &     & $\Lambda$ (1820) & $5/2^+$  & **** &                 &  &          & $\Xi_c$ (2790)       &  {\color{red} $1/2^-$}& ***        &  &  &  \\
			&  &     & $\Lambda$ (1830) & $5/2^-$ & **** &                 &  &          & $\Xi_c$ (2815)       &  {\color{red} $3/2^-$}& ***        &  &  &  \\
			&  &     & $\Lambda$ (1890) & $3/2^+$  & **** &                 &  &          & {\color{red}$\Xi_c$ (2923) }      &  &{\color{red} **   }            &  &  &  \\
			&  &     & $\Lambda$ (2000) & \color{red}$1/2^-$ & *    &                 &  &          & {\color{red}$\Xi_c$ (2930)}       &  & {\color{red}**}               &  &  &  \\
			&  &     & \color{red}$\Lambda$ (2050) & \color{red}$3/2^-$ &\color{red} *    &                 &  &          & {\color{red}$\Xi_c$ (2970) }      &  & {\color{red}*** }             &  &  &  \\
			&  &     & \color{red}$\Lambda$ (2070) & \color{red}$3/2^+$  &\color{red} *    &                 &  &          & {\color{red}$\Xi_c$ (3055)  }     &  & {\color{red}*** }             &  &  &  \\
			&  &     &\color{red} $\Lambda$ (2080) & \color{red}$5/2^-$ &\color{red} *    &                 &  &          &{\color{red} $\Xi_c$ (3080) }      &  & {\color{red}***    }          &  &  &  \\
			&  &     &\color{red} $\Lambda$ (2085) &\color{red} $7/2^+$  &\color{red} **   &                 &  &          & {\color{red}$\Xi_c$ (3123)   }    &  & {\color{red}*}               &  &  &  \\
			&  &     & $\Lambda$ (2100) & $7/2^-$ & **** &                 &  &          & $\Omega^0_c$          &     {\color{red}$1/2^+$}& ***         &  &  &  \\
			&  &     & $\Lambda$ (2110) & $5/2^+$  & ***  &                 &  &          & {\color{red}$\Omega_c$ (2770)$^0$ }  &  {\color{red} $3/2^+$}& {\color{red}*** }        &  &  &  \\
			&  &     & $\Lambda$ (2325) & $3/2^-$ & *    &                 &  &          & {\color{red}$\Omega_c$ (3000)$^0$}  &  & {\color{red}***}              &  &  &  \\
			&  &     & $\Lambda$ (2350) & $9/2^+$  & ***  &                 &  &          & {\color{red}$\Omega_c$ (3050)$^0$ }  &  & {\color{red}*** }             &  &  &  \\
			&  &     & $\Lambda$ (2585) &       & **   &                 &  &          & {\color{red}$\Omega_c$ (3065) $^0$}   &  & {\color{red}***  }            &  &  &  \\
			&  &     &                                              &  &  &  &  &  & {\color{red}$\Omega_c$ (3090)$^0$}   &  & {\color{red}***  }            &  &  &  \\
			&  &     &                                              &  &  &  &  &  & {\color{red}$\Omega_c$ (3120)$^0$}   &  &  {\color{red}***   }          &  &  &  \\
			
			\hline	
		\end{tabular}
	}
	\caption{Comparison of all baryon states from the PDG 2002 \cite{ParticleDataGroup:2002ivw} to PDG 2020 \cite{Zyla:2020zbs}. Modifications have been marked in red.}
	\label{tab:int:baryon_table_pdg}
\end{table}